\documentclass{aa}
\usepackage{soul}
\usepackage{afterpage}
\usepackage[varg]{txfonts}
\usepackage{supertabular}
\usepackage[normalem]{ulem}
\usepackage{csquotes}
\usepackage{units}
\usepackage{mathtools}
\usepackage{float}
\usepackage[caption = false]{subfig}
\usepackage{graphicx}
\usepackage{amsmath}

\bibpunct{(}{)}{;}{a}{}{,} 
\newcommand{\lk}{\mathcal{L}} % to be used inside math mode

\begin{document}

\title{Characterization of Gamma-Ray Burst prompt emission spectra down to soft X-rays}

\author{G. Oganesyan\inst{\ref{inst1}}
\and L. Nava\inst{\ref{inst2},\ref{inst3},\ref{inst4}}
\and G. Ghirlanda\inst{\ref{inst2}}
\and A. Celotti\inst{\ref{inst1}}
}

\institute{SISSA, via Bonomea 265, I--34136 Trieste, Italy; \email{\href{mailto:goganesy@sissa.it}{goganesy@sissa.it}}\label{inst1}
\and
INAF -- Osservatorio Astronomico di Brera, via Bianchi 46, I--23807 Merate (LC), Italy \label{inst2}
\and
INAF -- Osservatorio Astronomico di Trieste, via G.B. Tiepolo 11, I--34143 Trieste, Italy \label{inst3}
\and
INFN -- Istituto Nazionale di Fisica Nucleare, Sezione di Trieste, via Valerio 2, I-34127, Trieste, Italy \label{inst4}
}

%abstract 300 words max
\abstract
{
Detection of prompt emission by {\it Swift}-XRT provides a unique tool to study how the prompt spectrum of gamma-ray bursts (GRBs) extends down to the soft X-ray band. 
This energy band is particularly important for prompt emission studies, since it is towards low energies that the observed spectral shape is in disagreement with the synchrotron predictions.
Unfortunately, the number of cases where XRT started observing the GRB location during the prompt phase is very limited.
In this work, we collect a sample of 34 GRBs and perform joint XRT+BAT spectral analysis of prompt radiation, extending a previous study focused on the 14 brightest cases. 
{\it Fermi}-GBM observations are included in the analysis when available ({11 cases}), allowing the characterization of prompt spectra from soft X-rays to MeV energies.
In 62\% of the spectra, the XRT data reveal a hardening of the spectrum, well described by introducing an additional, low-energy power-law segment (with index $\alpha_1$) into the empirical fitting function. The break energy below which the spectrum hardens has values between { 3\,keV and 22\,keV}. A second power-law ($\alpha_2$) describes the spectrum between the break energy and the peak energy. The mean values of the photon indices are $\langle\alpha_1\rangle=-0.51$ ($\sigma={ 0.24}$) and $\langle\alpha_2\rangle={ -1.56}$ ($\sigma=0.26$). These are consistent, within one $\sigma$, with the synchrotron values in fast cooling regime. As a test, if we exclude XRT data from the fits we find typical results: the spectrum below the peak energy is described by a power law  with $\langle\alpha\rangle={-1.15}$. This shows the relevance of soft X-ray data in revealing prompt emission spectra consistent with synchrotron spectra. 
Finally, we do not find any correlation between the presence of the X-ray break energy and the flux, fluence, or duration of the prompt emission.
}


\maketitle

%---------------------------------------------------------------------
%==========================  INTRODUCTION ============================
\section{Introduction}\label{sec:intro}
The origin of gamma-ray burst (GRB) prompt emission is far from being fully understood. 
The mechanisms responsible for powering and launching the jet, the processes entailing jet energy dissipation, and the dominant radiative mechanisms responsible for the observed hard X-ray/$\gamma$-ray emission have not
yet been clearly identified.
An improved characterization of prompt spectra might be a good starting point for a reconsideration of the problem, since spectra carry the imprints of the properties of the emitting region (such as its location, bulk Lorentz factor, magnetic field strength and configuration, and particle acceleration efficiency).
All this information hidden in the spectra may help to discriminate among different theoretical scenarios of energy dissipation and jet composition.

Synchrotron radiation from a non-thermal population of ultra-relativistic electrons is expected to be a prime candidate for prompt radiation \citep{Rees94,Katz94,Tavani96,Sari96,Sari98}. 
However, the observed GRB prompt spectra are found to have, on average, a photon index $\langle\alpha\rangle\sim -1$ harder than the value $\alpha = -1.5$ expected from fast cooling synchrotron radiation \citep{Preece98,Frontera00,Ghirlanda02,Kaneko06,Sakamoto11,Nava11,goldstein12,gruber14,Lien16}. 

This inconsistency does not directly exclude the synchrotron mechanism as a viable radiation process for prompt radiation.
Synchrotron emission can still produce harder spectra in some tuned configurations. One of the possibilities is that either the cooling frequency $\nu_{\rm c}$ or the self-absorption frequency $\nu_{\rm sa}$ are close to the characteristic synchrotron frequency $\nu_{\rm m}$ \citep{Daigne11}. 
Moreover, if the low-energy part of the synchrotron spectrum is modified by 
the energy-dependent inverse Compton scattering in Klein-Nishina regime, a harder spectral shape (up to $\alpha=-2/3$) can be observed \citep{Derishev01,Nakar09,Daigne11}.
While these models assume a constant magnetic field, a magnetic field with a dependence on the radius and/or on the distance from the shock front can also harden the shape of the synchrotron spectrum  \citep{Peer06,Derishev07,Uhm14}. 
Finally, synchrotron spectra can have photon indices much harder than $-1.5$ if the pitch angles of the emitting electrons are distributed anisotropically \citep{Lloyd00,medvedev00}. 
Different scenarios require different conditions at the source.
It is then extremely important to explore whether these proposed scenarios represent a viable solution and are supported by observational evidence.

In order to gain a better understanding of the spectral shape below the $\nu F_\nu$ peak energy, \cite[][(hereafter O17]{Oganesyan2017} suggested to take advantage of those rare cases where the prompt radiation can be studied down to 0.5\,keV. 
This was done by selecting GRBs from {\it The Neil Gehrels Swift Observatory} (hereafter {\it Swift}) for which part of the prompt emission is observed also by the X-ray Telescope (XRT), in addition to the Burst Alert Telescope (BAT).
Time-resolved XRT-BAT joint spectral analysis of prompt emission was performed for 14~GRBs. 
The inclusion of soft X-ray data revealed a hardening of the spectral shape towards low energies, well described by adding a break (typically located between 2 and 20\,keV) and an additional power-law segment to the fitting function below such break energy.
The average values of the photon indices below and above the break energy were found to be very close (i.e., consistent within 1$\sigma$) to the expectations from synchrotron radiation ($\alpha_1^{\rm syn}=-0.67$ and $\alpha_2^{\rm syn}=-1.5$) in a scenario where the break energy corresponds to the cooling break.
In this work, we extend the work of O17 considering fainter GRBs, for which time-resolved analysis cannot be performed.
We selected 20 additional GRBs observed simultaneously by XRT and BAT, having significant signal to perform at least time-integrated analysis (one spectrum for each GRB).
We also address the question of why, in the usual situation (i.e., when data below 10\,keV are not available), the typical low-energy photon index has a value $\alpha\sim-1$.
We compare the sample of GRBs with low-energy breaks with the sample showing no hint for X-ray hardening and with the more general population of {\it Swift} GRBs, with the aim of correlating the presence of the break to other observables. The presence of the break seems independent from the fluence, flux, and duration, or a combination of these quantities.
The paper is organized as follows. 

In \S~\ref{sec:sample}, we define the sample-selection criteria. The procedures for data extraction and analysis are presented in \S~\ref{sec:data} and \S~\ref{sec:analysis}, respectively. In \S~\ref{sec:results}, we report the results of our analysis and discuss our results in \S~\ref{sec:conclusions}.

%================================= SAMPLE SELECTION =========================
\section{Sample selection}\label{sec:sample}
The full sample of GRBs with significant emission detected simultaneously by XRT and BAT includes 77 GRBs (as of January 2016, see O17 for details).
Time-resolved spectral analysis in at least four time bins can be performed only in 14 GRBs and the results of this analysis have been reported in O17. 
Spectral breaks between $\sim$\,2 and 20\,keV were found in 67\% of the 128 time-resolved spectra.

In this work, to further explore the occurrence of this spectral feature in GRBs' prompt emission spectra, we enlarge the sample by including fainter sources.
We relax the requirement of performing time-resolved analysis and select all cases with enough signal for a joint XRT+BAT time-integrated spectral analysis. 
More specifically, we consider all cases where the BAT signal-to-noise ratio (S/N) is larger than 30 (in the time interval where significant signal is detected simultaneously by XRT and BAT). 
This requirement is satisfied by 20 additional GRBs that, together with the 14 GRBs included in the analysis of O17, form a sample of 34 GRBs. 
In 11 cases (out of 34),  {\it Fermi} Gamma-ray Burst Monitor (GBM)  data are also available and have been included in the spectral analysis. For two additional GRBs, GBM data are available but have not been included because of inconsistencies between different NaI detectors.
For 17 GRBs, the redshift has been measured, and ranges from 0.73 to 5.91.
The list of GRBs and their redshift can be found in Table~\ref{tab:nh}.
%--------------------------------------------------------------------

%================================= DATA EXTRACTION =========================
\section{Data extraction}\label{sec:data}
For each GRB, we analyze the spectrum integrated over the time where significant signal is detected both by BAT and XRT. When GBM data are included, we re-define the edges of the time interval to match the coarser temporal resolution of GBM CSPEC data (see Sect.~\ref{sec:fermi-gbm}).

A detailed description of how data from the different instruments have been extracted and processed can be found in O17. We summarize here the main steps of the procedure.

%-----------------  BAT  -----------------
\subsection{Swift-BAT}
We downloaded the BAT event files from the Swift data archive\footnote{\url{http://heasarc.gsfc.nasa.gov/cgi-bin/W3Browse/swift.pl}}. 
BAT spectra and light curves have been extracted using the latest version of the {\sc heasoft} package (v6.17). 
The background-subtracted mask-weighted BAT light curves have been extracted in the energy range 15-150\,keV using the {\tt FTOOLS} {\tt batmaskwtevt} and {\tt batbinevt} tasks. 
BAT spectral files have been produced using the {\tt batbinevt} task and have been corrected through the {\tt batupdatephakw} and {\tt batphasyserr} to include systematic errors. 
We used {\tt batdrmgen} to generate response matrices for time intervals before, during, and after the satellite slew. 
The latest calibration files (CALDB release 2017-05-20) have been adopted.

%-----------------  XRT  -------------------
\subsection{Swift-XRT}
The XRT light curves have been retrieved from the Swift Science Data Center, provided by the University of Leicester\footnote{\url{http://www.swift.ac.uk/xrt_curves/}} \citep{Evans_09}.  
We downloaded the XRT event files from the \textit{Swift}-XRT archive\footnote{\url{http://www.swift.ac.uk/archive/}} and extracted source and background spectra in each time-bin with the {\tt xselect} tool.
In order to avoid possible pile-up effects, we removed the central region of the XRT images (see O17 for details) following the procedure suggested by \cite{Romano_06}. 
Ancillary response files have been generated using the task {\tt xrtmkarf}.
All the channels below 0.5\,keV were excluded from the spectral analysis.

%-----------------  FERMI  -------------------
\subsection{Fermi-GBM}\label{sec:fermi-gbm}
The GBM is composed of 12 sodium iodide (NaI) and two bismuth germanate (BGO) scintillation detectors \citep{meegan09}.
Two NaI (in the range 8-1000\,keV) and one BGO detector (300\,keV-40\,MeV) 
have been used for the spectral analysis. We excluded channels in the range 30-40\,keV due to the presence of the Iodine K-edge at 33.17\,keV.
The BGO is included only if the detected signal is above the background noise during the time of interest.
The extraction of spectra has been performed using the \textsc{gtburst} tool. \footnote{\url{http://fermi.gsfc.nasa.gov/}}
We selected pre- and post-burst data to model the background, and fit it with a energy- and time-dependent polynomial.

%=================================  SPECTRAL ANALYSIS  ============================
\section{Spectral analysis}\label{sec:analysis}
Spectral analysis has been performed with XSPEC  (v12.9.1).
Different likelihoods have been used for different detectors: CSTAT for XRT data, Gaussian for BAT, and PGSTAT for GBM data. The list of tested models and the procedure applied to select the best fit model are discussed in the following sections.

Following the same procedure adopted in O17, we have introduced multiplicative factors between different instruments, in order to account for uncertainties in their cross-calibration. 
For cases with XRT and BAT data, we leave the calibration constant free to vary between 0.9 and 1.1. When GBM data are also available, we freeze the calibration constant between XRT and BAT and adopt a calibration constant for the GBM spectrum, free to vary between 0.9 and 1.1.
We do not find significant correlation (Pearson correlation coefficient R$>$0.5 and p-value $<$0.001) between the photon index below the break energy and the normalization constant.
In the following sections, we summarize how metal absorption has been treated, which spectral models have been tested, and how the best fit is chosen among all the spectral models that provide a reasonable fit to the data.
%--------------------------------------------------------------
\begin{figure}[ht!]
\centering
\includegraphics[width=1.0\columnwidth,trim=0 0.2cm 0.2cm -0.33cm]{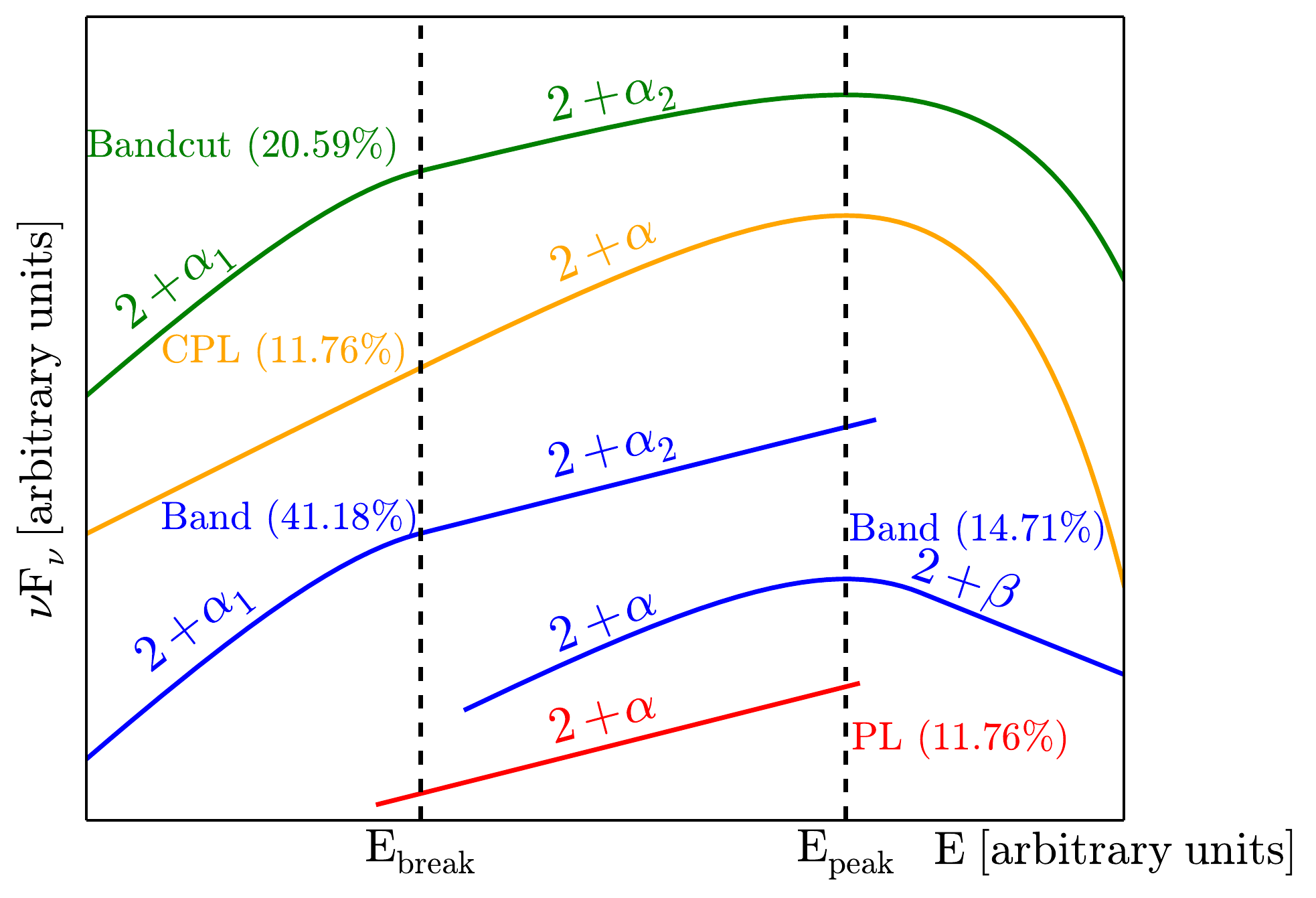}
\caption{\label{fig:sketch} Spectral models (in $\nu F_{\nu}$ representation) tested in this work. 
From top to bottom: a Band function with a high-energy exponential cutoff (Bandcut, green),
a Cutoff power-law model (CPL, orange), 
a Band function (Band, blue) and a single power-law model (PL, red). 
The Band model can describe two different cases: either both spectral indices are larger than $-2$, or the second one is smaller than $-2$.
In all models, we use the letter $\alpha$ for spectral segments where the $\nu F_\nu$ flux increases with energy (and distinguish between $\alpha_1$ and $\alpha_2$ in case two increasing segments, separated by a spectral break, are present), and $\beta$ to refer to a decreasing spectral segment.
The number within brackets next to the model name refers to the percentage of cases for which each model provides the best fit to the data.}
\end{figure}
%--------------------------------------------------------------

\subsection{Treatment of the absorption}
We take into account Galactic and intrinsic absorption by applying the multiplicative {\tt tbabs} and {\tt ztbabs} models \citep{wilms00} in XSPEC, respectively. 
Galactic absorption by neutral hydrogen in the direction of the GRB is estimated from \cite{Kalberia_05}.
To account for intrinsic absorption within the host galaxy, we apply the method described in O17: we infer the intrinsic column density of neutral hydrogen from a late time X-ray spectrum taken during the power-law decay phase of the afterglow emission, provided that no significant spectral evolution is evident at that time \citep{Butler07}. 
The inferred value is used as fixed input value for the intrinsic column density, $N_{\rm H}$, in the joint XRT and BAT (and eventually GBM) spectral analysis.
For GRBs without redshift, late-time X-ray spectra are fitted by applying the {\tt tbabs} model only.
The values of the intrinsic $N_{\rm H}$ and the late time interval used to constrain $N_{\rm H}$ are reported in Table~\ref{tab:nh}. 
For the 14 GRBs already included in the sample studied in O17, some values of $N_{\rm H}$ might differ from those reported in O17. The reason is that we choose CSTAT likelihood to fit the late-time X-ray spectra.

\subsection{Spectral models}
The standard models generally used to fit prompt spectra (PL, CPL, Band and smoothly broken PL) have the possibility to describe at most one change of the spectral slope, typically corresponding to a peak energy $E_{\rm peak}$ of the $\nu F_\nu$ spectrum.
In order to capture an additional change in the slope, we first need to introduce an appropriate empirical fitting function.
We consider a Band function modified to include a high-energy exponential cutoff (see the model named Bandcut in Fig~\ref{fig:sketch}).
This model was introduced in \cite{Zheng}. 
In this model, $E_{\rm peak}$ is located around the high-energy exponential cutoff, while the additional break feature that we want to describe is located at the smooth connection between the two low-energy PL segments.

The Bandcut model is defined as follows: 
%----------------------------------------------
\begin{equation}    \label{eq:bcpl}
N_{\rm E}^{\rm Bandcut}\propto \begin{cases}
E^{\alpha_{1}}e^{-\frac{E}{E_1}} \quad\quad\quad\quad\quad\quad{\rm for} \quad\quad E\le E_{\rm break}\\
  \\
\left[\frac{E_1E_2}{E_2-E_1}(\alpha_{1}-\alpha_{2})\right]^{\alpha_{1}-\alpha_{2}} e^{\alpha_{1}-\alpha_{2}} E^{\alpha_{2}}e^{-\frac{E}{E_2}} \\ 
\quad\quad\quad\quad\quad\quad\quad\quad\quad\quad\quad {\rm for}\quad\quad E>E_{\rm break}
\end{cases}
,\end{equation}
%----------------------------------------------
where $E_{\rm break}=\frac{E_1E_2}{E_2-E_1}(\alpha_1-\alpha_2)$. The peak energy is defined by $E_{\rm peak}=E_2(2+\alpha_2)$. 
The introduction of this model represents a difference as compared to the analysis presented in O17, where a broken power-law with an exponential cutoff was used. 
The difference is then in the description of the shape around the break energy (sharp break in O17 and smooth break in this work).

We also modeled the spectra with functions including three PL segments.
However, we did not succeed in constraining the photon index above $E_{\rm peak}$ for any of the spectra in our sample.
Therefore, the final set of tested models includes PL, CPL, Band, and Bandcut.
All these models are shown schematically in Fig.~\ref{fig:sketch}. 
We note that a Band model is found to describe two different situations: standard cases where a peak energy is present, and cases where the $\nu F_\nu$ flux increases with energy over the full spectral range, but with a change in slope that identifies a break energy $E_{\rm break}$.
According to the notation introduced in O17, we adopt the following terminology when referring to photon indices: 
%-------------------------------------------
\begin{enumerate}
\item The letter $\alpha$ refers to the photon index of spectral segments increasing in $\nu F_\nu$ (i.e., photon indices larger than -2). 
If there are two consecutive segments (separated by a break) with photon indices larger than -2, we call them $\alpha_{1}$ and $\alpha_{2}$ (below and above the break, respectively).
\item The letter $\beta$ is used to refer to a part of the spectrum that is decreasing in $\nu F_\nu$ (i.e., $\beta<-2$).
\end{enumerate}
%-------------------------------------------

%--------------------------------------------------------------------
\begin{figure}
\centering
 \includegraphics[width=0.94\columnwidth,trim=0 0.12cm 0 0]{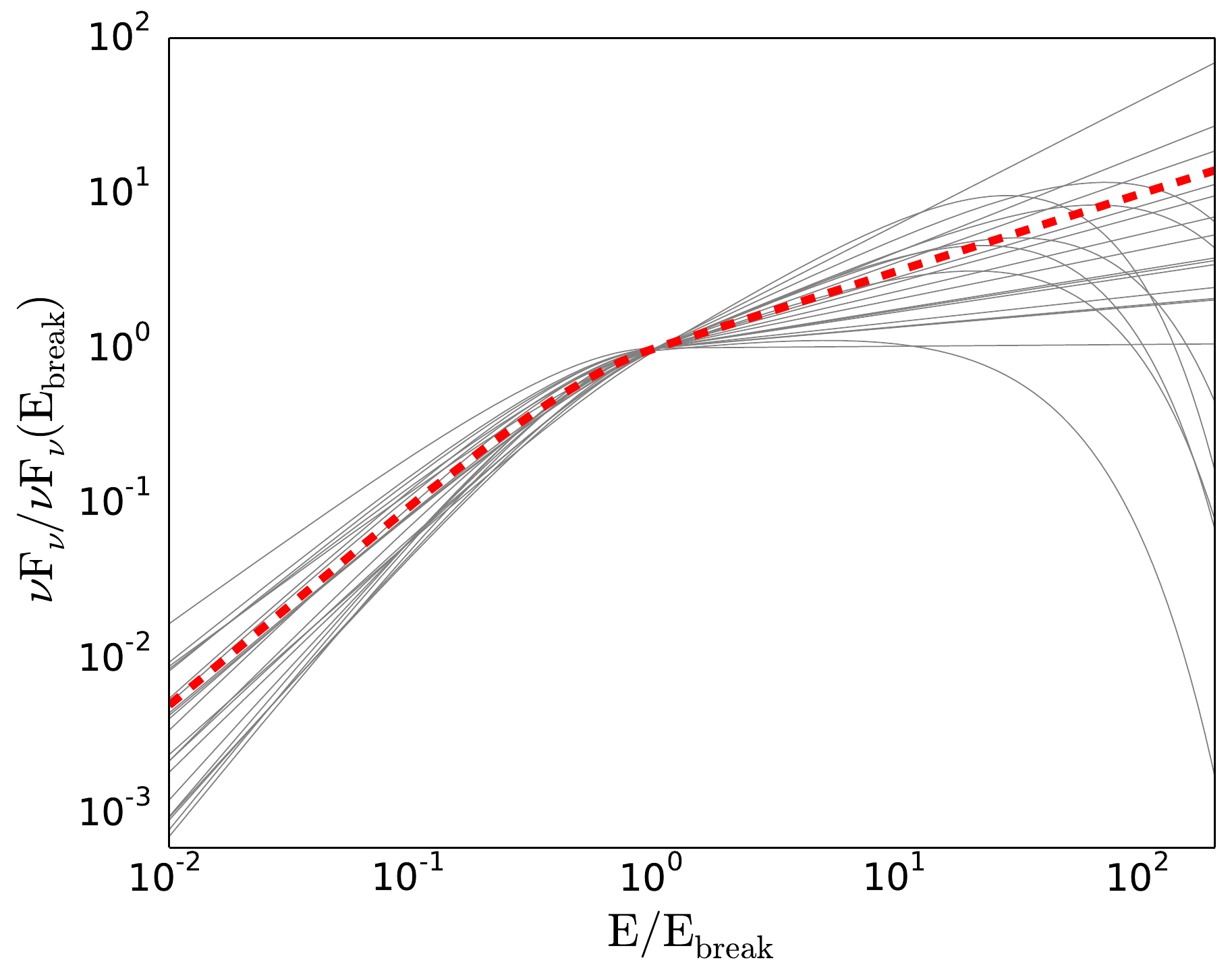}
\caption{\label{fig:averaged} Best fit models (in  $\nu F_{\nu}$ representation) for all cases where the break energy is found (gray lines), i.e., all cases where the best fit model is either a Band function continuously rising in $\nu F_\nu$ or a Bandcut (see Fig.~\ref{fig:sketch}). A Band model with photon indices equal to those expected from fast cooling synchrotron radiation ($\alpha_1^{\rm syn}=-2/3$ and $\alpha_2^{\rm syn}=-3/2$) is shown with a dashed, red curve.}
\end{figure}
%--------------------------------------------------------------------

%--------------------------------------------------------------------
%[htbp]
\begin{figure*}
%trim = left, bottom, right, up
\includegraphics[scale=0.26,trim=-1.6cm 0 0 0]{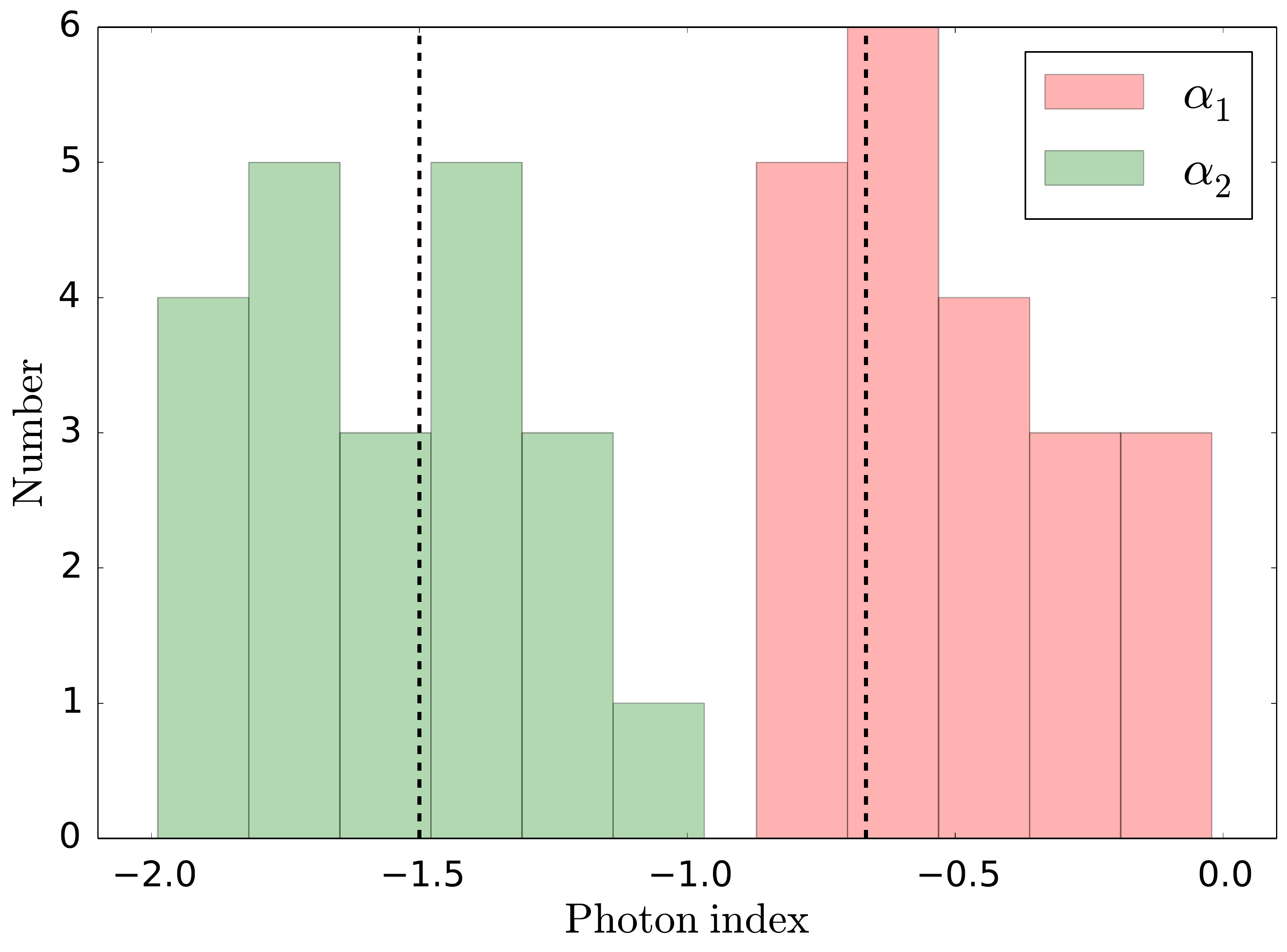}  
\includegraphics[scale=0.26,trim=1.5cm 0.3cm -1.5cm 0]{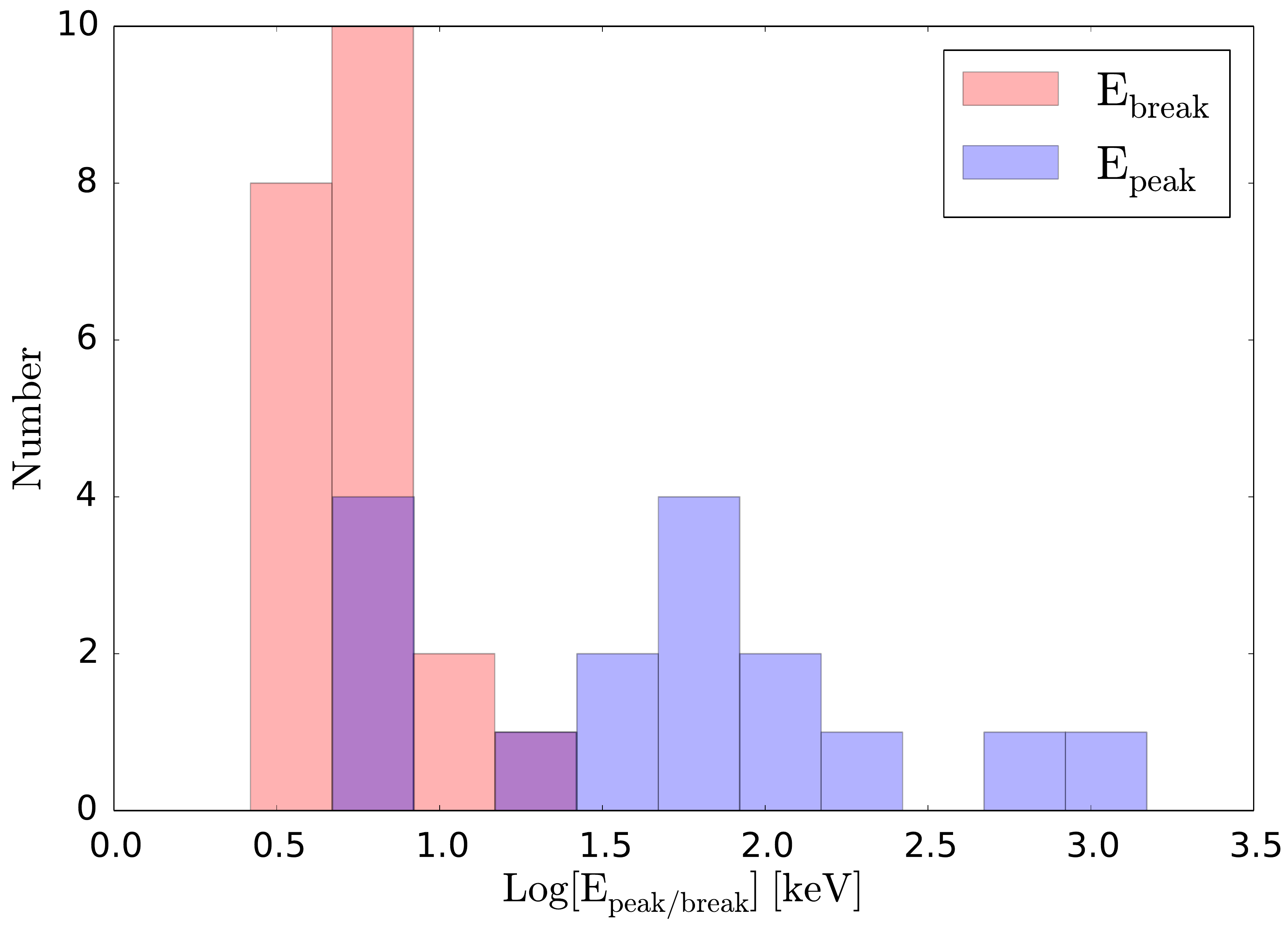}  
\includegraphics[scale=0.26,trim=0 0 0 -0.6cm]{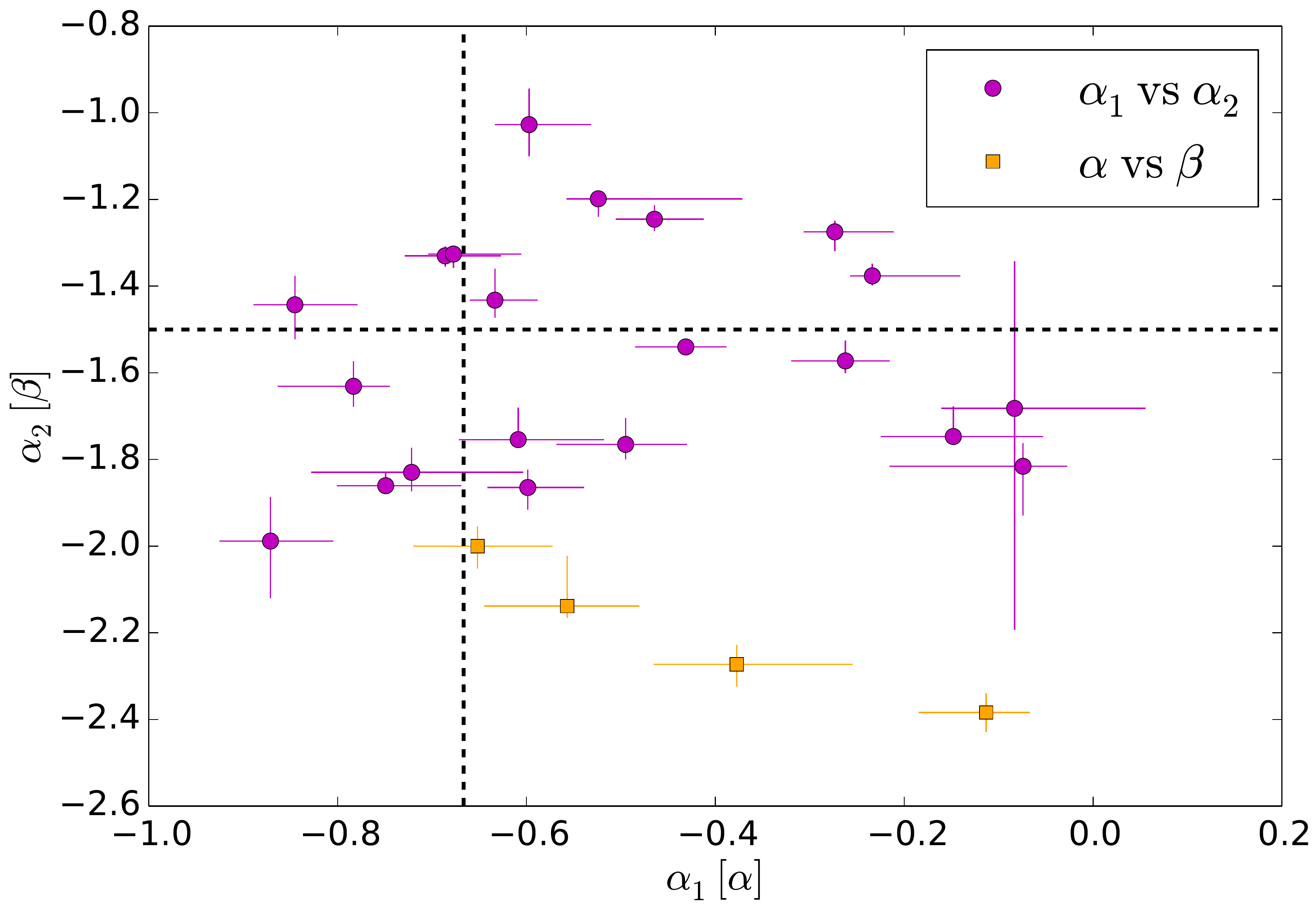} 
\includegraphics[scale=0.26,trim=0 0.2cm 0 -0.8cm]{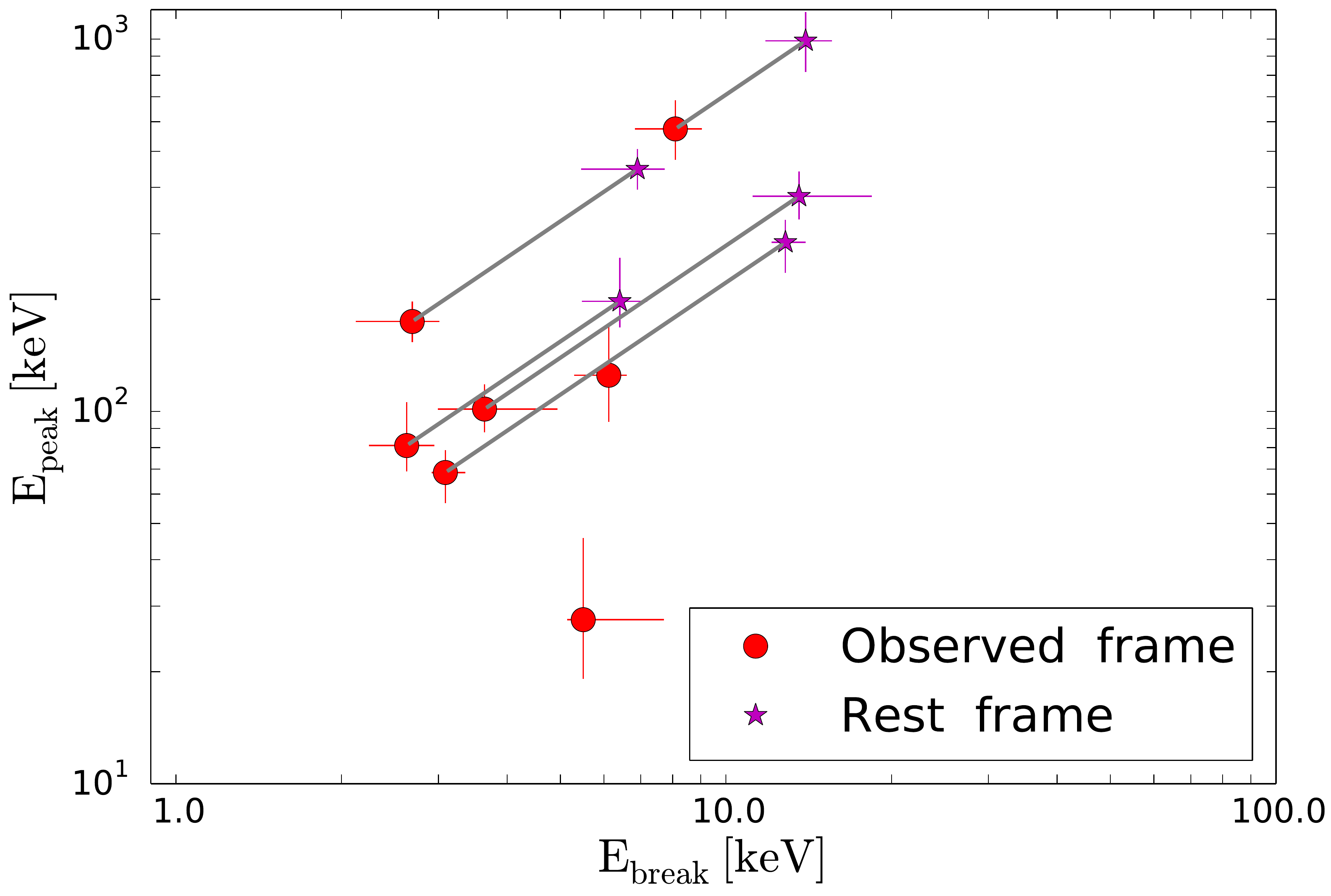} 
\caption{Best fit parameters resulting from the spectral analysis of the full sample. Left-hand panels: photon indices. Right-hand panels: peak and break energies.
Top left: distribution of $\alpha_1$ (red) and $\alpha_2$ (green), representing the indices below and above the break energy.
Bottom left: $\alpha_1$ vs. $\alpha_2$ (circles) and $\alpha$ vs. $\beta$ for GRBs with a spectrum modeled by a Band function with second index $<-2$ (squares). 
In both left-hand panels, the values for $\alpha_1$ and $\alpha_2$ predicted from fast cooling synchrotron emission are drawn as dashed lines. In the bottom panel they should be used as reference lines for the circle symbols only. 
Top right: $E_{\rm peak}$ (blue) and $E_{\rm break}$ (red) distributions. 
Bottom right: $E_{\rm peak}$ vs. $E_{\rm break}$ for spectra where both features can be constrained (Bandcut model). Circles refer to the observer frame, while stars are used for the rest frame, for those GRBs with measured redshift. The positions before and after cosmological redshift correction are connected with a solid line.
\label{fig:best_fit_param}}
\end{figure*}
%--------------------------------------------------------------------

\subsection{Selection of the best fit model}
From the joint usage of different likelihoods (CSTAT for XRT data, Gaussian for BAT and PGSTAT for GBM data) it is possible to derive an overall likelihood from the product of the single likelihoods obtained separately for each instrument.
For each GRB, we then derive the overall likelihood  for all tested models. In practice, we define the likelihood ($\lk$) as the sum of $\chi^2$, CSTAT, and PGSTAT given by the best fit in XSPEC. To identify the best model, we compare pairs of models and we associate to a given improvement (i.e., a $\delta\lk\,=\,\lk_{modelB} - \lk_{modelA}$) a chance probability, and select the most complex model only if the chance probability is less than $1\%$. The association between  a given $\delta\lk$ and its chance probability has been obtained by performing simulations, as described in the following.

For each GRB, we fit a simple model (e.g., PL) to the spectrum.
Then, we simulate $10^3$ fake spectra using as input this model and its best-fit parameters obtained from the fit of the real spectrum. 
We re-fit the fake spectra using both the input model and using more complex models (e.g., BPL). As a result, we have a distribution of $\delta\lk\,=\,\lk_{\rm PL}-\lk_{\rm BPL}$. An example of $\delta\lk$ distribution from spectra simulated with a PL model and re-fitted with PL and i) CPL model and ii) BPL model is presented in Appendix~ \ref{app:a} (left- and right-hand panels in Fig.~\ref{fig:sim}, respectively).
The distribution allows to associate a chance probability to a given $\delta\lk$: large improvements correspond to a small probability of being obtained by chance.

A more complex model is preferred over a simpler one whenever the improvement derived from the fit of real spectra corresponds to a probability of less than $1\%$ of being a chance improvement. In the example in Fig.~\ref{fig:sim}, the two panels show the distributions of $\delta\lk_{\rm PL-CPL}$ and $\delta\lk_{\rm PL-BPL}$ for simulated spectra.
The $\delta\lk$ obtained when the real GRB spectrum is fitted with PL and CPL is 59, and is 162 when comparing fits with PL and BPL models (as reported in the figure title). 
As can be seen from the comparison with the red horizontal line (corresponding to $1\%$ chance probability), the probability that such improvements are obtained by chance are much smaller than $1\%$. Both CPL and BPL models are in this example preferred over the PL. The whole procedure is repeated replacing the PL model with more complex models, until the best fit model is found.

%=========================  RESULTS  ==========================
\section{Results}\label{sec:results}

We fit the 34 time-integrated spectra (one for each GRB) with all the models (PL, CPL, Band, and Bandcut, see Fig.~\ref{fig:sketch}) and define the best-fit model according to the method explained in the previous section.
The results are reported in Table~\ref{tab:table} and Fig.~\ref{fig:lc+sp}.
For each GRB, the table reports the time interval used for the spectral analysis, the best fit model, the best fit parameters, the average energy flux (in the energy range 0.5\,keV - 10\,MeV), the $\lk$ and d.o.f., and the instruments included in the analysis. 
Figure~\ref{fig:lc+sp} shows, for each GRB, the XRT, BAT (and eventually GBM) light curve (with the time interval used for spectral analysis highlighted ) and the photon spectrum obtained with the best fit model.

The best fit model is a PL in 4 cases (GRB 070721B, GRB 080906, GRB 100413A and GRB 130606A), a CPL in 5 cases, a Band model with a $\nu F_\nu$ peak in 4 cases, a Band model with both photon indices $>-2$ in 14 cases, and a Bandcut model in the remaining 7 cases. 
This means that 62\% of prompt emission spectra in our sample (i.e., 21 out of 34 GRBs) display a low-energy spectral break $E_{\rm break}$ separating two power-law segments with photon indices $>-2$.

We first focus on these 21 cases and show the best-fit model in $\nu F_{\nu}$ units (gray curves) in Fig.~\ref{fig:averaged}.
For reference, we also plot a Band model with photon indices $\alpha_1^{\rm syn}=-0.67$ and $\alpha_2^{\rm syn}=-1.5$ (dashed, red curve).
As can be seen, the dashed red curve is on average a good representation of the observed spectra. 

Figure~\ref{fig:best_fit_param} shows the distribution of the photon indices $\alpha_1$ and $\alpha_2$ (upper left-hand panel) and the relation between them (circled symbols in the bottom left-hand panel).
In both panels, the dashed lines mark the values expected for $\alpha_1^{\rm syn}$ and $\alpha_2^{\rm syn}$ in the case of the fast-cooling regime.
A Gaussian fit to the distributions returns $\langle\alpha_1\rangle=-0.51$ ($\sigma=0.24$) and  $\langle\alpha_2\rangle=-1.56$ ($\sigma=0.26$). 
These values are consistent within 1$\sigma$ with the values found in O17 and with the synchrotron values. 
The bottom left-hand panel in Fig.~\ref{fig:best_fit_param} also reports $\alpha$ versus $\beta$ values for those spectra best modeled by a peaked Band model (square symbols).

The right-hand panels in Fig.~\ref{fig:best_fit_param} summarize the results on $E_{\rm peak}$ and $E_{\rm break}$.
A Gaussian fit to the $E_{\rm break}$ logarithmic distribution returns $\langle \log(E_{\rm break}) \rangle=0.74$ ($\sigma=0.20$), in agreement within 1$\sigma$ with the results from O17. 
The $E_{\rm peak}$ distribution (blue histogram) is wide and flat, and values range from 5 to 915 \,keV.
The bottom right-hand panel of Fig.~\ref{fig:best_fit_param} shows $E_{\rm peak}$ versus $E_{\rm break}$ for the subsample of spectra for which both features are constrained (i.e., for spectra best modeled by a Bandcut function). 
Red circles refer to energies in the observer frame, while purple stars refer to the rest frame values obtained after redshift correction (for GRBs with known redshift).
No hint of a correlation between $E_{\rm peak}$ and $E_{\rm break}$ is found.

%--------------------------------------------------------------------
\subsection{Origin of the typically observed value $\alpha=-1$}\label{sec:test}
The typical value $\alpha\sim-1$ describing the part of the spectrum below $E_{\rm peak}$ was inferred from studies of prompt spectra down to 8-25\,keV, mainly from BATSE, {\it Swift}-BAT, and {\it Fermi}-GBM.
In O17 and in this work, we found that when soft X-ray data are available and require a model including a low-energy spectral break, the part of the spectrum immediately below the peak energy is described by a value $\alpha_2\sim-1.5$, softer than $\alpha$ and consistent with the synchrotron theory.
These results, in apparent contradiction, seem to suggest that fit results depend on the extension of the energy range over which observations are available and/or on the shape of the function used to model the data.
Another explanation is that the subsample of events studied here is somehow peculiar and not representative of the whole population.

\begin{figure}
\centering
%\vskip -1.8 truecm 
 \includegraphics[width=1.0\columnwidth,trim=0 0.3cm 0 0]{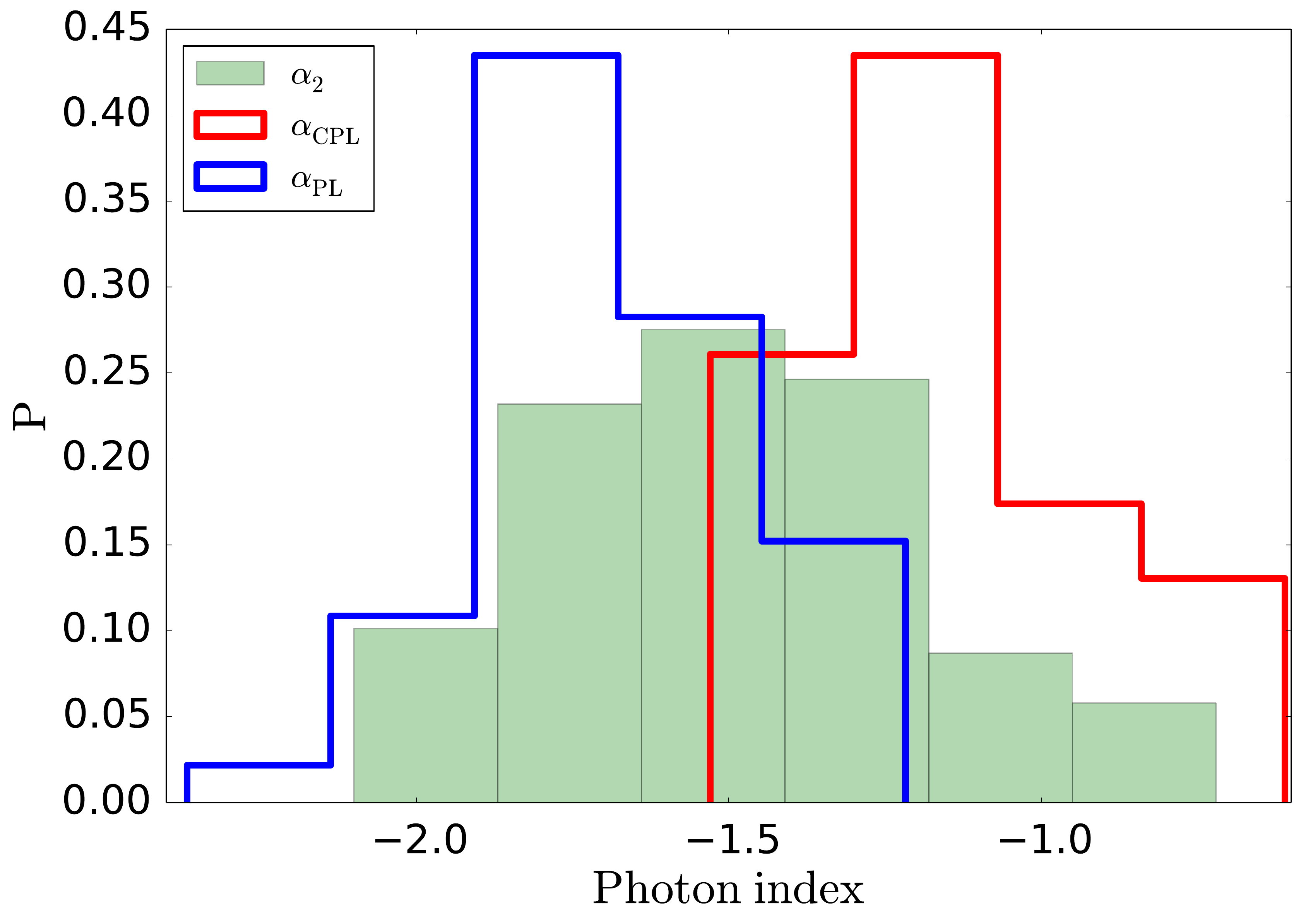}
% \vskip -2.5 truecm
\caption{\label{fig:no_XRT_test} Distribution of photon indices describing the spectrum below the peak energy, for the merged sample studied in O17 and this paper. The green histogram shows the indices $\alpha_2$ describing the spectrum between $E_{\rm break}$ and $E_{\rm peak}$.
The blue and red histograms show the distribution of $\alpha$ for the same sample of spectra, obtained when the analysis is performed without including XRT. 
The red histogram denotes cases where the best model is a CPL, and the blue histogram cases where the best-fit model is a single PL.}
\end{figure}

To test all these possibilities, we perform the following exercise.
We collect all the spectra (among those presented here and in O17) displaying a low-energy break, and plot their $\alpha_2$ distribution (green histogram) in Fig.~\ref{fig:no_XRT_test}.
Subsequently, for all these spectra, we re-do the spectral analysis by excluding XRT data. 
The best-fit model, after excluding XRT, is either a PL or a CPL.
The distributions of the photon indices are shown in Fig.~\ref{fig:no_XRT_test}, separately for $\alpha_{\rm PL}$ (blue histogram) and $\alpha_{\rm CPL}$ (red histogram).
We find $\langle \alpha_{\rm PL}\rangle=-1.70$ ($\sigma= 0.23$) and $\langle\alpha_{\rm CPL}\rangle=-1.15$ ($\sigma=0.21$).
A Kolmogorov-Smirnov (KS) test between the distributions of $\rm \alpha_{PL}$ and $\rm \alpha_{CPL}$
has a probability of $\rm {\bf 3} \times 10^{-9}$ that the two populations are drawn from the same parent distribution.
A similar separation between $\alpha_{\rm PL}$ and $\alpha_{\rm CPL}$ and similar mean values are found in the population of BAT  \citep{Lien16}, GBM \citep{gruber14}, and BATSE bursts \citep{Kaneko06}.
We conclude that when XRT data are removed, the best fit model has a shape similar to the typical shape of the general GRB population.
The KS tests for $\alpha_{\rm PL}-\alpha_2$ and $\alpha_{\rm CPL}-\alpha_2$ give probabilities of $3 \times 10^{-3}$ and $9 \times 10^{-5}$, respectively. 
This shows that this is not a peculiar subsample of GRBs: when XRT data are excluded, the best-fit parameters are in full agreement with the general population.

\begin{figure}[ht!]
\centering
%\vskip -1.8 truecm 
 \includegraphics[scale=0.48]{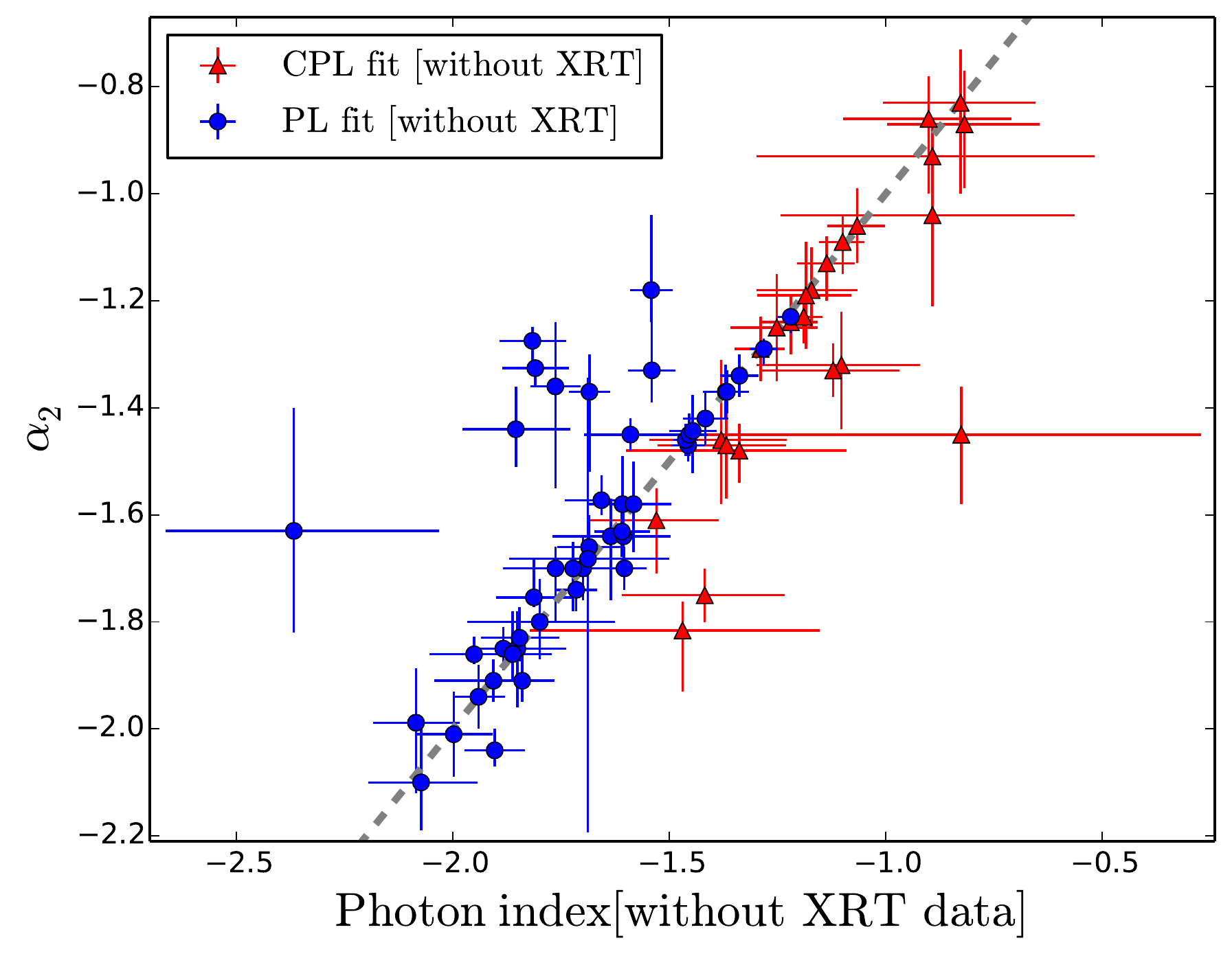}
% \vskip -2.5 truecm
\caption{\label{fig:a2_vs_a} Comparison between the photon index $\alpha_2$ and the photon index $\alpha$ derived after removing XRT data from the spectral analysis. In this last analysis, the best-fit model is either a PL (and the index is called $\alpha_{\rm PL}$, blue points) or a CPL ($\alpha_{\rm CPL}$, red points). The identity line is drawn as a dashed gray line.}
\end{figure}

In Fig.~\ref{fig:a2_vs_a}, the values of $\alpha_2$ are shown versus the values of $\alpha$ derived after excluding XRT data.
Different colors and symbols are used to distinguish between cases in which the best-fit model after XRT exclusion is a PL (blue) or a CPL (red).
First of all, we note the separation between red and blue points, which was already evident from Fig.~\ref{fig:no_XRT_test}.
Most of the points are consistent within 1$\sigma$ with the equality line (dashed gray line). However, almost half of the CPL fits return a harder value of $\alpha$. 
On the contrary, PL fits tend to return softer spectra.

The extension of the energy range and the introduction of a function with a low-energy break (necessary for a good description of the overall spectral shape) have then a strong impact on the inferred value of the photon index describing the spectral shape at energies below the spectral peak energy.
The overall result is that when XRT data are available, the part of the spectrum below $E_{\rm peak}$ is described by a theoretically motivated value, but when XRT data are removed and the spectrum is modeled with a CPL function, the best fit photon index describing the part below $E_{\rm break}$ has a harder value, that if taken as face value leads to the opposite conclusion: the inconsistency of observed spectra with synchrotron radiation.

%------------------------------------------------------
%------------------------------------------------------

%------------------------------------------------------
%-------------------------------------------------------

%=======================================================
\subsection{Comparison with the full BAT catalog}
\begin{figure}
\includegraphics[width = 0.50\textwidth]{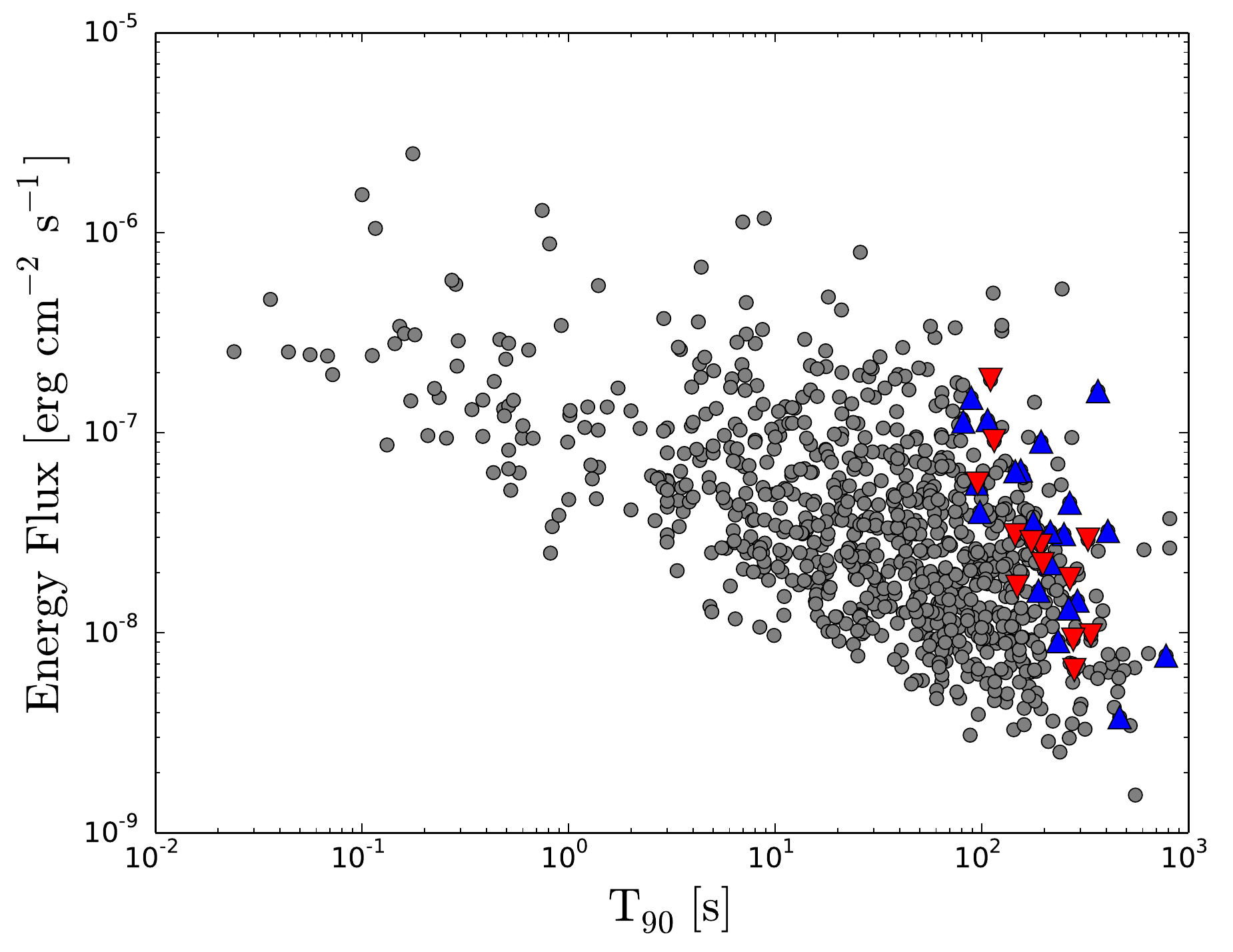}  %\hspace{1.1em}
\includegraphics[width = 0.50\textwidth]{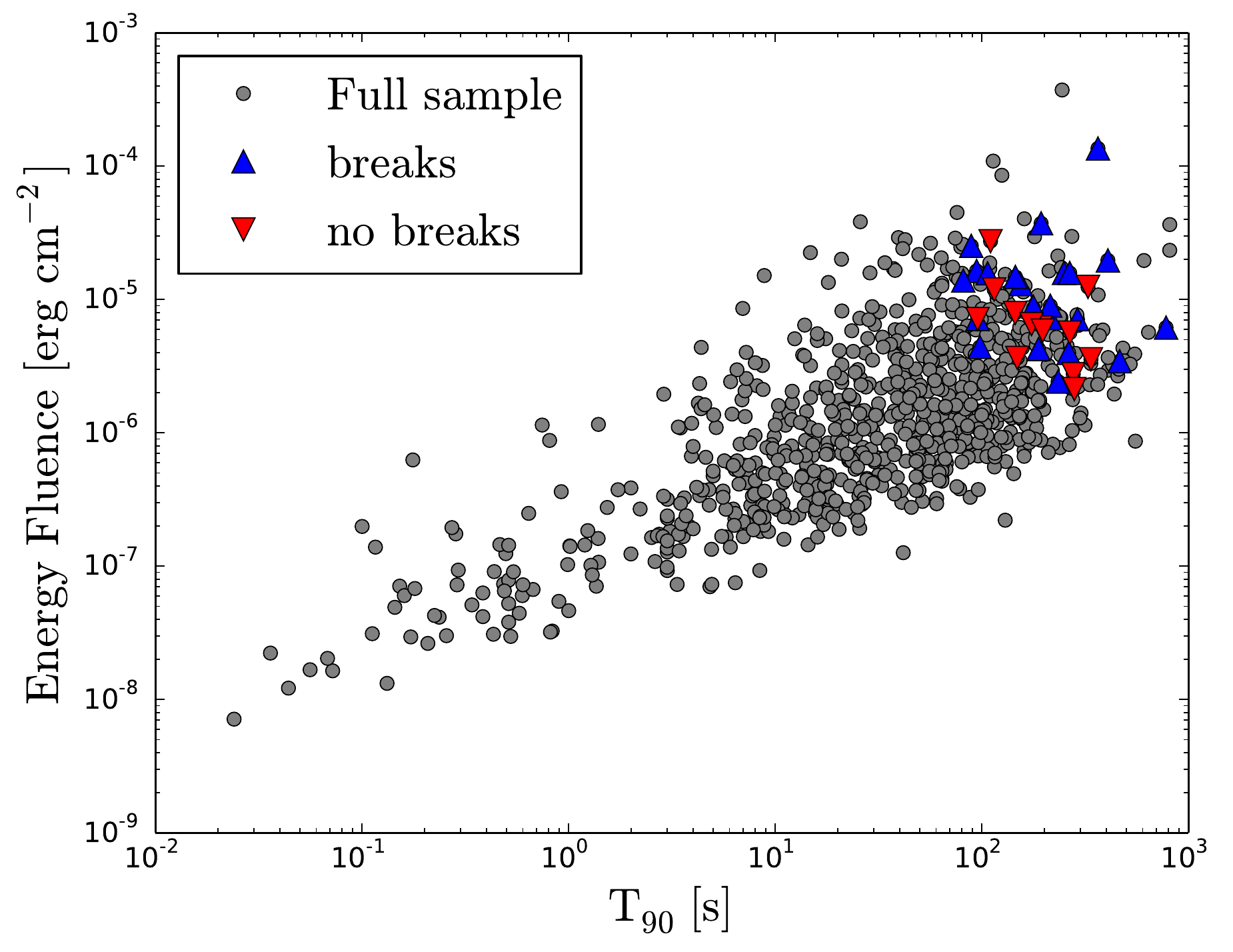} 
\caption{Comparison in terms of flux, fluence, and $T_{90}$ between the sample studied in this work (triangles) and the full catalog of BAT GRBs, from \cite{Lien16} (gray circles). Our sample is divided into two classes: GRBs with a low-energy break (blue triangles) and GRBs without a low-energy break (red upside-down triangles).
Upper panel: average flux vs. $T_{90}$. Lower panel: fluence vs. $T_{90}$.
\label{fig:flux+fluence}}
\end{figure}

Figure~\ref{fig:flux+fluence} shows the average energy-flux versus $T_{90}$ (upper panel), and the fluence versus $T_{90}$ (bottom panel) for our sample (blue and red triangles) and for a large sample of {\it Swift}-BAT GRBs (gray circles, from \citealt{Lien16}). 
The values of fluences and fluxes are integrated in the BAT energy range 15-150\,keV. 
The sample of GRBs studied in this paper is clearly biased towards long prompt emission durations.
This reflects the slew time required by the satellite to place the BAT source within the XRT field of view: prompt emission can be observed with the XRT only if it lasts longer than the typical slewing time.
In some cases however, the $T_{90}$ does not reflect the duration of the main emission episode, since the large $T_{90}$ duration is caused by the presence of a precursor, while the main emission (detected by the XRT) has a more standard duration (the light curves of all 34 GRBs can be seen in Fig.~\ref{fig:lc+sp}).
Limiting the comparison to GRBs with a similar duration, we note that the whole range of fluxes of the full sample is spanned also by our sample.
To look for differences within our sample, we mark GRBs with a low-energy break (blue triangles) and GRBs with no evidence of a break (red upside-down triangles).
The two different subsamples do not display any relevant difference in terms of flux, fluence, duration, or a combination of these quantities. 
The question of whether GRBs with spectral breaks in the soft X-ray band have some characteristics that distinguish them from GRBs without X-ray breaks remains therefore an open question.
%-------------------------------------------------------
%-------------------------------------------------------
%=======================   CONCLUSIONS   =====================

\section{Conclusions}\label{sec:conclusions}
We studied a sample of 34~GRBs for which the prompt emission (or part of it) was detected simultaneously by XRT and BAT.
We performed time-integrated joint spectral analysis over the time interval where signal above background is observed in both instruments.
In particular, since the signal in BAT is, in general, fainter, we required a  BAT signal-to-noise ration (S/N) $>$30 to guarantee a reliable spectral analysis.

Our results confirm the results obtained by \cite{Oganesyan2017} on a smaller sample, and can be summarized as follows.
\begin{itemize}
\item  62\% of the prompt spectra display a change in slope at low energy, (between 3 and 22\,keV, observer frame). In other words, the data in the soft X-ray band do not lie on the power-law extrapolation of the Band spectrum describing the 10\,keV-MeV data. The change in slope can be well described by adding a break into the fitting function and an additional power-law segment below the break.
\item The spectral indices $\alpha_1$ and $\alpha_2$ below and above the break energy have a distribution centered around the values $\langle\alpha_1\rangle=-0.51$ ($\sigma=0.24$) and $\langle\alpha_2\rangle=-1.56$ ($\sigma=0.26$), consistent within 1$\sigma$ with the values predicted in case of fast cooling synchrotron radiation. 
\item The value of the spectral index describing the part of the spectrum below the peak energy is sensitive to the inclusion of low-energy data and to the fitting function. If XRT data are included in the analysis and if the break is modeled, the average value is around -1.5, consistent with synchrotron radiation. If XRT data are removed and a CPL model is used to describe the spectrum, the photon index is harder, leading to the opposite conclusion of an inconsistency with synchrotron radiation.
\item GRBs with low-energy breaks share similar observational properties in terms of flux and fluence as compared to those without a break in their soft X-ray band.   
\end{itemize}

The average values of the photon indices allow us to speculate about a possible synchrotron origin of the observed spectrum. 
In such a scenario, the break energy $E_{\rm break}$ would correspond to the cooling frequency, and the peak energy $E_{\rm peak}$ to the typical synchrotron frequency.
The ratio between the two characteristic energies ranges from $\sim5$ to $\sim 71$, implying a ratio $\gamma_{\rm m}/\gamma_{\rm c}\sim 2-8$, where $\gamma_{\rm m}$ is the typical energy of the particles injected by the acceleration process, and $\gamma_{\rm c}$ is the cooling Lorentz factor.

For these values of $\gamma_{\rm m}/\gamma_{\rm c}$, particle cooling is still very efficient \citep{Daigne11}.
Such a regime has been extensively discussed in the literature, and is often referred to as a moderately fast cooling regime. 
Different scenarios have been proposed to achieve such a situation.
Adiabatic cooling effects have been widely discussed by \cite{Daigne11}, and can explain the large cooling frequencies if the dissipation takes place at large radii ($>$10$^{15}$\,cm) in a region characterized by a relatively weak magnetic field (10-100\,G in the comoving frame) and moving with large bulk Lorentz factor ($\Gamma>$\,400).
A variation of the magnetic field as a function of the distance from the shock front \citep{Peer06,Derishev07} can also lead to cooling timescales larger than the one inferred from typical magnetic field values.
Another possibility is to modify the standard assumption on particle acceleration, and invoke slow particle heating \citep{asano09} or particle reacceleration \citep{kumar08,beniamini13}.

The comparisons between the photon indices derived from the empirical model and the predictions of the 
synchrotron scenario represent a first consistency check, but they do not prove the validity of the synchrotron interpretation. 
Several studies 
\citep{Beloborodov2013,Axelsson2015,Yu2015,Vurm2016} have argued that the observed spectral width around the peak energy is narrower than the one characterizing synchrotron spectra. 
A more detailed, theory-driven analysis is now required in order to assess the validity of the synchrotron interpretation.

\begin{acknowledgements}
LN acknowledges funding from the European Union's Horizon 2020 Research and Innovation programme under the Marie Sk\l odowska-Curie grant agreement n.\,664931.
This work made use of public {\it Fermi}-GBM data and data supplied by the UK Swift Science Data Centre at the University of Leicester.
This research has made use of data and software provided by the High Energy Astrophysics Science Archive Research Center (HEASARC), which is a service of the Astrophysics Science Division at NASA/GSFC and the High Energy Astrophysics Division of the Smithsonian Astrophysical Observatory.
\end{acknowledgements}

\bibliographystyle{aa} % style aa.bst
\bibliography{references} % your references Yourfile.bib

%=========================  APPENDIX  ===========================
\begin{appendix}
\onecolumn 
\section{Simulations}\label{app:a}
%-------------------------------------------------------
We report an example of simulations performed to select the best-fit model. The example refers to the joint XRT+BAT+GBM spectrum of GRB~080928. For this spectrum, the difference between the likelihoods obtained with PL and CPL models is 59, and the difference between PL and BPL likelihoods is 162.

The best fit parameters of the PL fit are used as input to simulate $10^3$ fake spectra, which are then re-fitted with CPL and BPL models.
For these simulated spectra, the figure shows the distributions of $\delta\lk_{\rm PL-CPL}$ and $\delta\lk_{\rm PL-BPL}$. The horizontal red line marks the $1\%$ threshold used to define the best fit model.

\begin{figure}[ht!]
\includegraphics[width = 0.90\textwidth]{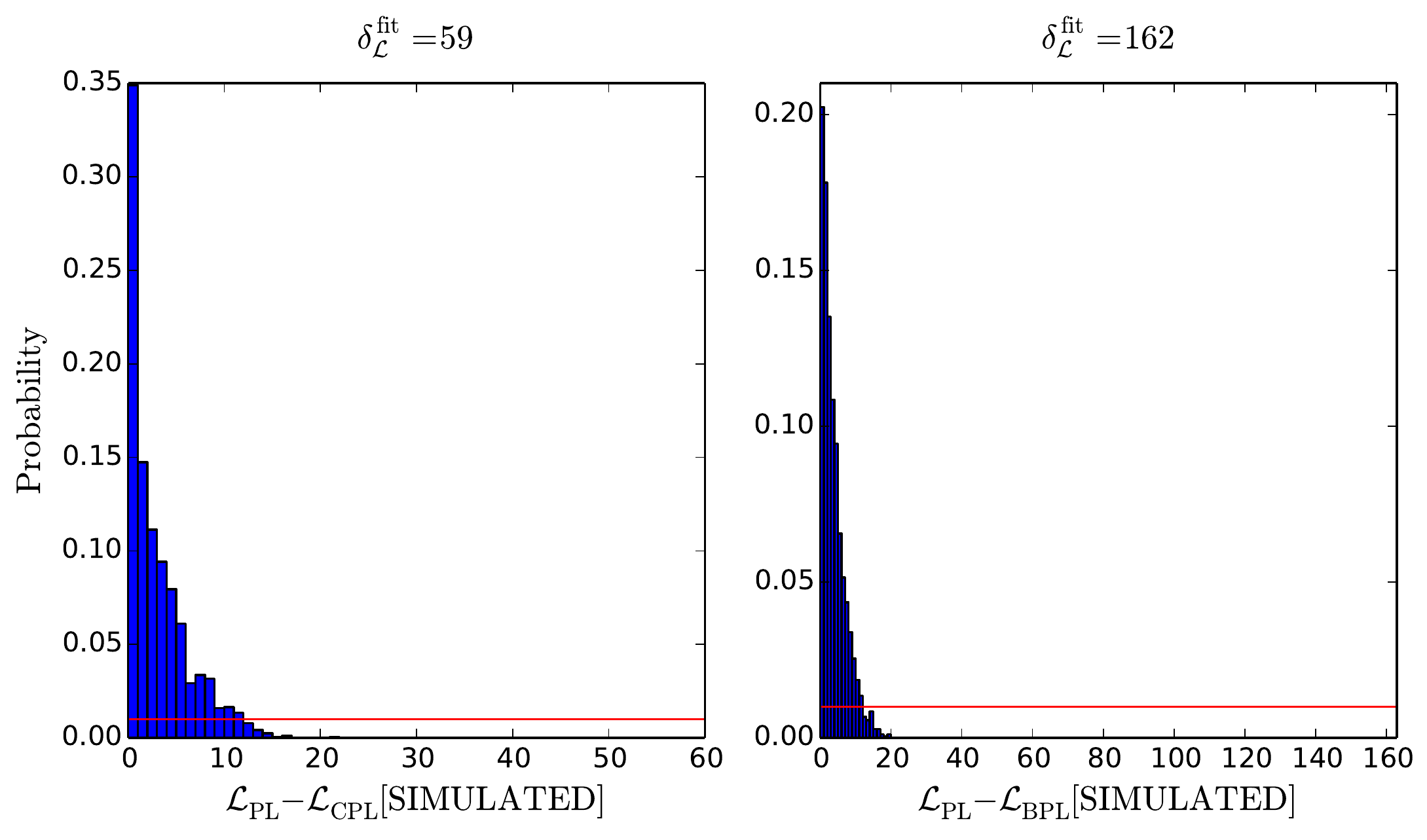}  %\hspace{1.1em} 
\caption{Left panel: distribution 
of the improvement of the CPL fit with respect to the PL fit, for 1000 fake spectra. All fake spectra have been simulated adopting the best fit model obtained fitting a PL model to the observed spectrum of GRB~080928. 
A $\delta\lk>11$ corresponds to a probability $<1\%$ to have a stochastic improvement (horizontal red line in the figure).
The $\delta\lk$ obtained when the real spectrum is fitted with a PL and a CPL is $\delta\lk_{\rm PL-CPL}=59$, corresponding to a chance probability much smaller than $1\%$. Right panel: As in the left panel, but the PL fits are compared to BPL fits. Also in this case, the improvement $\delta\lk=162$ exceeds the critical value of 12. 
\label{fig:sim}}
\end{figure}
%-------------------------------------------------------
\clearpage

\section{Tables}
\begin{table}[ht!]
\LTcapwidth=0.6\textwidth 
\caption{List of the 34 GRBs analyzed in this work. The GRB's name and redshift are reported in the first and second columns. 
The third column lists the values of the intrinsic $N_{\rm H}$, derived from spectral analysis of late time XRT observations (see Sect.~\ref{sec:analysis} for the method used to estimate $N_{\rm H}$ for GRBs with and without measured redshift).  The late time interval (LTI, from the BAT trigger time) chosen for the estimate of $N_{\rm H}$ can be found in the last column.}
\label{tab:nh}
\centering
\begin{tabular}{l c c c} 
\hline\hline \\[-0.2cm]
GRB & Redshift     &  $N_{\rm H}$ &  LTI  \\[0.1 cm]
&  & $10^{22}$\,cm$^{-2}$& $10^4$\,s \\[0.1 cm]
\hline
      &       &      &      \\ 
060510B      &  $4.940$     &   $0.00$    & $0.05-62.41$     \\ [0.1 cm]
060814   & $1.92$    & $1.93$ &  $16.83-137.78$\\ [0.1 cm]
061121   & $1.314$   & $0.82$ &  $3.46-9.25$\\ [0.1 cm]
070616   & ...           &  $0.49$ & $0.46-37.11$\\ [0.1 cm]
070721B      &  $3.626$     & $0.00$     & $0.02-13.95$     \\ [0.1 cm]
080906      &  ...     &  $0.34$    &  $0.54-73.69$    \\ [0.1 cm]
080928      &   $1.69$    &  $0.62$    & $0.42-3.34$     \\ [0.1 cm]
081008      &   $1.969$    & $0.69$     &  $0.06-1.77$    \\ [0.1 cm]
090709A      &  ...     &  $0.27$    &  $0.41-6.27$    \\ [0.1 cm]
090715B      &  $3.000$     & $2.02$     &  $0.02-164.88$    \\ [0.1 cm]
100413A      &  ...     &  $0.25$    & $0.63-17.46$     \\ [0.1 cm]
100619A & ...           & $0.45$ & $5.34-100.59$\\ [0.1 cm]
100704A      &  ...     & $0.37$     &  $1.26-131.81$    \\ [0.1 cm]
100725B & ...          &  $0.59$ & $2.18-80.35$\\ [0.1 cm]
100728A & $1.567$  &  $3.43$ & $0.50-68.29$\\ [0.1 cm]
100814A      & $1.440$      &  $0.25$    & $21.20-671.30$     \\ [0.1 cm]
100906A & $1.727$  & $0.94$ &  $1.06-46.86$\\ [0.1 cm]
110102A & ...            & $0.19$ & $1.04-24.32$ \\ [0.1 cm]
110119A      &   ...    &  $0.16$    & $0.49-26.09$     \\ [0.1 cm]
110205A & $2.22$   &  $0.70$ & $0.14-38.29$\\ [0.1 cm]
111103B      &  ...     &  $0.35$    & $4.55-30.06$     \\ [0.1 cm]
111123A      &  $3.152$     &  $4.81$    & $0.42-4.06$     \\ [0.1 cm]
121123A & ...           &   $0.05$ & $1.66-13.91$\\ [0.1 cm]
121217A      &  ...    &  $0.63$    & $2.87-547.65$     \\ [0.1 cm]
130514A      &  ...     &  $0.36$    & $1.15-38.81$     \\ [0.1 cm]
130606A      &  $5.913$     &  $9.12$    & $0.53-1.26$     \\ [0.1 cm]
130907A & $1.238$  & $1.05$  &  $0.76-238.41$\\ [0.1 cm]
140108A & ...           &   $1.70$ & $1.05-43.16$ \\ [0.1 cm]
140206A & $2.73$   & $2.03$ &  $2.12-8.71$\\ [0.1 cm]
140323A      & ...      &  $0.39$    & $2.82-9.84$     \\ [0.1 cm]
140512A & $0.725$  & $0.44$ &  $2.79-32.94$\\ [0.1 cm]
141031A  & ...      & $0.30$     &  $4.57-84.70$    \\ [0.1 cm]
150724A      & ...      &  $0.55$    & $0.13-13.26$     \\ [0.1 cm]
151021A      & $2.330$      & $2.40$     & $0.49-67.93$     \\ [0.1 cm]
\hline
\end{tabular}
\end{table}

\clearpage

%%%%TABLE_WITH_PARAMETERS_OF_THE_BEST_FIT_MODEL%%%%%%%%%
{\tiny 
\LTcapwidth=1.0\textwidth 
\begin{longtable}{lcccccccccl}
\caption{\label{tab:table} Results of the spectral analysis. The table lists the GRB name, the best fit model name (PL = power-law, CPL = cutoff power-law, Band =  Band function, Bandcut = Band function with a high energy cutoff), the best fit parameters (columns 4 to 8, for a definition see Fig.~\ref{fig:sketch}), the average flux, the overall likelihood $\lk$, and the degrees of freedom (d.o.f.). The last column reports the instruments included in the spectral analysis: X = XRT, B = BAT, G = GBM.} \\

\hline\hline\\ [-0.1 cm]
GRB & Time interval &Model & $\alpha_1$ & $E_{\rm break}$ & $\alpha / \alpha_2$ &   $E_{\rm peak}$  & $\beta$ & Flux & $\lk$ \,(d.o.f.) & Instr. \\ [0.1 cm]
           &                 &       &                  & keV  &                  &   keV   &  & $10^{-7}$erg\,s$^{-1}$\,cm$^{-2}$ & & \\ [0.1 cm]
\hline \\[-0.1 cm]
060510B &[127.00,325.00]& Band  & ${ -0.61_{-0.06}^{+0.09}}$ & ${   6.15_{-1.08}^{+1.21}}$  & ${   -1.75_{-0.02}^{+0.07}}$ &  &  & $ {   [0.22 - 0.79]}$ & {  1052(1002)} & X+B  \\ [0.1 cm] 

060814 &[77.50,200.00] &Band  & ${   -0.15_{-0.08}^{+0.09}}$ & ${   3.37_{-0.37}^{+0.34}}$  & ${   -1.75_{-0.01}^{+0.07}}$ &  &  & $ {   [0.40 - 1.45]}$ & {  848(1002)} & X+B  \\ [0.1 cm] 

061121 &[62.00,110.00]& Band  & ${   -0.23_{-0.02}^{+0.09}}$ & ${   4.54_{-0.40}^{+0.53}}$  & ${   -1.38_{-0.02}^{+0.03}}$ &  &  & $ {   [5.32 - 76.78]}$ & {  794(1002)} & X+B  \\  [0.1 cm] 
 
070616 &[138.00,615.00]& Bandcut  & ${   -0.63_{-0.03}^{+0.05}}$ & ${   6.12_{-0.82}^{+0.48}}$  & ${   -1.43_{-0.04}^{+0.07}}$ & ${   125.10_{-31.35}^{+44.05}}$ & & ${   0.64_{-0.08}^{+0.10}}$ & {  968(1001)} & X+B  \\ [0.1 cm] 

070721B & [311.00,361.00] &PL  &  &  & ${  -1.20_{-0.05}^{+0.08}}$ &  &  & ${  [0.09 - 2.60]}$ & {  721(1004)} & X+B  \\   [0.1 cm] 

080906 & [78.00,95.00]& PL  &  &  & ${  -1.72_{-0.04}^{+0.09}}$ &  &  & ${  [0.11 - 0.43]}$ & {  700(1004)} & X+B  \\ [0.1 cm] 

080928 & [197.93,256.30]& Band  & ${   -0.72_{-0.11}^{+0.12}}$ & ${   4.79_{-1.20}^{+0.87}}$  & ${   -1.83_{-0.04}^{+0.06}}$ &  &  & $ {   [0.66 - 1.13]}$ & {  1054(1233)} & X+B+G  \\ [0.1 cm] 

081008 & [94.00,193.00]& Band  & ${   -0.75_{-0.05}^{+0.08}}$ & ${   4.34_{-0.63}^{+0.22}}$  & ${   -1.86_{-0.02}^{+0.03}}$ &  &  & $ {   [0.28 - 0.74]}$ & {  900(1002)} & X+B  \\  [0.1 cm] 
 
090709A & [75.00,101.00]& Band & ${   -0.85_{-0.04}^{+0.07}}$ & ${   22.06_{-4.19}^{+12.08}}$  & ${   -1.44_{-0.08}^{+0.07}}$ &  &  & $ {   [1.07 - 12.55]}$ & {  897(1002)} & X+B  \\ [0.1 cm] 

090715B & [53.00,295.00]& CPL  &  &  & ${   -1.56_{-0.03}^{+0.03}}$ & ${   27.81_{-7.40}^{+3.10}}$  &  & ${   0.13_{-0.00}^{+0.02}}$ & {  909(1003)} & X+B  \\ [0.1 cm] 

100413A & [147.00,228.00]& PL &  &  & ${  -1.48_{-0.04}^{+0.01}}$ &  &  & ${  [0.25 - 2.38]}$ & {  902(1004)} & X+B  \\ [0.1 cm] 
 
100619A & [80.68,100.13]& Band  & ${   -0.60_{-0.04}^{+0.06}}$ & ${   6.95_{-0.85}^{+0.71}}$  & ${   -1.86_{-0.05}^{+0.04}}$ &  &  & $ {   [3.30 - 5.40]}$ & {  1147(1218)} & X+B+G  \\ [0.1 cm] 

100704A & [93.37,197.82]& Band  &  &  & ${  -0.38_{-0.09}^{+0.12}}$ & ${  4.70_{-0.56}^{+0.50}}$  & ${  -2.27_{-0.05}^{+0.04}}$ & ${  0.50_{-0.17}^{+0.38}}$ & {  1476(1234)} & X+B+G  \\ [0.1 cm] 
 
100725B & [89.49,229.78]& Band  &  &  & ${   -0.65_{-0.07}^{+0.08}}$ & ${   5.10_{-0.70}^{+0.57}}$  & ${   -2.00_{-0.05}^{+0.05}}$ & ${   1.10_{-0.31}^{+0.44}}$ & {  818(1002)} & {  X+B}  \\ [0.1 cm] 

100728A & [81.53,158.33]& Bandcut  & ${   -0.69_{-0.04}^{+0.06}}$ & ${   2.69_{-0.57}^{+0.33}}$  & ${   -1.33_{-0.02}^{+0.02}}$ & ${   174.41_{-20.88}^{+22.86}}$ & & ${   1.92_{-0.29}^{+0.49}}$ & {  1945(1352)} & X+B+G  \\ [0.1 cm] 

100814A& [94.70,180.71]& Bandcut  & ${   -0.68_{-0.03}^{+0.07}}$ & ${   2.63_{-0.38}^{+0.32}}$  & ${   -1.33_{-0.03}^{+0.02}}$ & ${   80.98_{-12.04}^{+24.92}}$ & & ${   0.35_{-0.05}^{+0.11}}$ & {  1418(1349)} & X+B+G  \\ [0.1 cm] 

100906A & [85.72,125.65]& Band  &  &  & ${  -0.11_{-0.07}^{+0.05}}$ & ${  6.48_{-0.36}^{+0.70}}$  & ${  -2.38_{-0.05}^{+0.04}}$ & ${  1.36_{-0.43}^{+0.32}}$ & {  1274(1234)} & X+B+G  \\ [0.1 cm] 

110102A & [195.17,290.40]& Band  & ${   -0.43_{-0.05}^{+0.04}}$ & ${   5.49_{-0.29}^{+0.54}}$  & ${   -1.54_{-0.01}^{+0.01}}$ &  &  & $ {   [1.29 - 9.90]}$ & {  949(1002)} & X+B  \\ [0.1 cm] 

110119A & [162.56,214.79]& Band  & ${   -0.26_{-0.06}^{+0.05}}$ & ${   5.42_{-0.63}^{+0.70}}$  & ${   -1.57_{-0.03}^{+0.05}}$ &  &  & $ {   [0.90 - 2.52]}$ & {  1238(1233)} & X+B+G  \\  [0.1 cm] 

110205A & [160.00,350.00]& Band  & ${   -0.50_{-0.07}^{+0.07}}$ & ${   7.60_{-1.05}^{+2.75}}$  & ${   -1.77_{-0.03}^{+0.06}}$ &  &  & $ {   [0.65 - 2.36]}$ & {  926(1002)} & X+B  \\  [0.1 cm] 

111103B & [104.00,127.00]& CPL  &  &  & ${   -1.38_{-0.08}^{+0.06}}$ & ${   64.22_{-12.26}^{+47.80}}$  &  & ${   0.47_{-0.08}^{+0.02}}$ & {  756(1003)} & X+B  \\   [0.1 cm] 

111123A & [106.00,274.00] & Bandcut  & ${   -0.27_{-0.03}^{+0.06}}$ & ${   3.09_{-0.17}^{+0.27}}$  & ${   -1.27_{-0.04}^{+0.03}}$ & ${   68.55_{-11.81}^{+10.27}}$ & & ${   0.31_{-0.04}^{+0.08}}$ & {  994(1001)} & X+B  \\ [0.1 cm] 

121123A &  [193.15,299.65]& CPL &  &  & ${   -0.89_{-0.03}^{+0.02}}$ & ${   70.15_{-5.26}^{+4.52}}$  &  & ${   1.11_{-0.02}^{+0.08}}$ & {  1029(1003)} & X+B  \\  [0.1 cm] 

121217A & [717.30,767.48]& Band  & ${   -0.78_{-0.08}^{+0.04}}$ & ${   8.49_{-1.61}^{+1.65}}$  & ${   -1.63_{-0.05}^{+0.06}}$ &  &  & $ {   [1.20 - 2.97]}$ & {  1383(1233)} & X+B+G  \\ [0.1 cm] 

130514A & [96.00,158.00]& Band  &   ${   -0.87_{-0.05}^{+0.07}}$ & ${   8.86_{-1.82}^{+3.21}}$  & ${   -1.99_{-0.13}^{+0.10}}$ &  &  & $ {   [0.53 - 1.07]}$ & {  936(1002)} & X+B  \\  [0.1 cm] 

130606A & [117.00,166.00]& {  PL}  &  &  & ${  -1.15_{-0.02}^{+0.04}}$ &  &  & ${  [0.35 - 12.83]}$ & {  878(1004)} & X+B  \\ [0.1 cm] 

130907A & [71.00,87.00]&Band  & ${   -0.52_{-0.03}^{+0.15}}$ & ${   5.83_{-1.26}^{+1.15}}$  & ${   -1.20_{-0.04}^{+0.01}}$ &  &  & $ {   [6.87 - 203.28]}$ & {  919(1002)} & X+B  \\  [0.1 cm] 

140108A & [76.76,101.33]& {  CPL}  & & & ${  -1.28_{-0.02}^{+0.02}}$ & ${  915.40_{-236.34}^{+405.32}}$  &  & ${  8.40_{-0.50}^{+0.51}}$ & {  1209(1234)} & X+B+G  \\ [0.1 cm] 

140206A & [50.25,100.00]&Bandcut  & ${   -0.60_{-0.04}^{+0.07}}$ & ${   3.64_{-0.64}^{+1.30}}$  & ${   -1.03_{-0.07}^{+0.08}}$ & ${   101.49_{-13.54}^{+16.82}}$ & & ${   3.51_{-0.79}^{+0.45}}$ & {  844(1001)} & {  X+B}  \\ [0.1 cm] 

140323A & [101.52,122.00]&Bandcut  & ${   -0.07_{-0.14}^{+0.05}}$ & ${   5.50_{-0.36}^{+2.21}}$  & ${   -1.82_{-0.11}^{+0.05}}$ & ${   27.59_{-8.44}^{+18.19}}$ & & ${   0.74_{-0.40}^{+0.09}}$ & {  1210(1001)} & X+B+G  \\ [0.1 cm] 

140512A & [102.86,158.16]&Bandcut  & ${   -0.46_{-0.04}^{+0.05}}$ & ${   8.09_{-1.26}^{+0.95}}$  & ${   -1.25_{-0.03}^{+0.03}}$ & ${   573.93_{-100.43}^{+111.73}}$ & & ${   6.07_{-0.87}^{+1.31}}$ & {  1475(1338)} & X+B+G  \\ [0.1 cm] 

141031A & [857.00,893.00]&Band  & ${   -0.08_{-0.08}^{+0.14}}$ & ${   4.59_{-0.82}^{+0.54}}$  & ${   -1.68_{-0.51}^{+0.34}}$ &  &  & $ {   [0.25 - 1.15]}$ & {  934(1002)} & X+B  \\  [0.1 cm] 

150724A & [216.00,235.00]& {  CPL}  &  &  & ${   -1.50_{-0.11}^{+0.09}}$ & ${   21.13_{-3.62}^{+9.78}}$  &  & ${   0.31_{-0.06}^{+0.02}}$ & {  732(1003)} & X+B  \\  [0.1 cm] 

151021A & [95.00,128.00]&Band  &  &  & ${   -0.56_{-0.09}^{+0.08}}$ & ${   5.34_{-0.61}^{+0.75}}$  & ${   -2.14_{-0.03}^{+0.12}}$ & ${   1.05_{-0.40}^{+0.27}}$ & {  868(1002)} & X+B  \\  [0.1 cm] 
\hline
%\end{tabular}
\end{longtable}}

%------------------  LC and SPECTRA  -----------------
\clearpage
\section{Light curves and spectra}
\begin{figure}[b!]
%\vskip -1truecm
\caption{Light curves (left-hand panel) and spectra (middle and right-hand panel) for each GRB in the sample. The light-blue shaded area in the left-hand panels highlights the time interval chosen for the spectral analysis. The middle panels show the $\nu F_\nu$ spectra and the best fit model. We note that the data points in the $\nu F_\nu$ panels have been derived for a specific model, and should not be used to perform comparisons with a different model.
The right-hand panels show the count spectra.
In all panels, XRT data are shown in red, BAT in green, GBM-NaI in blue and light-blue, and GBM-BGO in purple.}
\label{fig:lc+sp}
{\includegraphics[width = 0.40\textwidth]{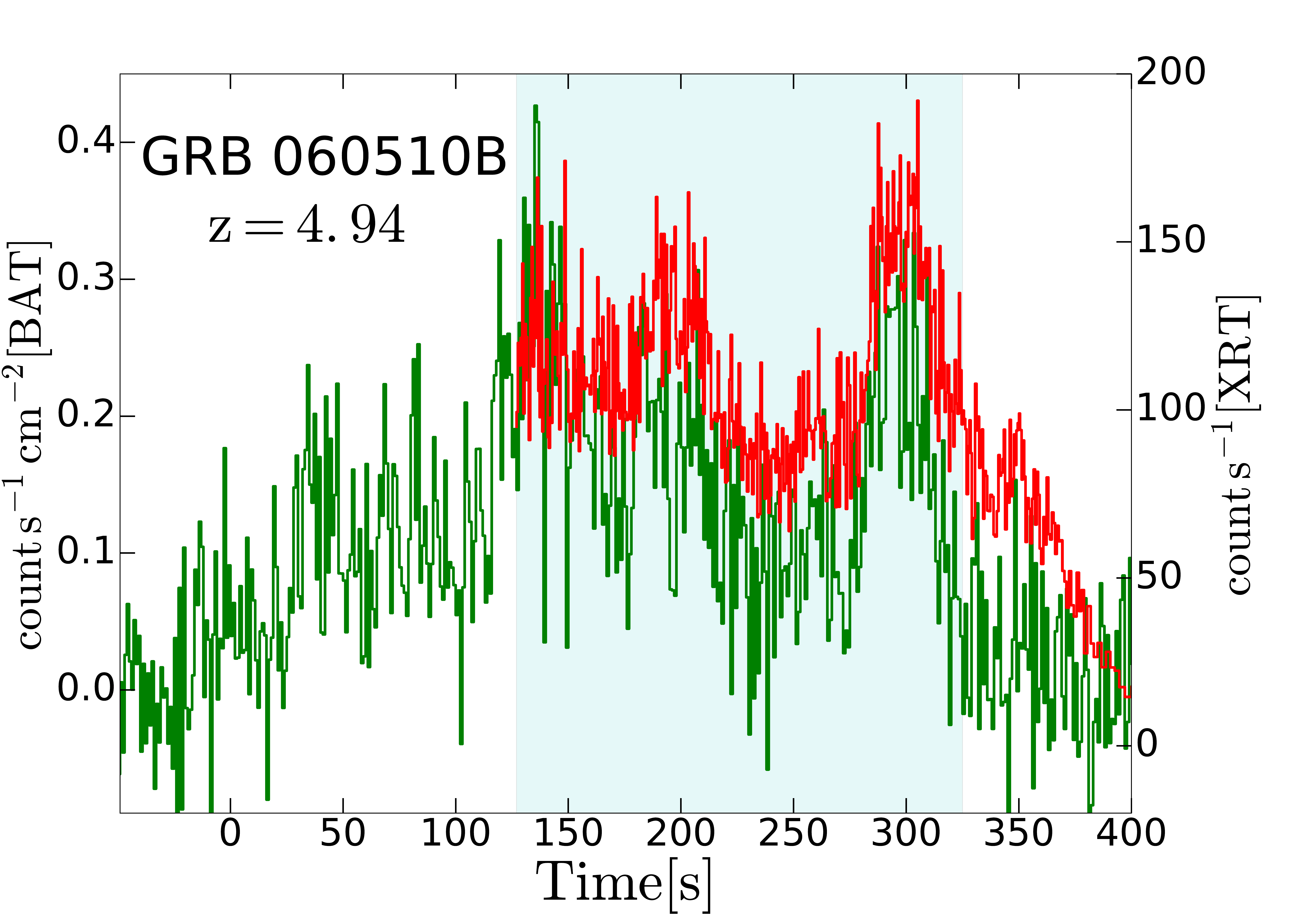} 
\includegraphics[width = 0.30\textwidth]{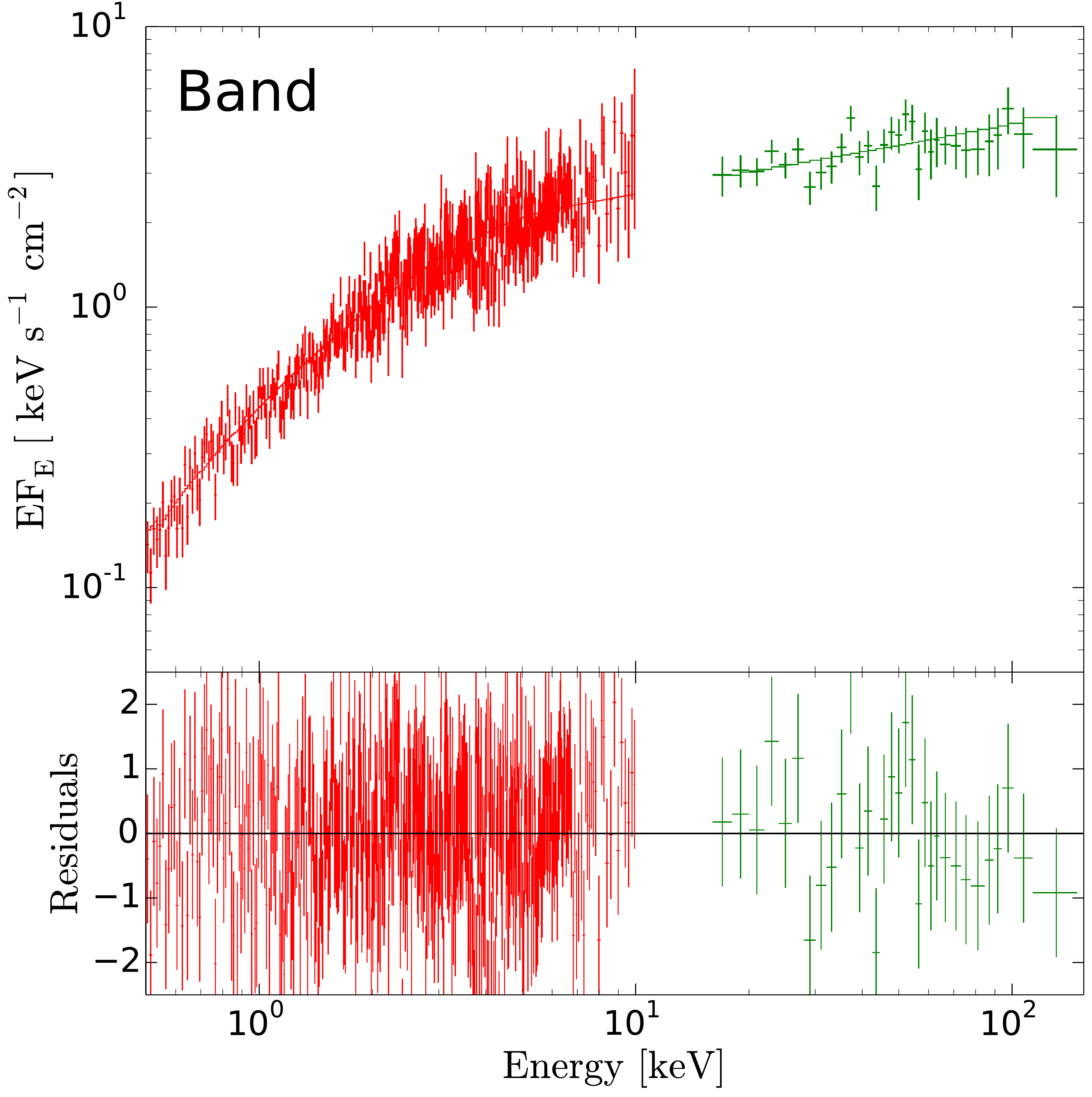}
\includegraphics[width = 0.30\textwidth]{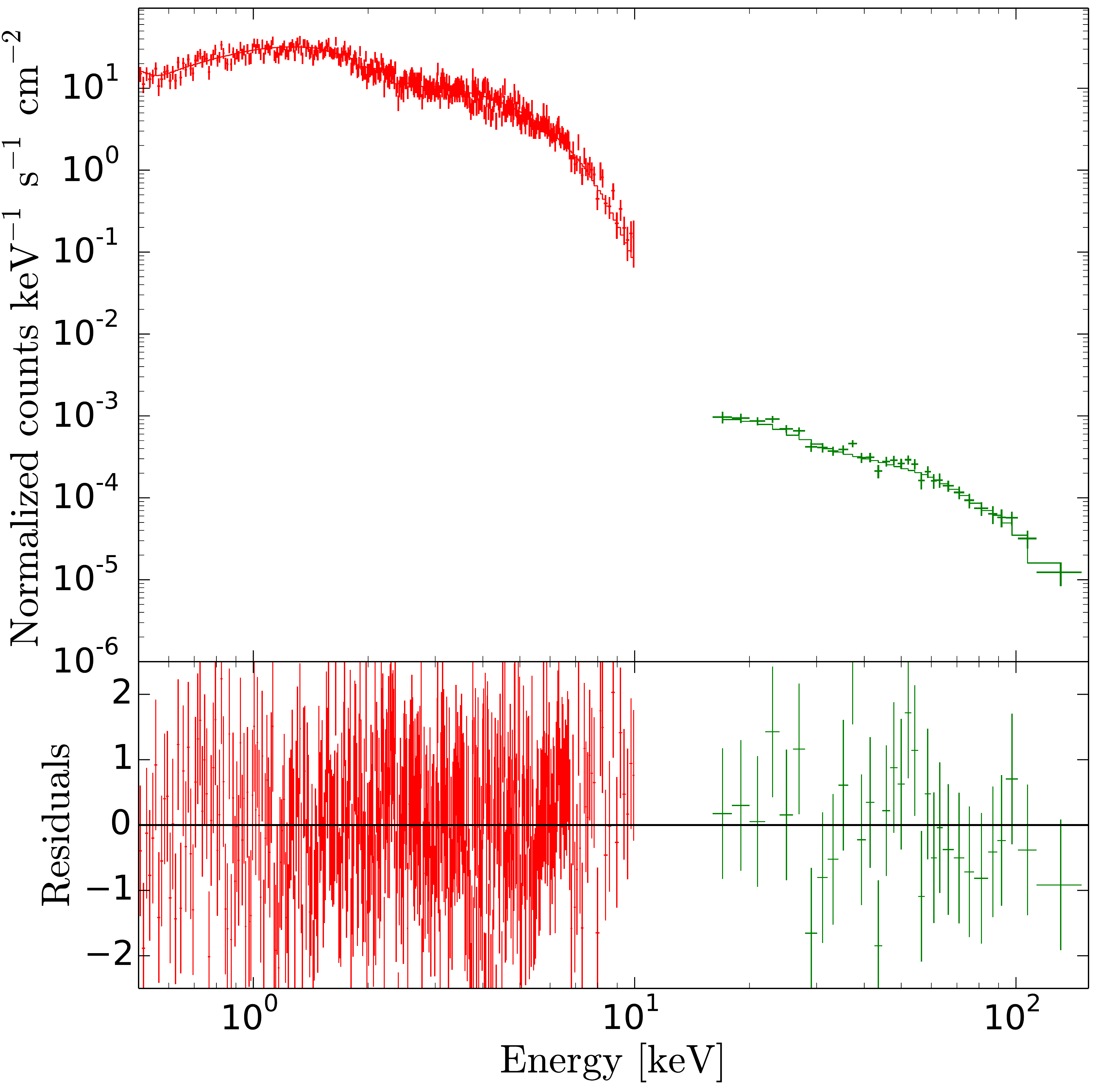}\\
\includegraphics[width = 0.40\textwidth]{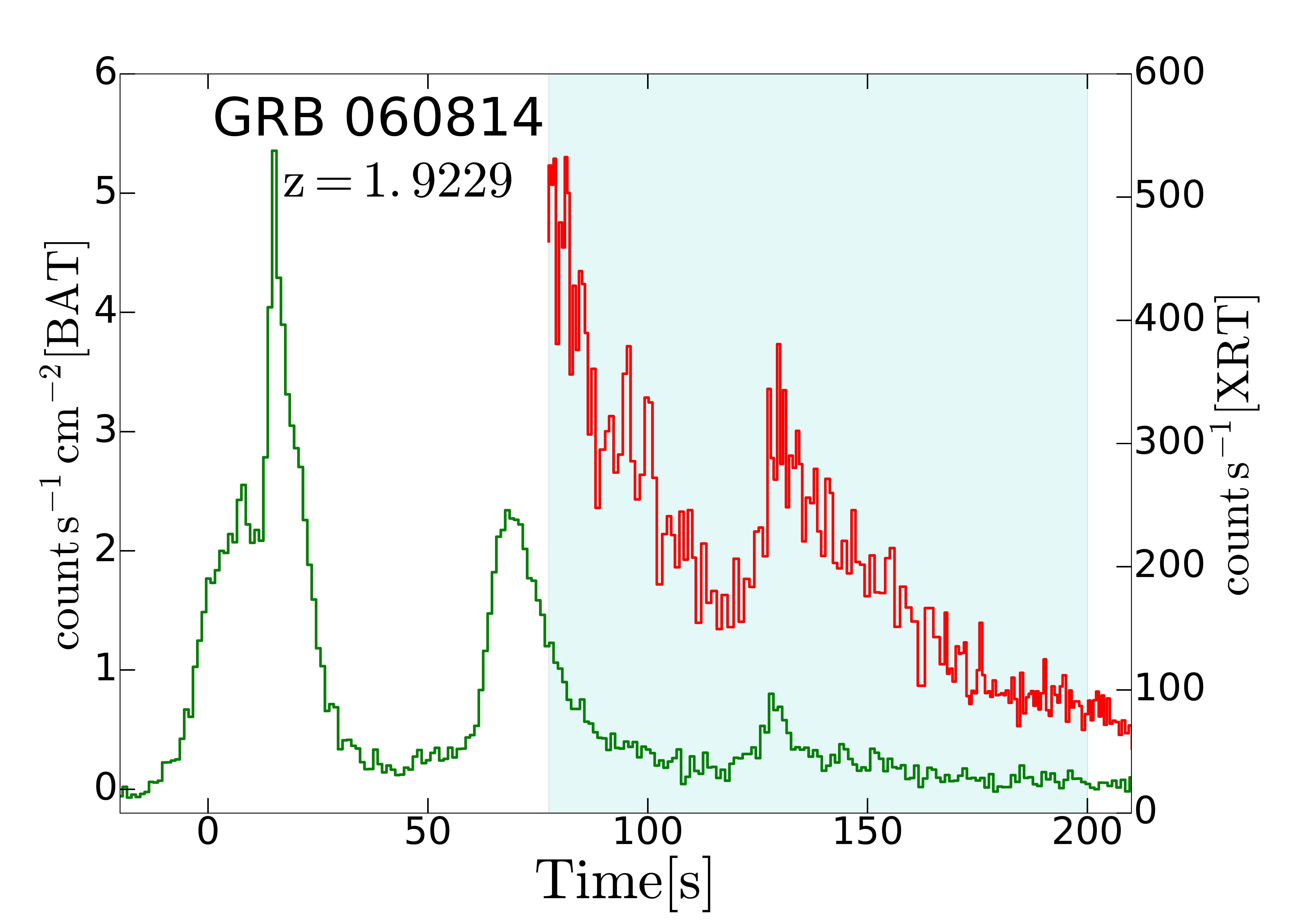} 
\includegraphics[width = 0.30\textwidth]{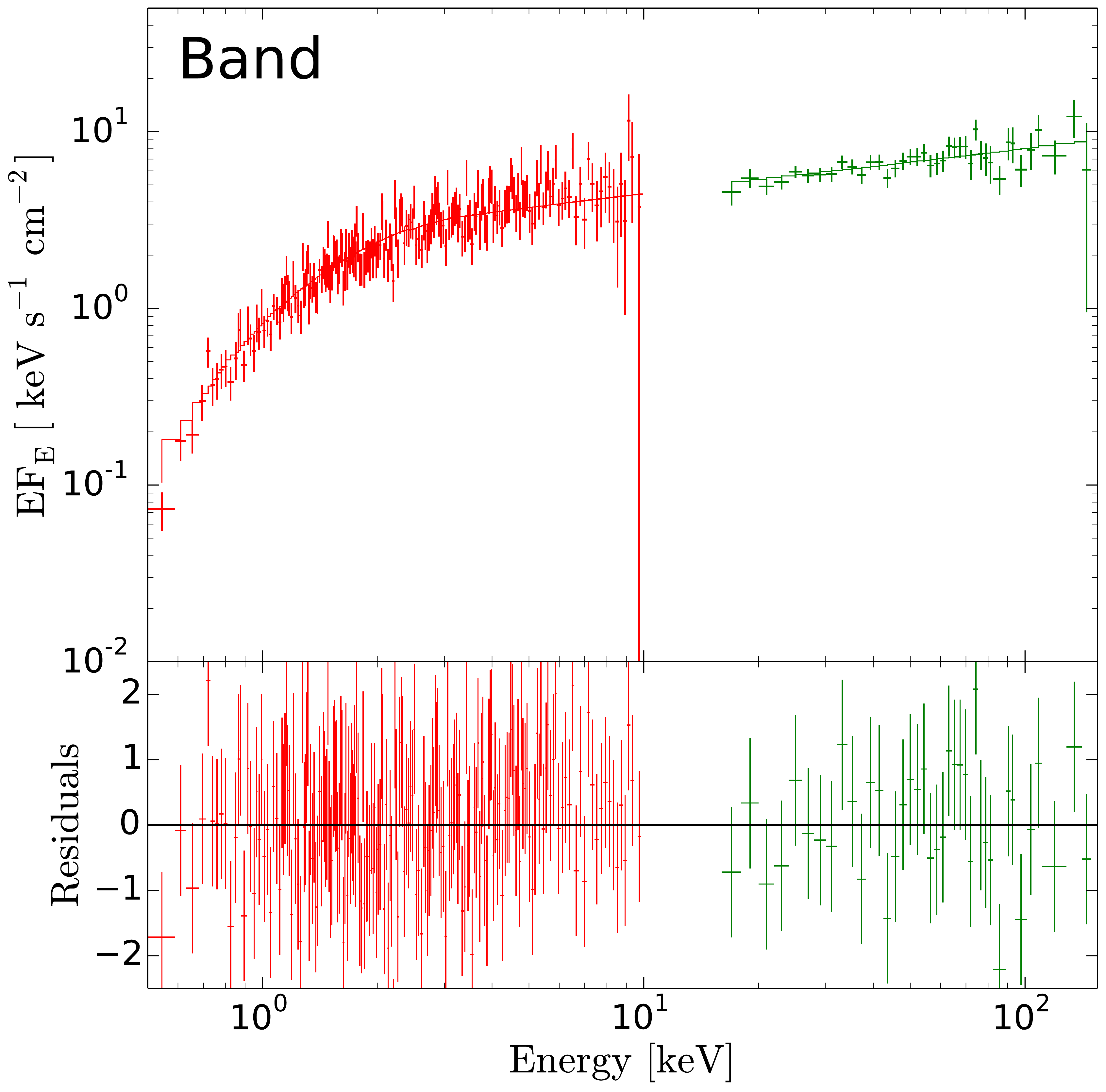}
\includegraphics[width = 0.30\textwidth]{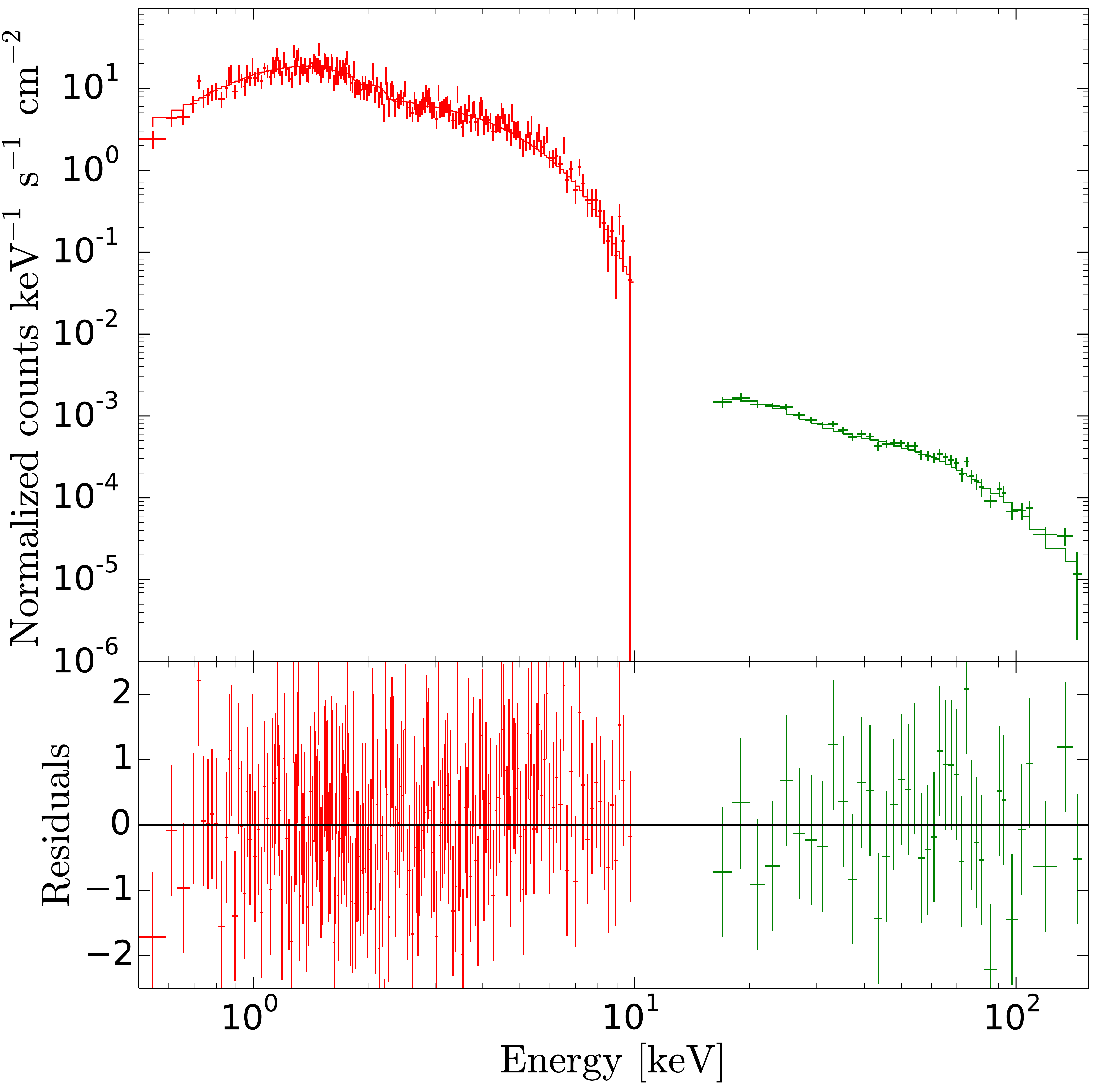}\\
\includegraphics[width = 0.40\textwidth]{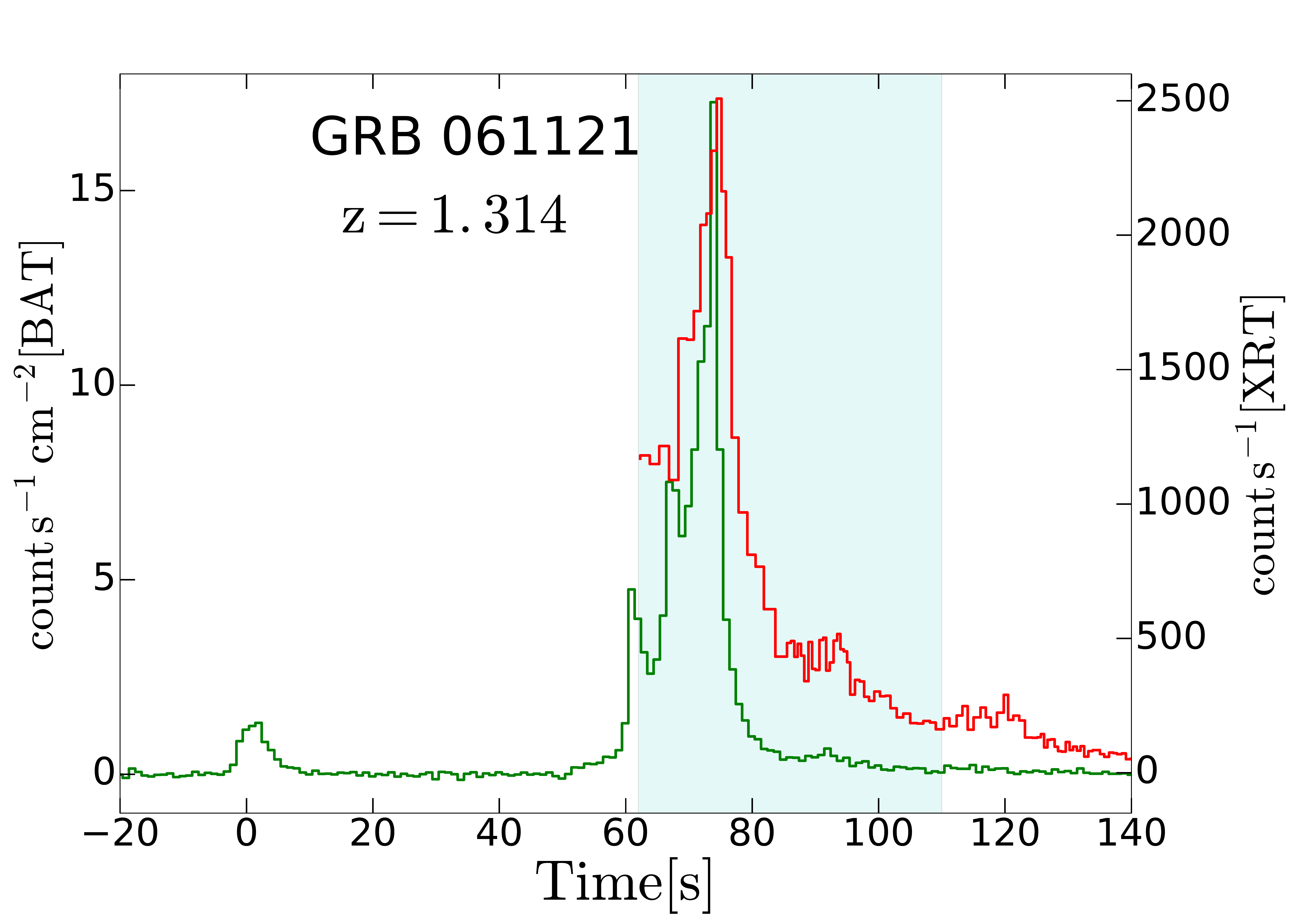}
\includegraphics[width = 0.30\textwidth]{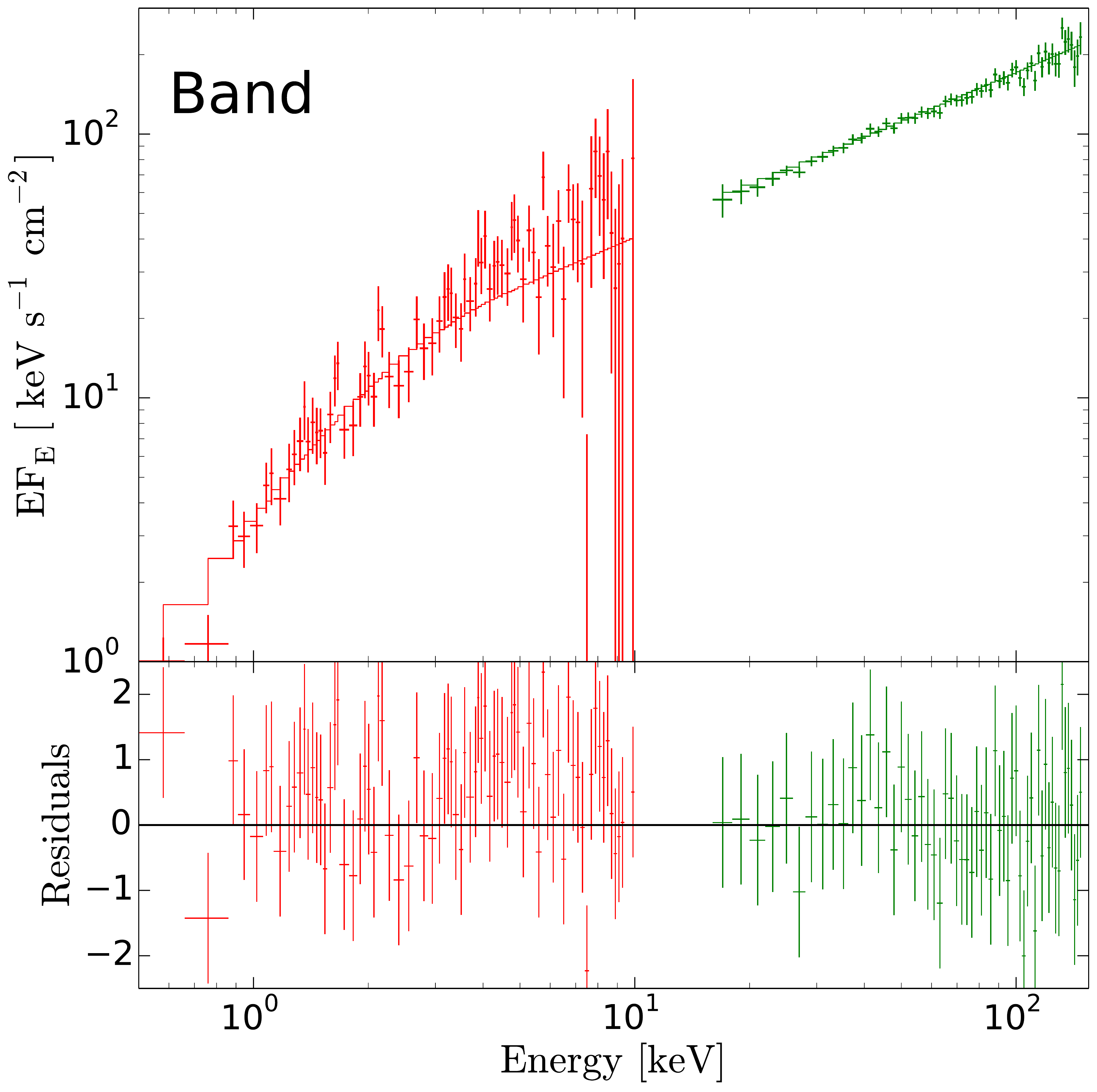}
\includegraphics[width = 0.30\textwidth]{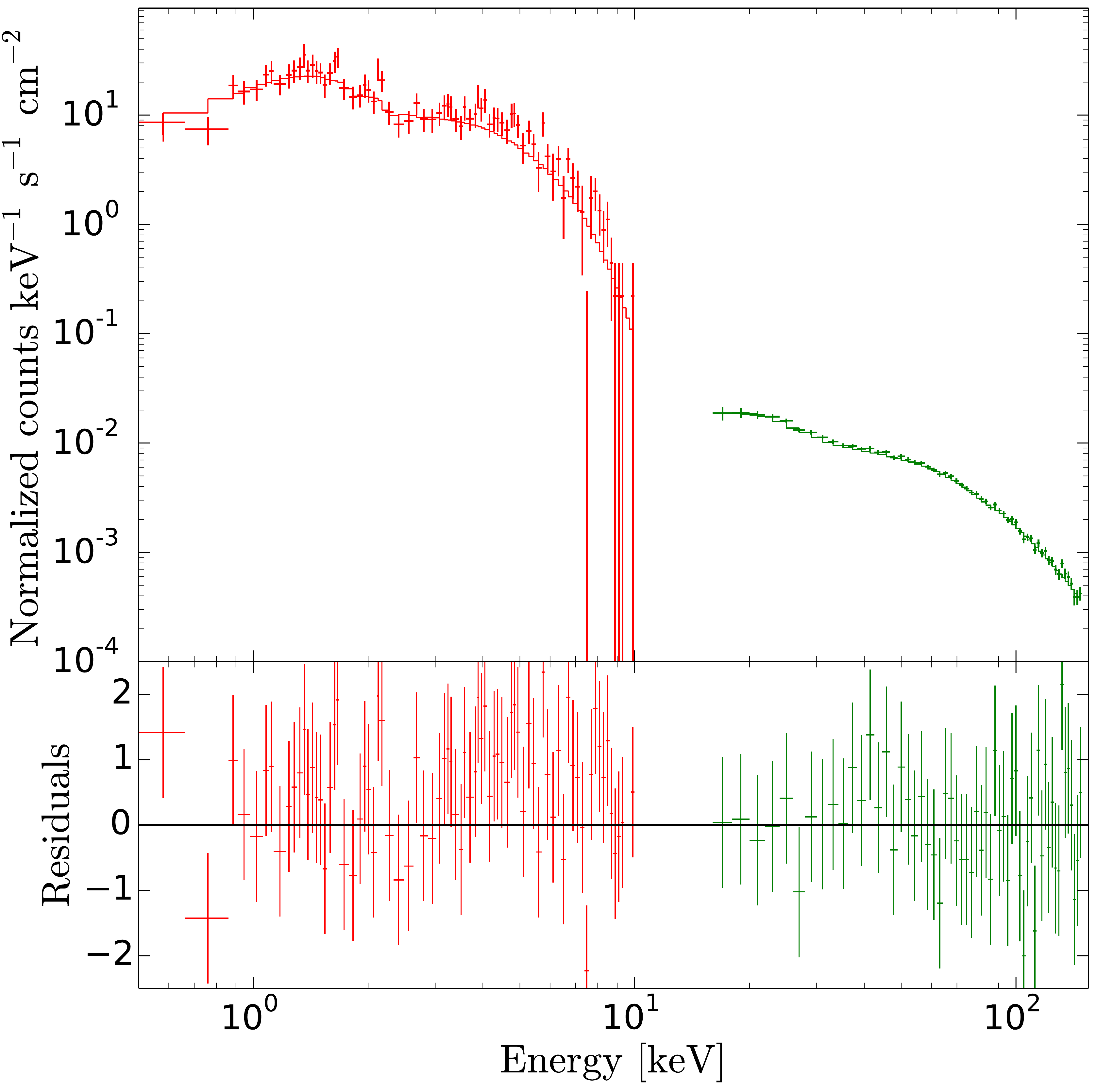}}
\end{figure}

\begin{figure}\ContinuedFloat
\includegraphics[width = 0.40\textwidth]{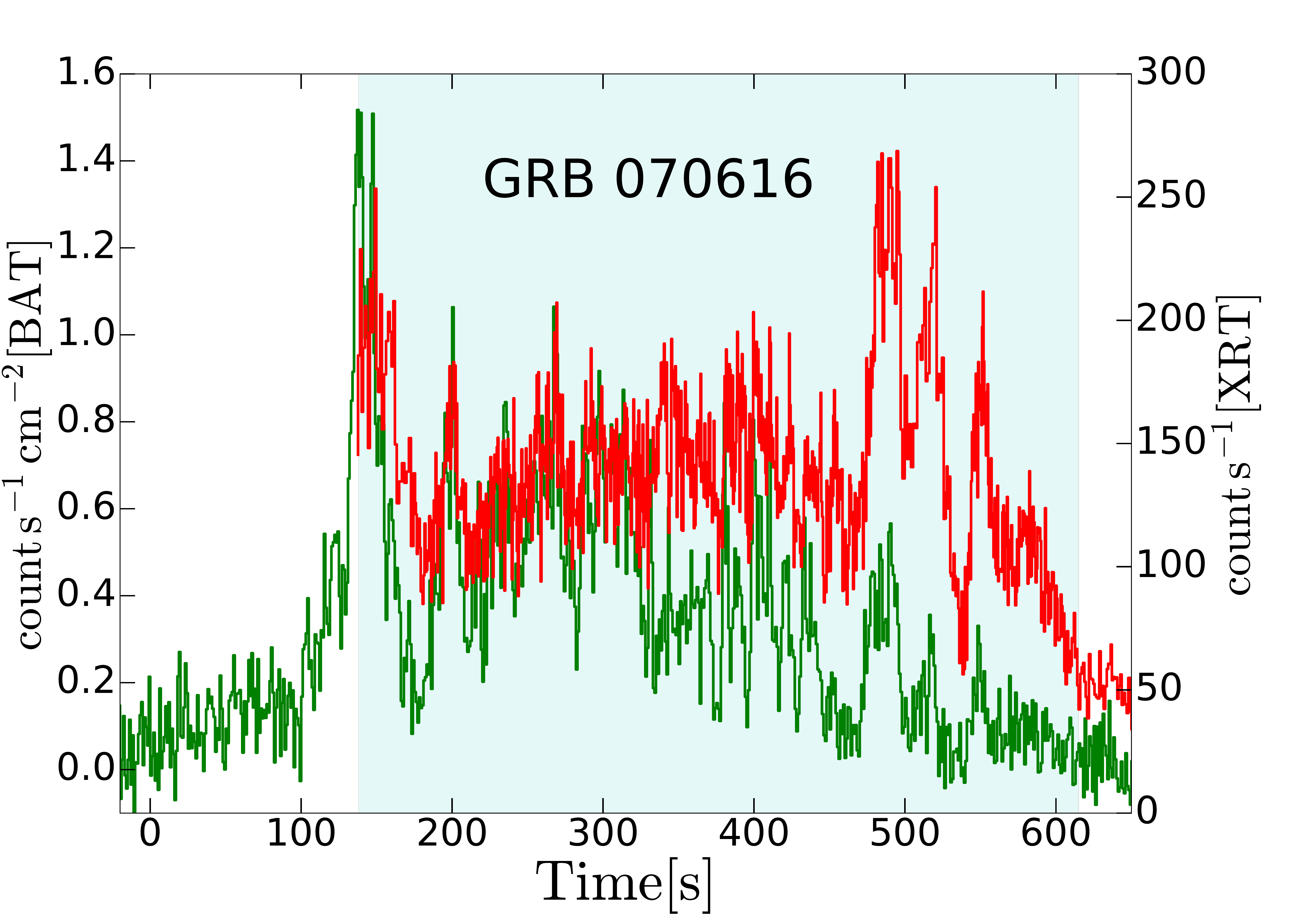} 
\includegraphics[width = 0.30\textwidth]{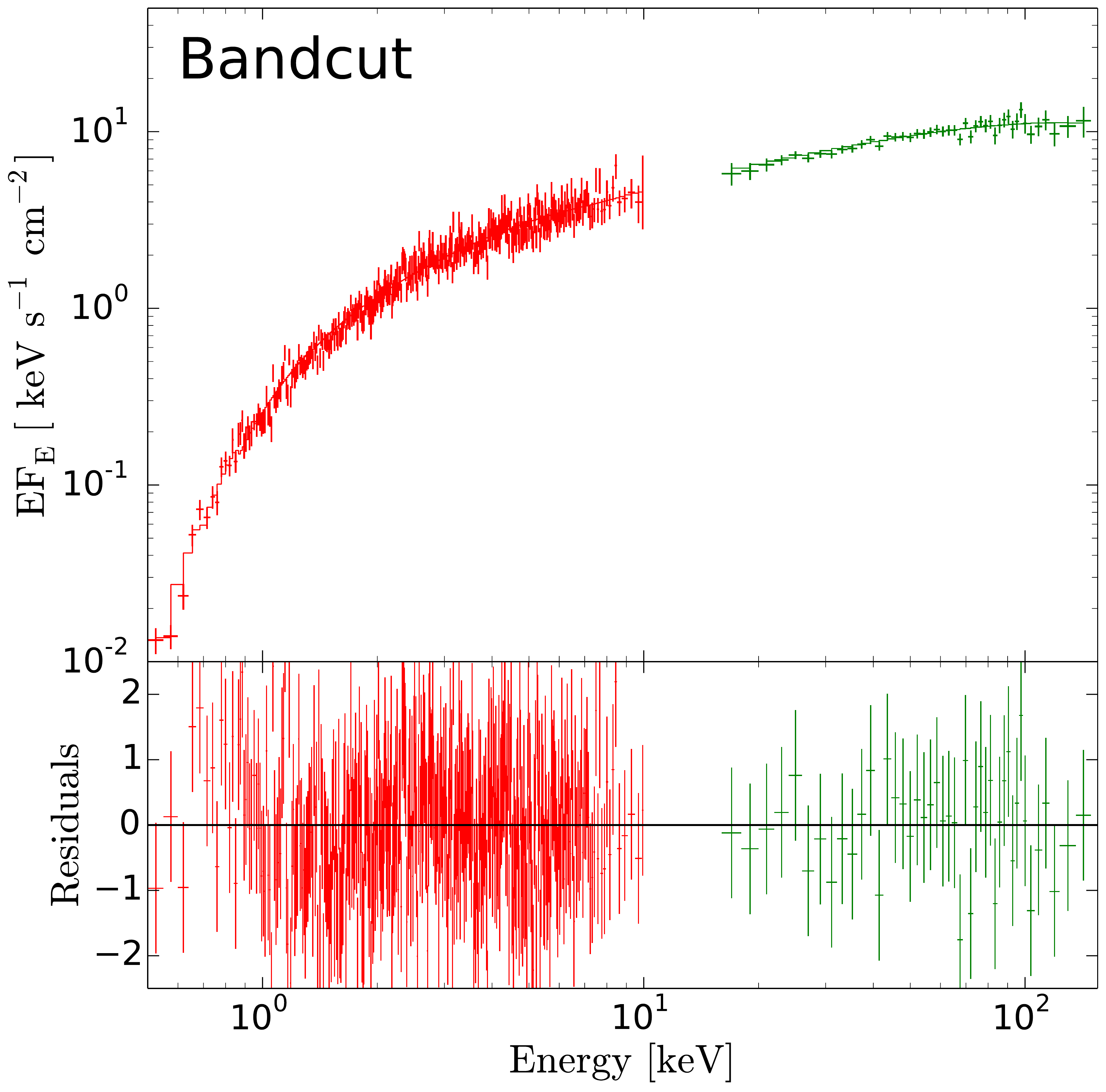}
\includegraphics[width = 0.30\textwidth]{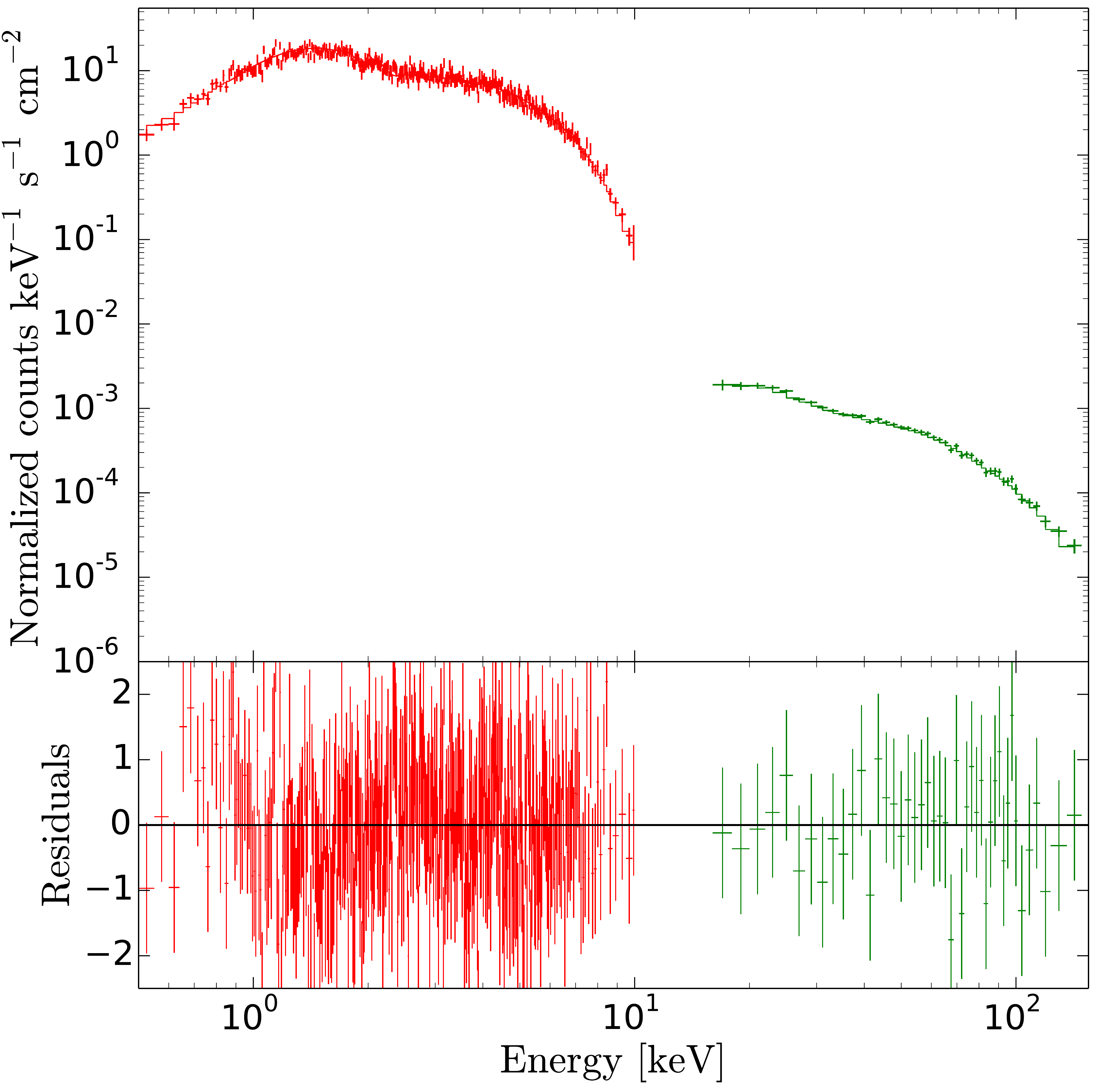} \\
\includegraphics[width = 0.40\textwidth]{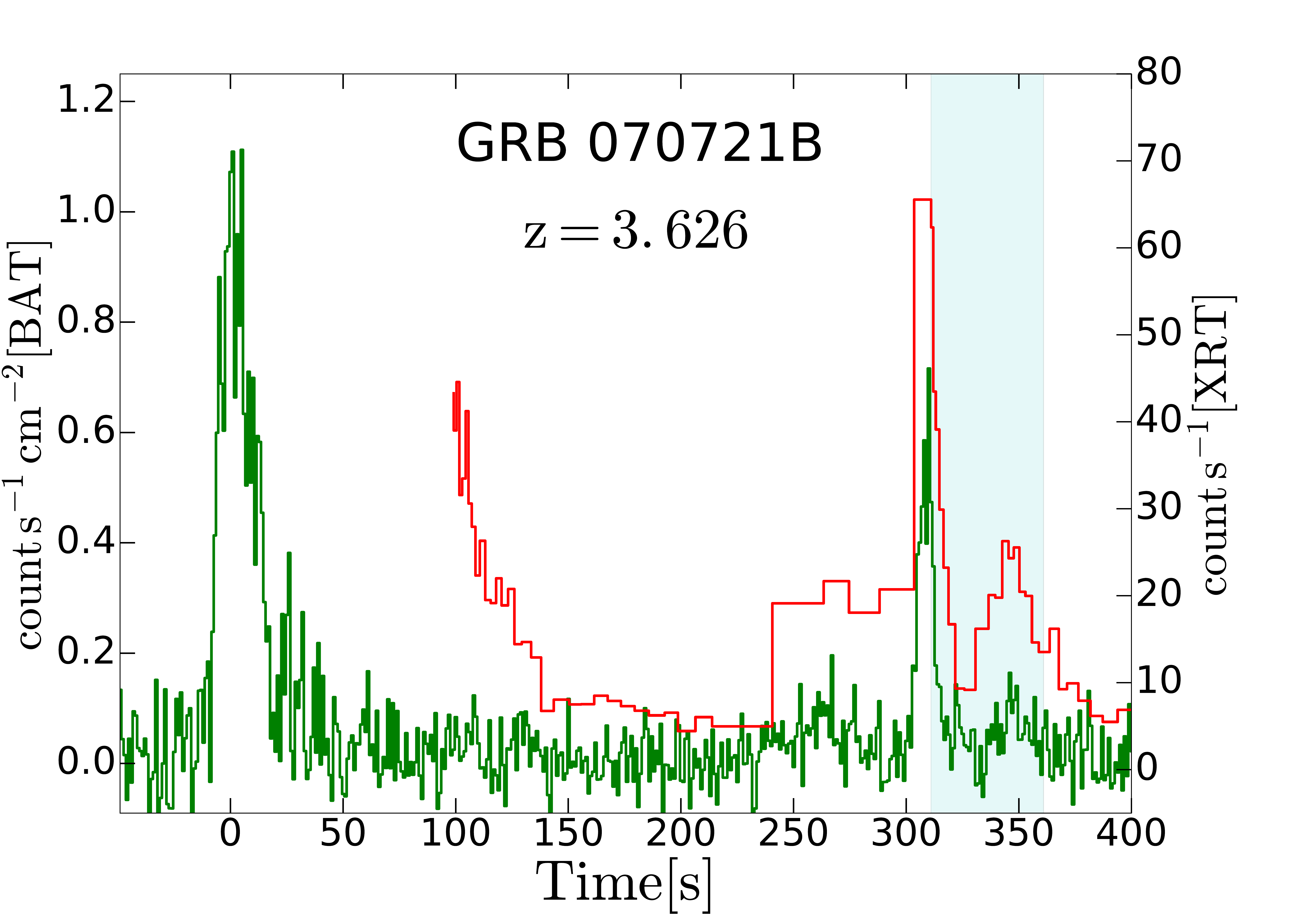} 
\includegraphics[width = 0.30\textwidth]{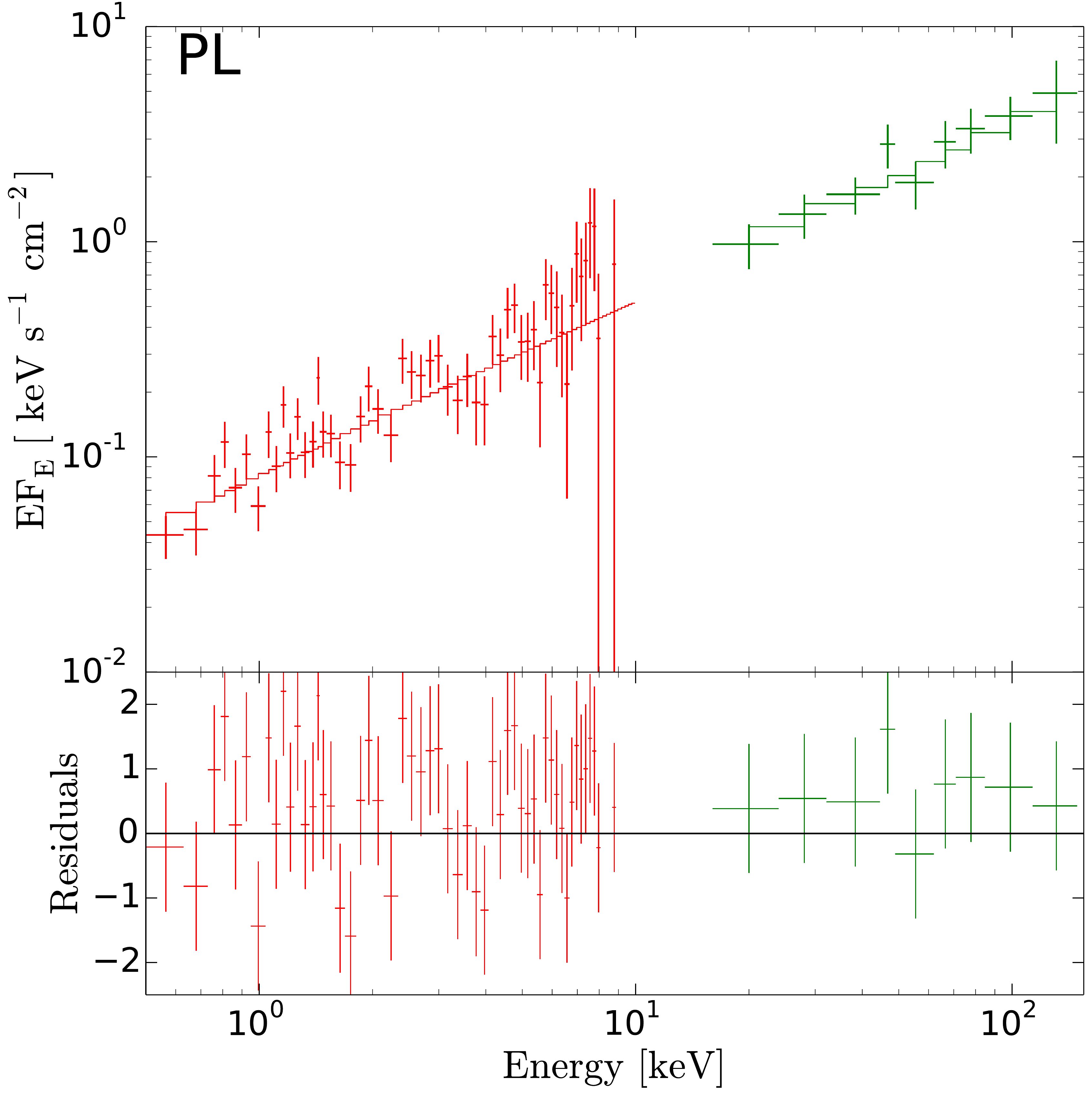} 
\includegraphics[width = 0.30\textwidth]{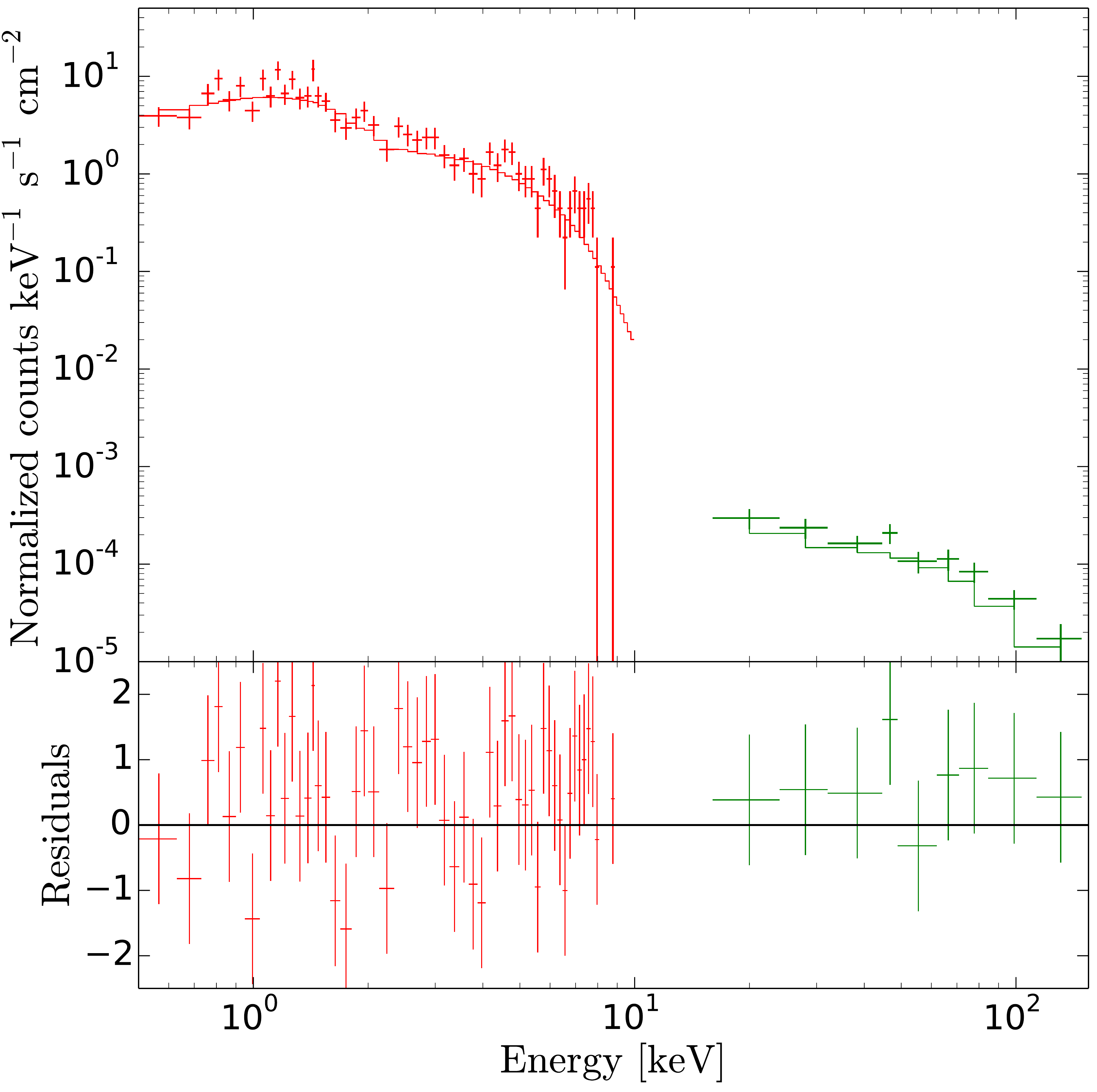} \\
\includegraphics[width = 0.40\textwidth]{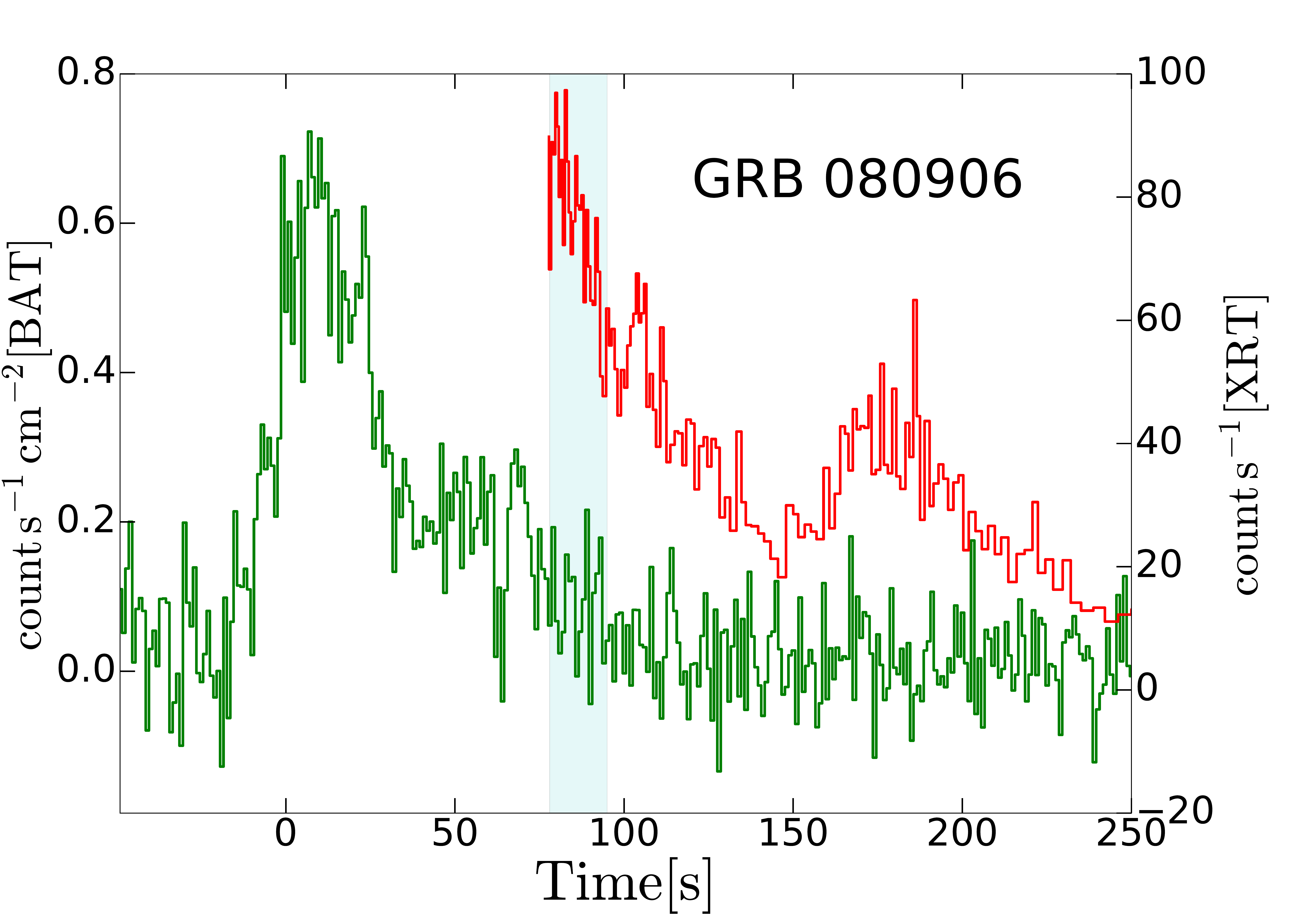}  
\includegraphics[width = 0.30\textwidth]{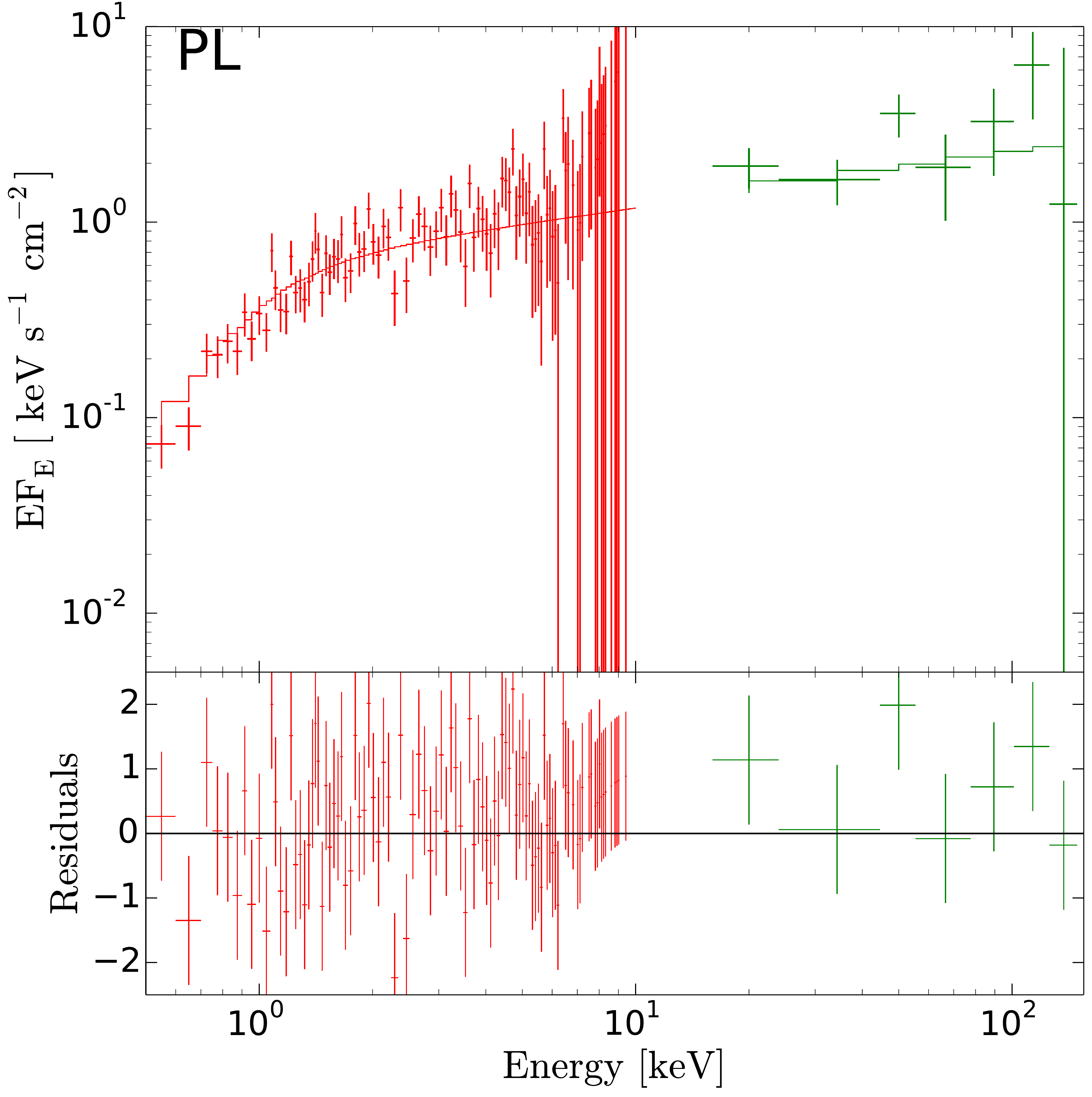} 
\includegraphics[width = 0.30\textwidth]{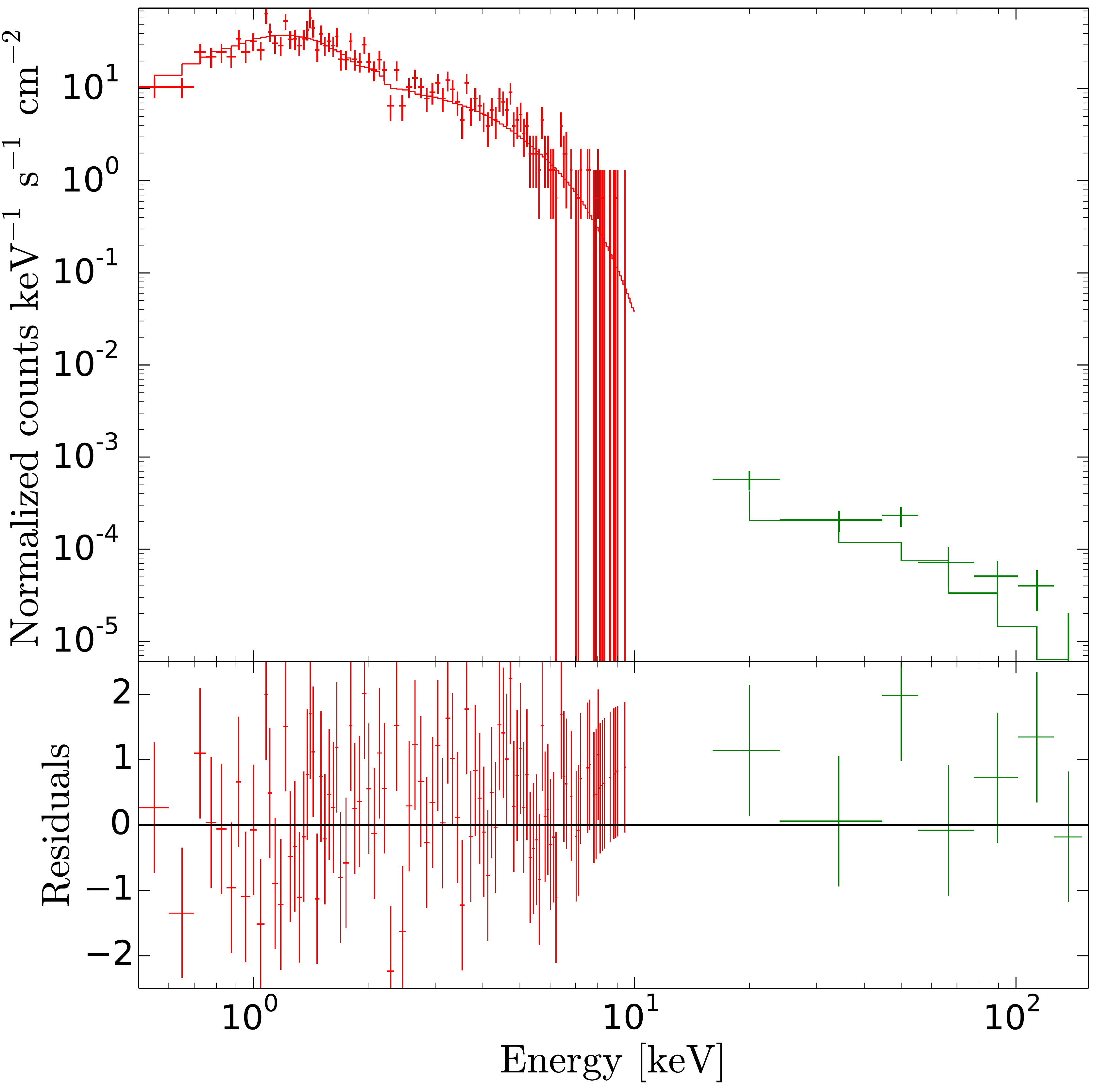} 

\end{figure}

\begin{figure}\ContinuedFloat
\includegraphics[width = 0.40\textwidth]{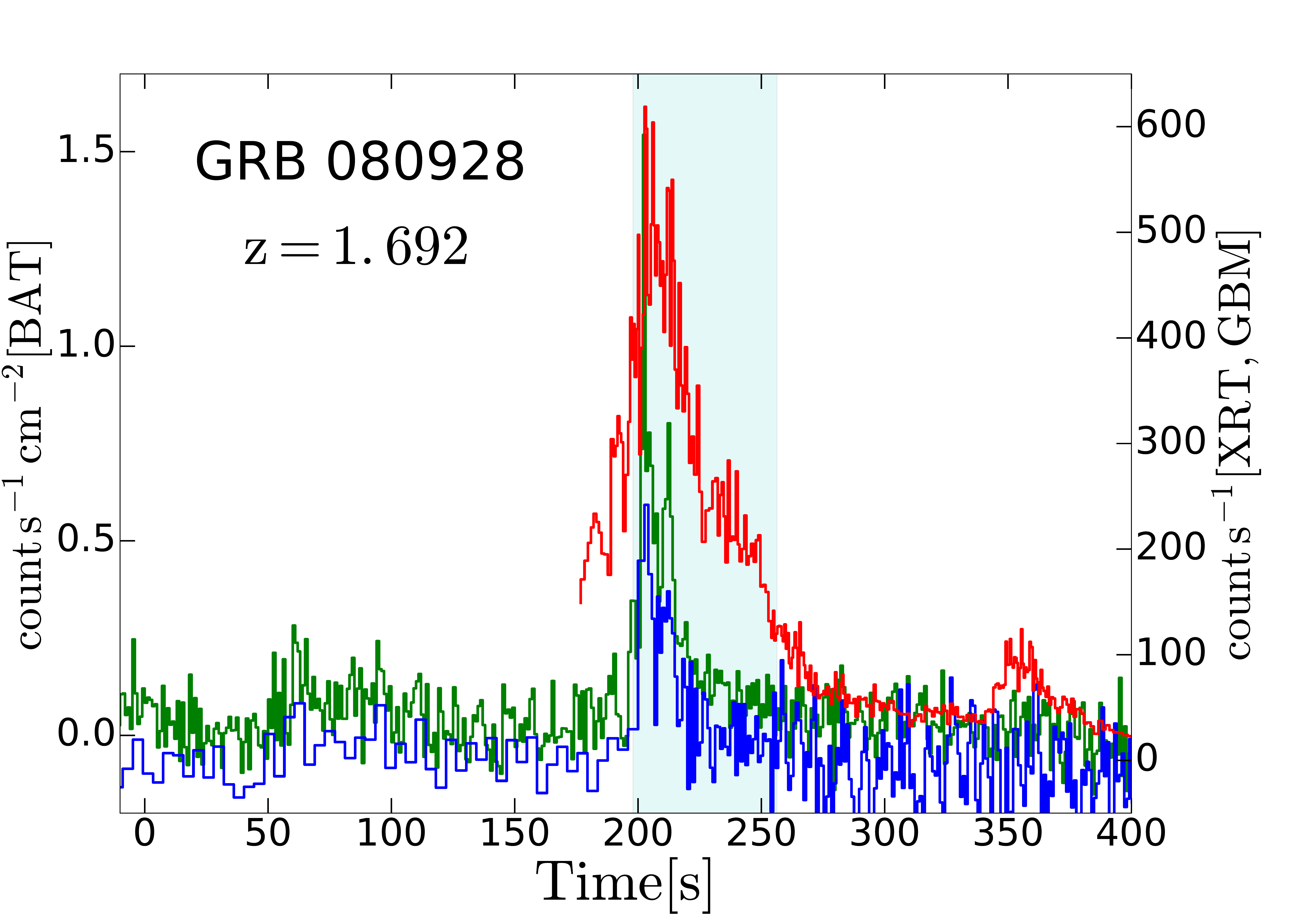} 
\includegraphics[width = 0.30\textwidth]{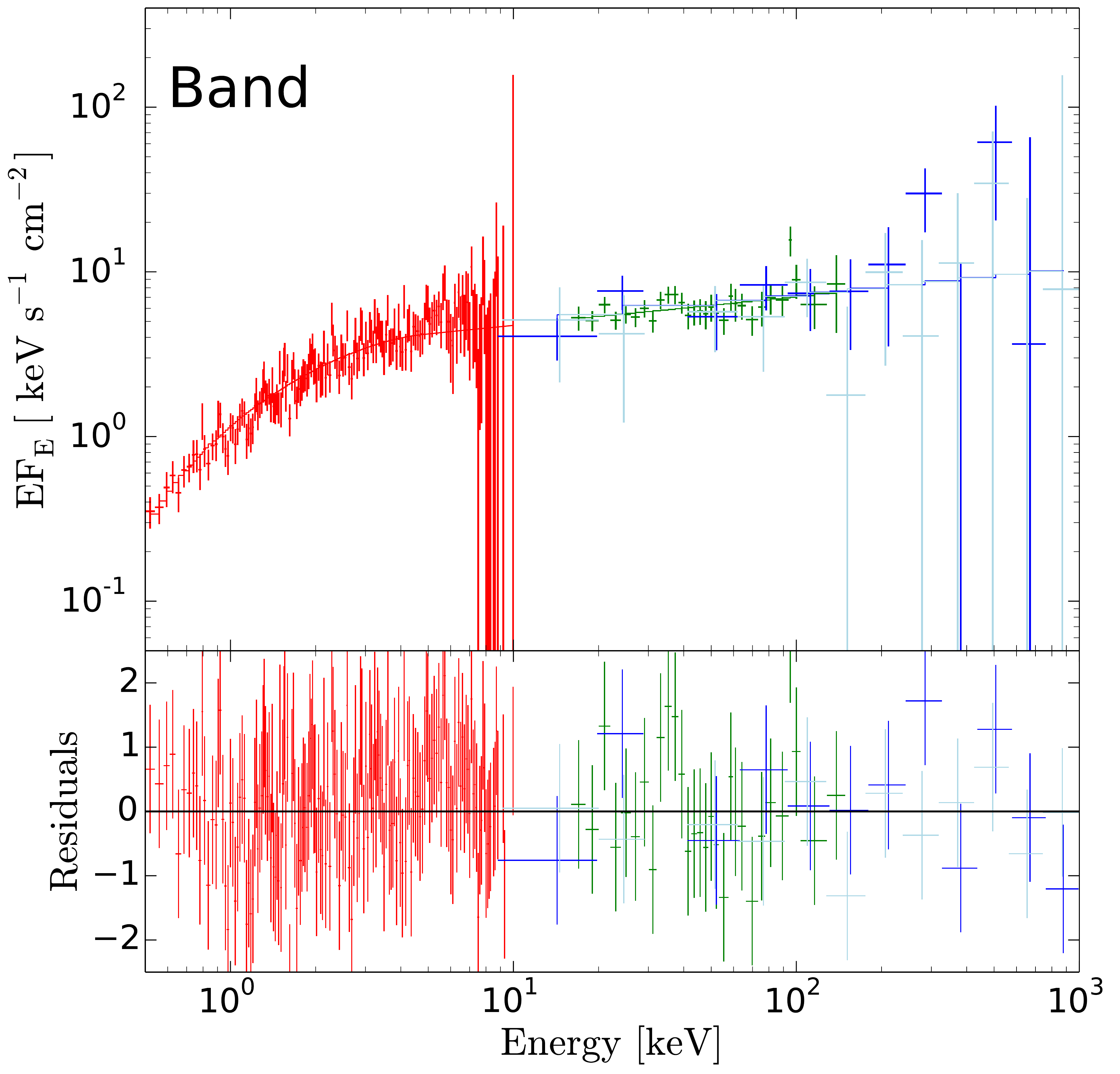} 
\includegraphics[width = 0.30\textwidth]{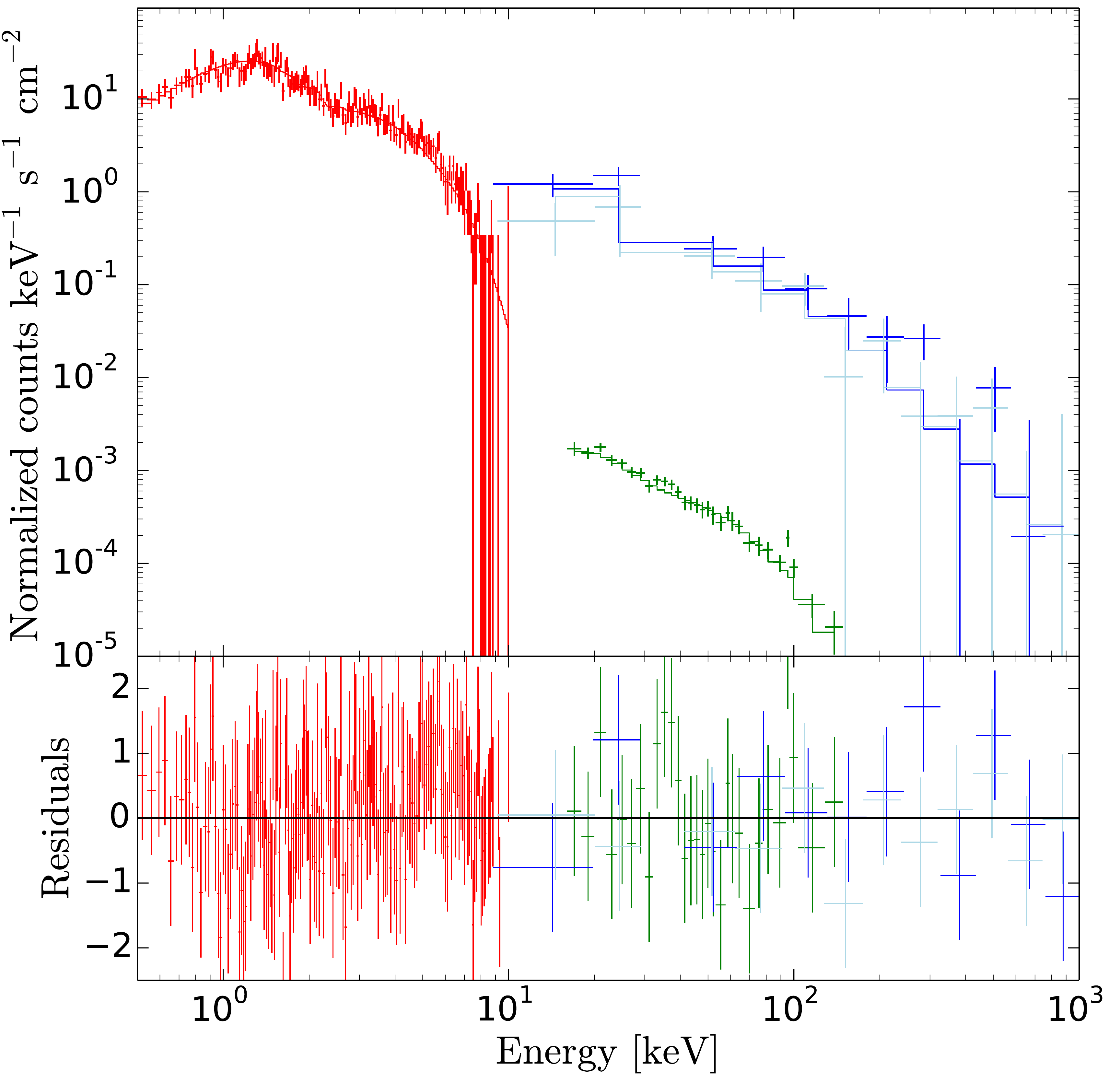} \\
\includegraphics[width = 0.40\textwidth]{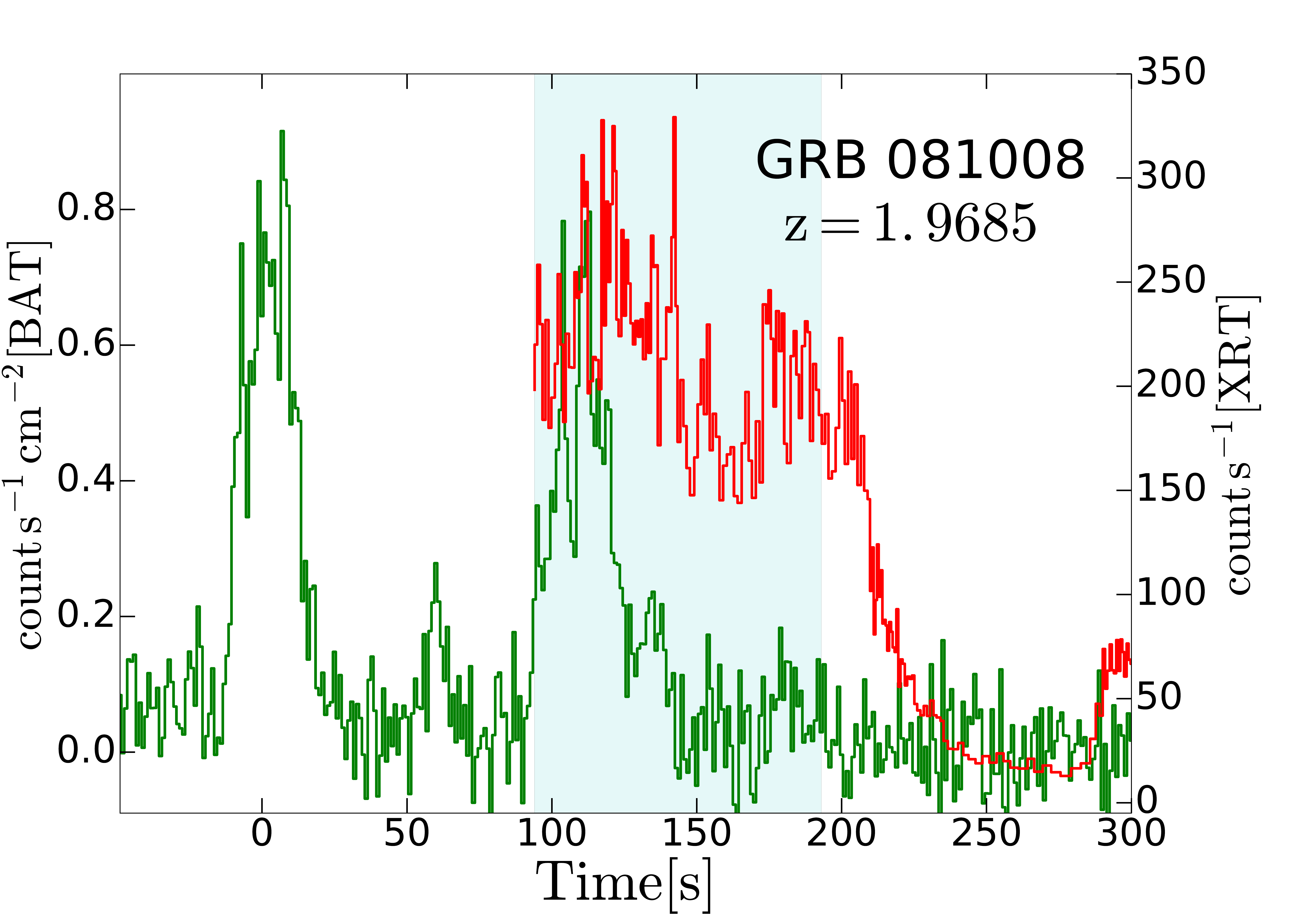} 
\includegraphics[width = 0.30\textwidth]{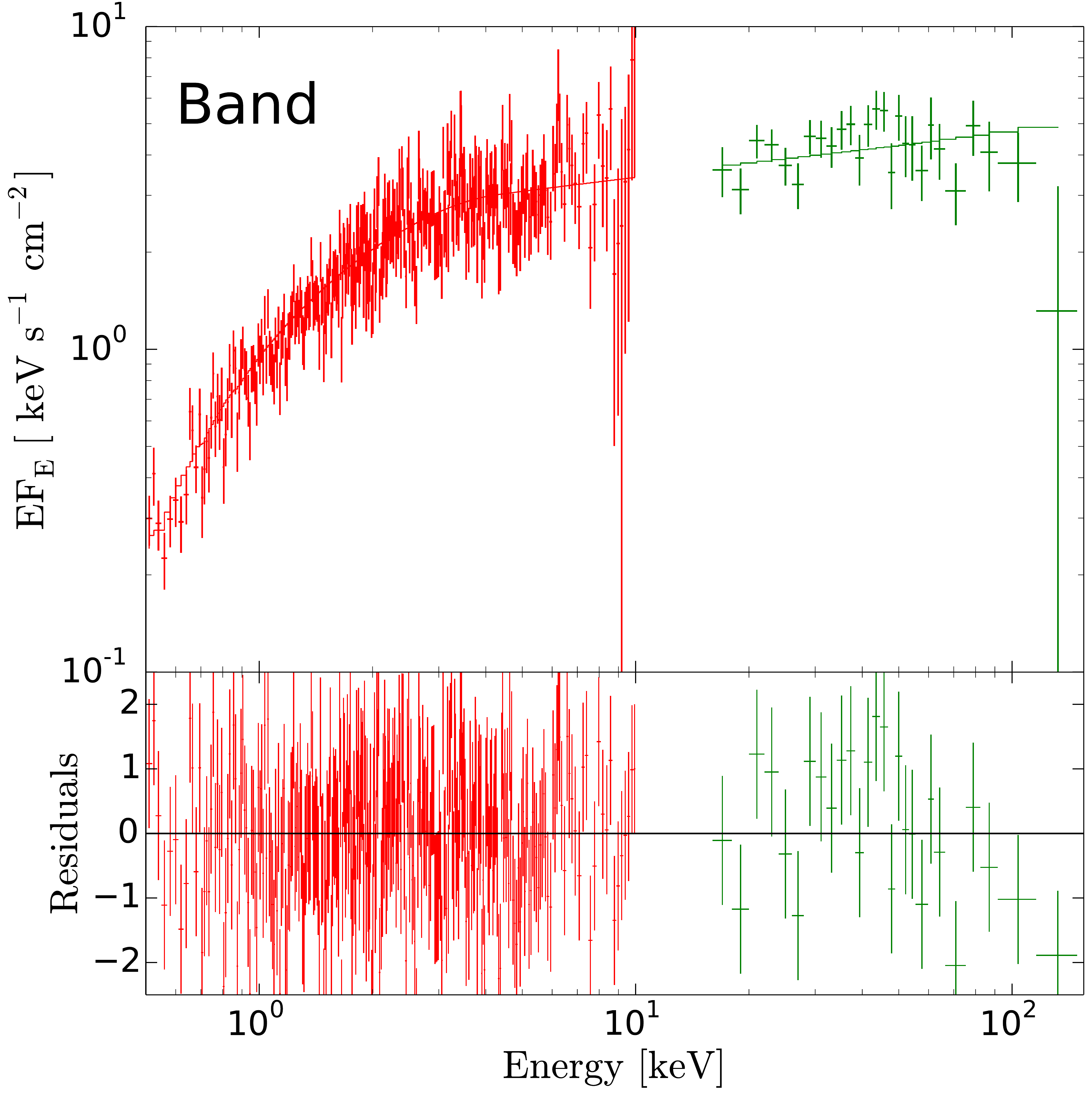} 
\includegraphics[width = 0.30\textwidth]{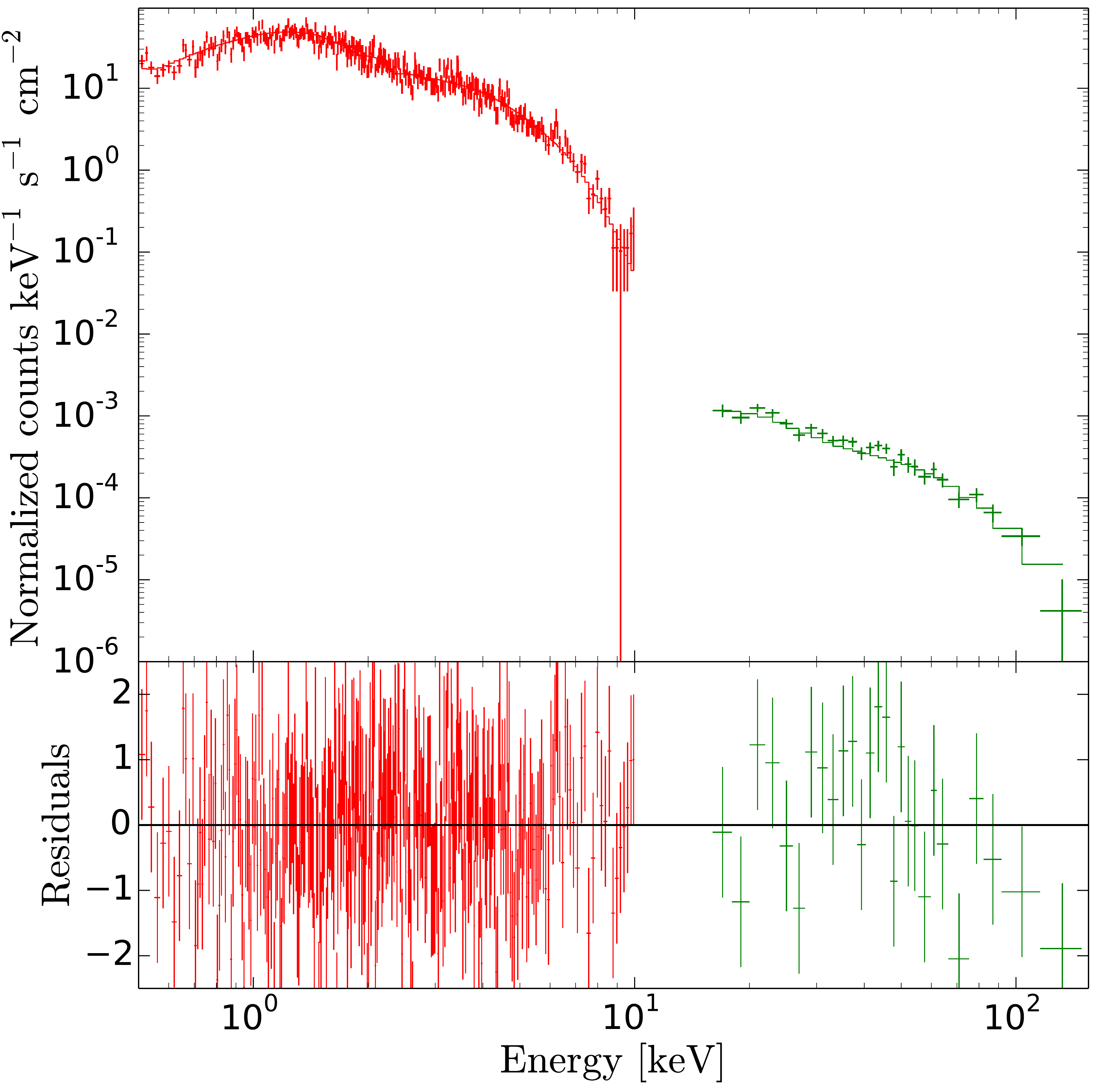} \\
\includegraphics[width = 0.40\textwidth]{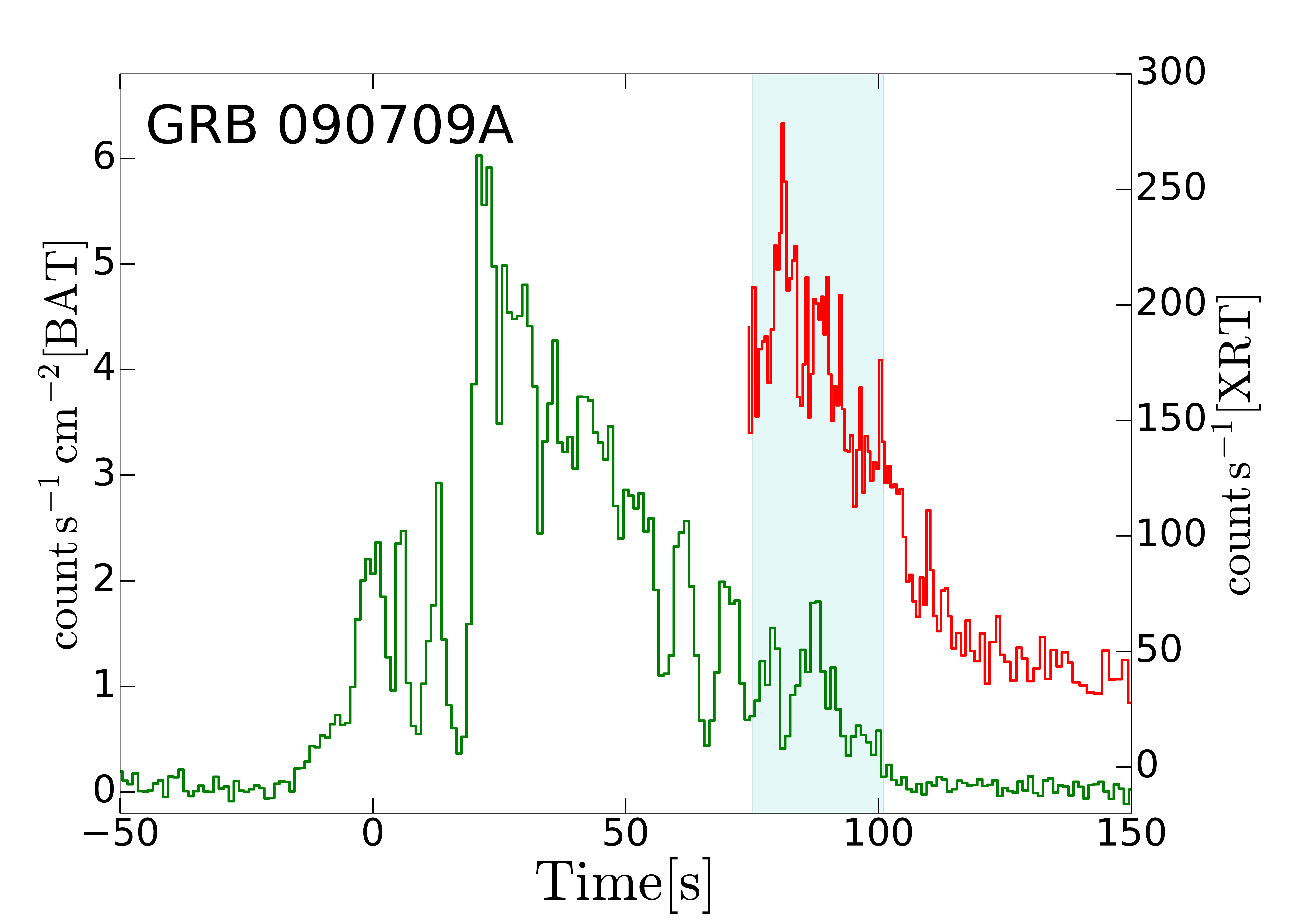} 
\includegraphics[width = 0.30\textwidth]{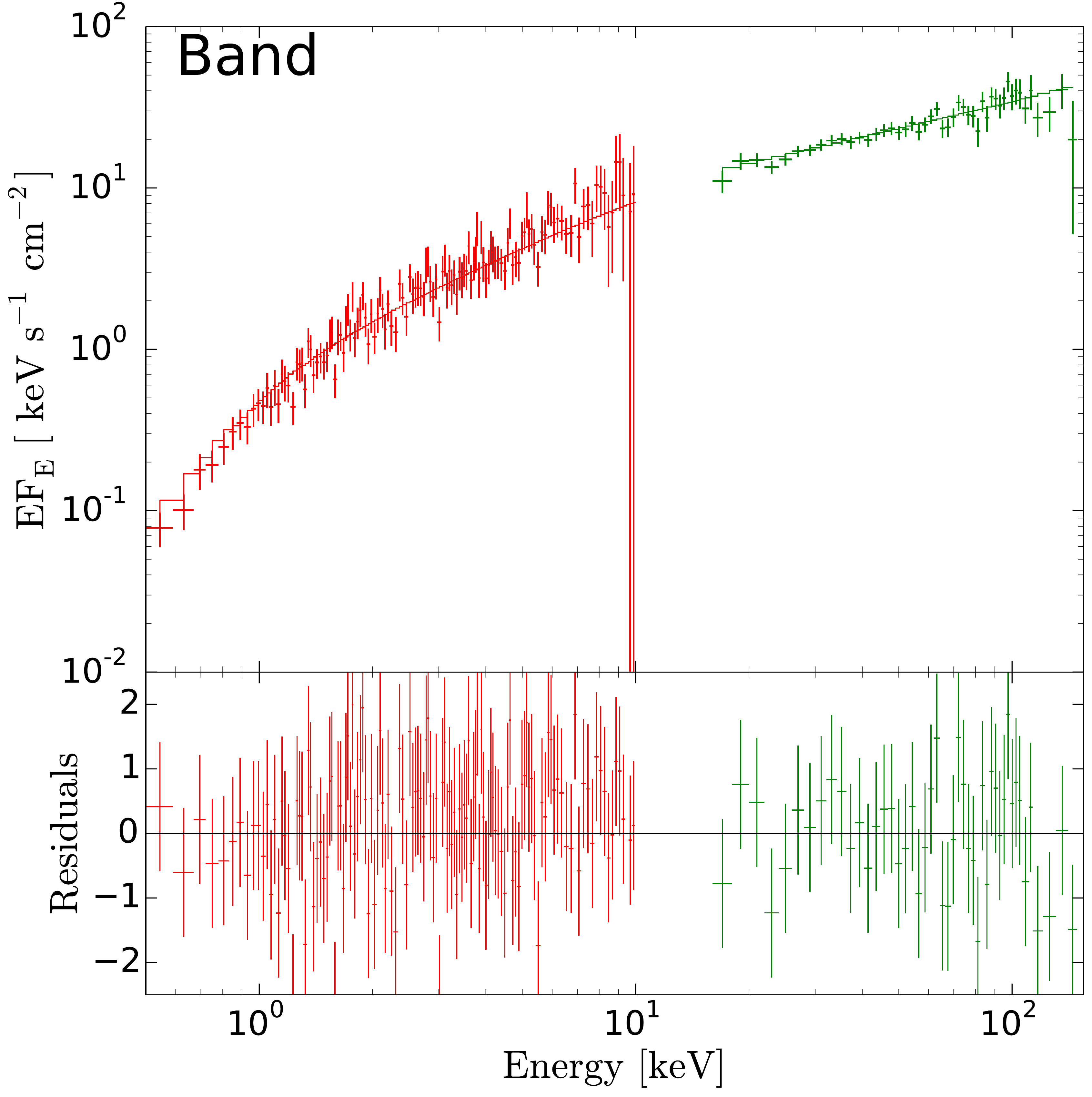} 
\includegraphics[width = 0.30\textwidth]{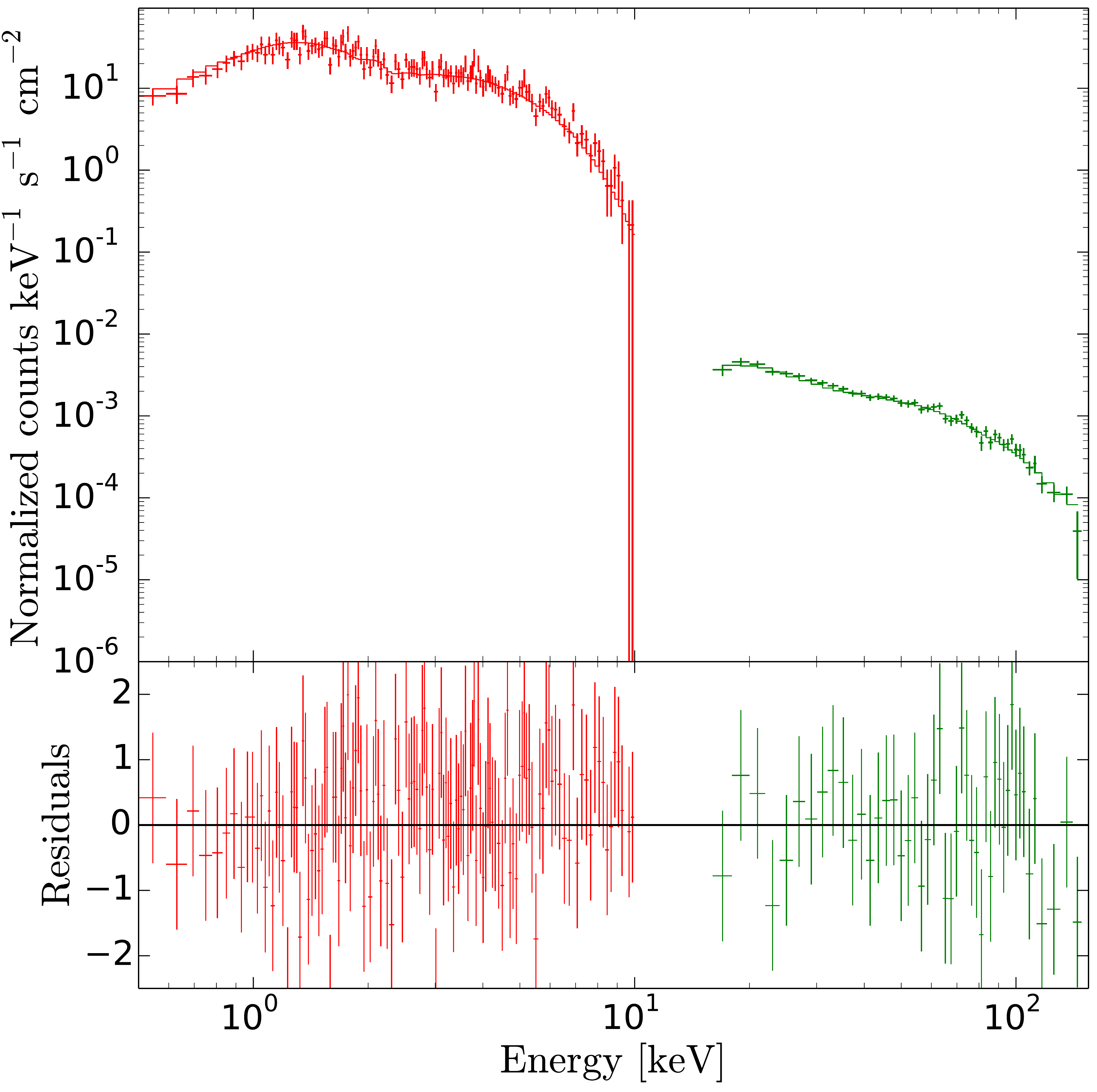} 
\end{figure}

\begin{figure}\ContinuedFloat
\includegraphics[width = 0.40\textwidth]{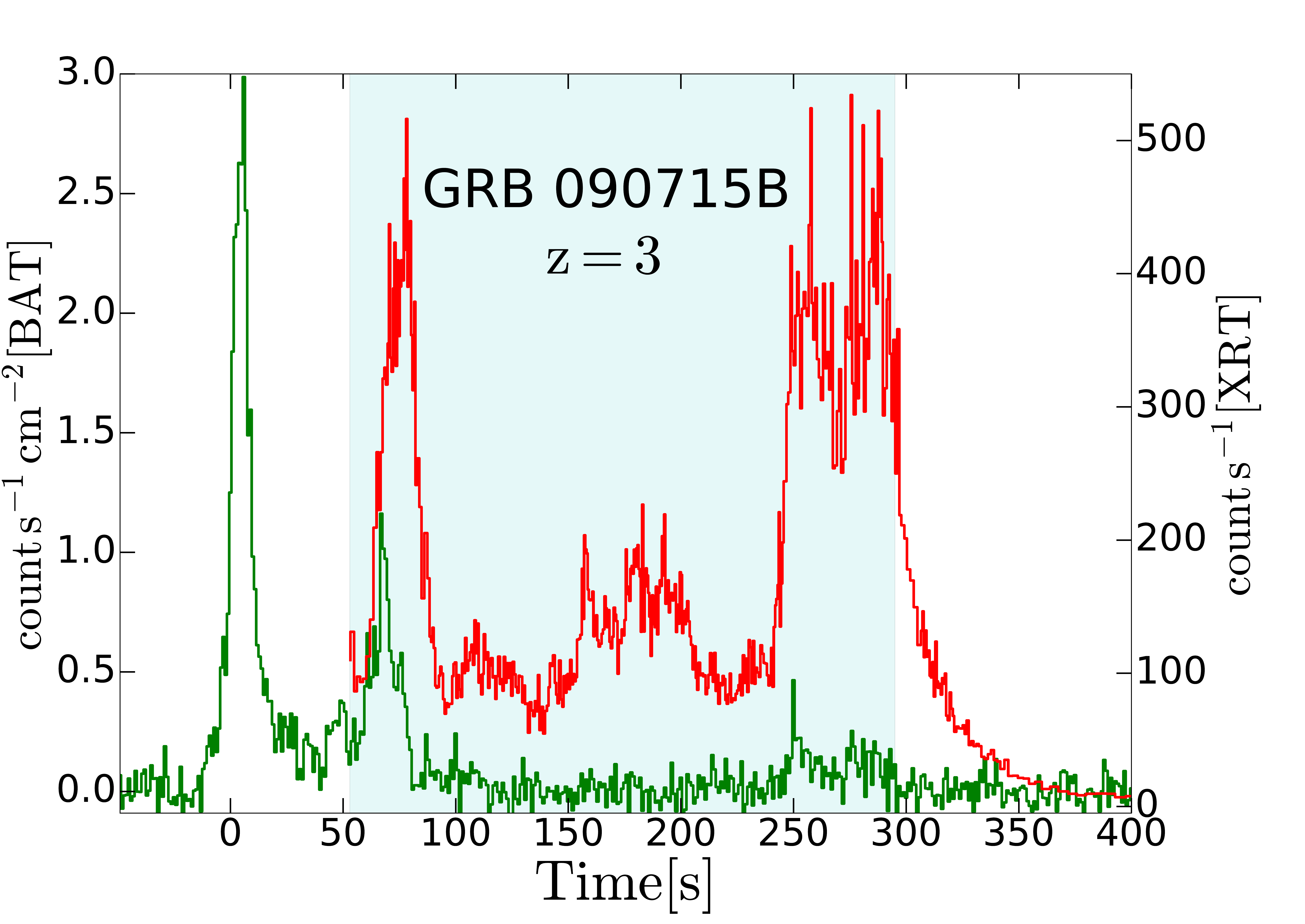} 
\includegraphics[width = 0.30\textwidth]{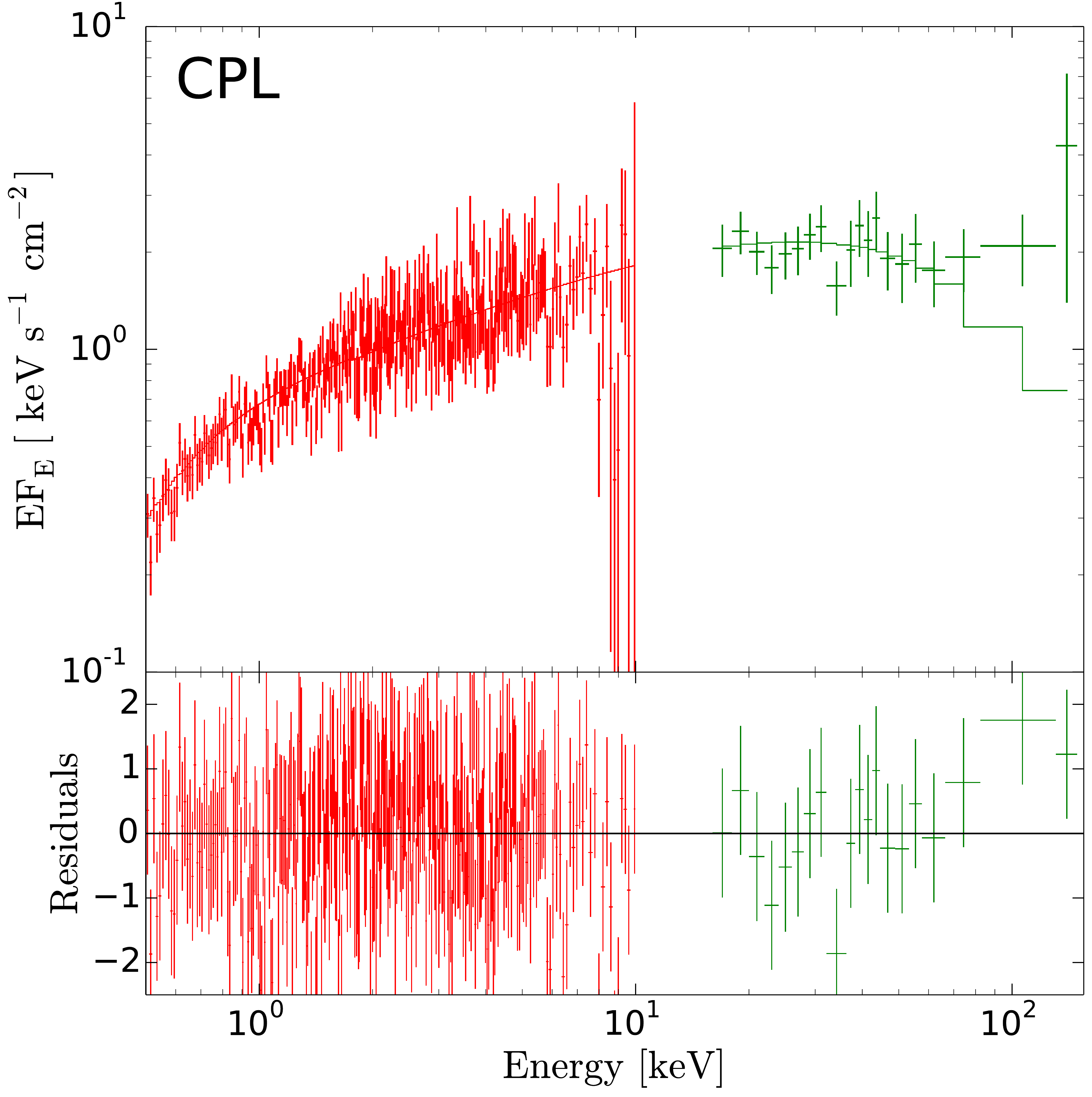} 
\includegraphics[width = 0.30\textwidth]{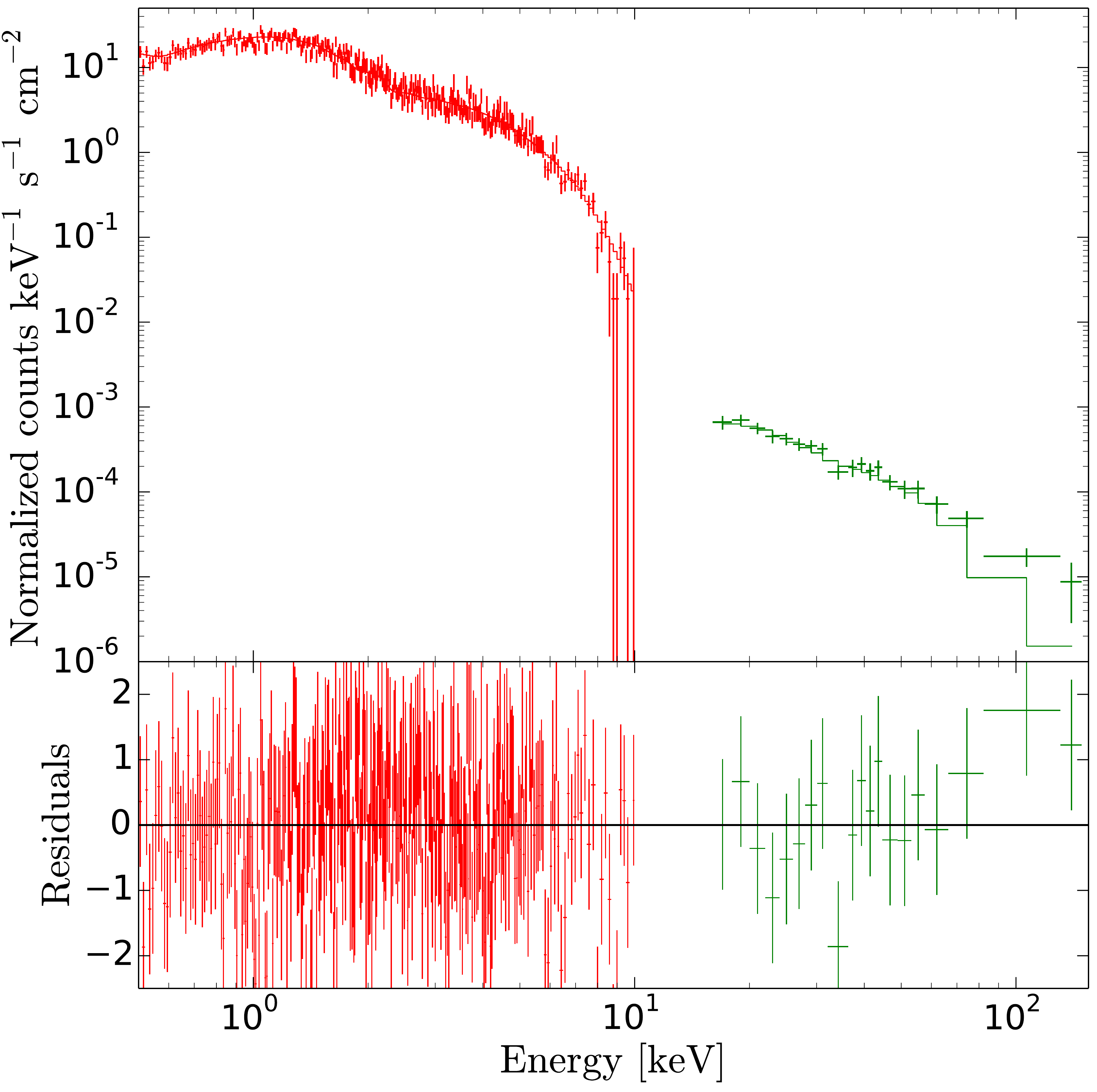} \\
\includegraphics[width = 0.40\textwidth]{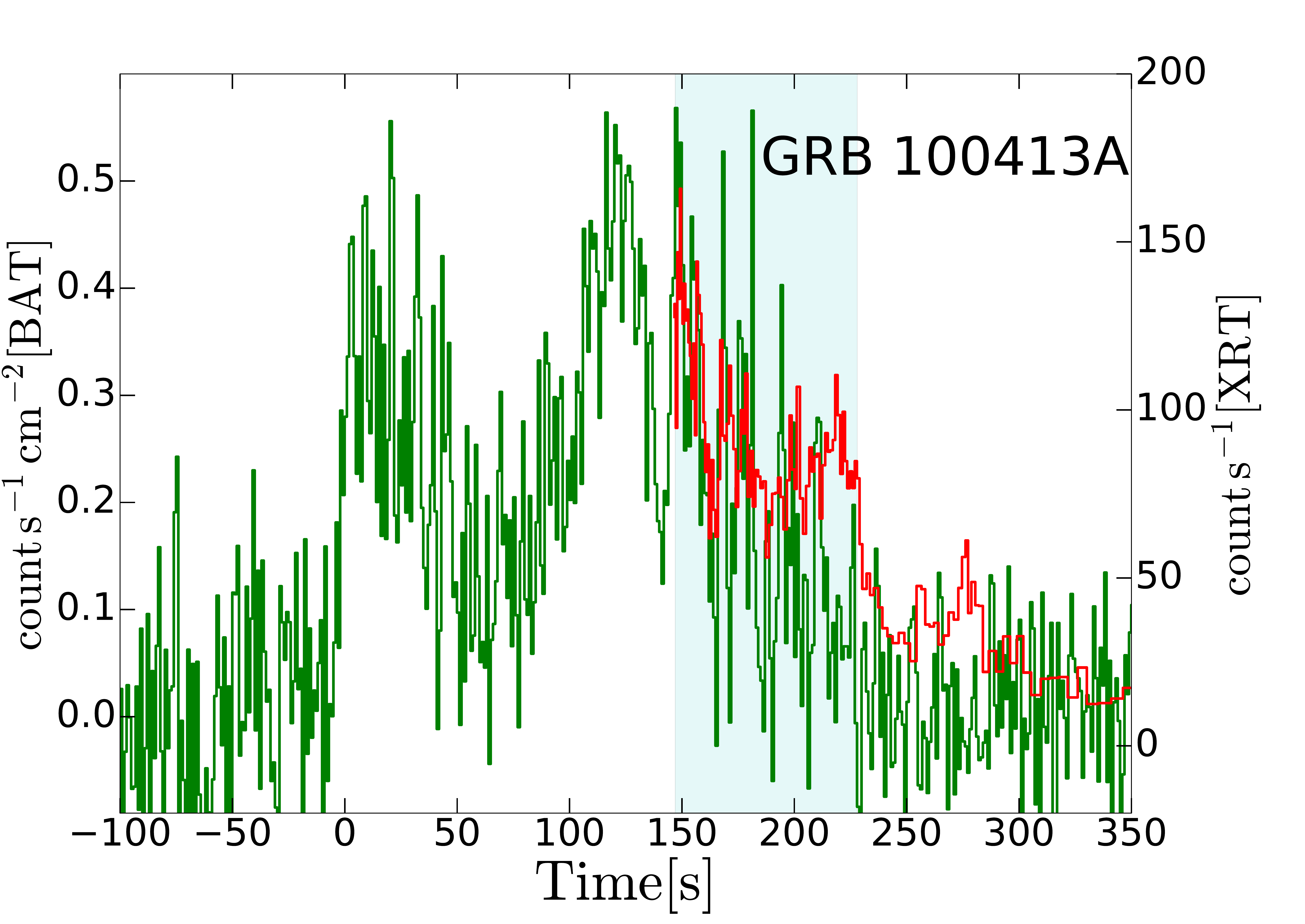} 
\includegraphics[width = 0.30\textwidth]{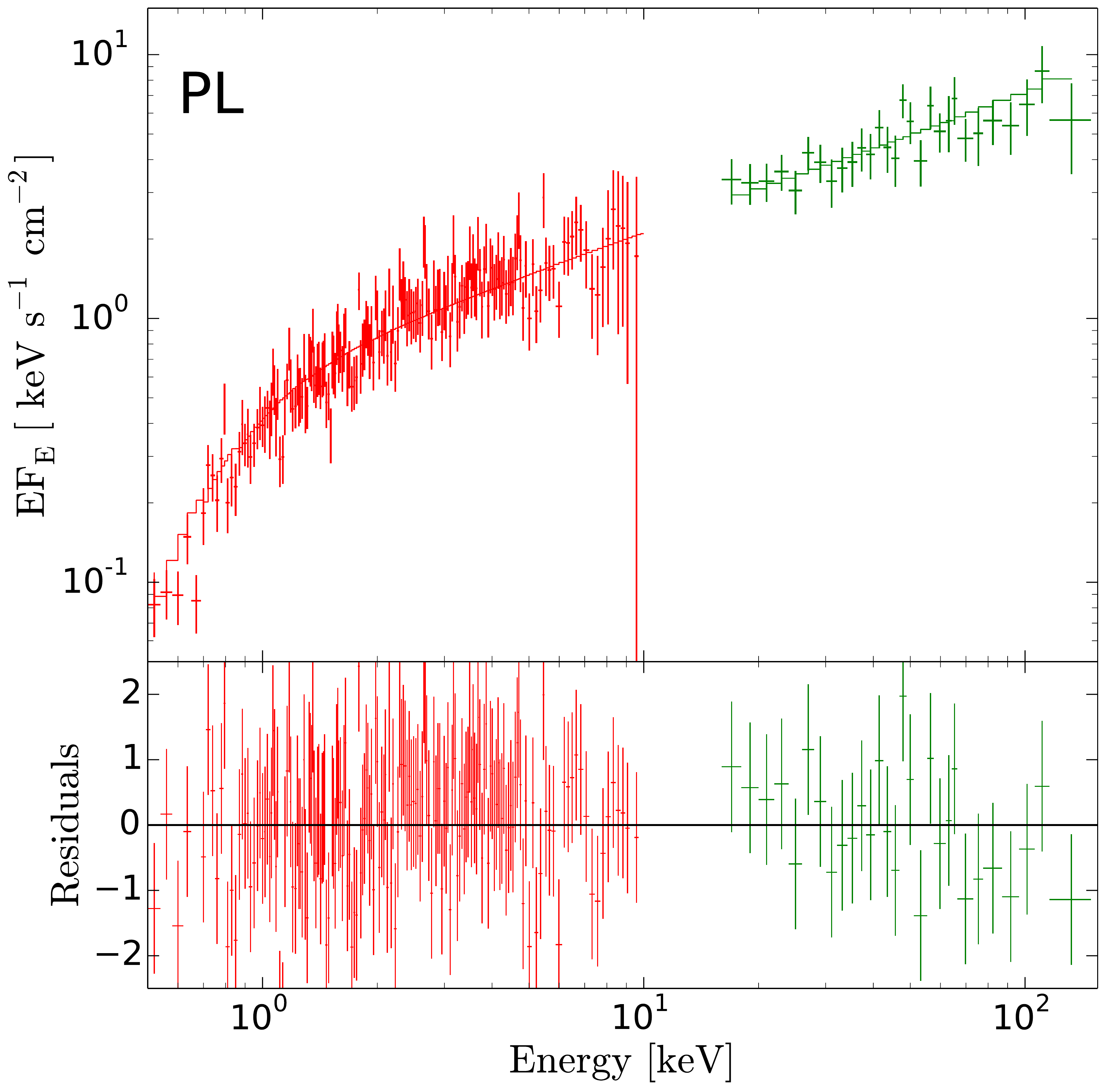} 
\includegraphics[width = 0.30\textwidth]{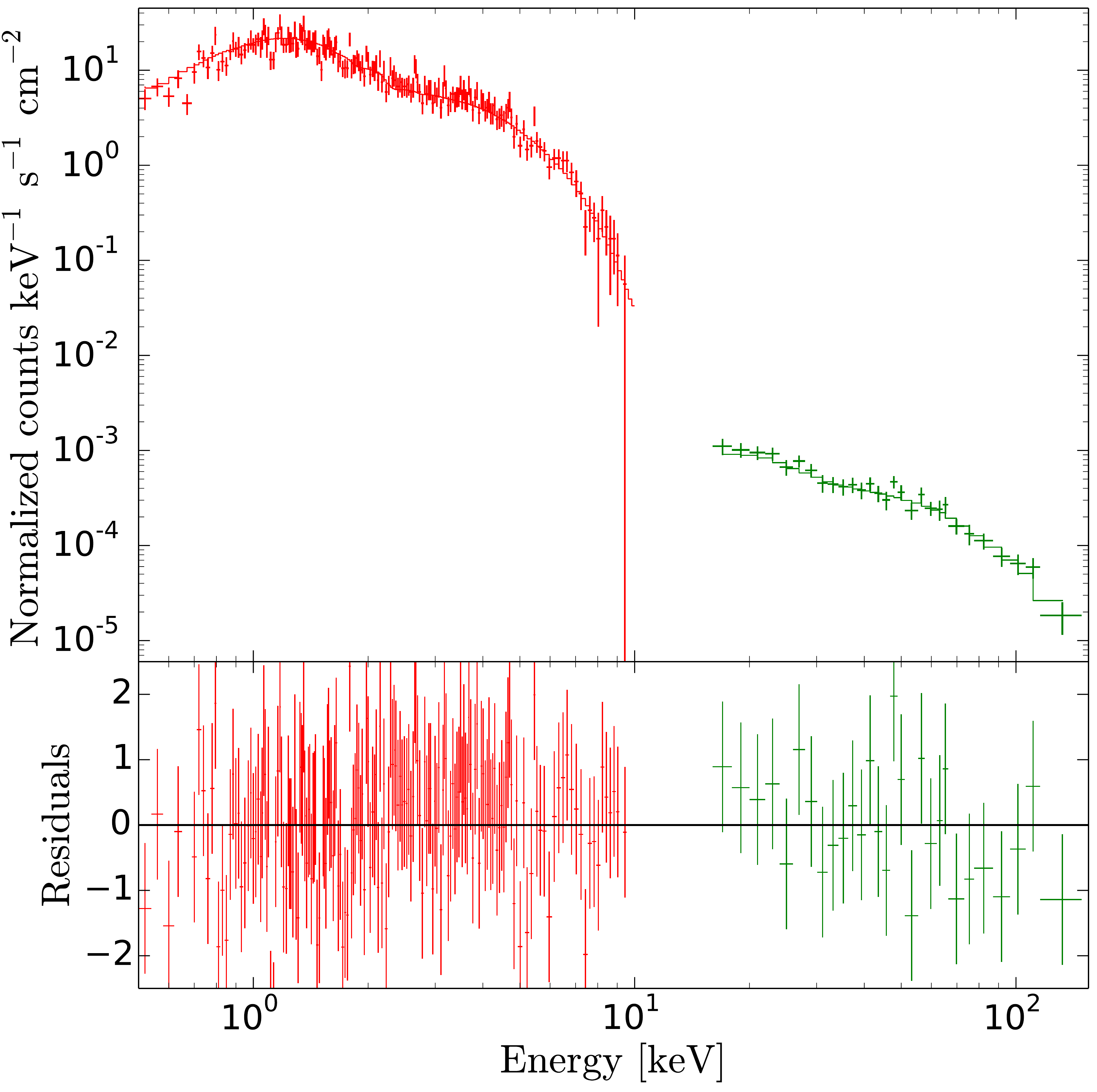} \\
\includegraphics[width = 0.40\textwidth]{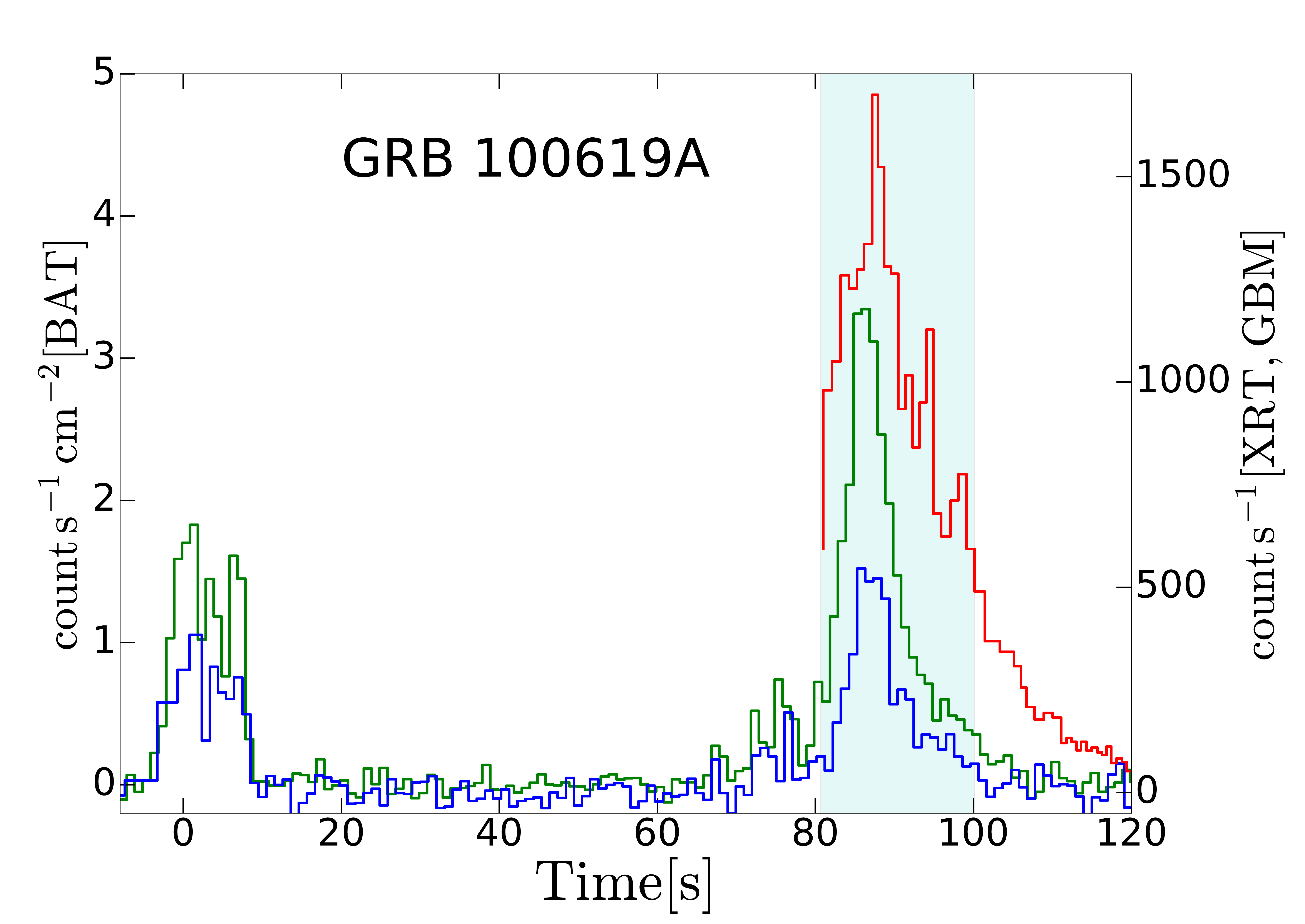} 
\includegraphics[width = 0.30\textwidth]{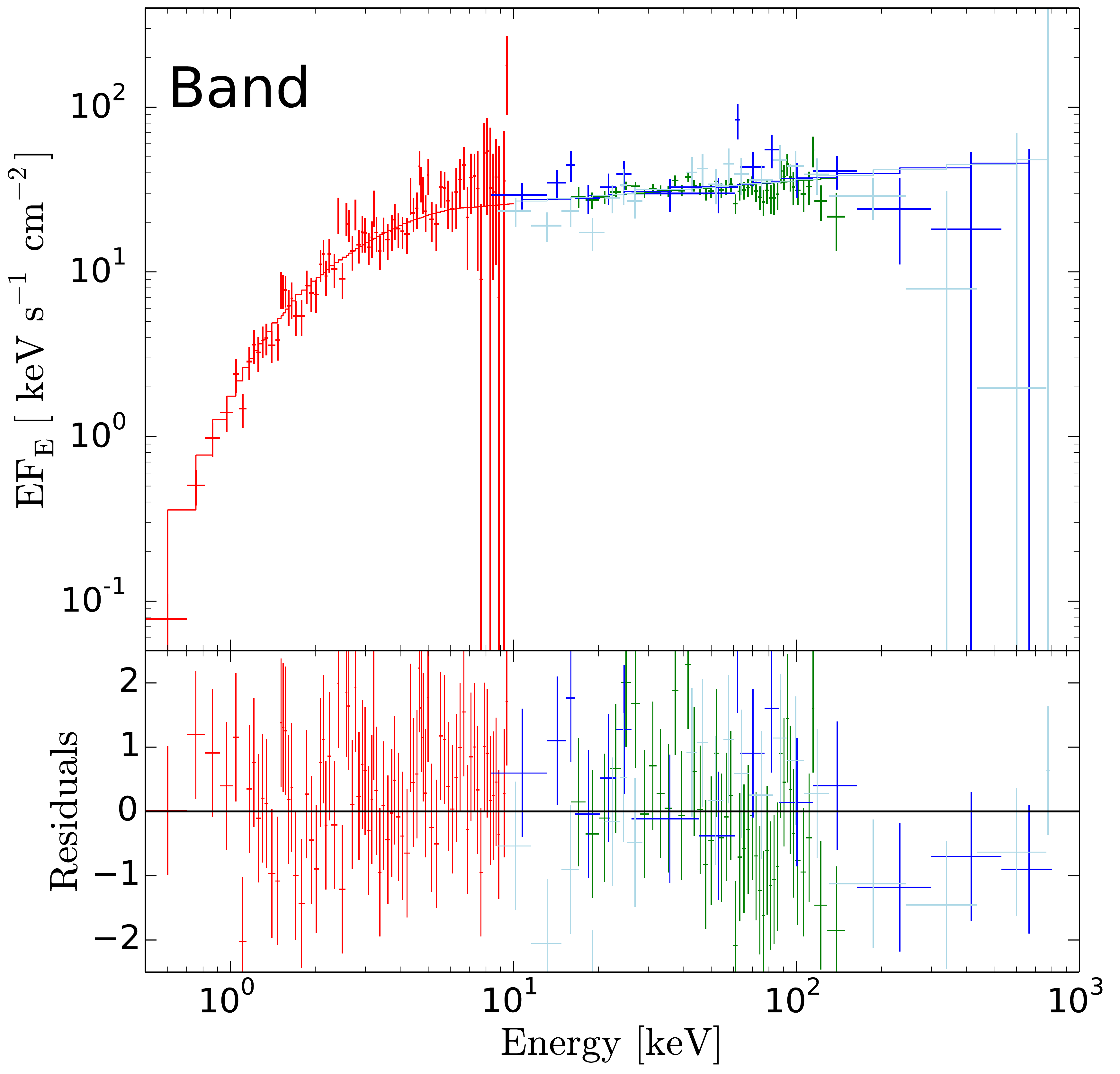} 
\includegraphics[width = 0.30\textwidth]{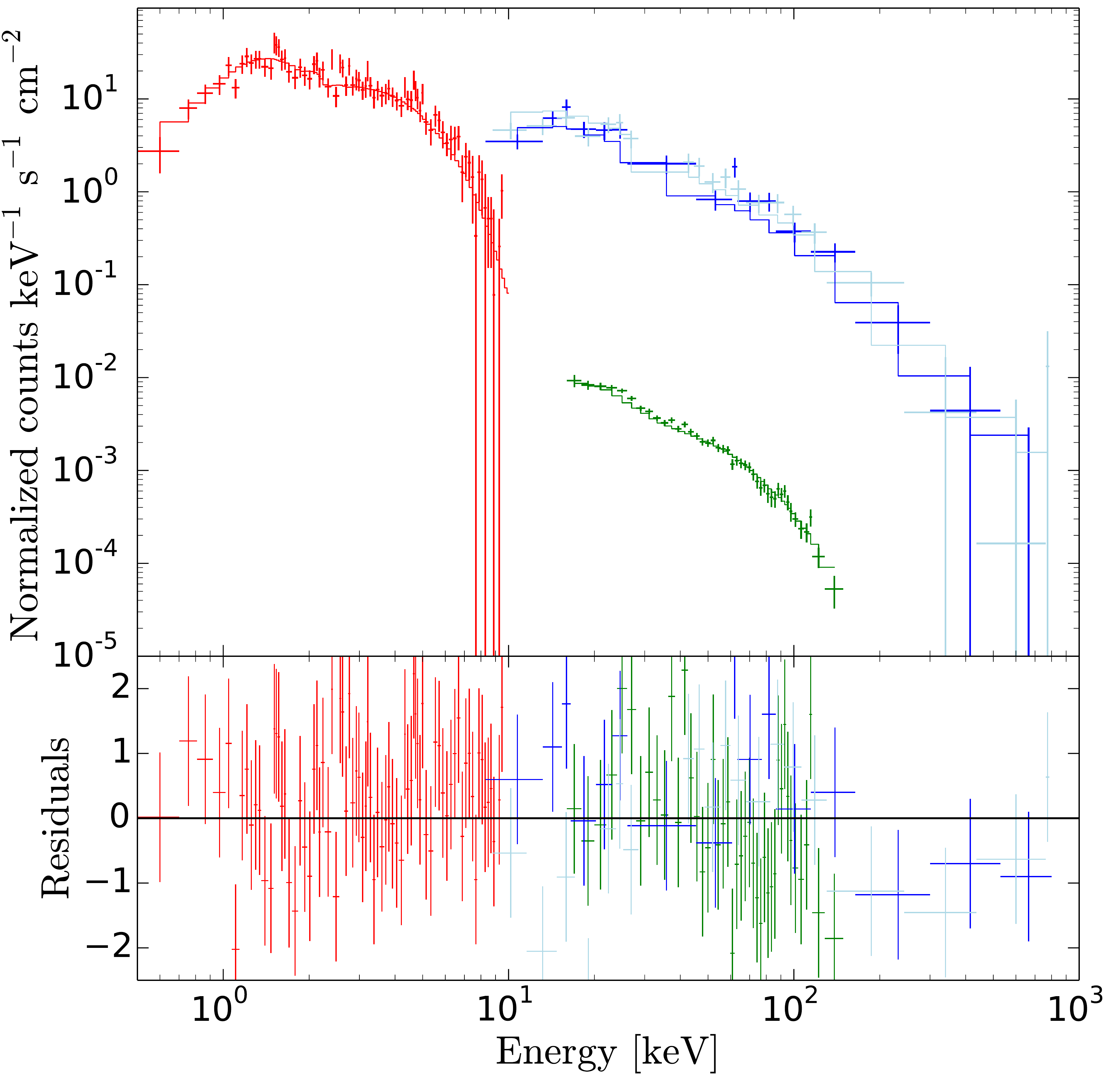} 
\end{figure}

\begin{figure}\ContinuedFloat
\includegraphics[width = 0.40\textwidth]{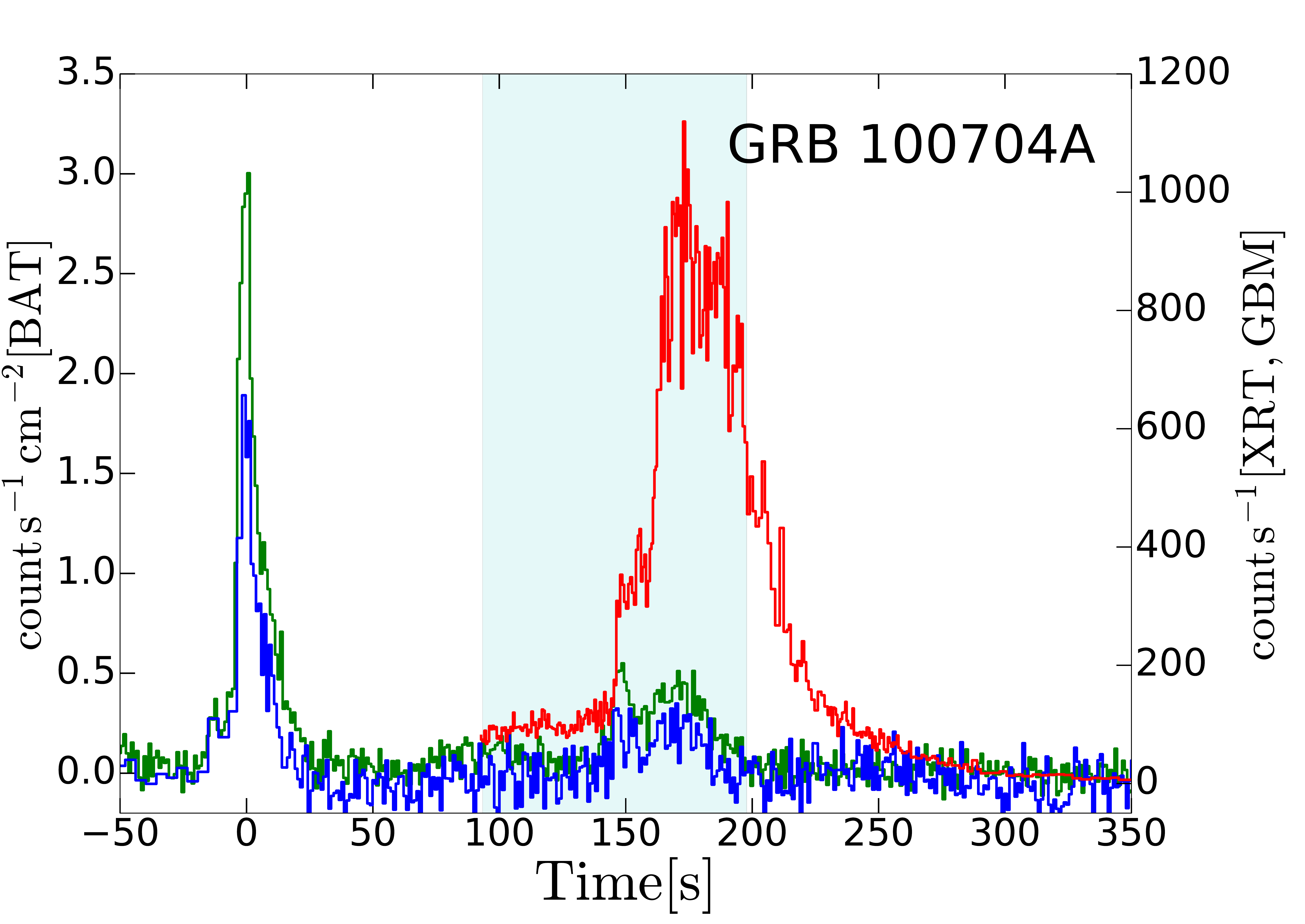} 
\includegraphics[width = 0.30\textwidth]{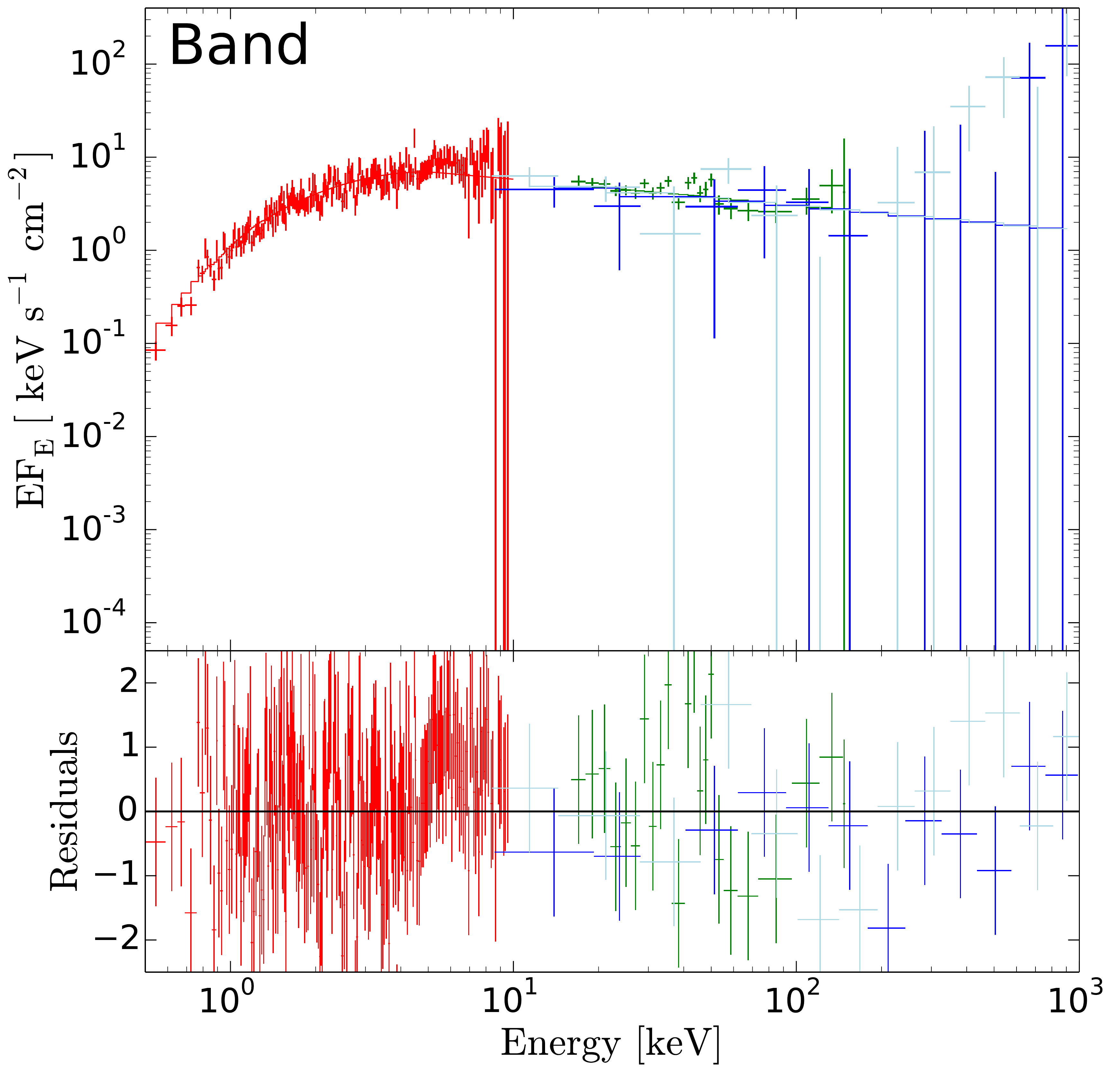} 
\includegraphics[width = 0.30\textwidth]{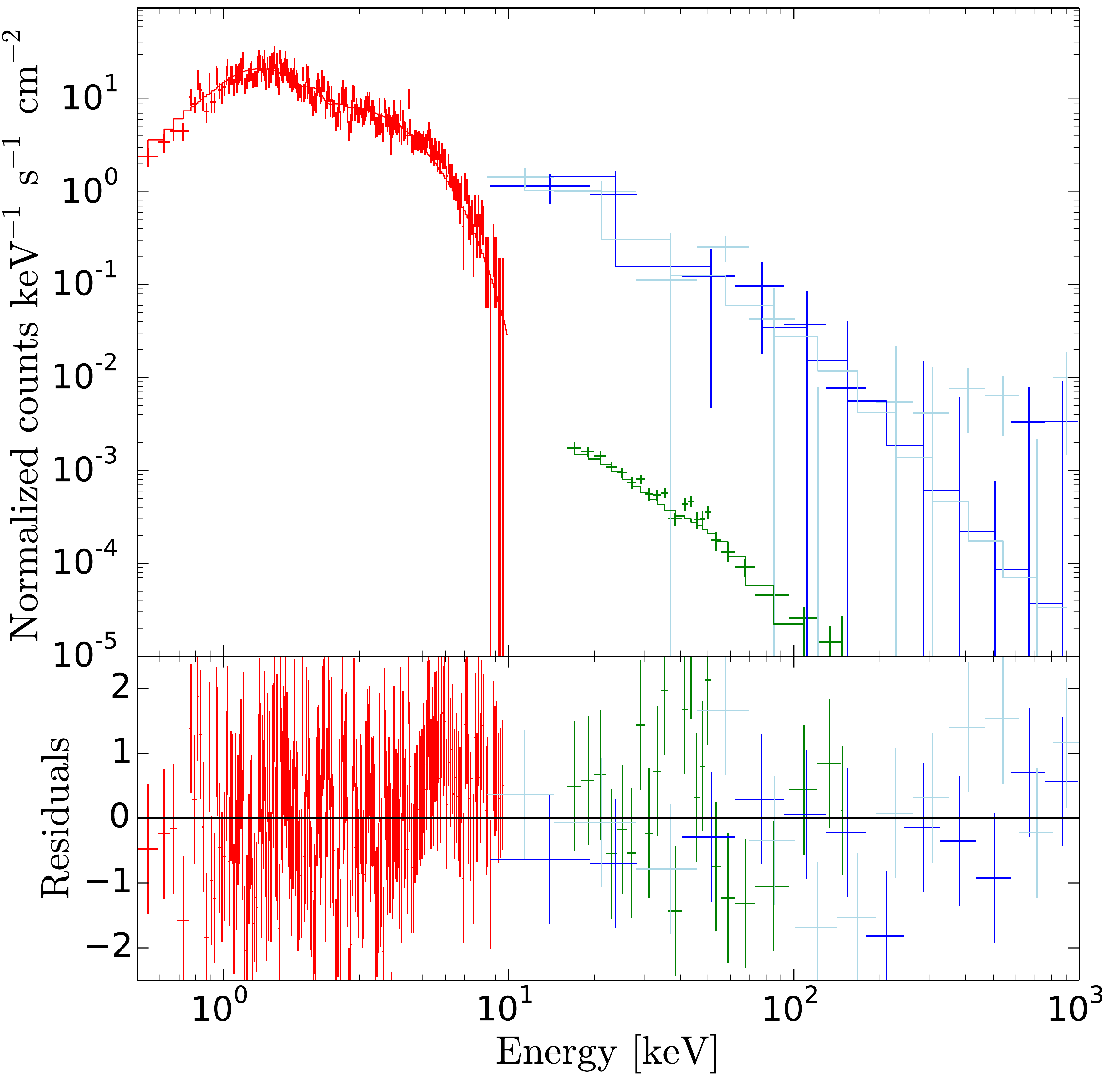} \\
\includegraphics[width = 0.40\textwidth]{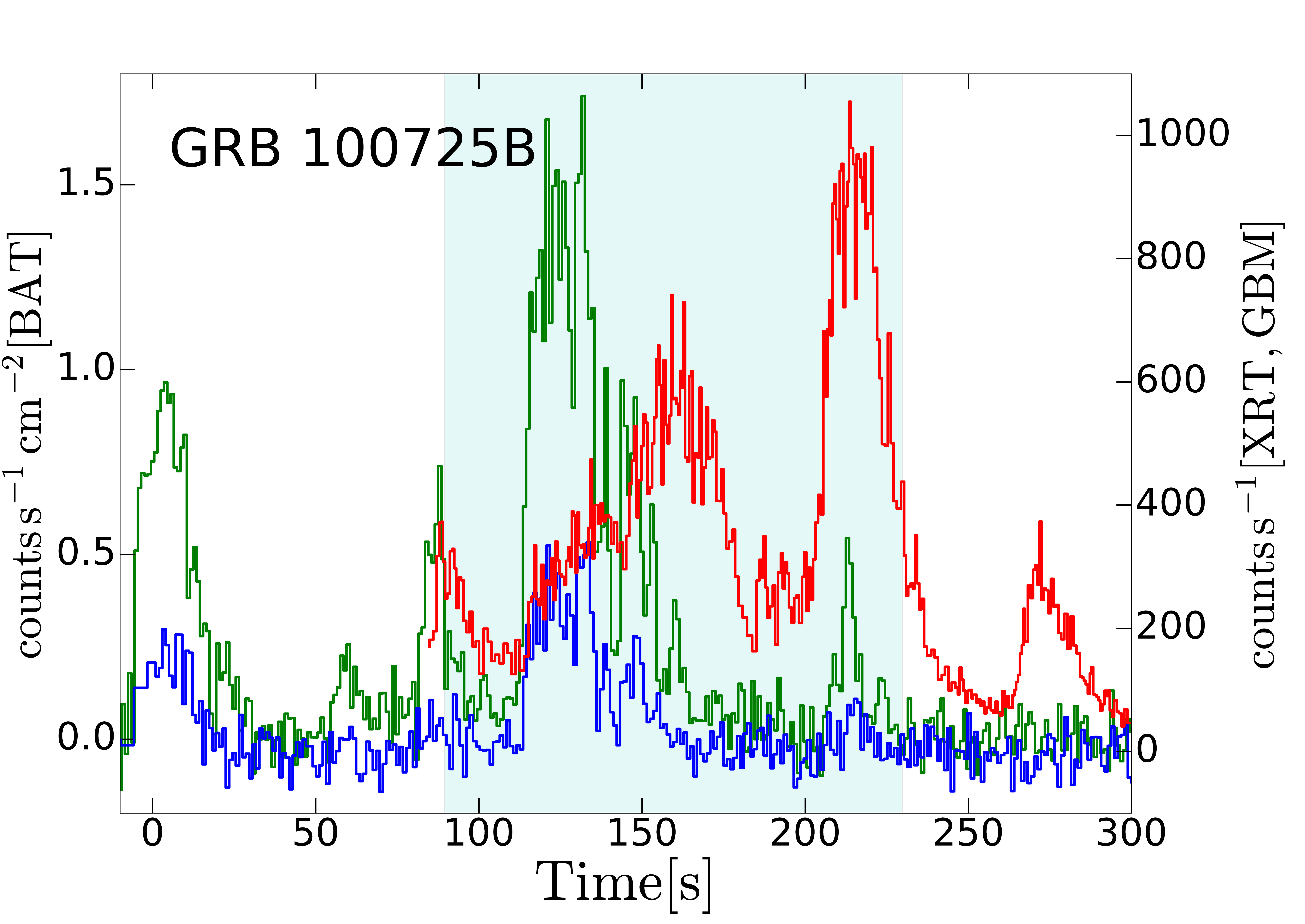} 
\includegraphics[width = 0.30\textwidth]{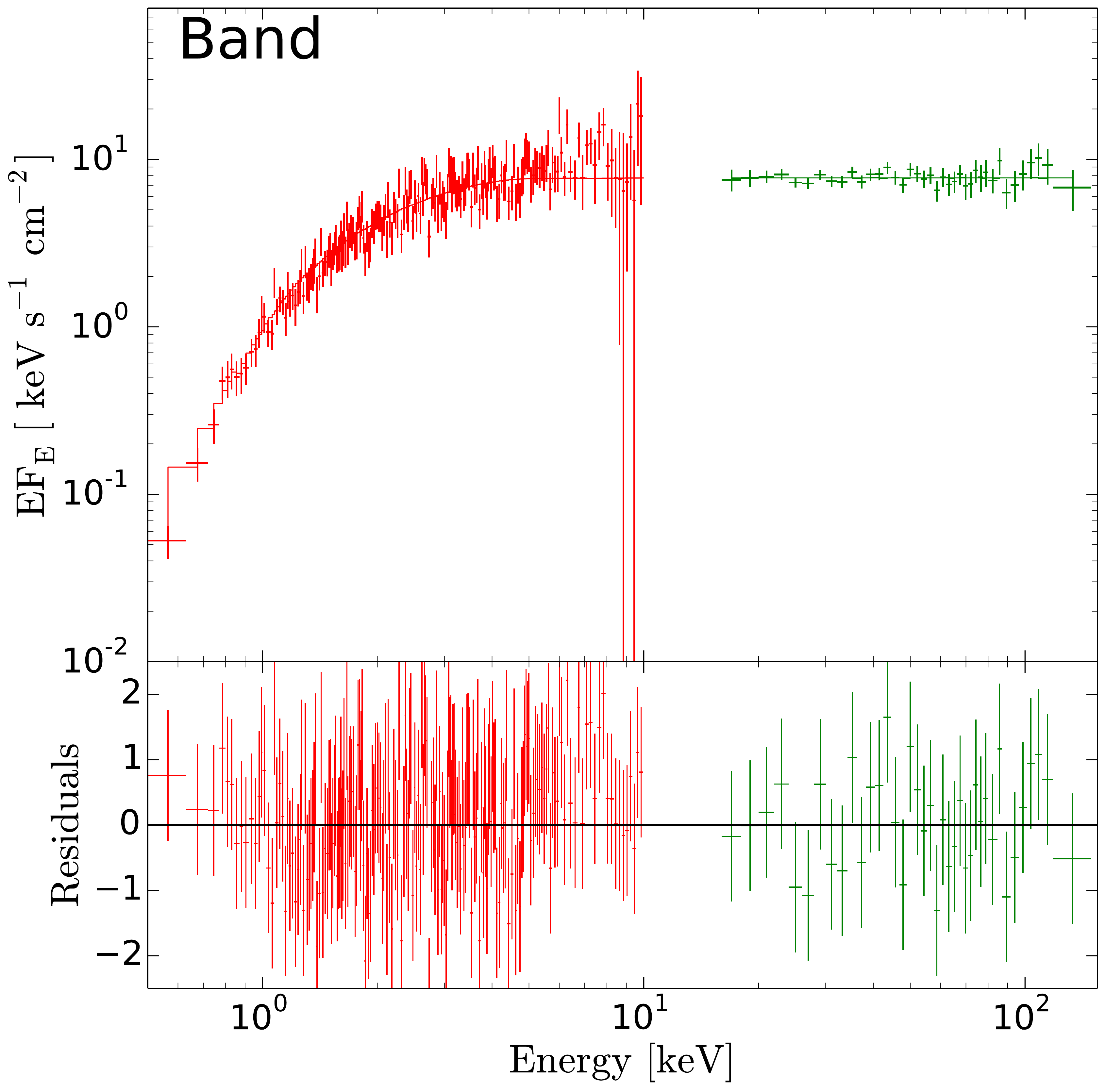} 
\includegraphics[width = 0.30\textwidth]{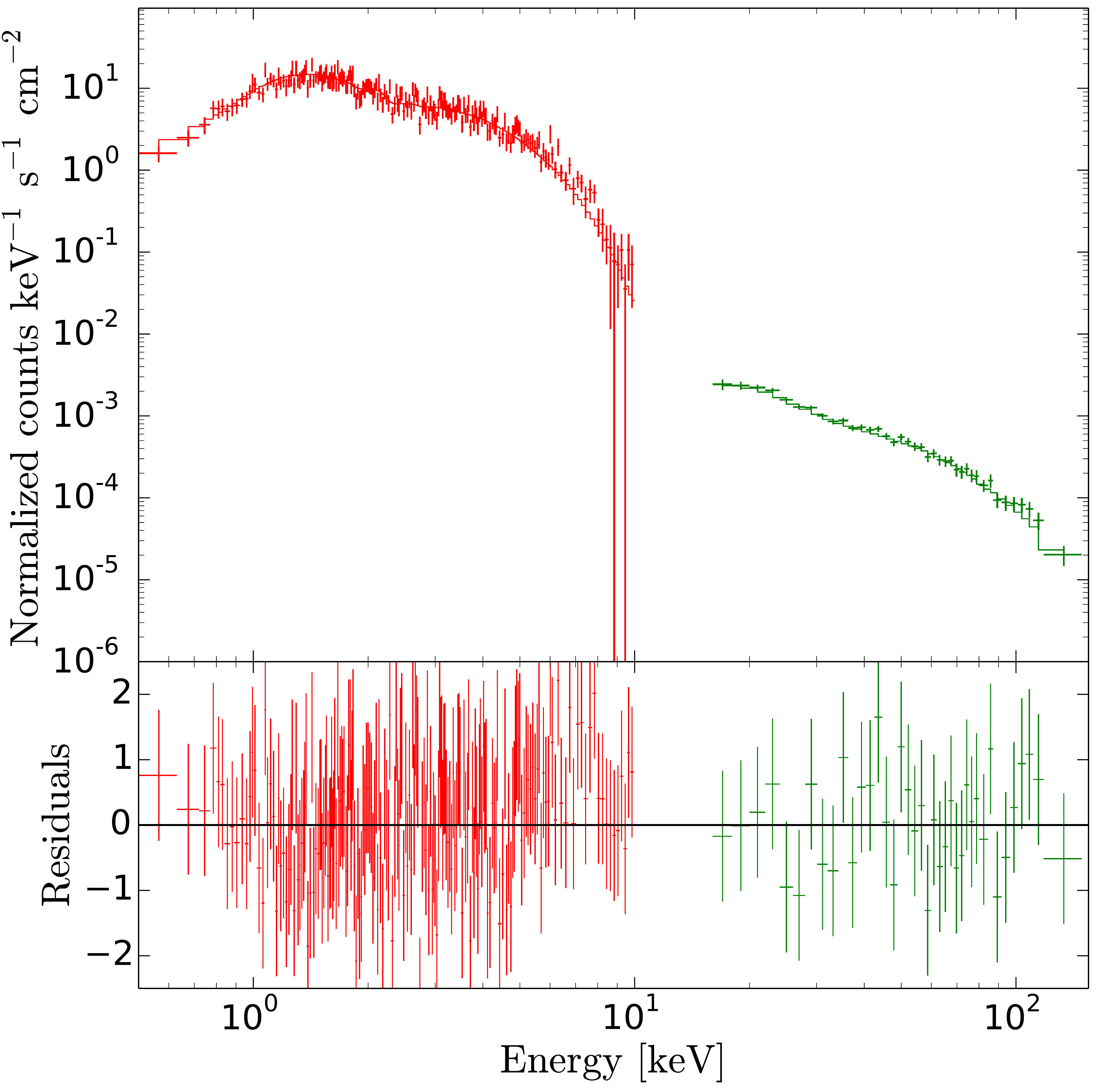} \\
\includegraphics[width = 0.40\textwidth]{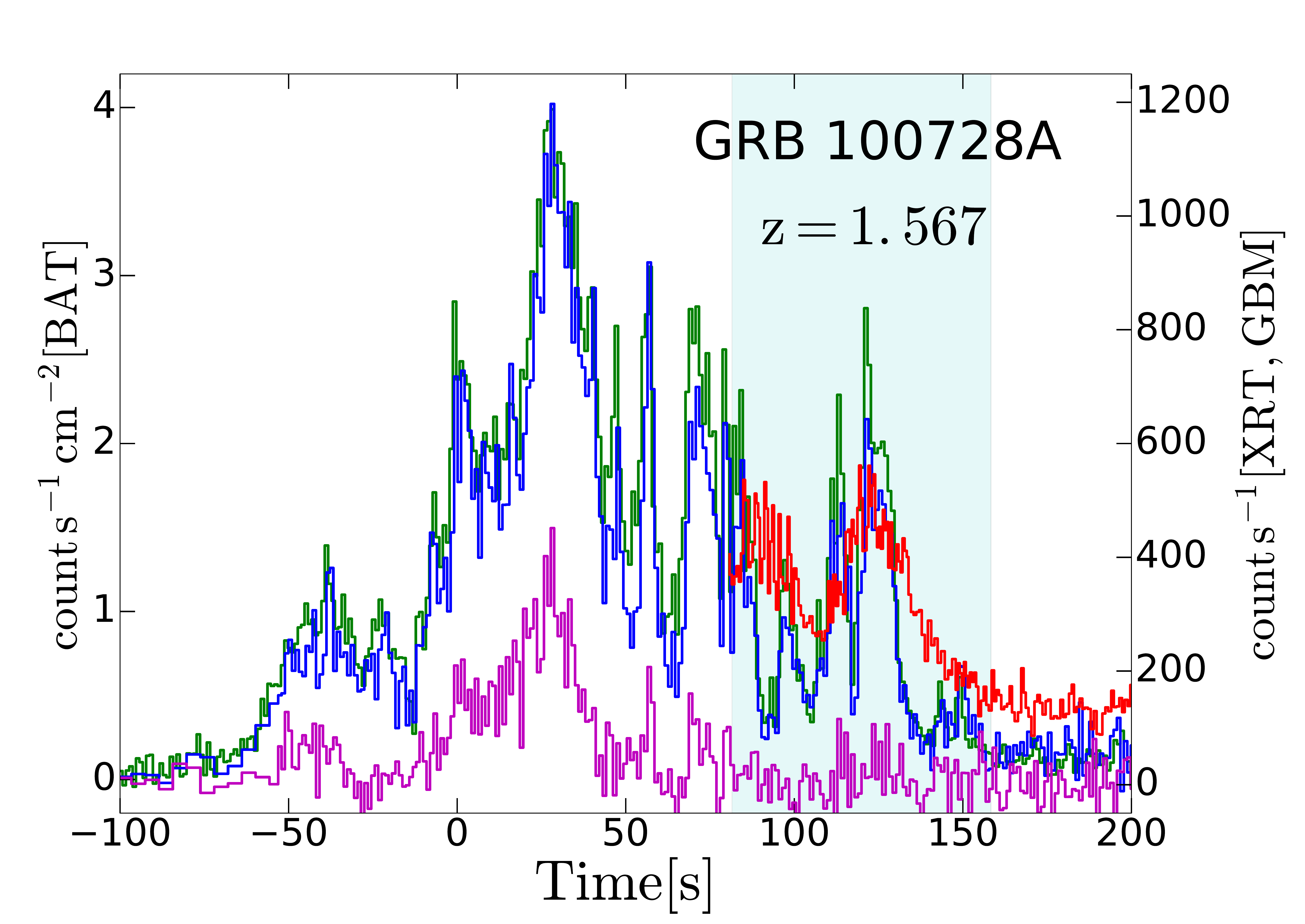} 
\includegraphics[width = 0.30\textwidth]{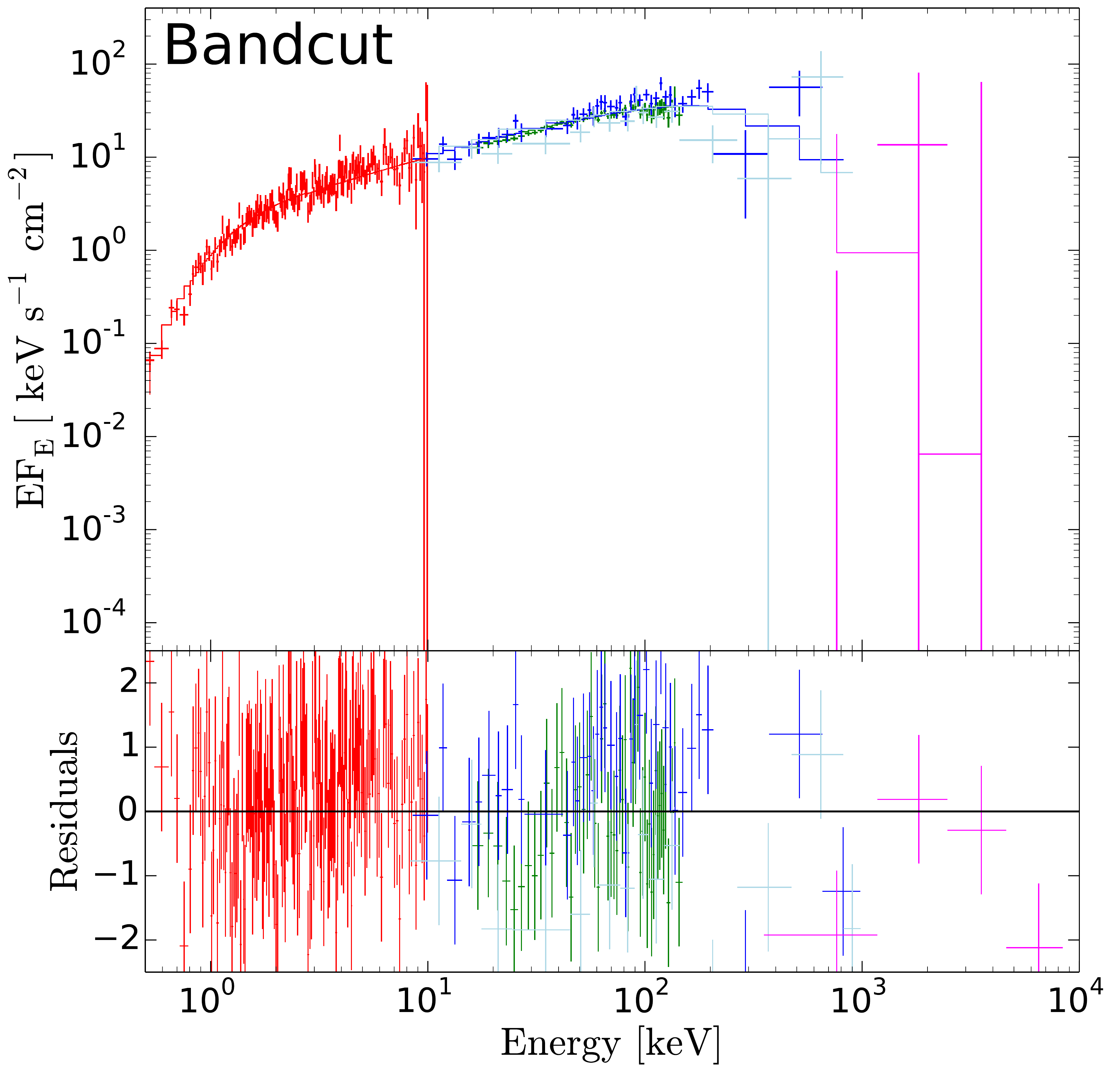} 
\includegraphics[width = 0.30\textwidth]{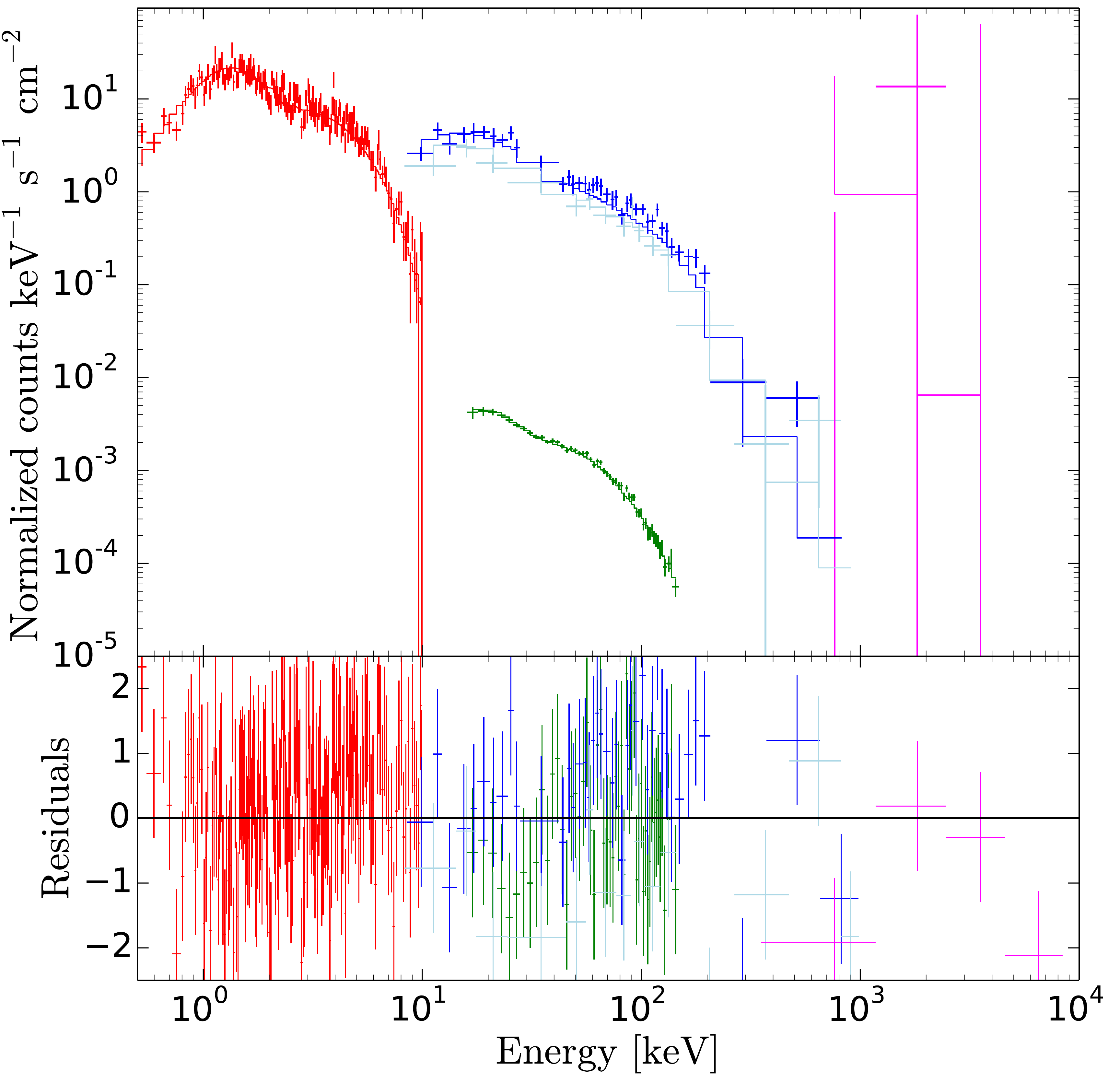} 
\end{figure}

\begin{figure}\ContinuedFloat
\includegraphics[width = 0.40\textwidth]{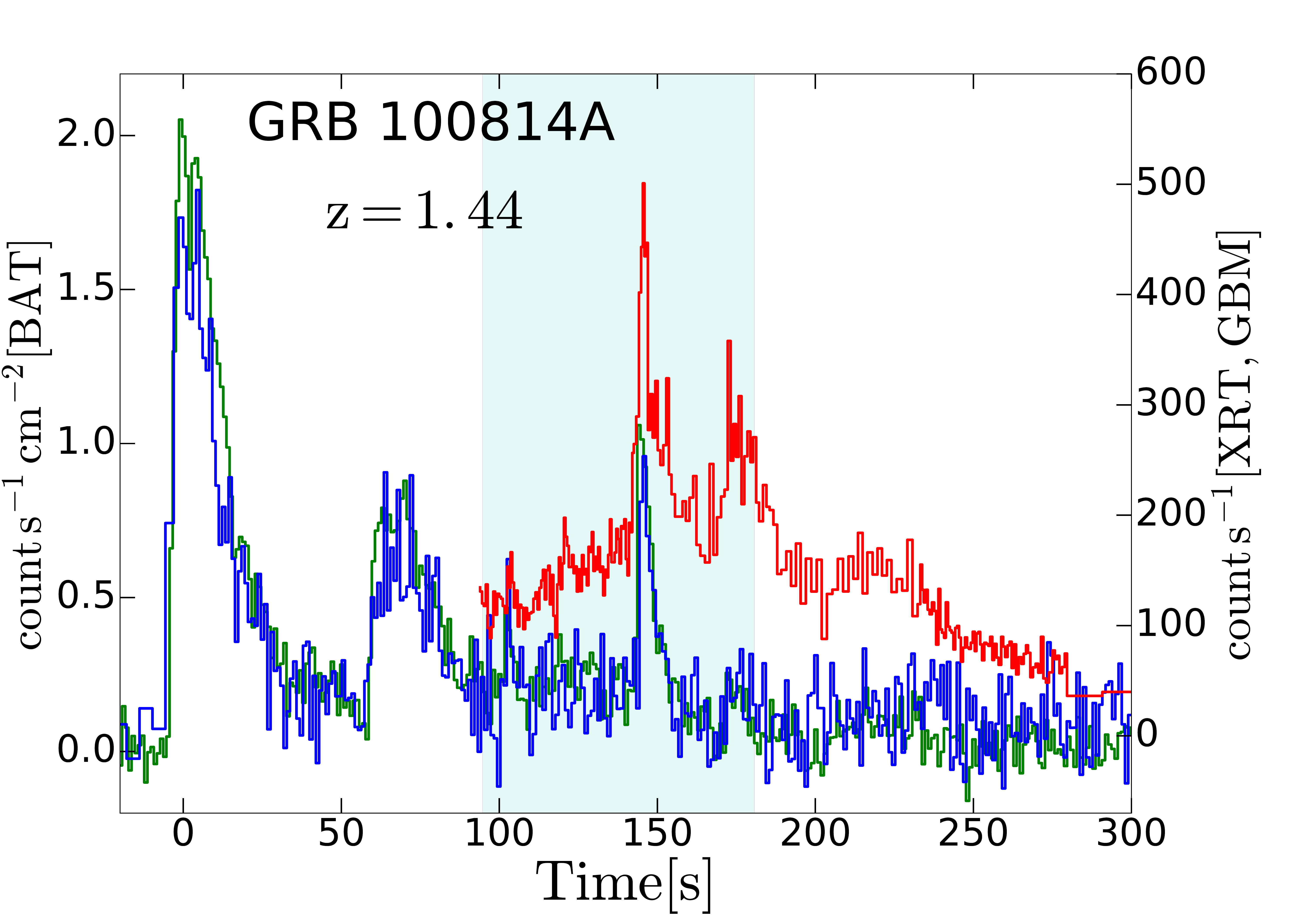} 
\includegraphics[width = 0.30\textwidth]{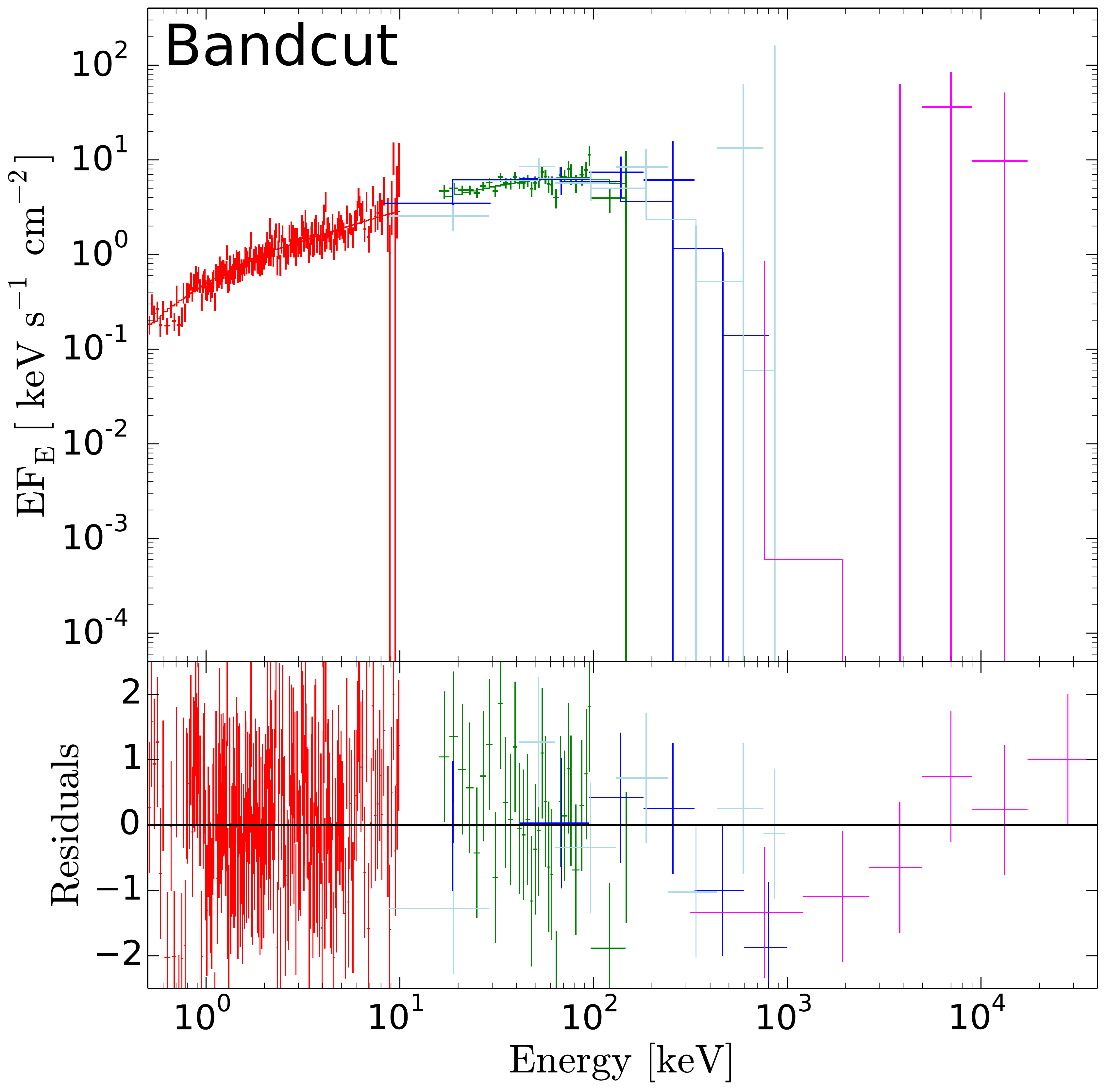} 
\includegraphics[width = 0.30\textwidth]{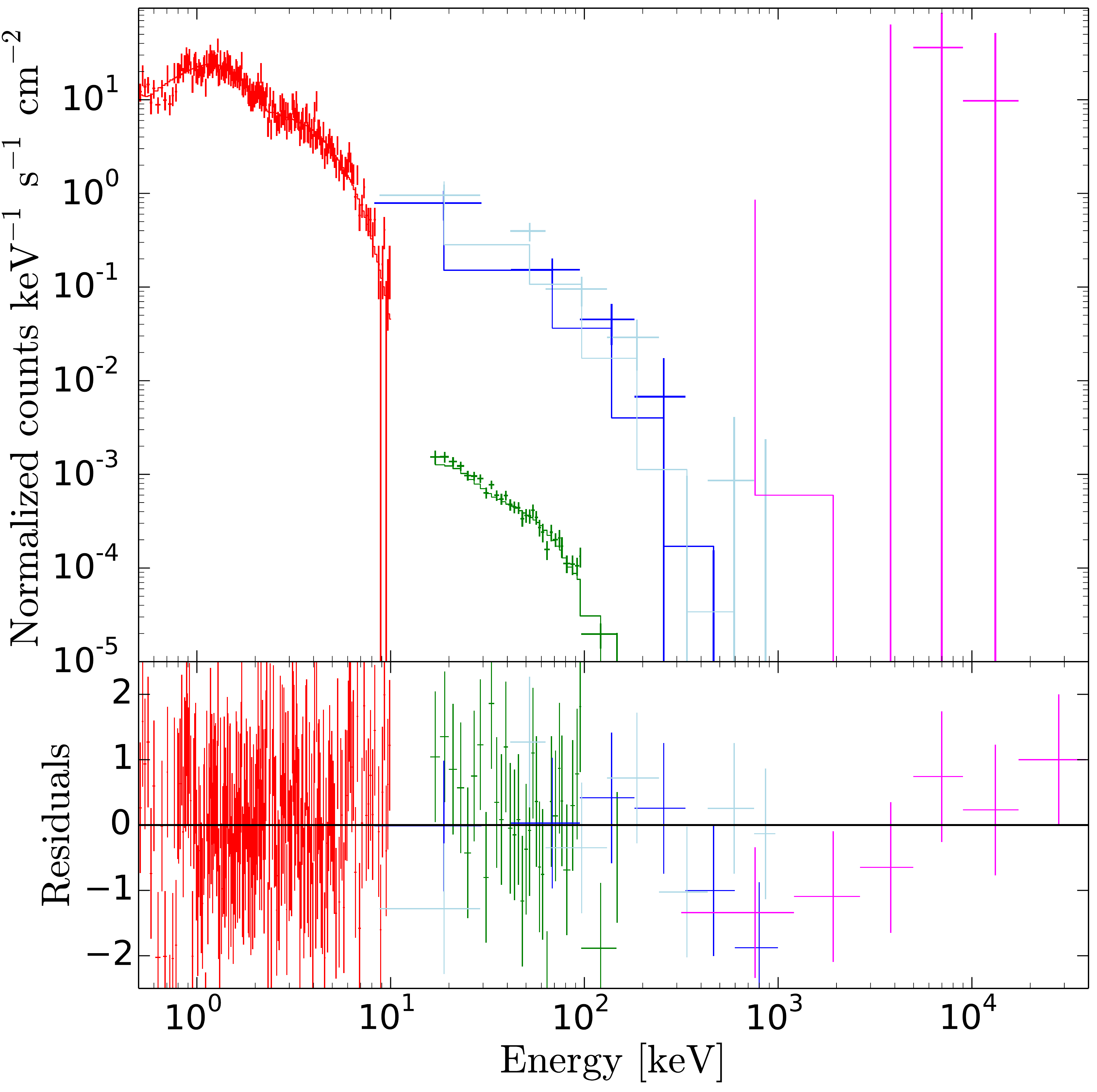} \\
\includegraphics[width = 0.40\textwidth]{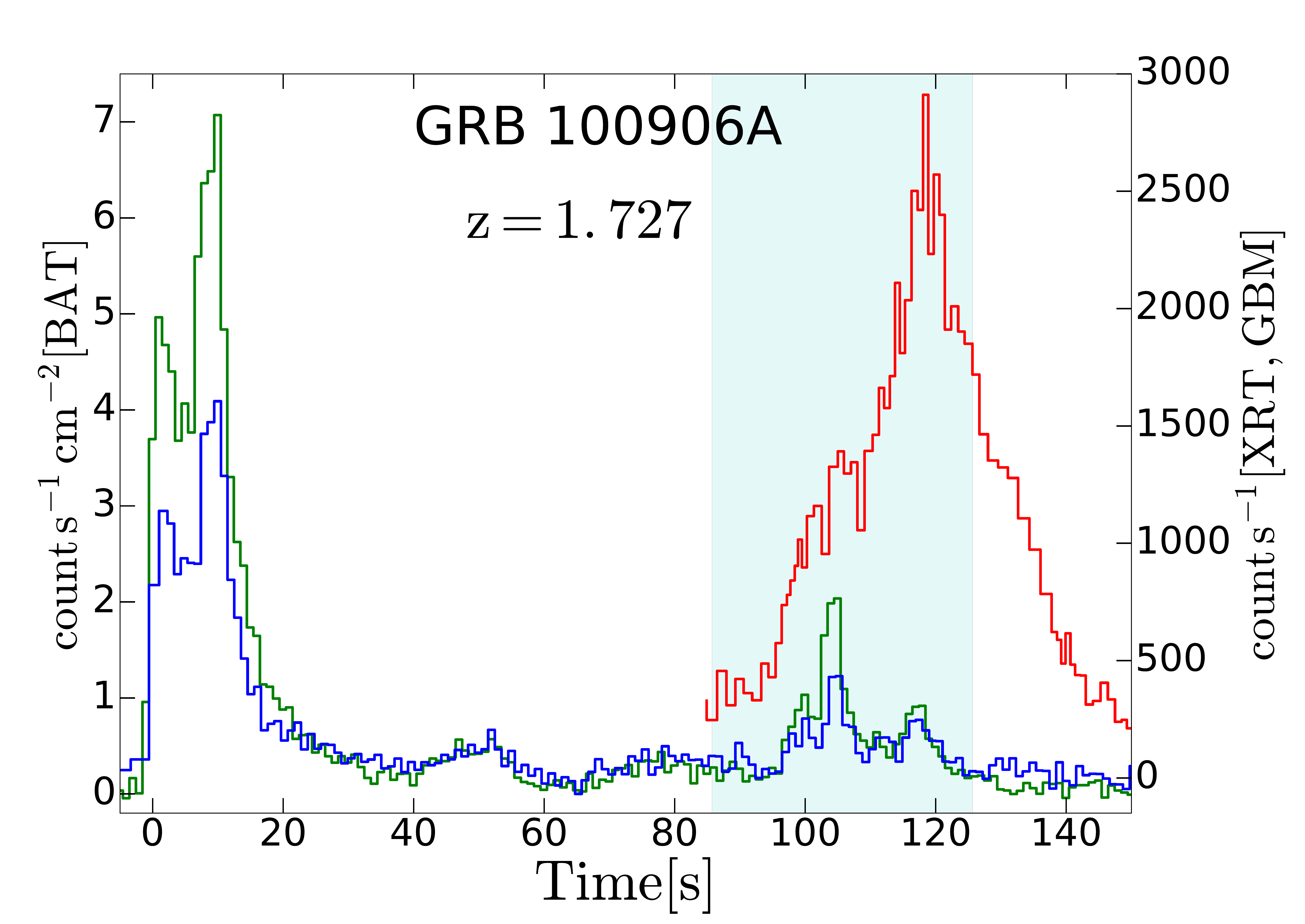} 
\includegraphics[width = 0.30\textwidth]{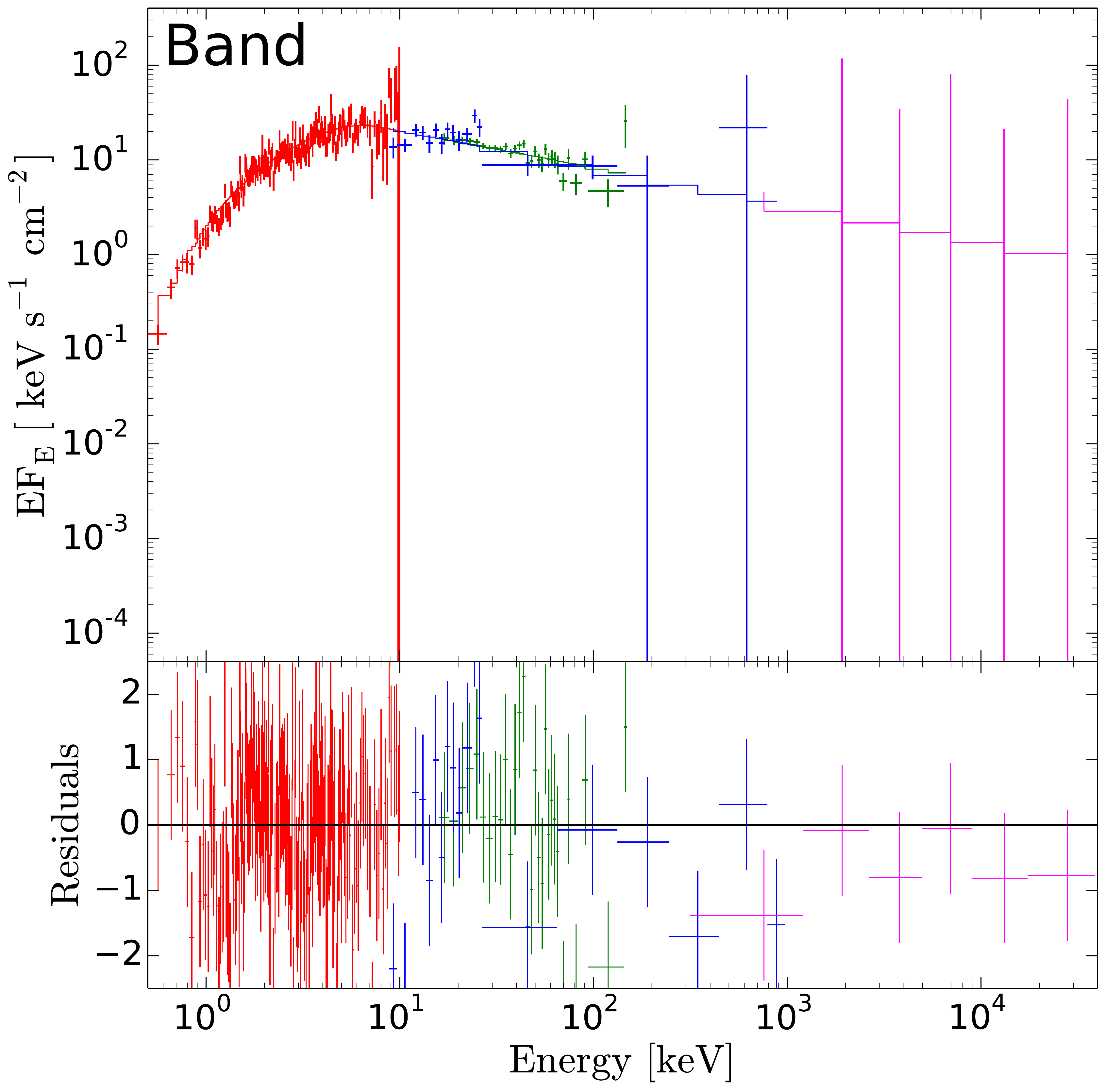} 
\includegraphics[width = 0.30\textwidth]{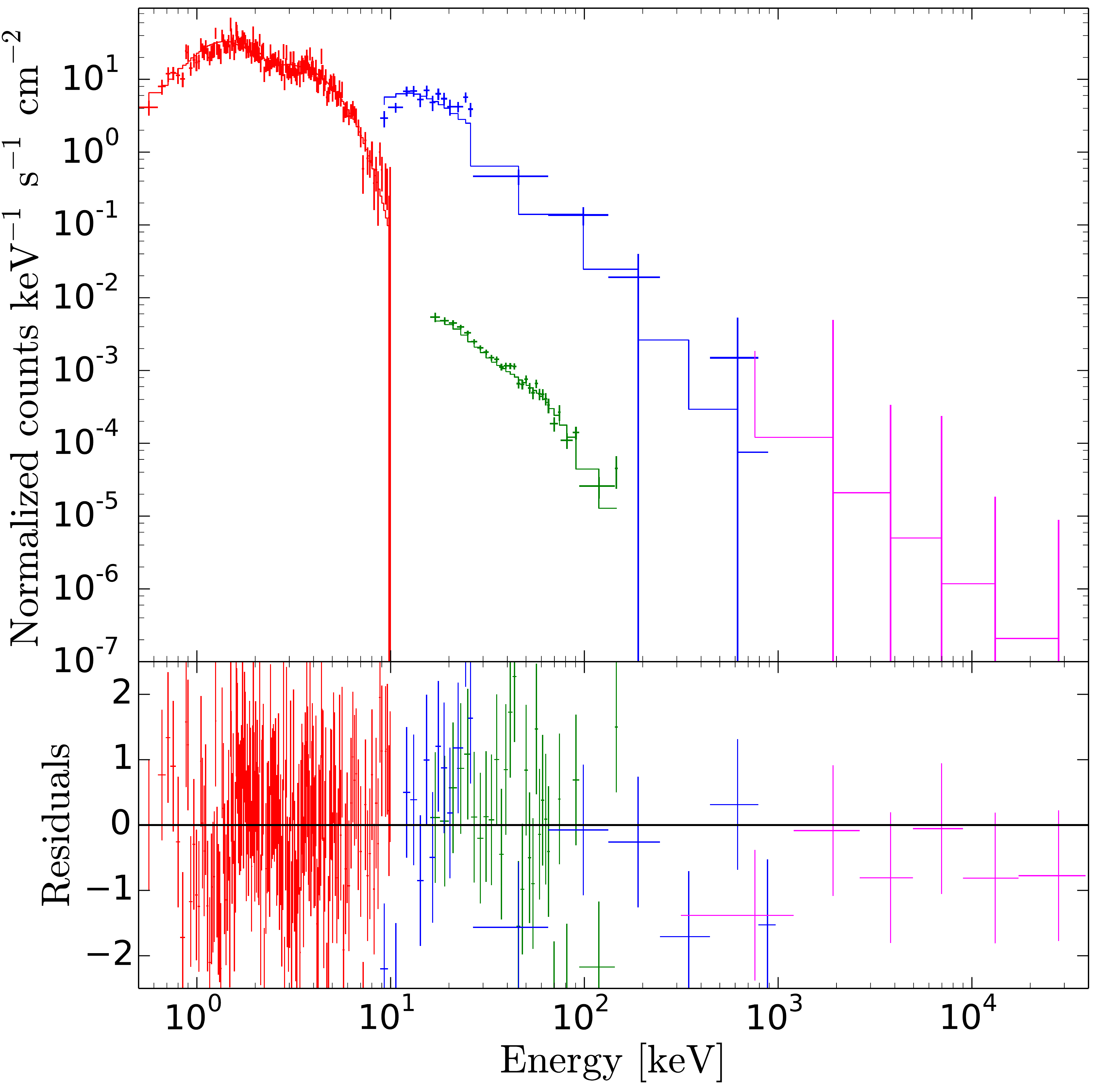} \\
\includegraphics[width = 0.40\textwidth]{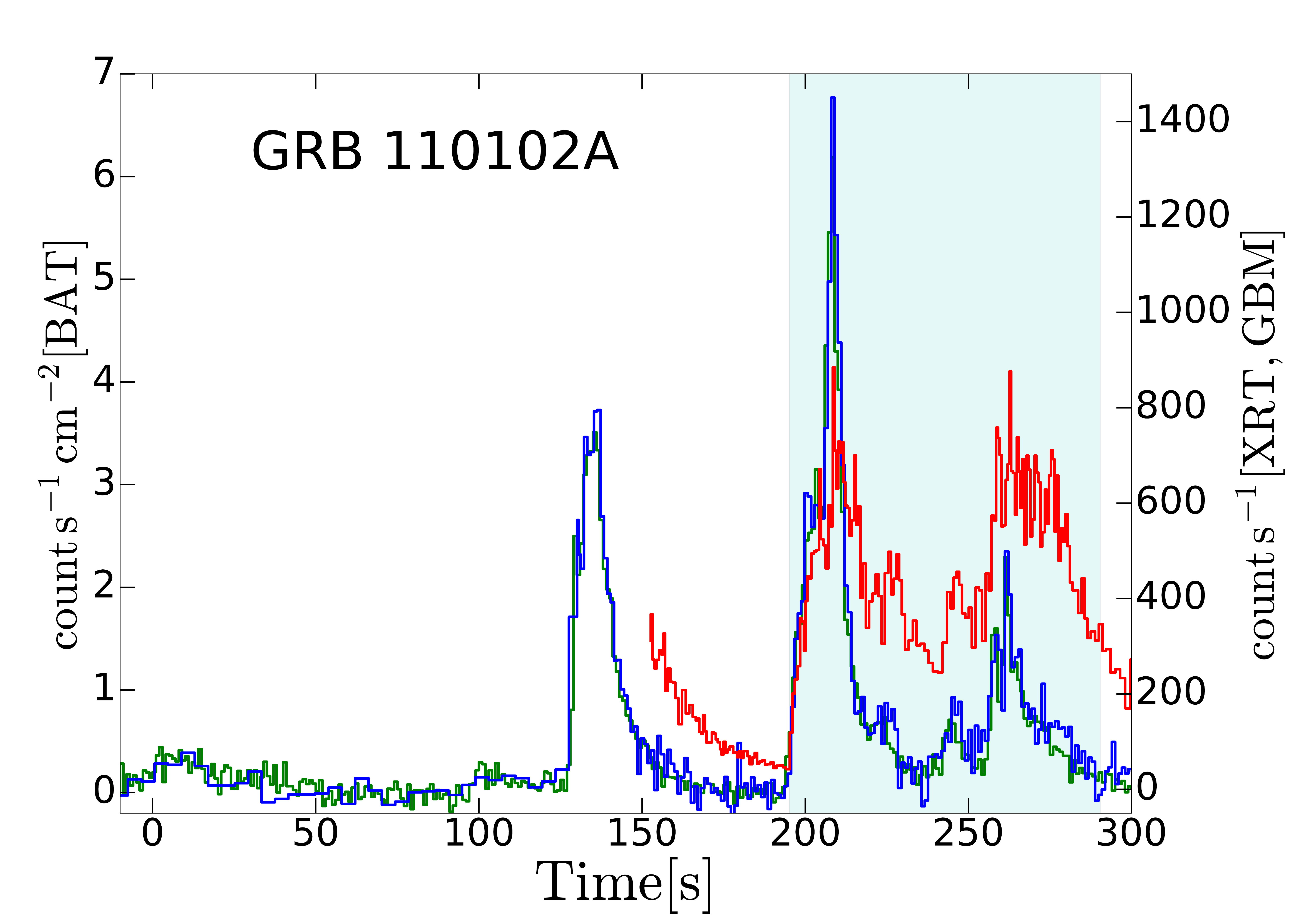} 
\includegraphics[width = 0.30\textwidth]{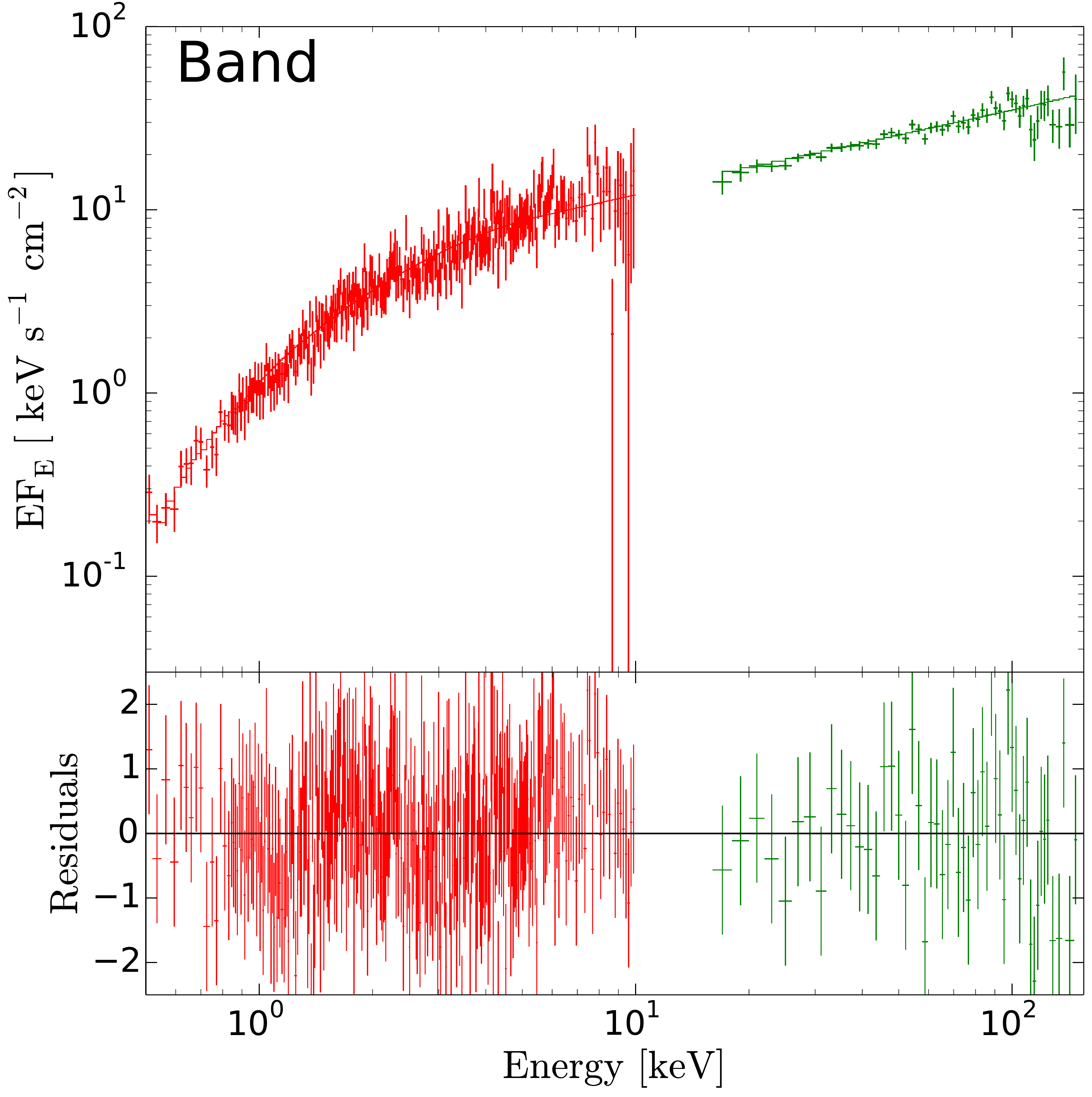} 
\includegraphics[width = 0.30\textwidth]{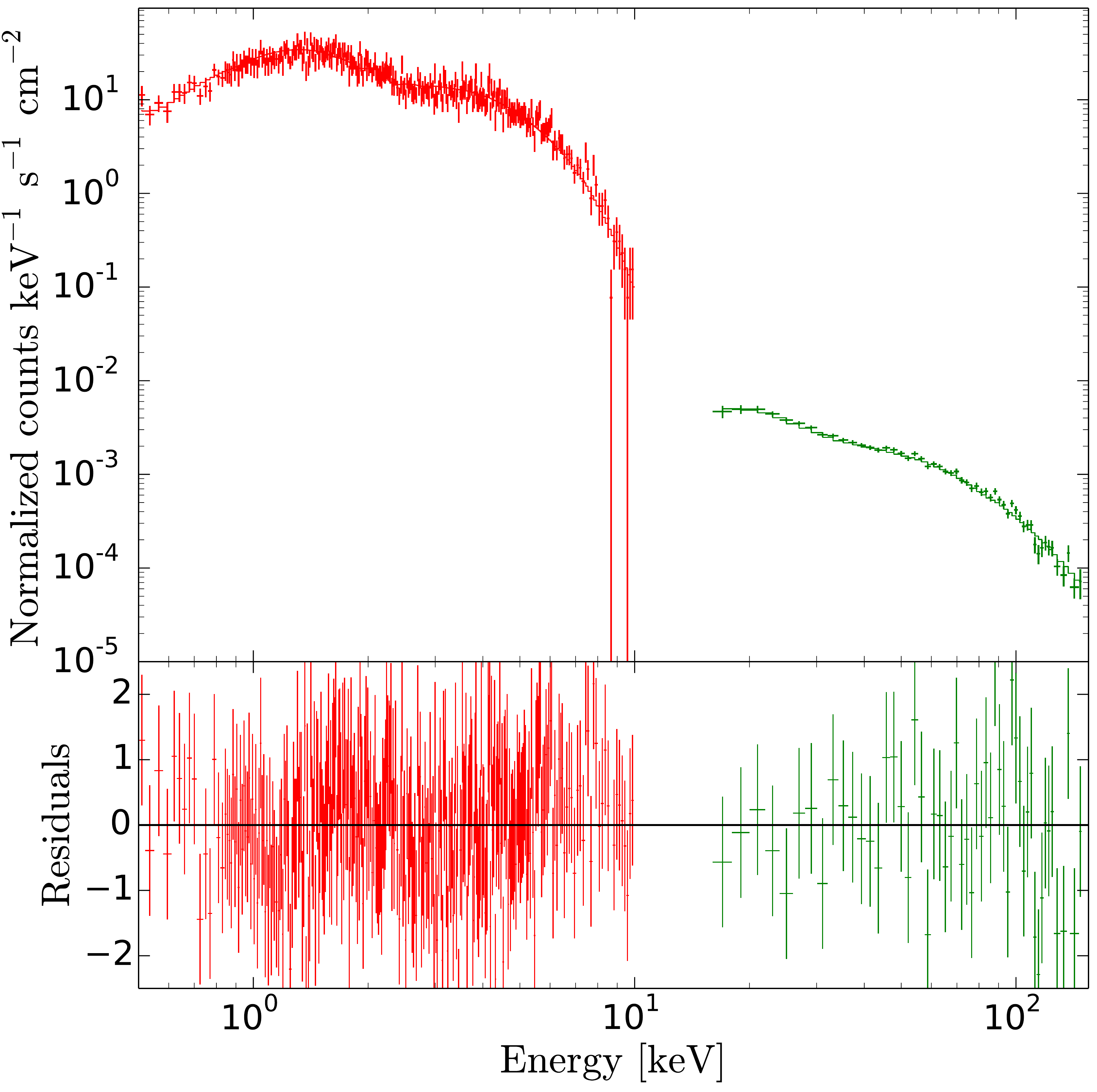} 
\end{figure}

\begin{figure}\ContinuedFloat
\includegraphics[width = 0.40\textwidth]{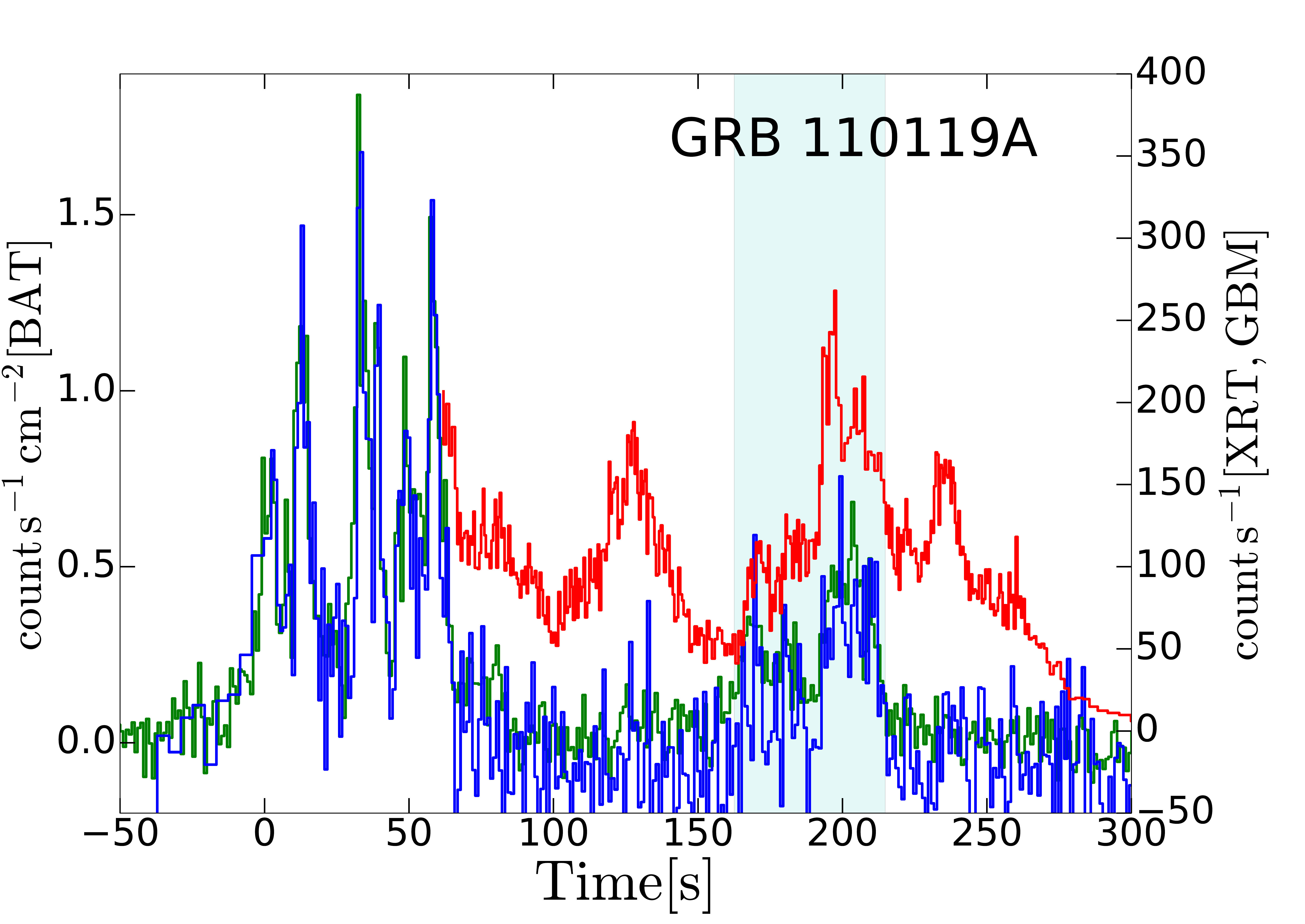} 
\includegraphics[width = 0.30\textwidth]{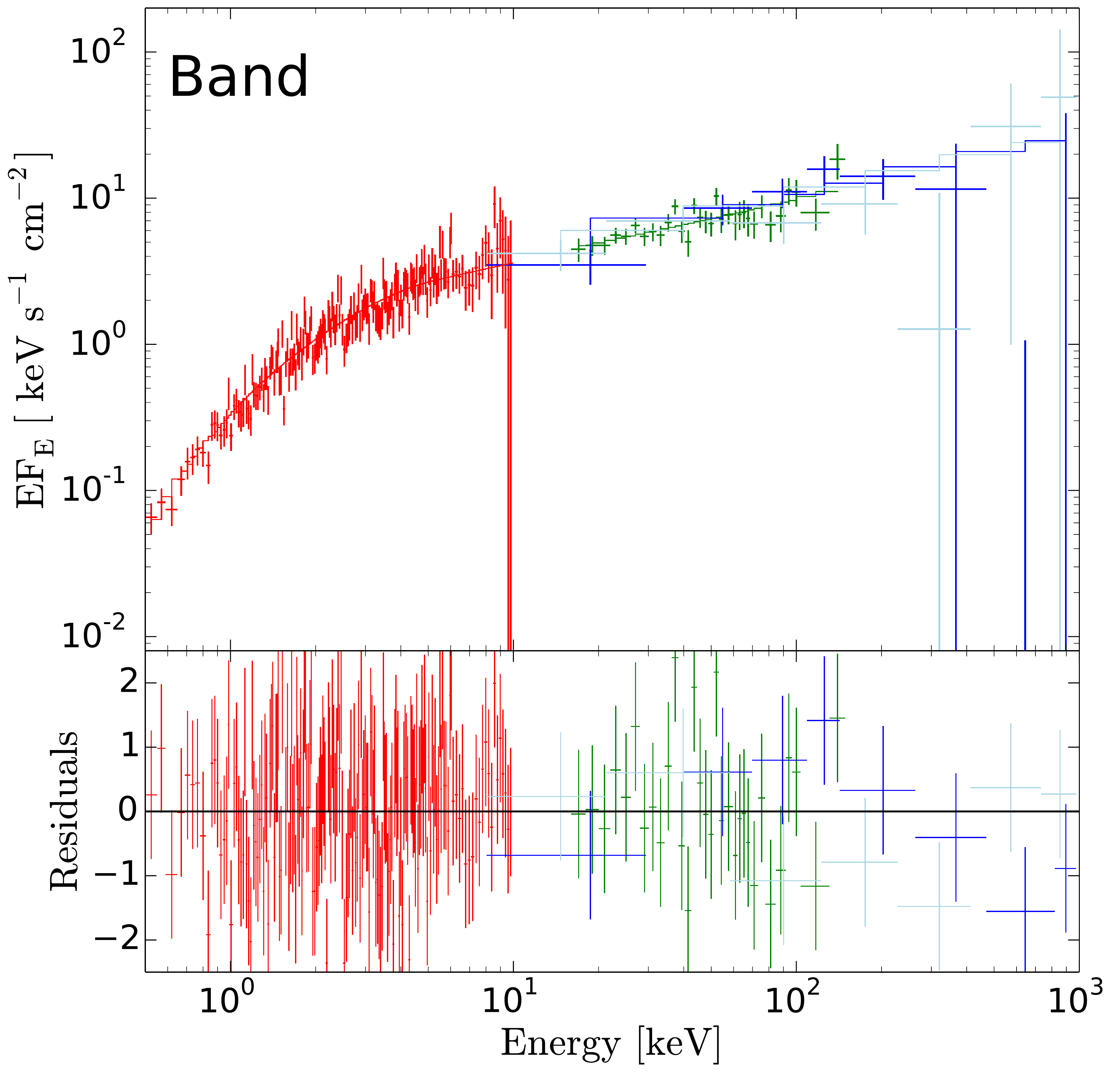} 
\includegraphics[width = 0.30\textwidth]{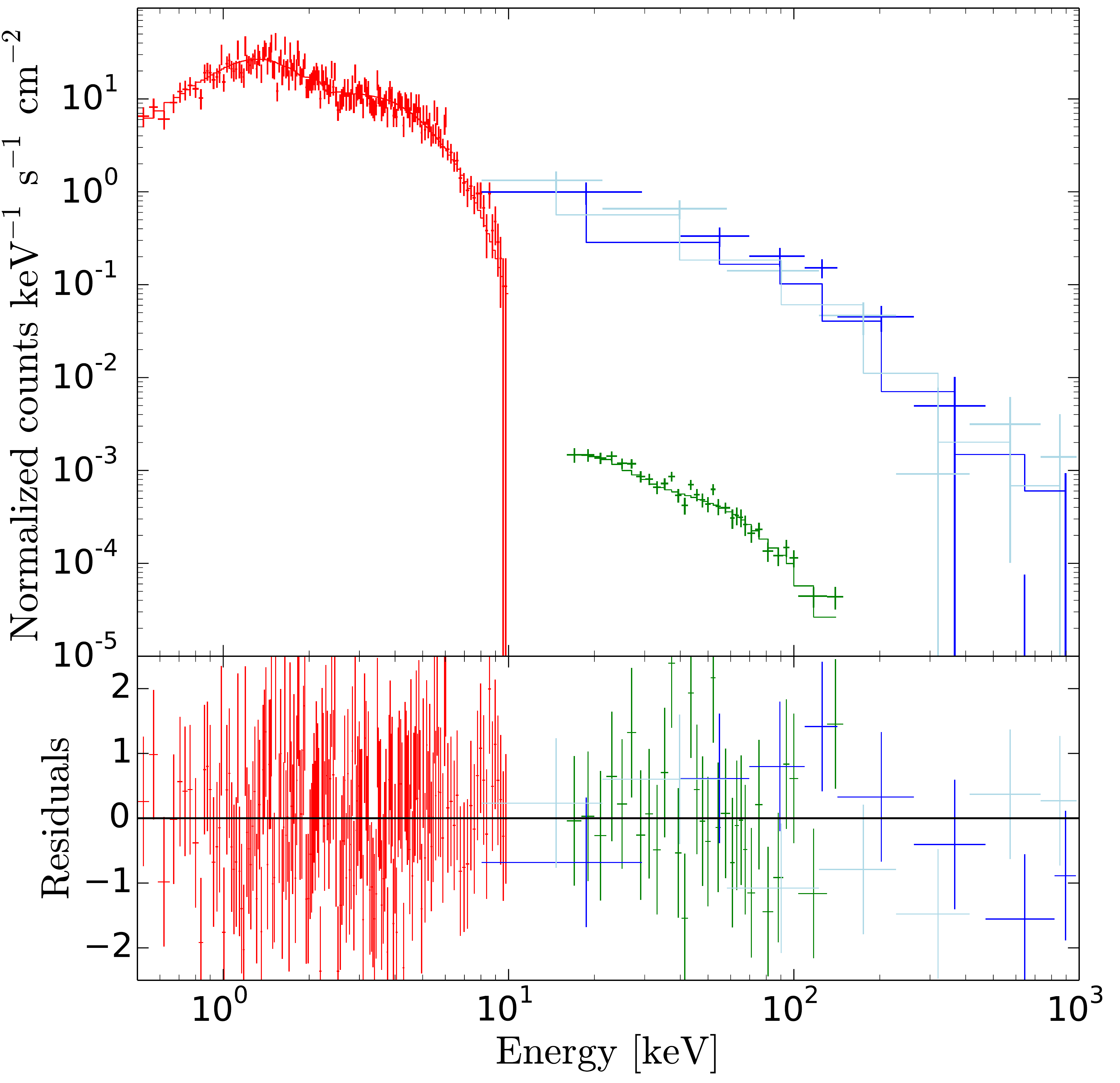} \\
\includegraphics[width = 0.40\textwidth]{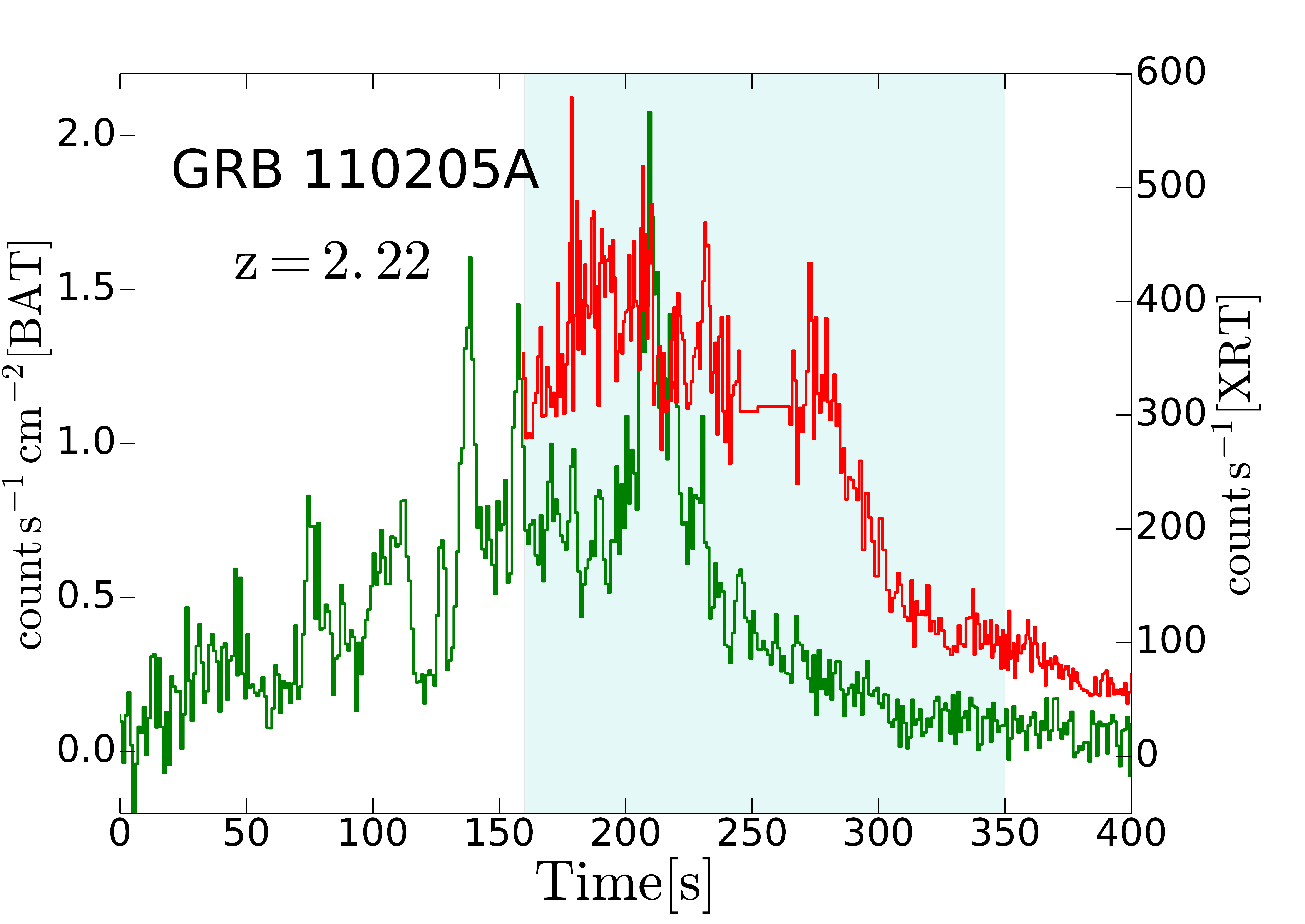} 
\includegraphics[width = 0.30\textwidth]{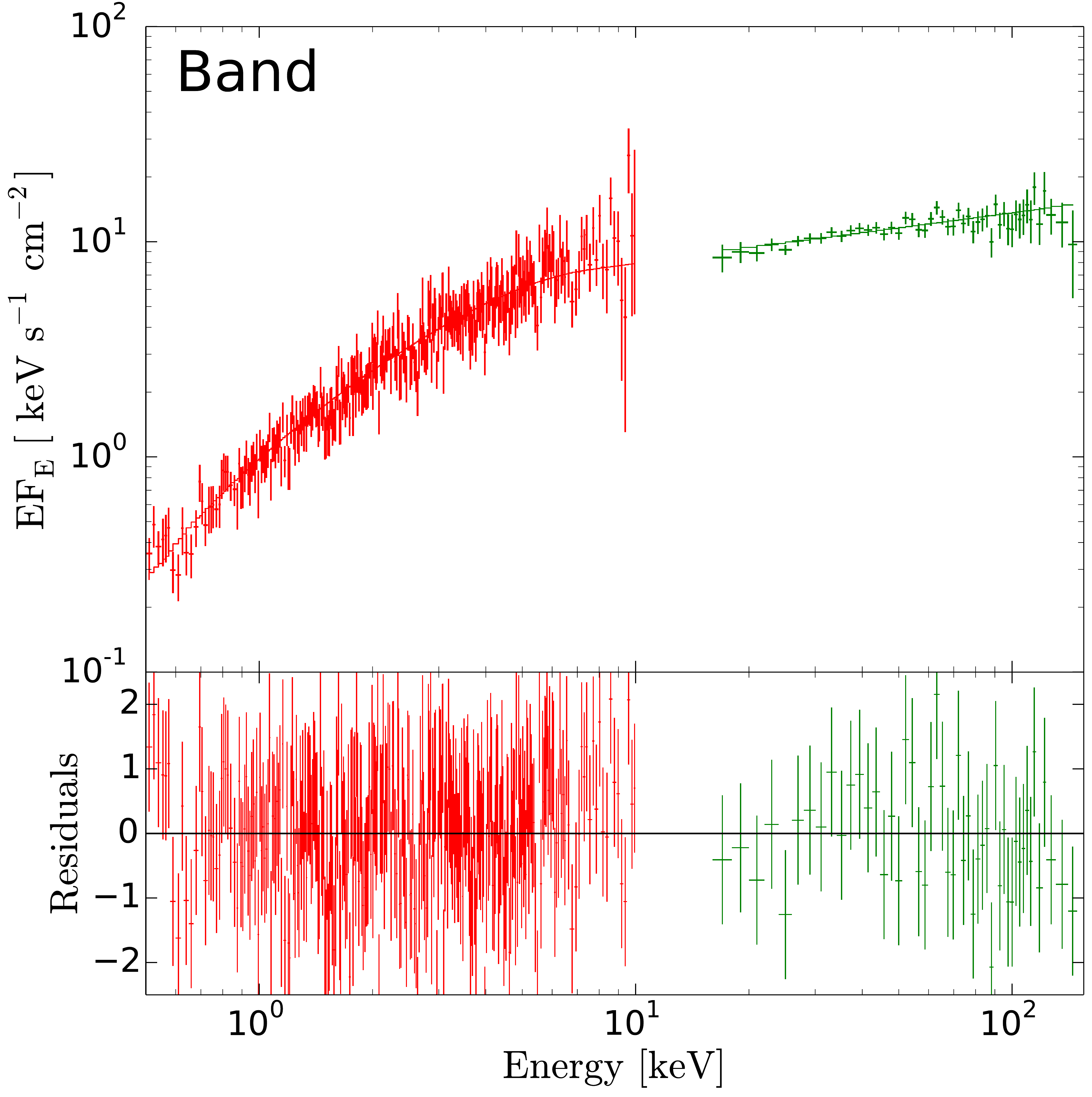} 
\includegraphics[width = 0.30\textwidth]{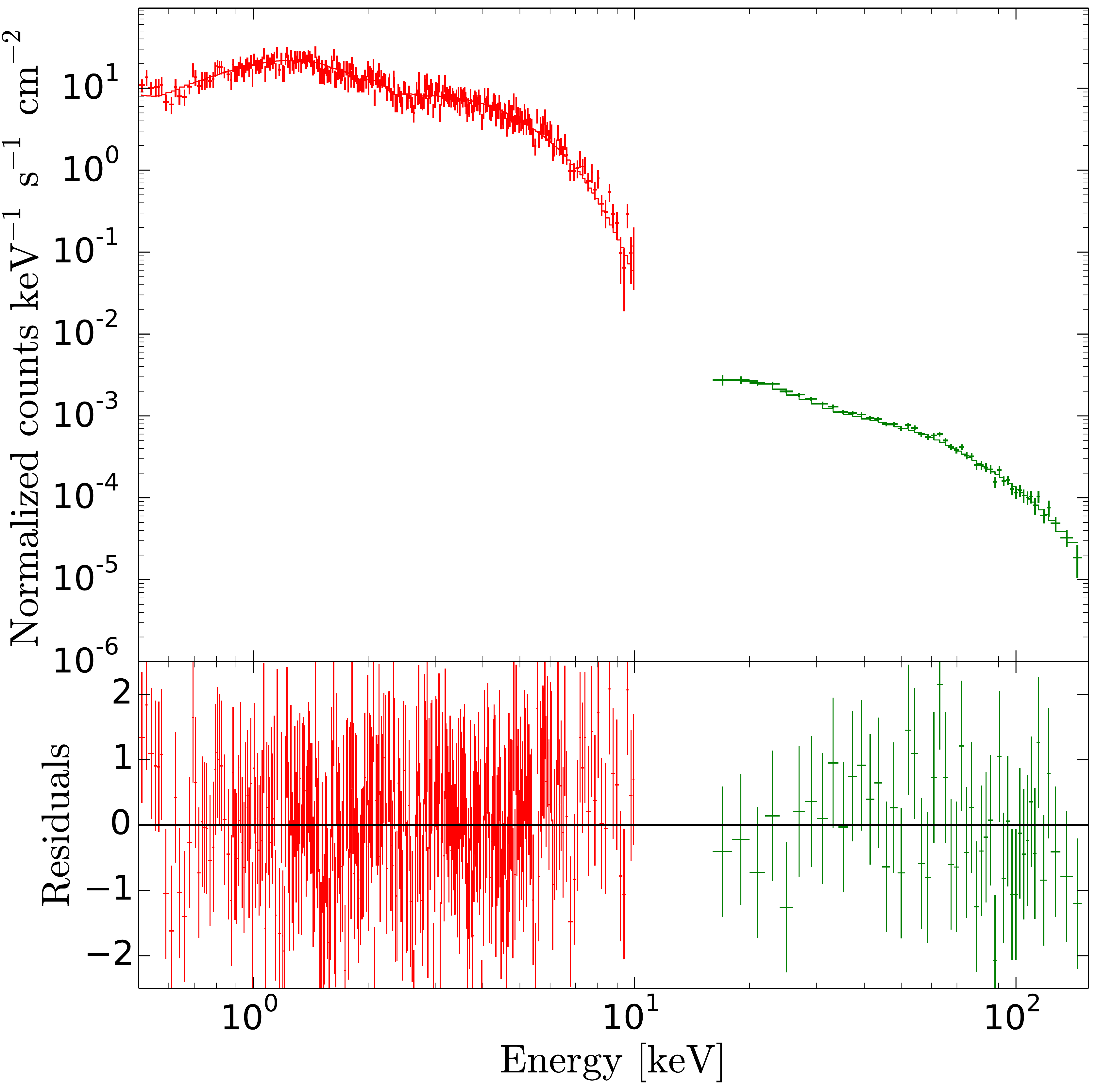} \\
\includegraphics[width = 0.40\textwidth]{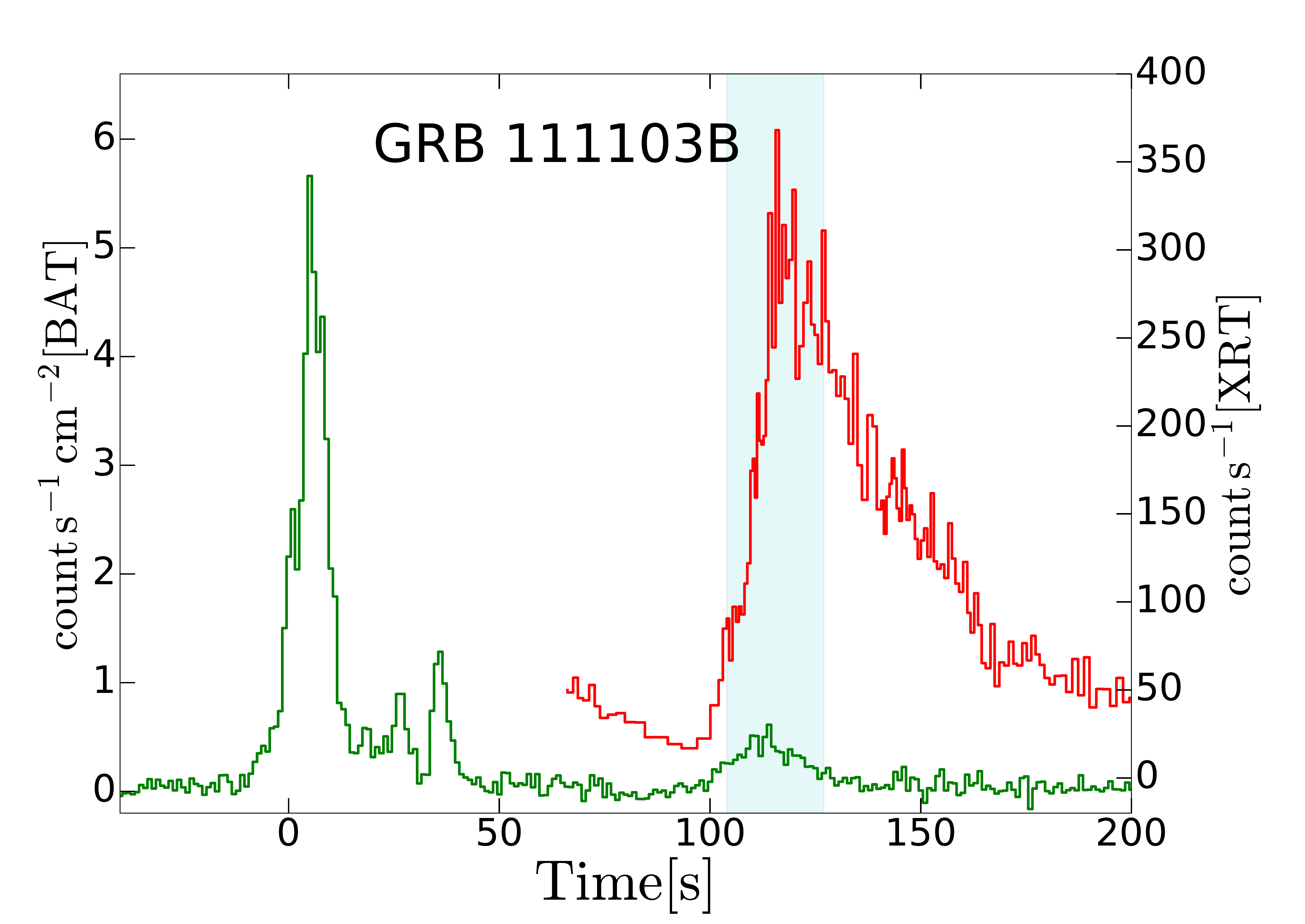} 
\includegraphics[width = 0.30\textwidth]{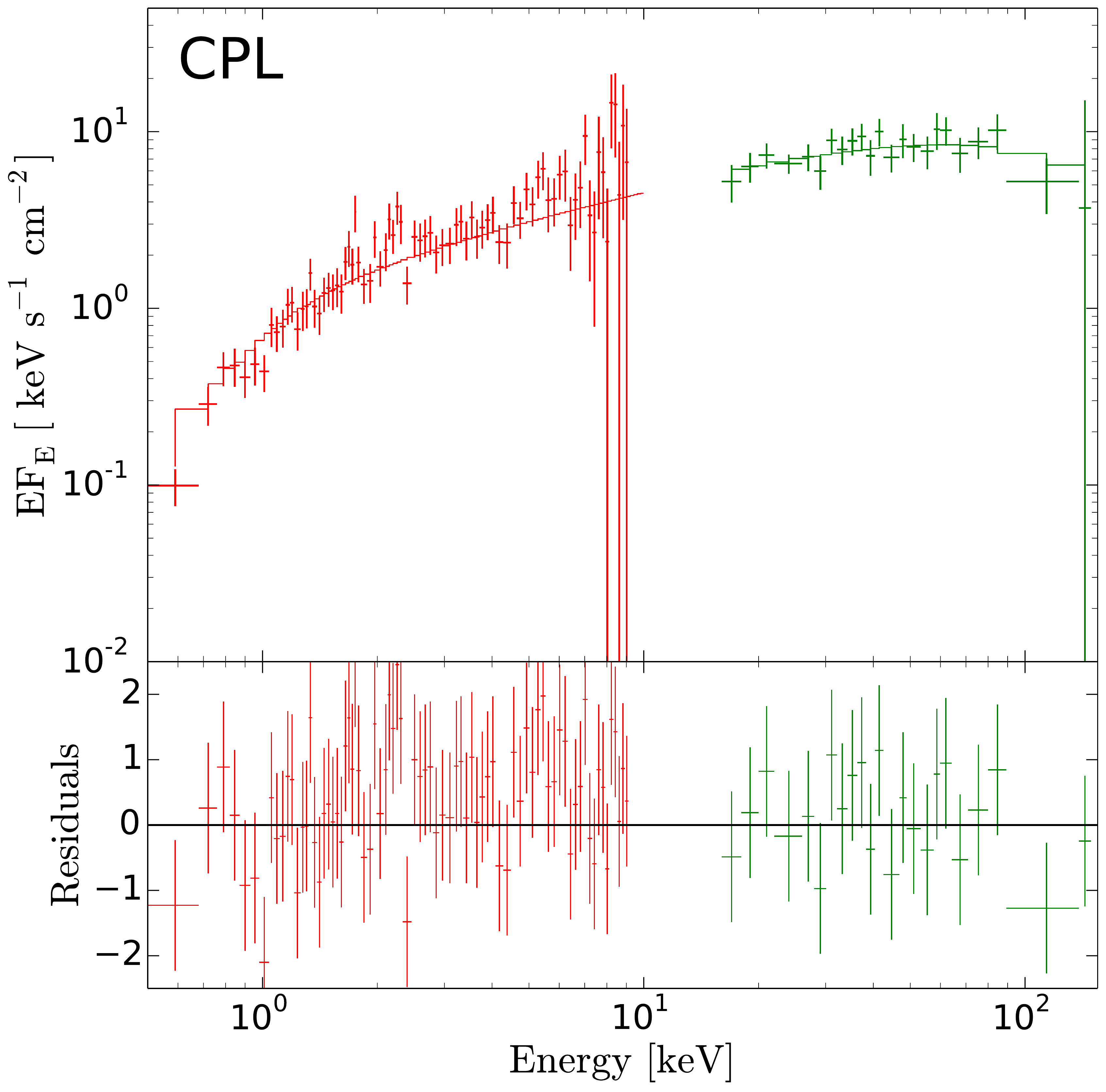} 
\includegraphics[width = 0.30\textwidth]{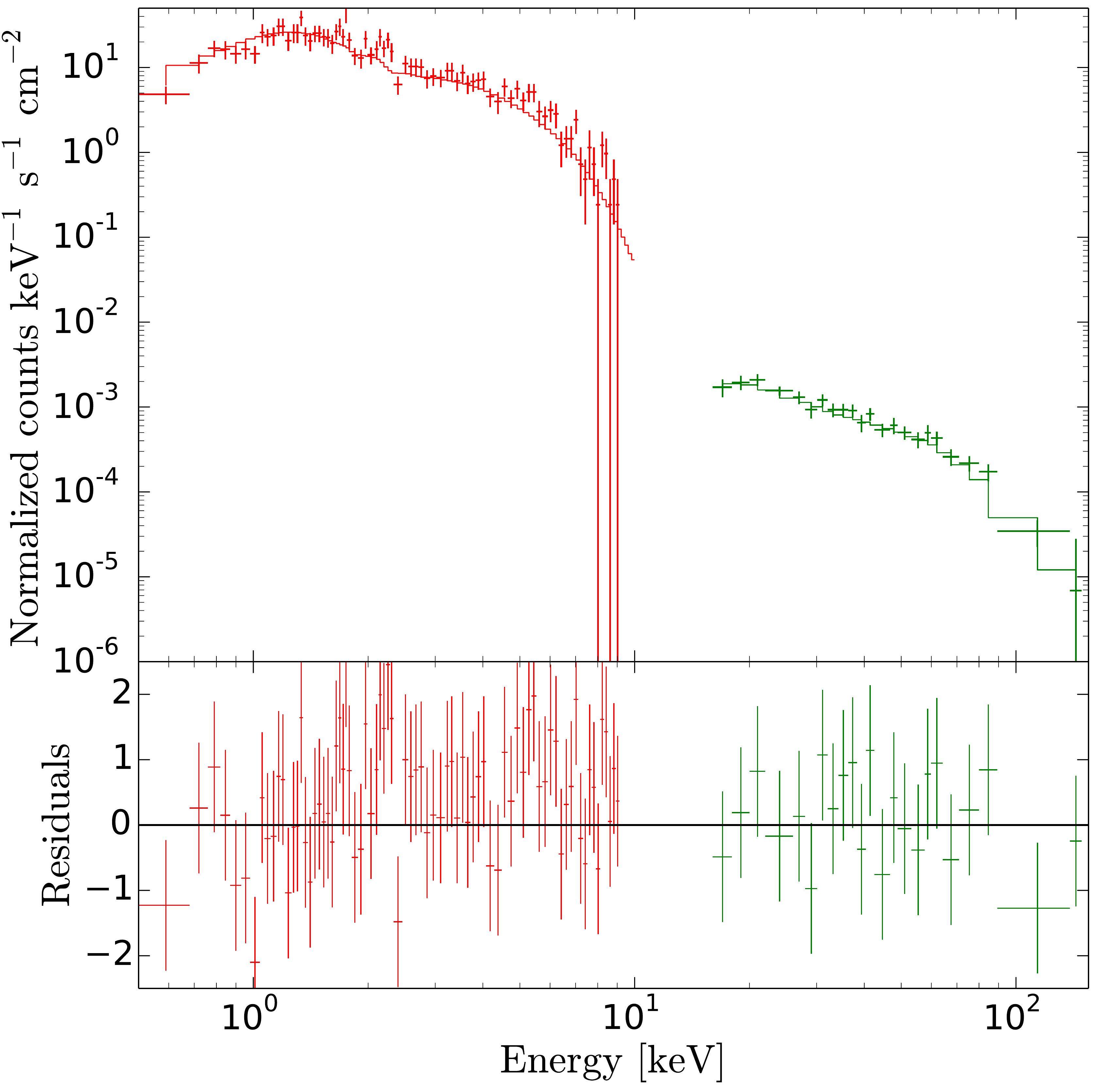} 
\end{figure}

\begin{figure}\ContinuedFloat
\includegraphics[width = 0.40\textwidth]{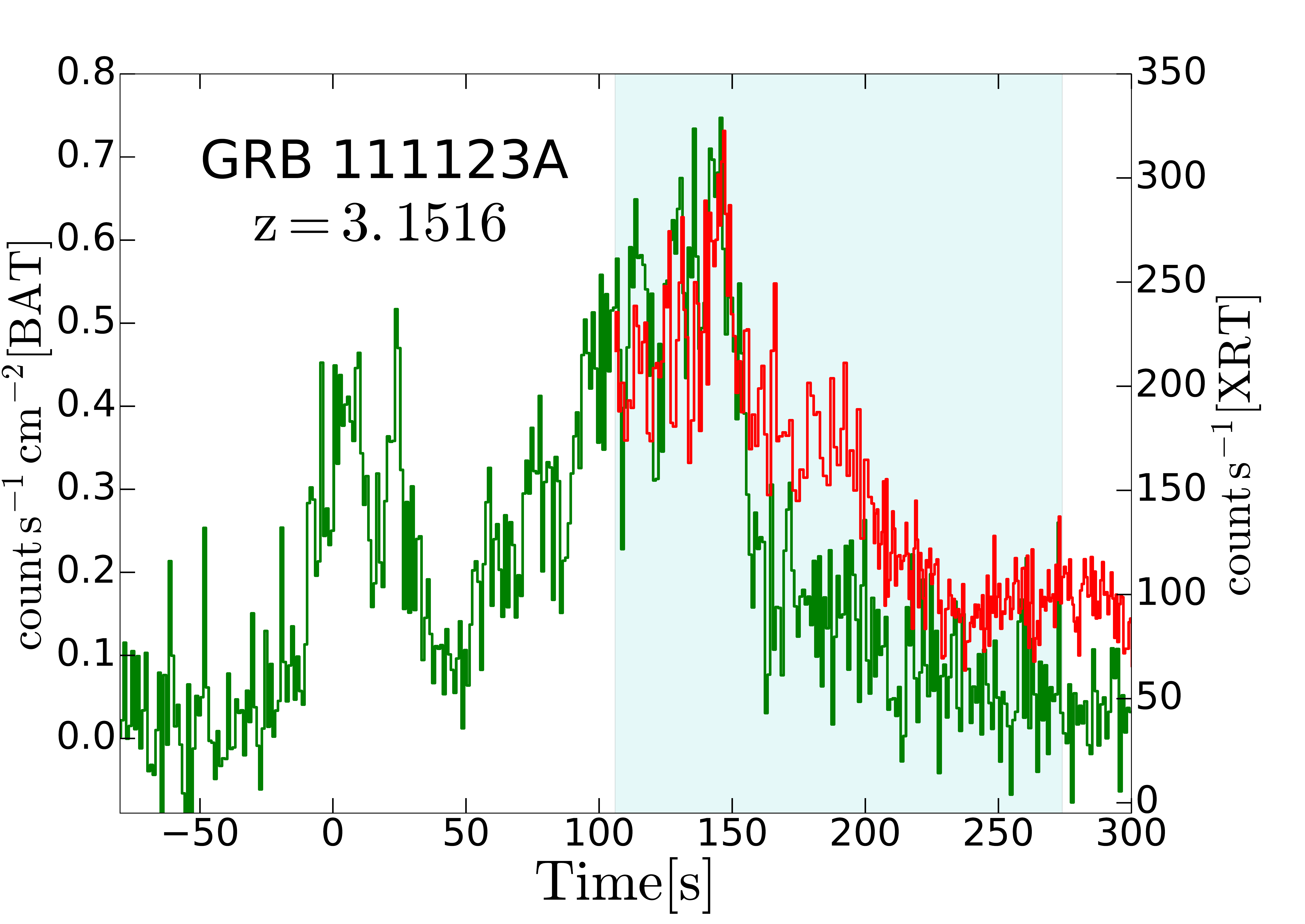}
\includegraphics[width = 0.30\textwidth]{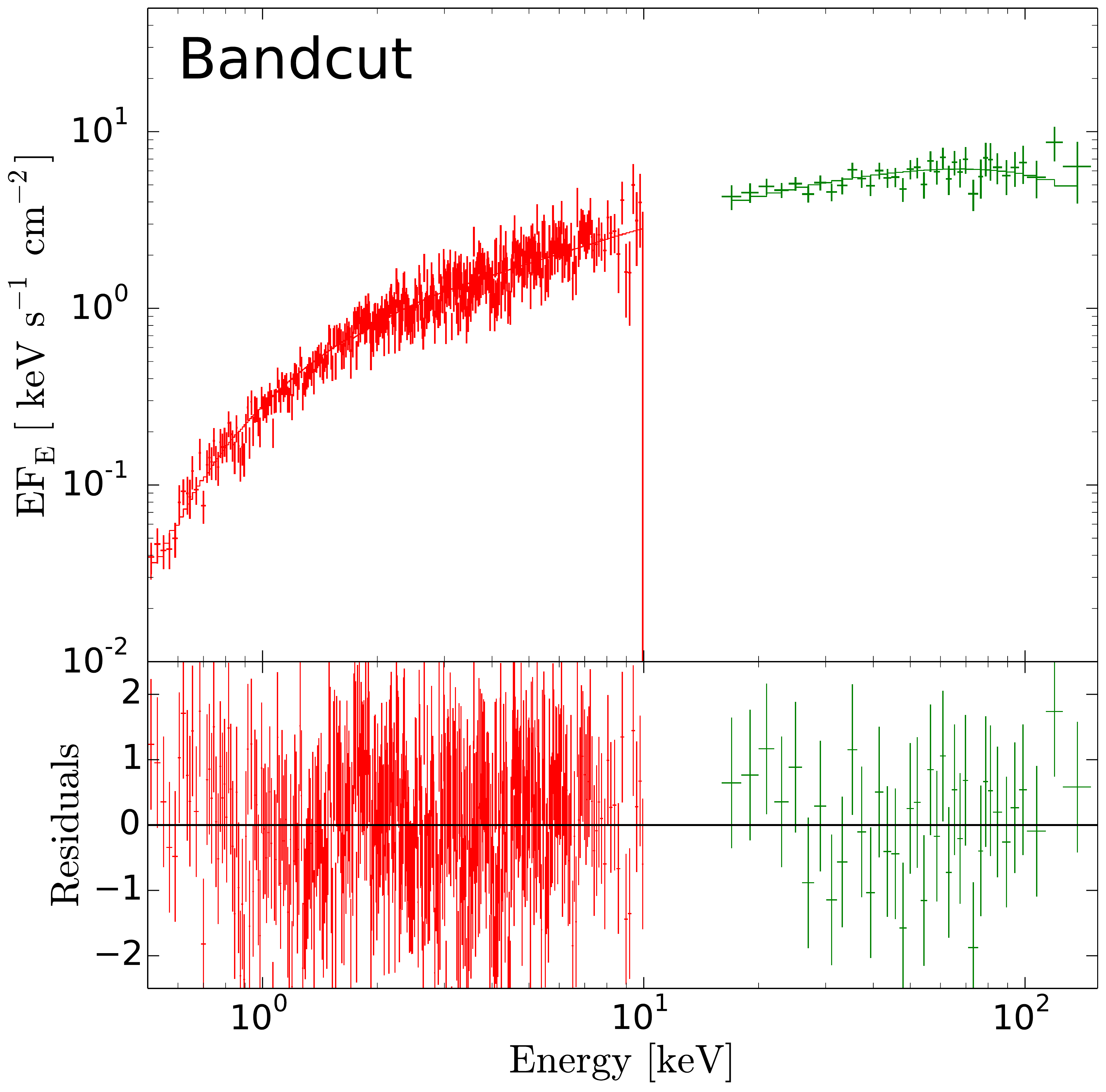} 
\includegraphics[width = 0.30\textwidth]{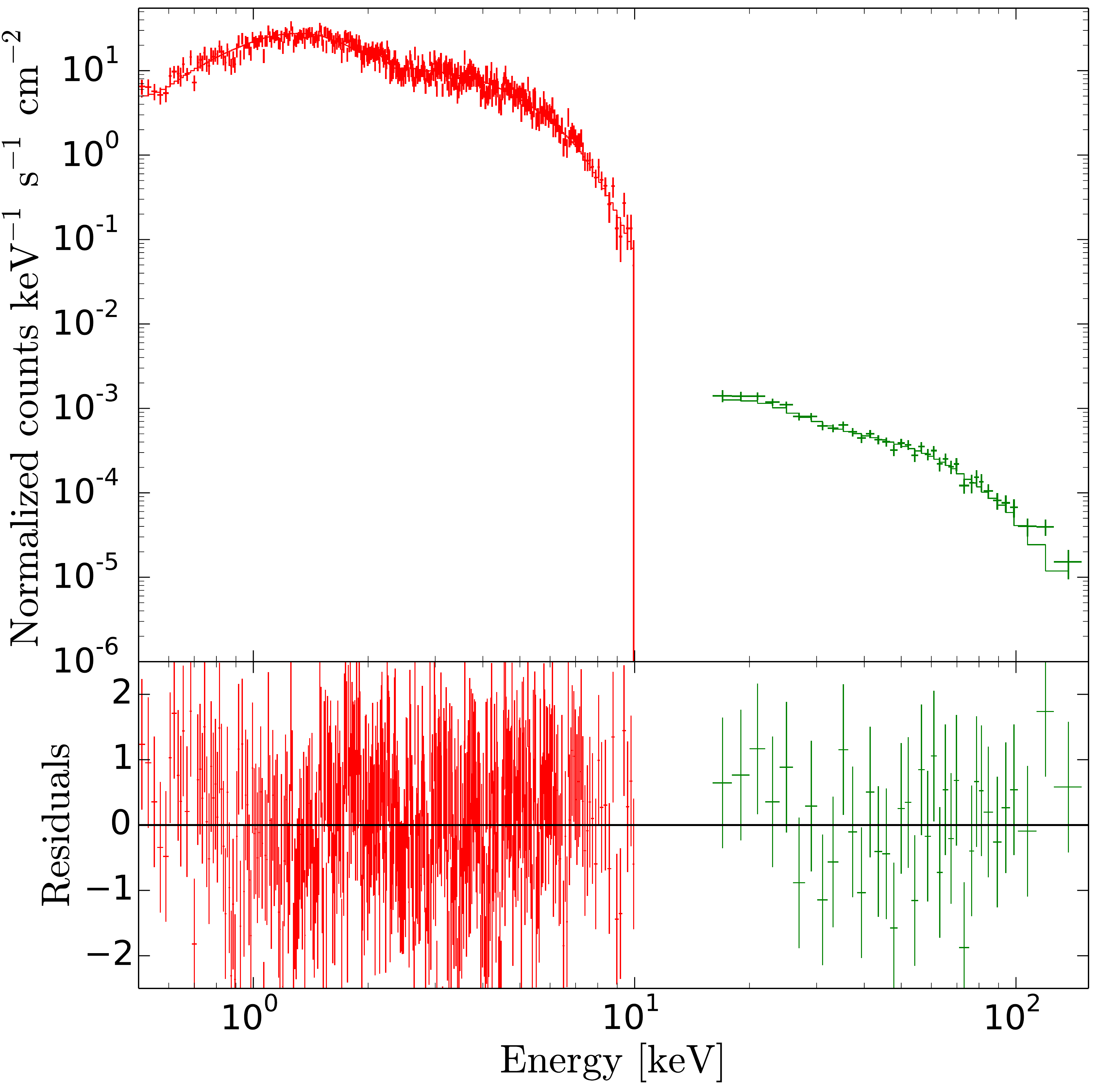} \\
\includegraphics[width = 0.40\textwidth]{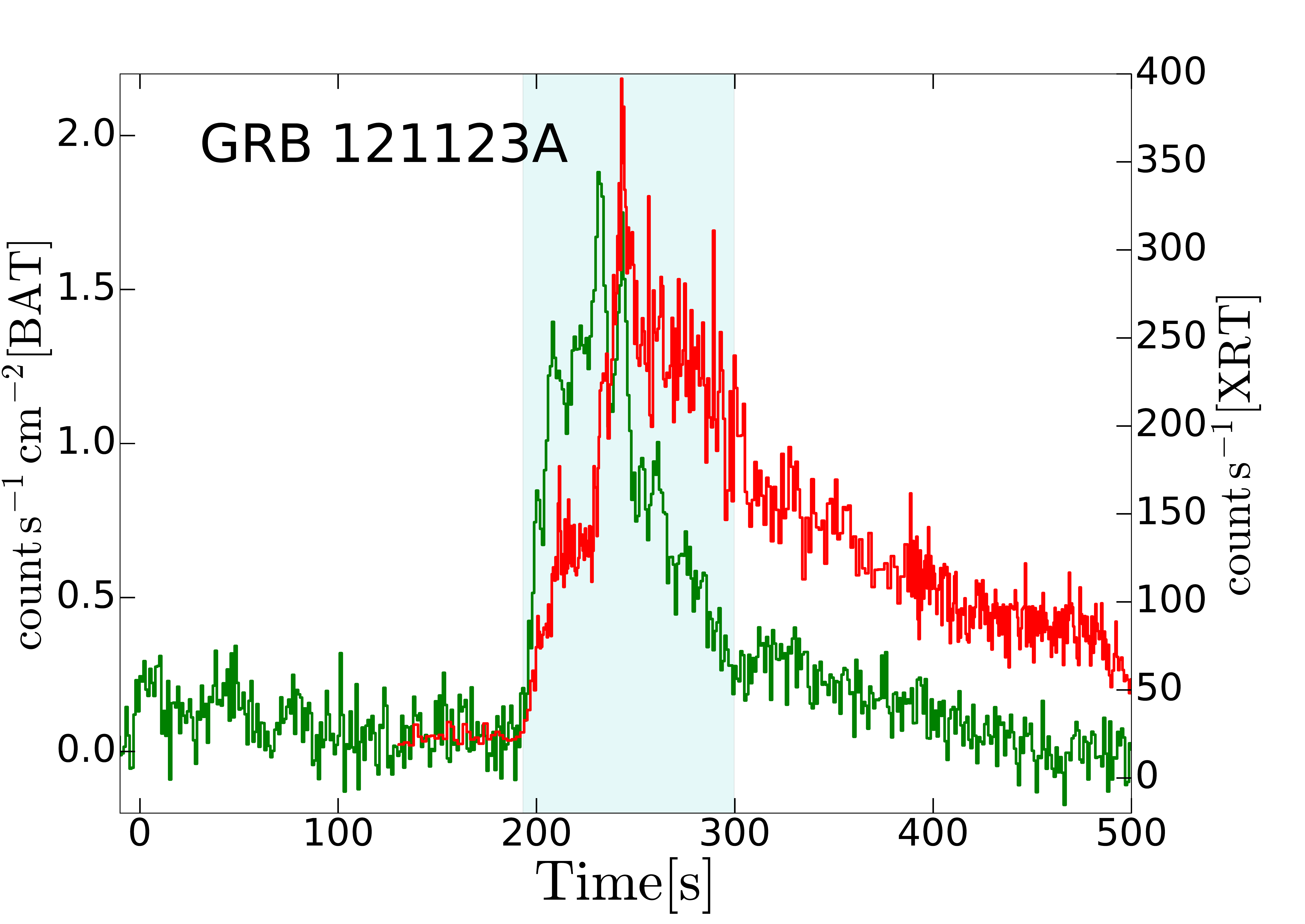} 
\includegraphics[width = 0.30\textwidth]{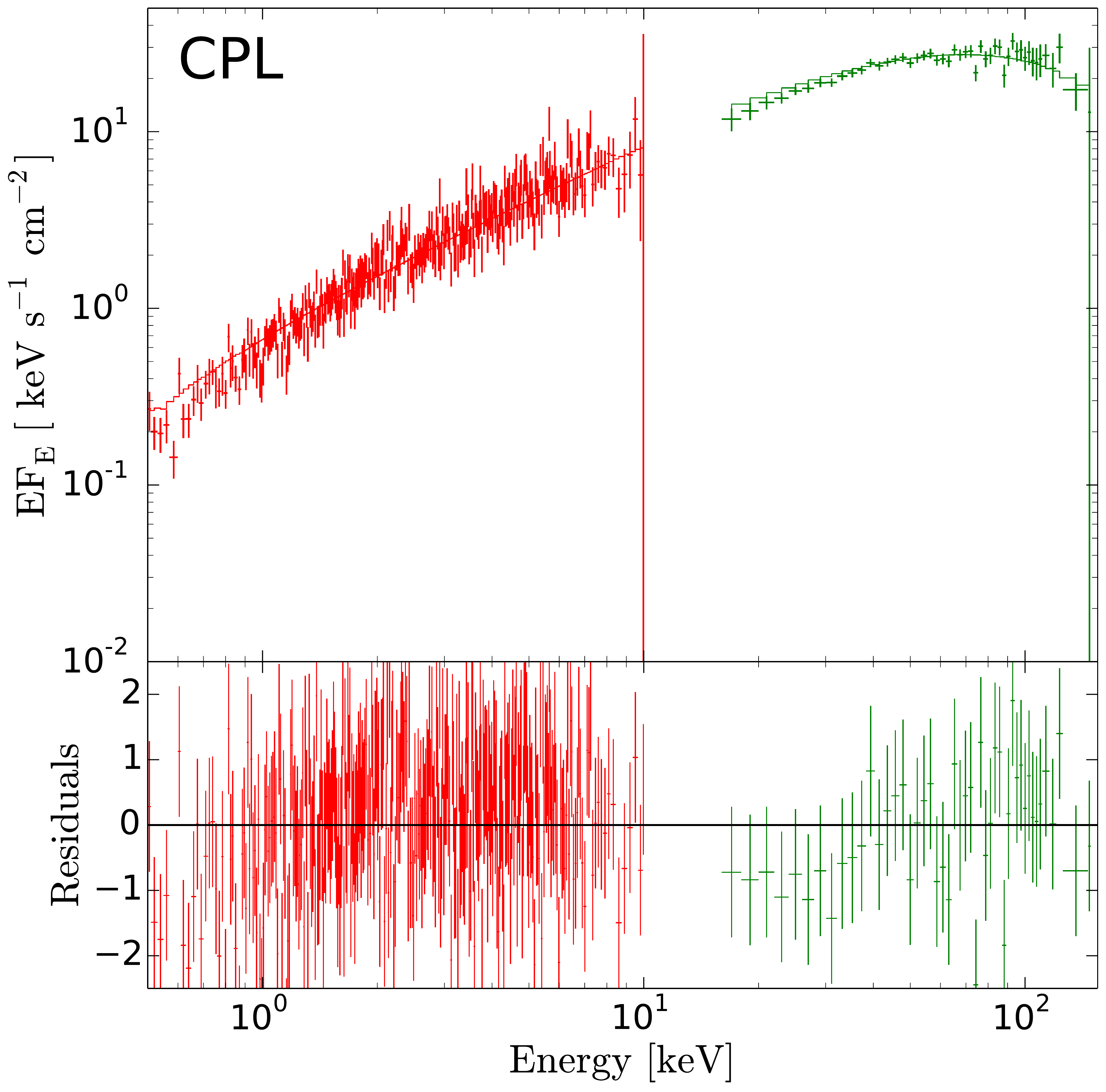} 
\includegraphics[width = 0.30\textwidth]{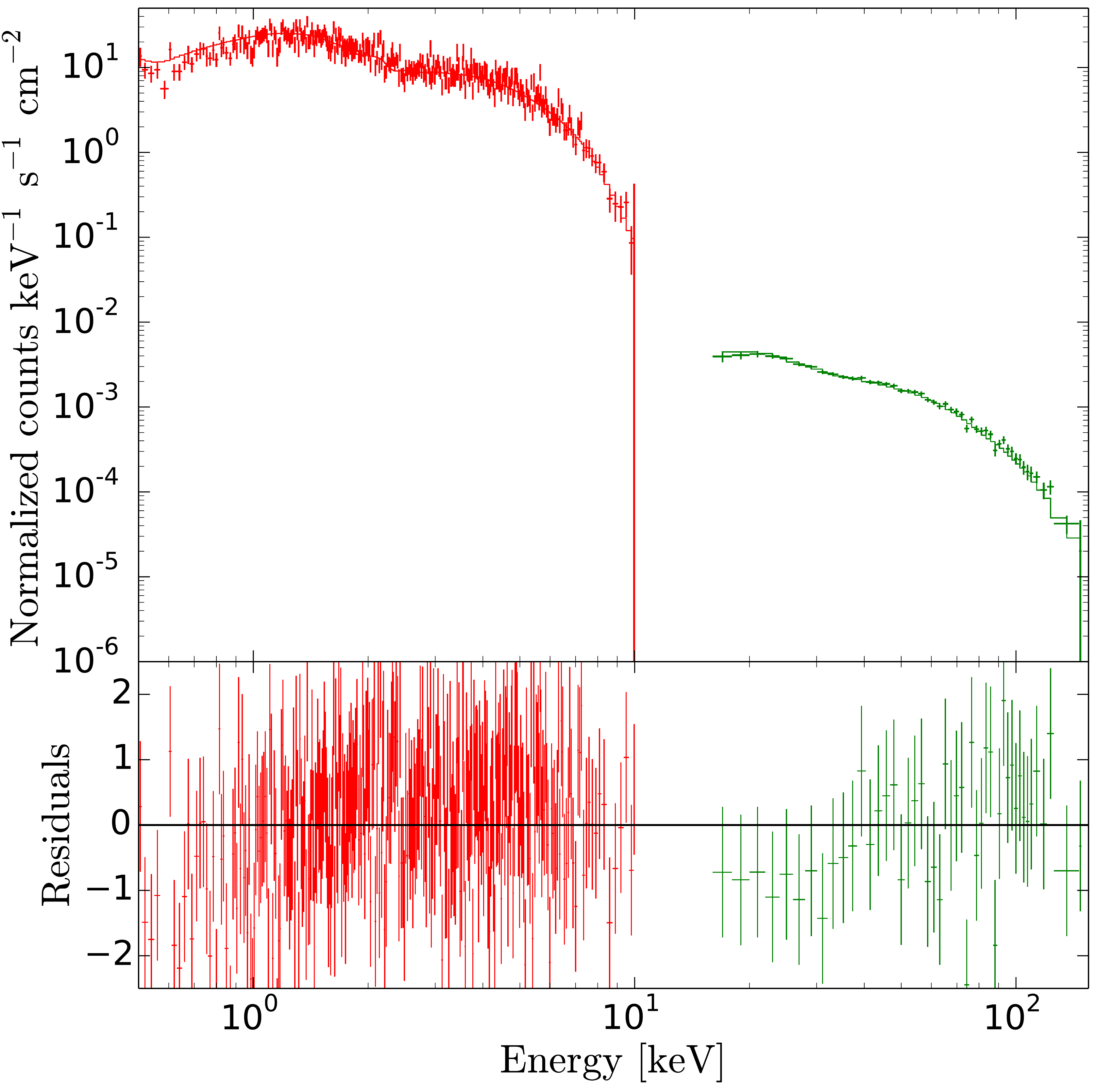} \\
\includegraphics[width = 0.40\textwidth]{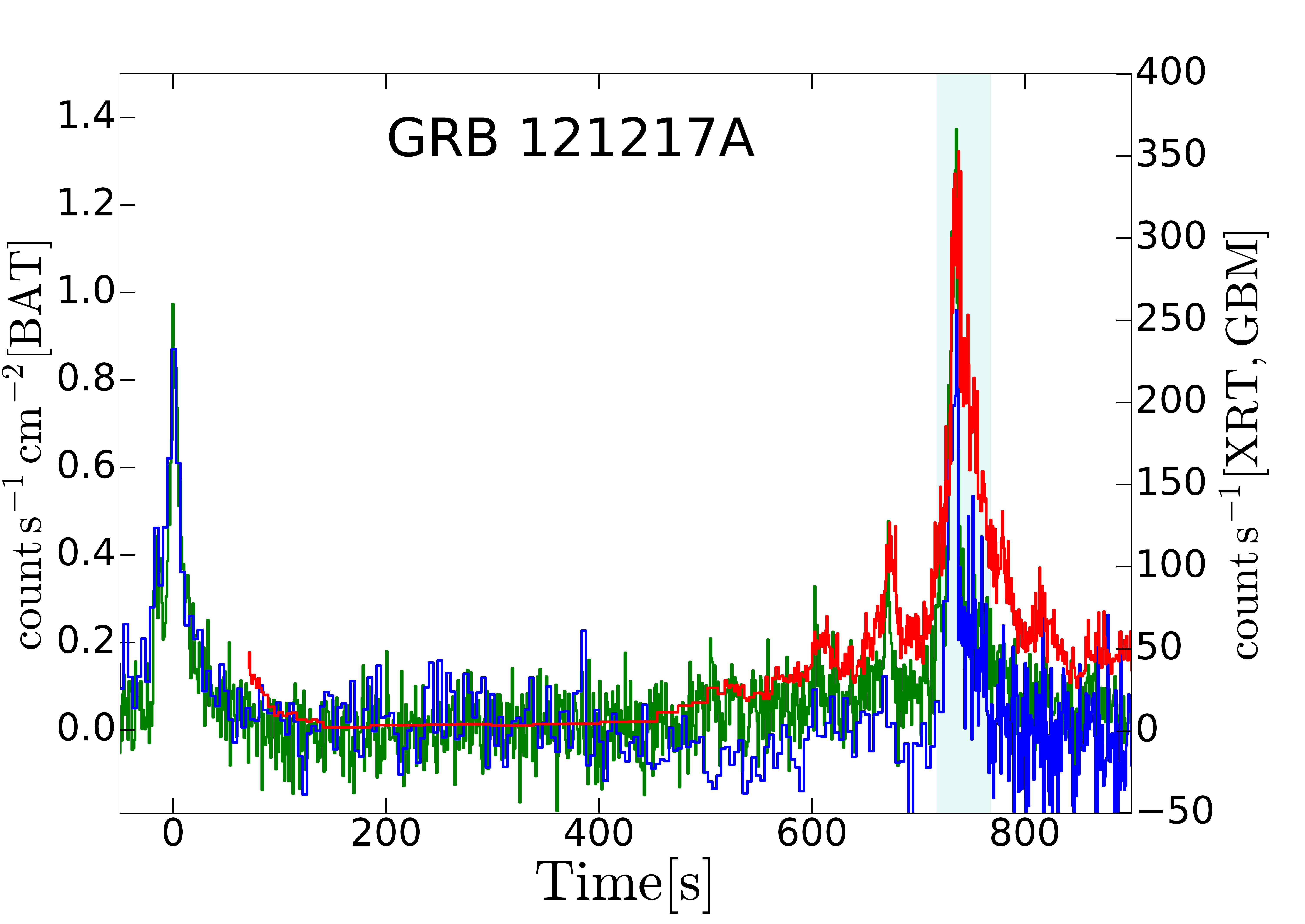} 
\includegraphics[width = 0.30\textwidth]{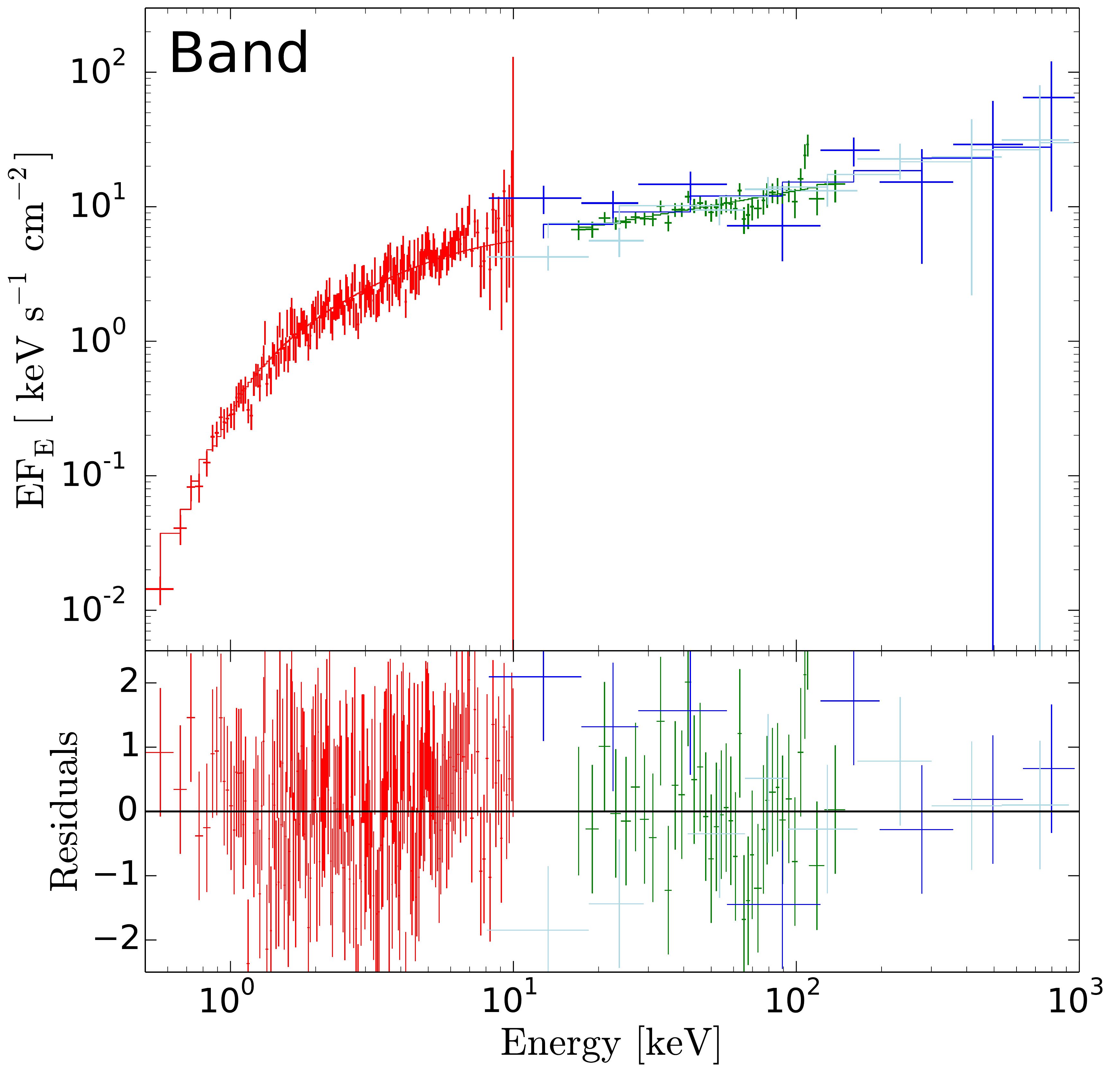} 
\includegraphics[width = 0.30\textwidth]{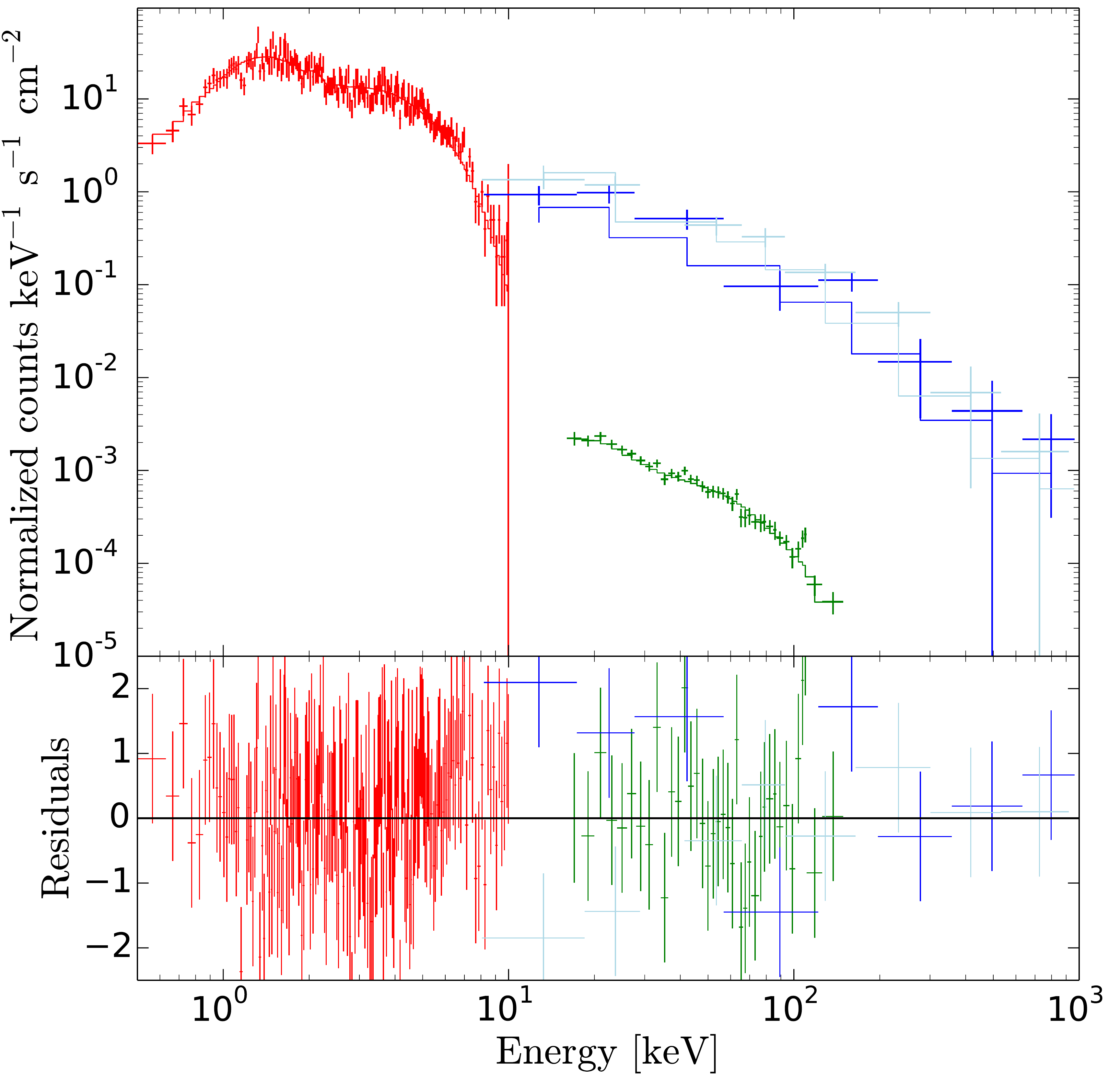} 
\end{figure}

\begin{figure}\ContinuedFloat
\includegraphics[width = 0.40\textwidth]{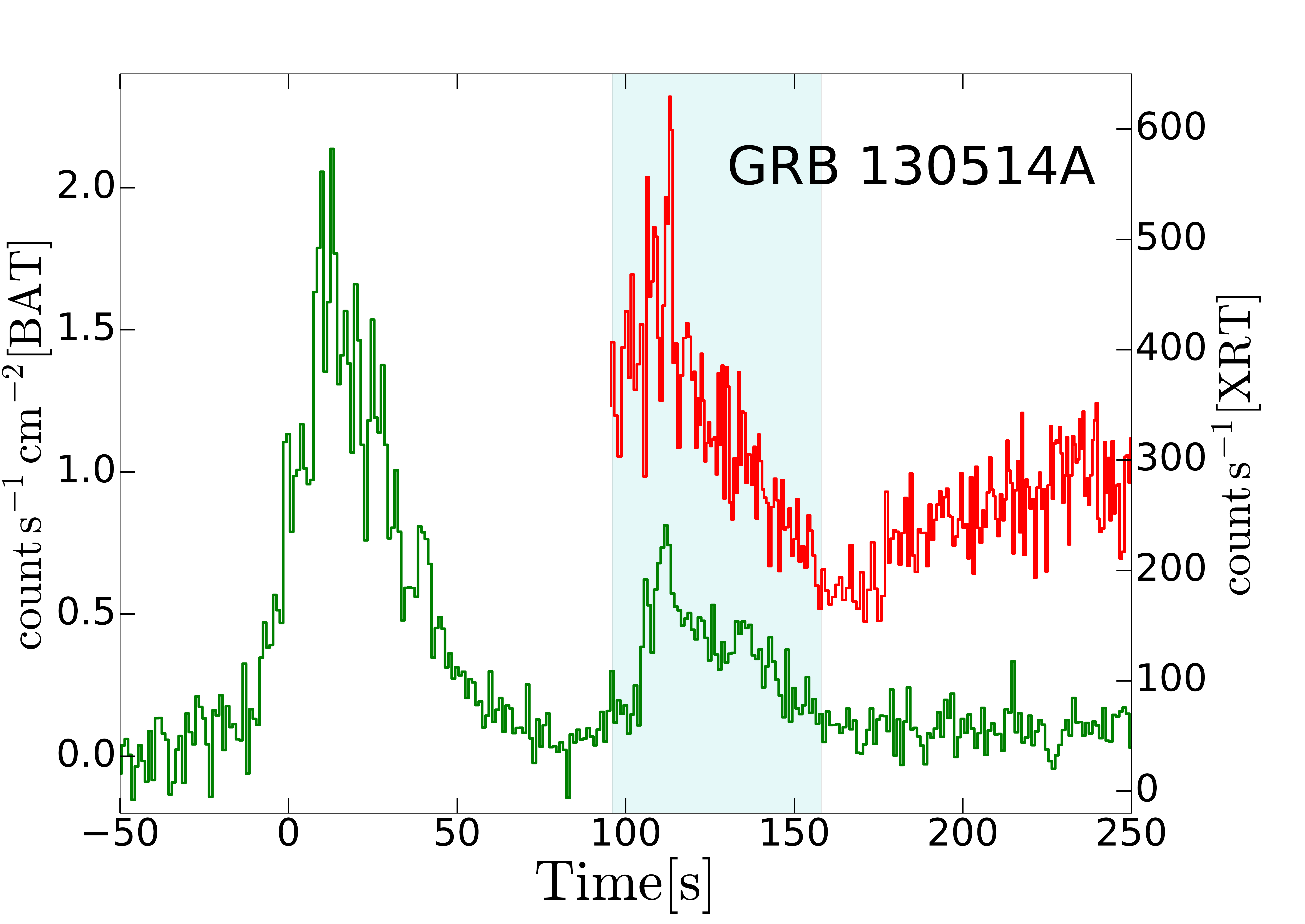}
\includegraphics[width = 0.30\textwidth]{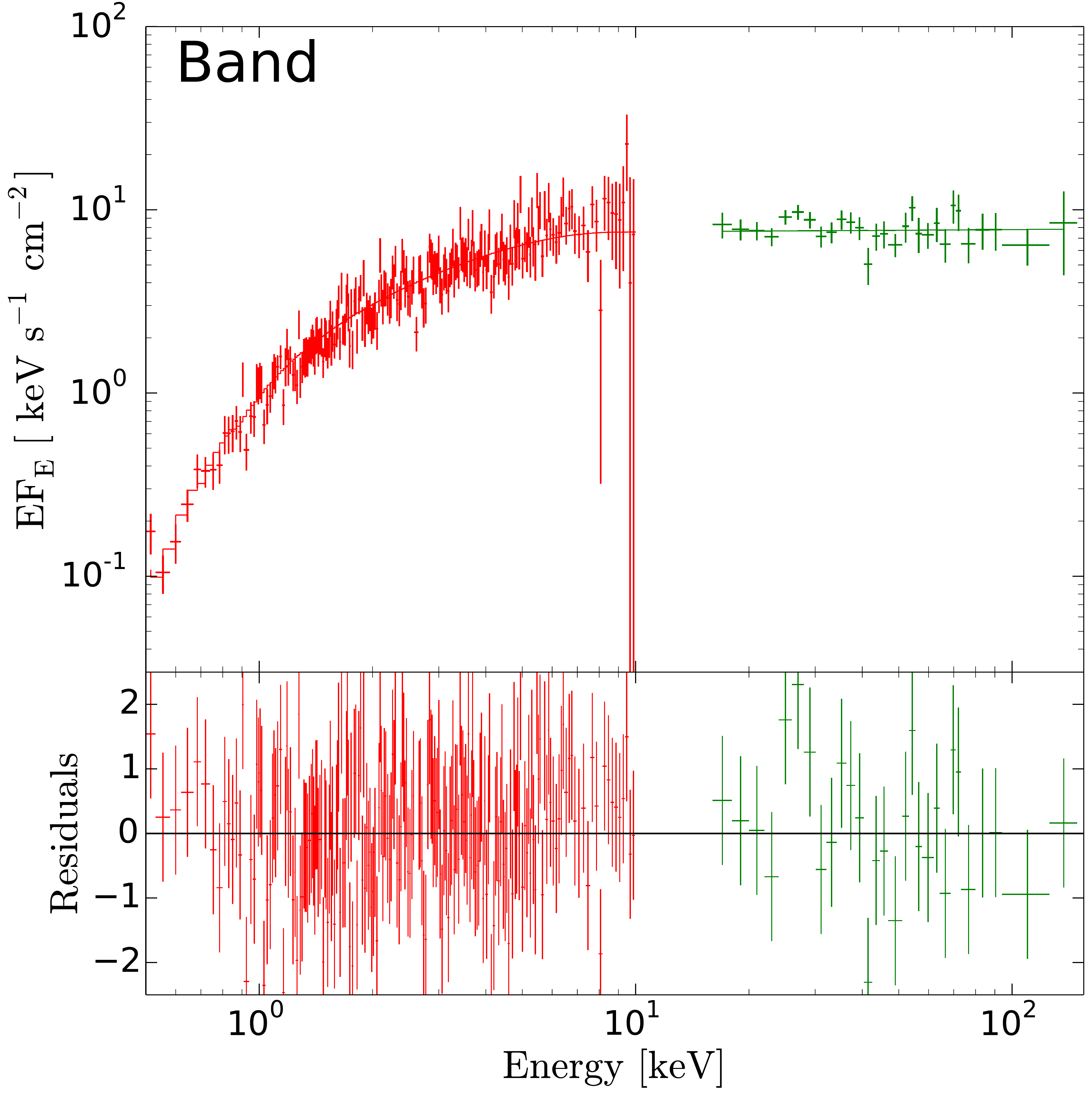} 
\includegraphics[width = 0.30\textwidth]{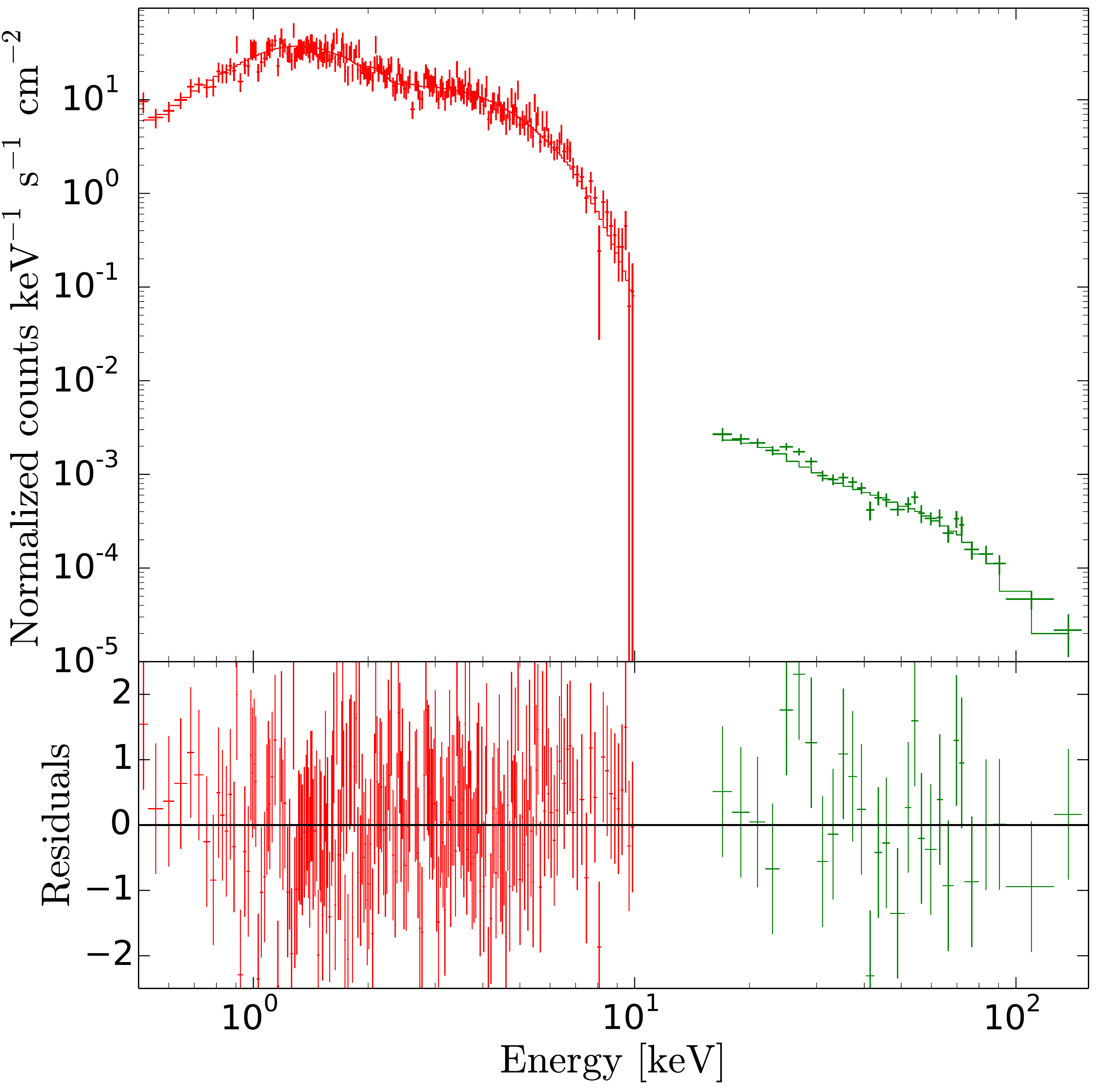} \\
\includegraphics[width = 0.40\textwidth]{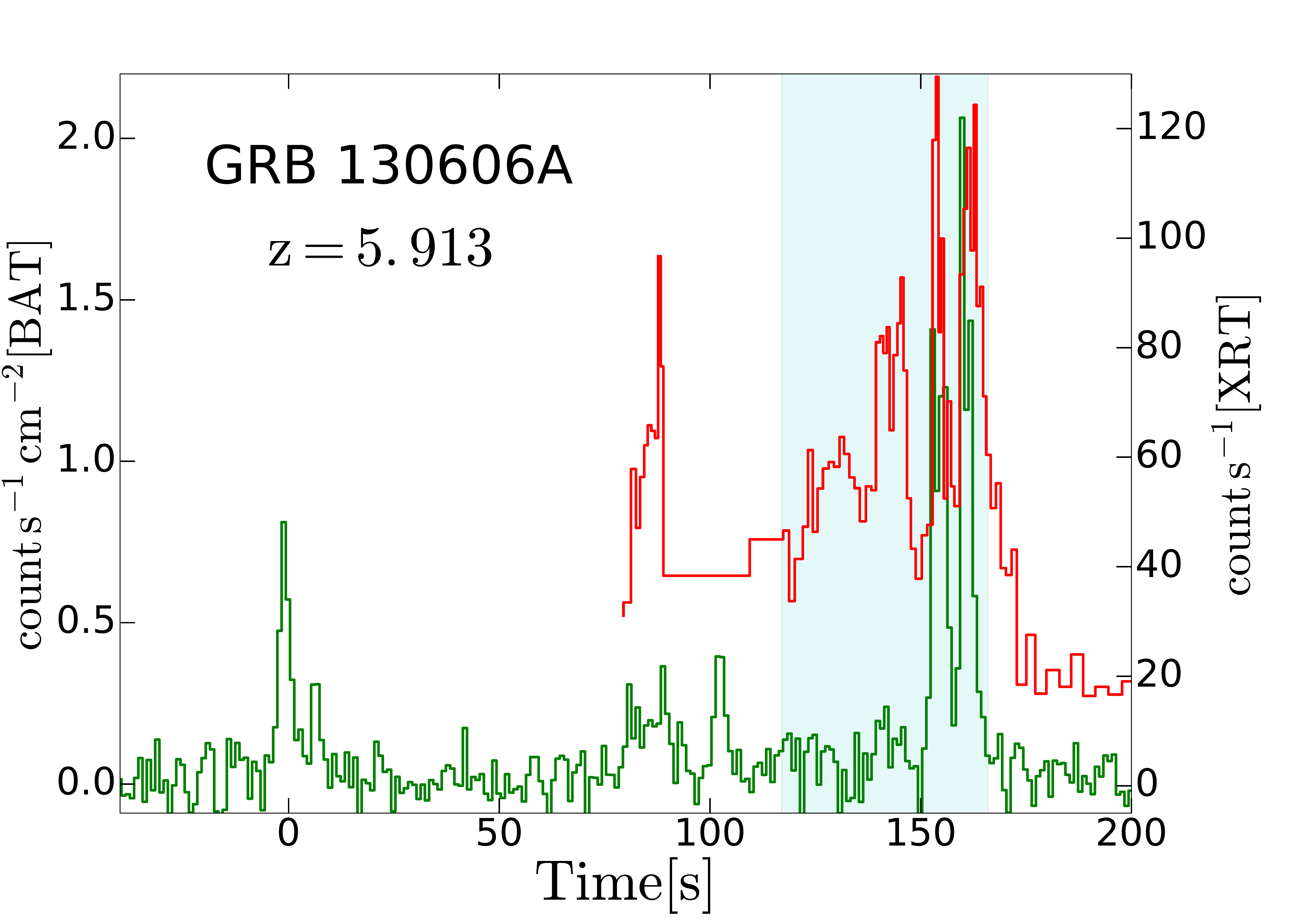} 
\includegraphics[width = 0.30\textwidth]{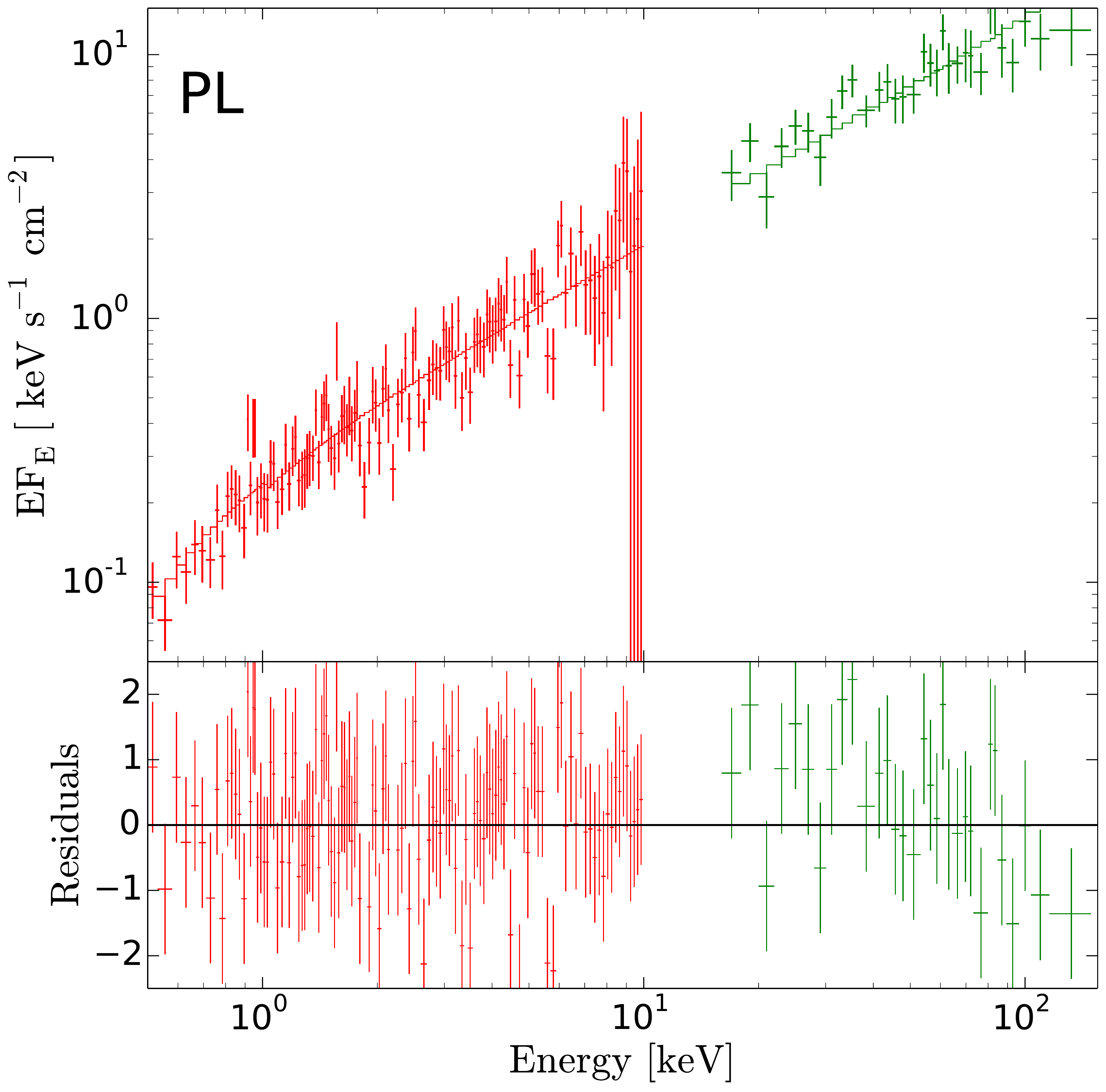} 
\includegraphics[width = 0.30\textwidth]{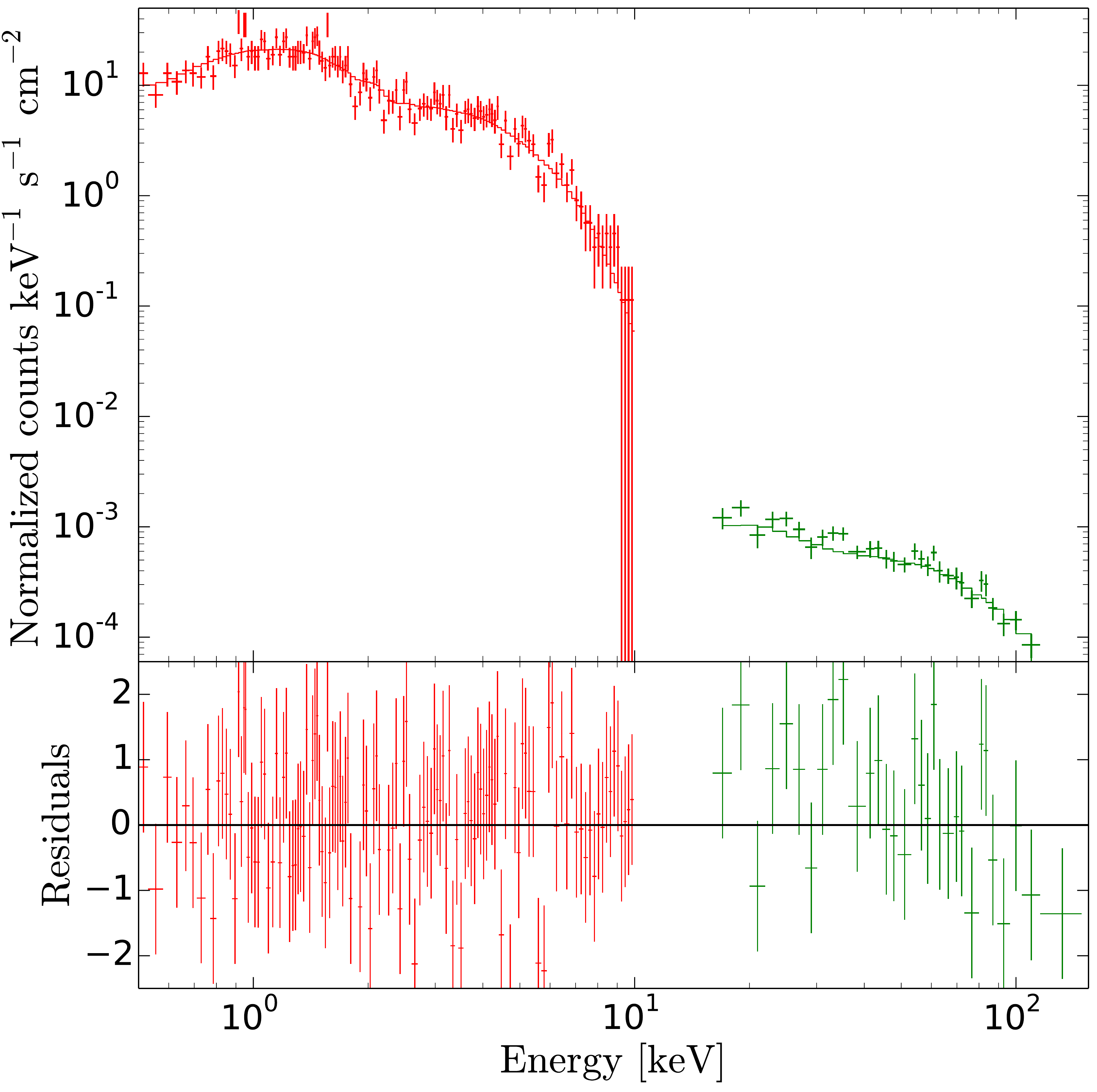} \\
\includegraphics[width = 0.40\textwidth]{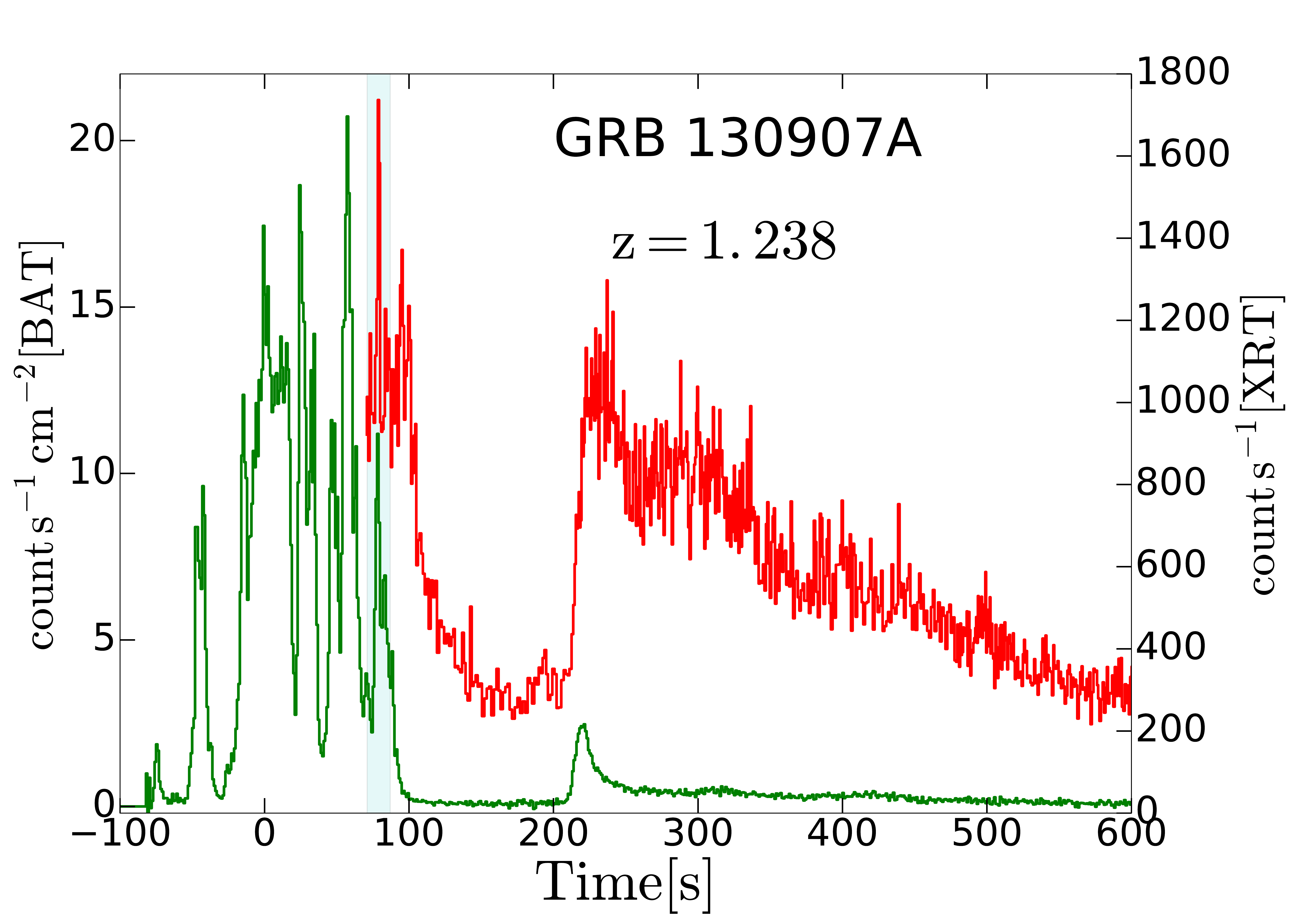} 
\includegraphics[width = 0.30\textwidth]{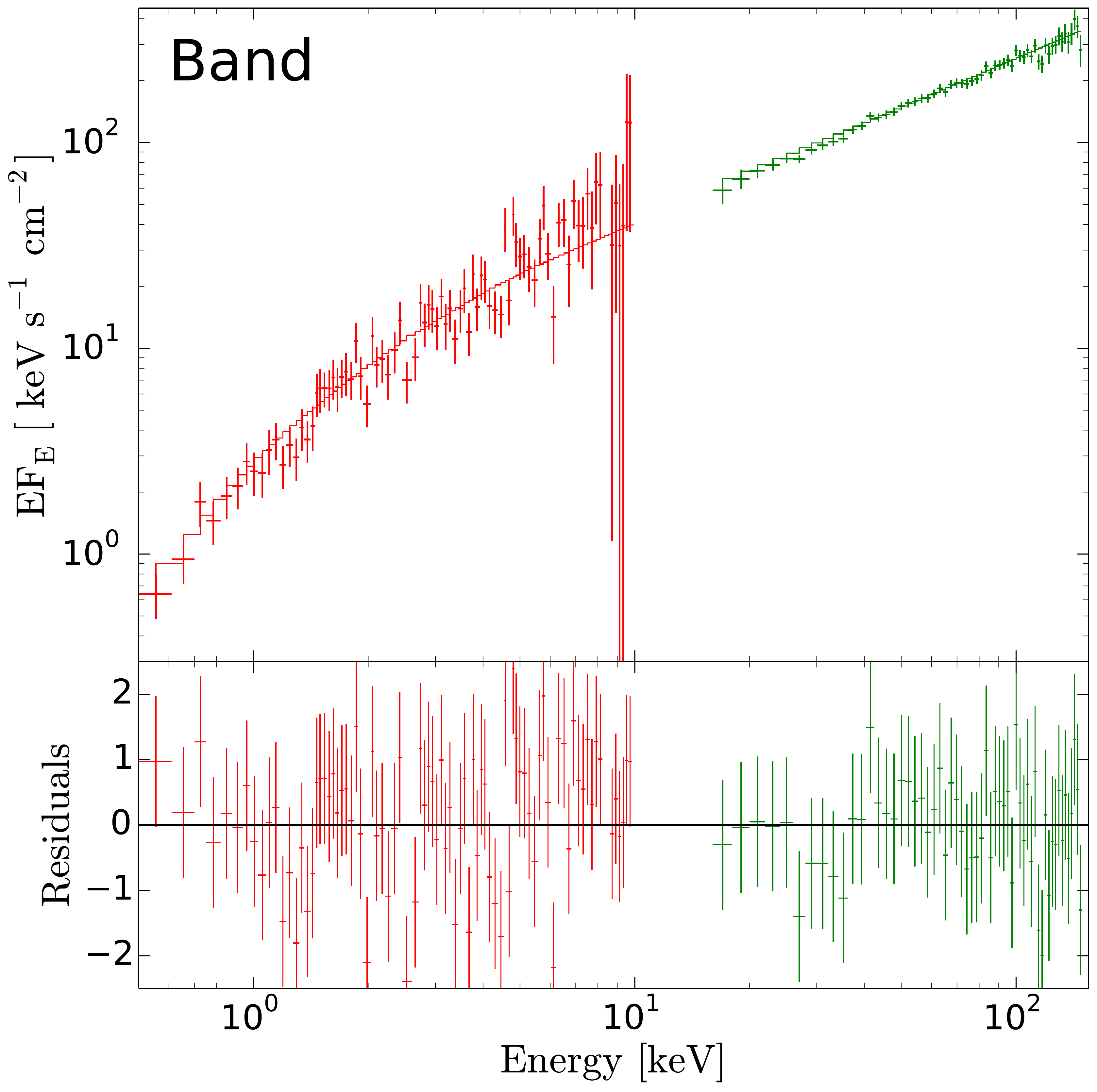} 
\includegraphics[width = 0.30\textwidth]{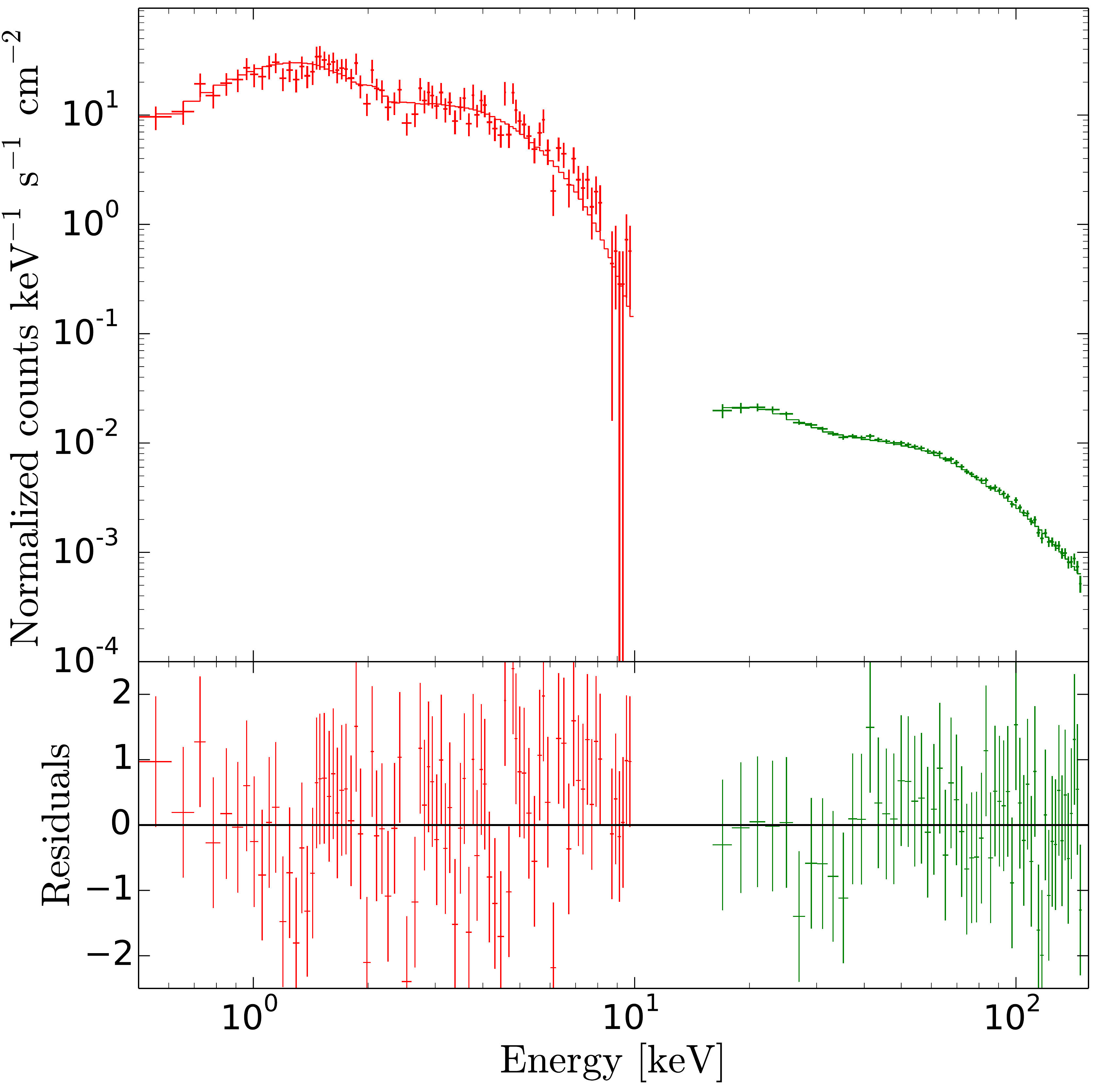} 
\end{figure}

\begin{figure}\ContinuedFloat
\includegraphics[width = 0.40\textwidth]{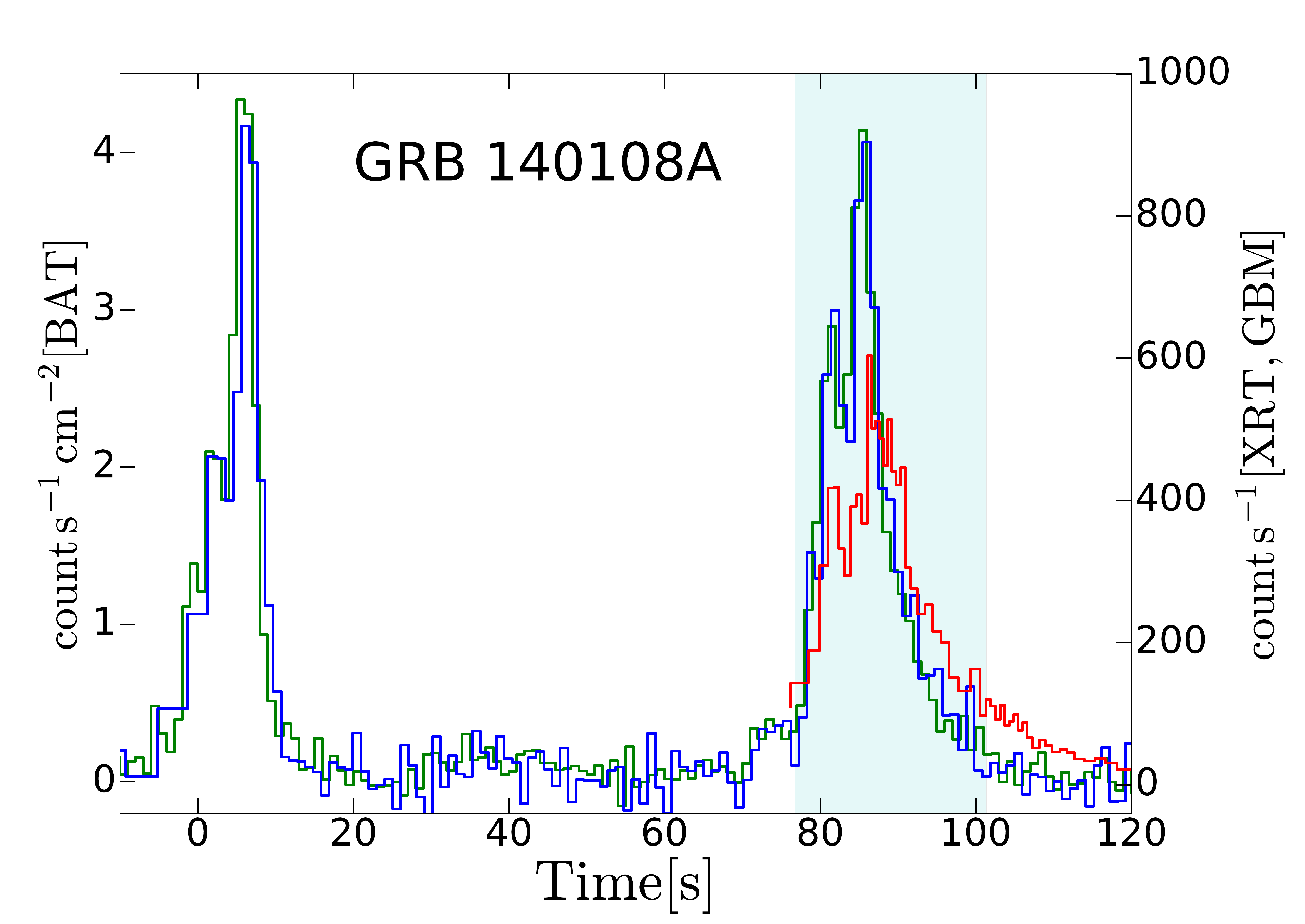}
\includegraphics[width = 0.30\textwidth]{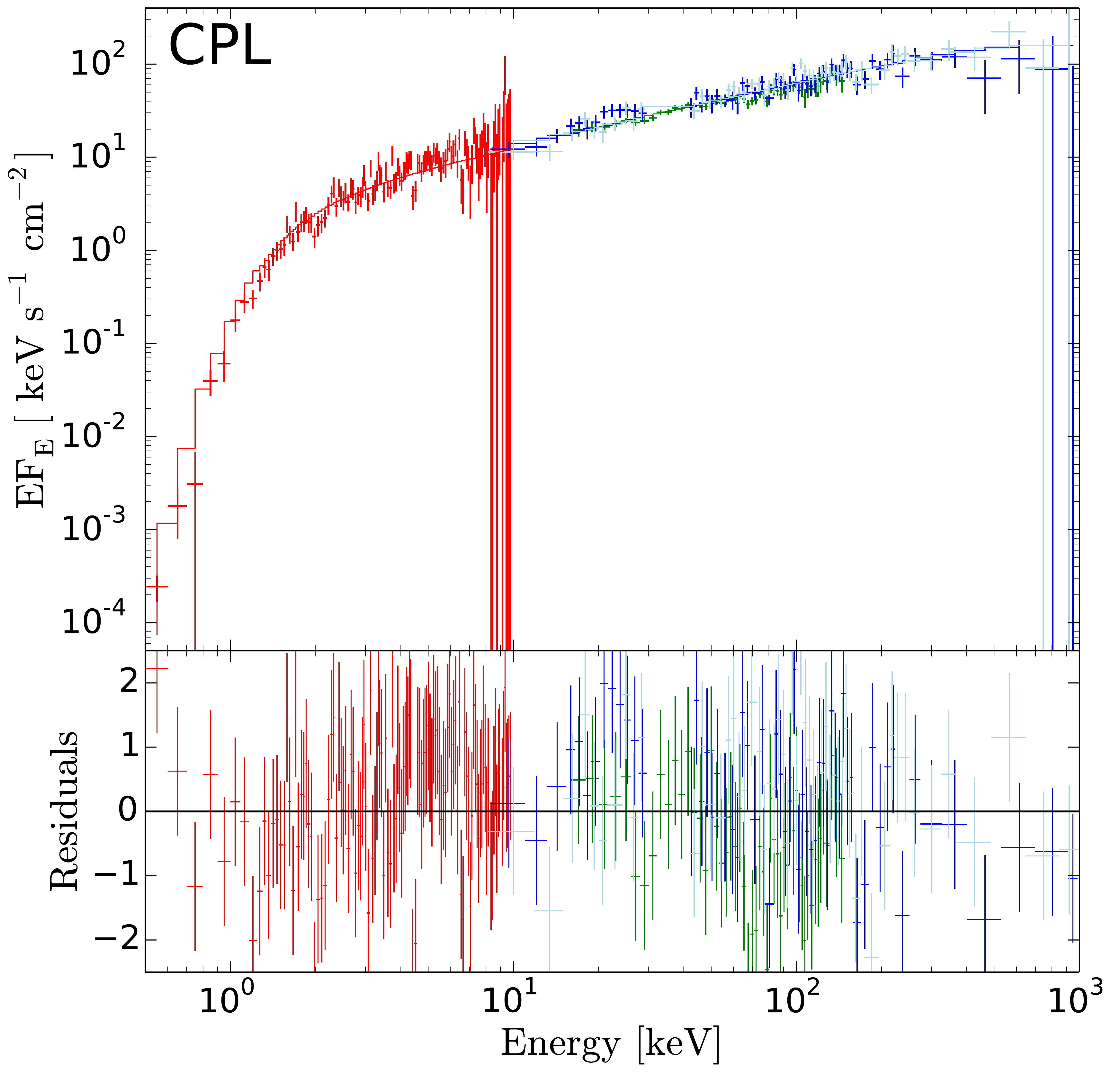} 
\includegraphics[width = 0.30\textwidth]{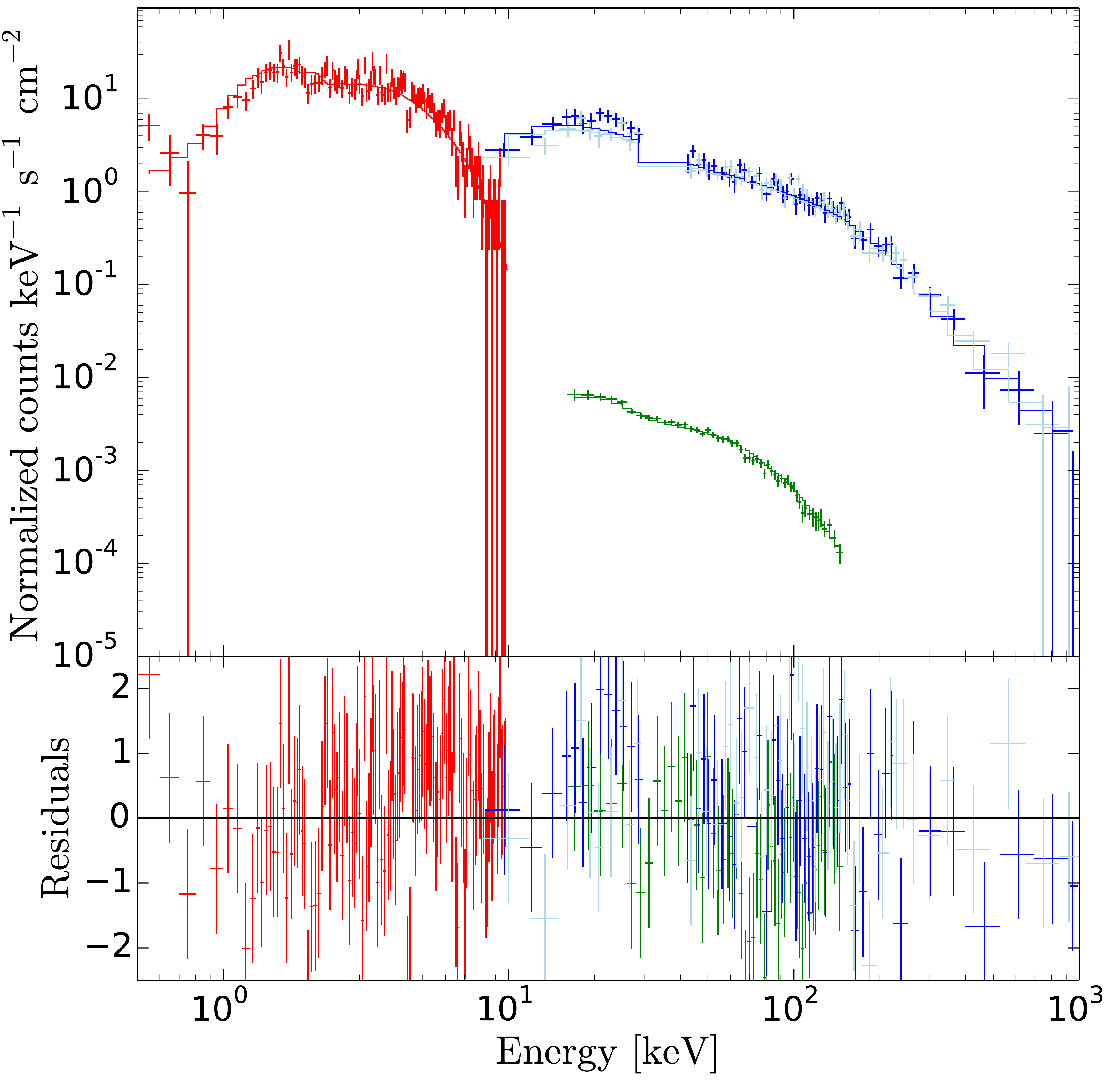} \\
\includegraphics[width = 0.40\textwidth]{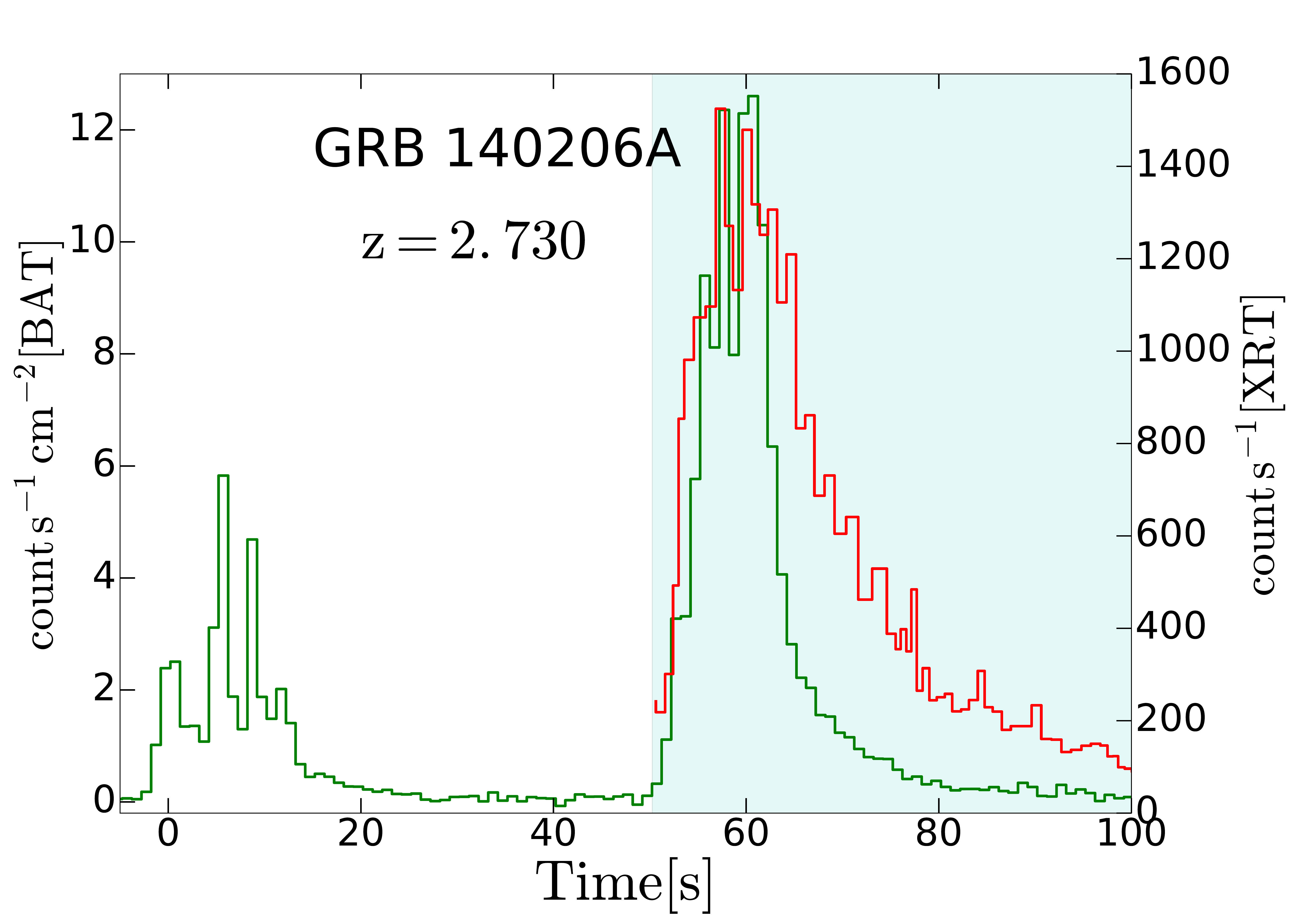}
\includegraphics[width = 0.30\textwidth]{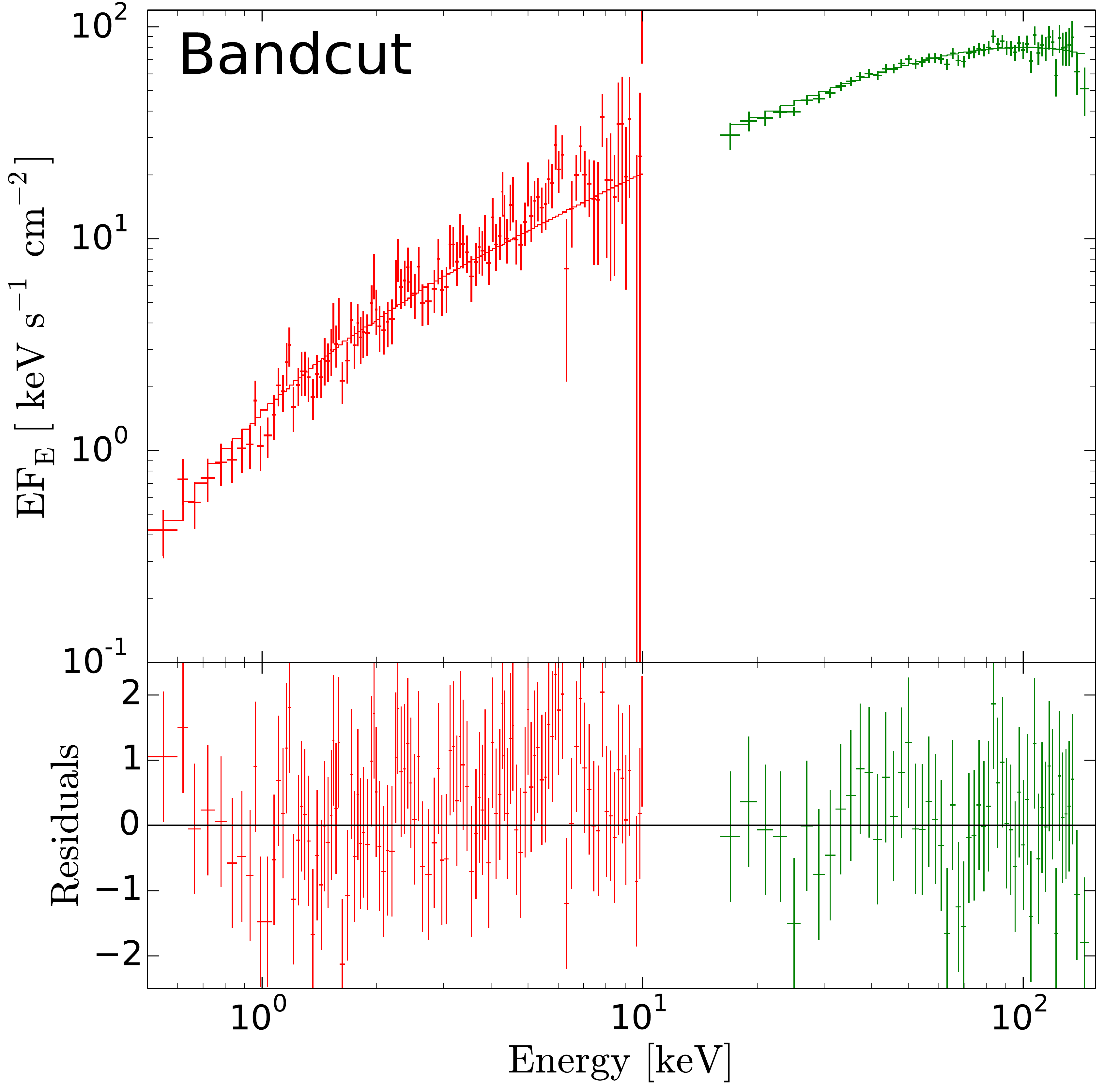} 
\includegraphics[width = 0.30\textwidth]{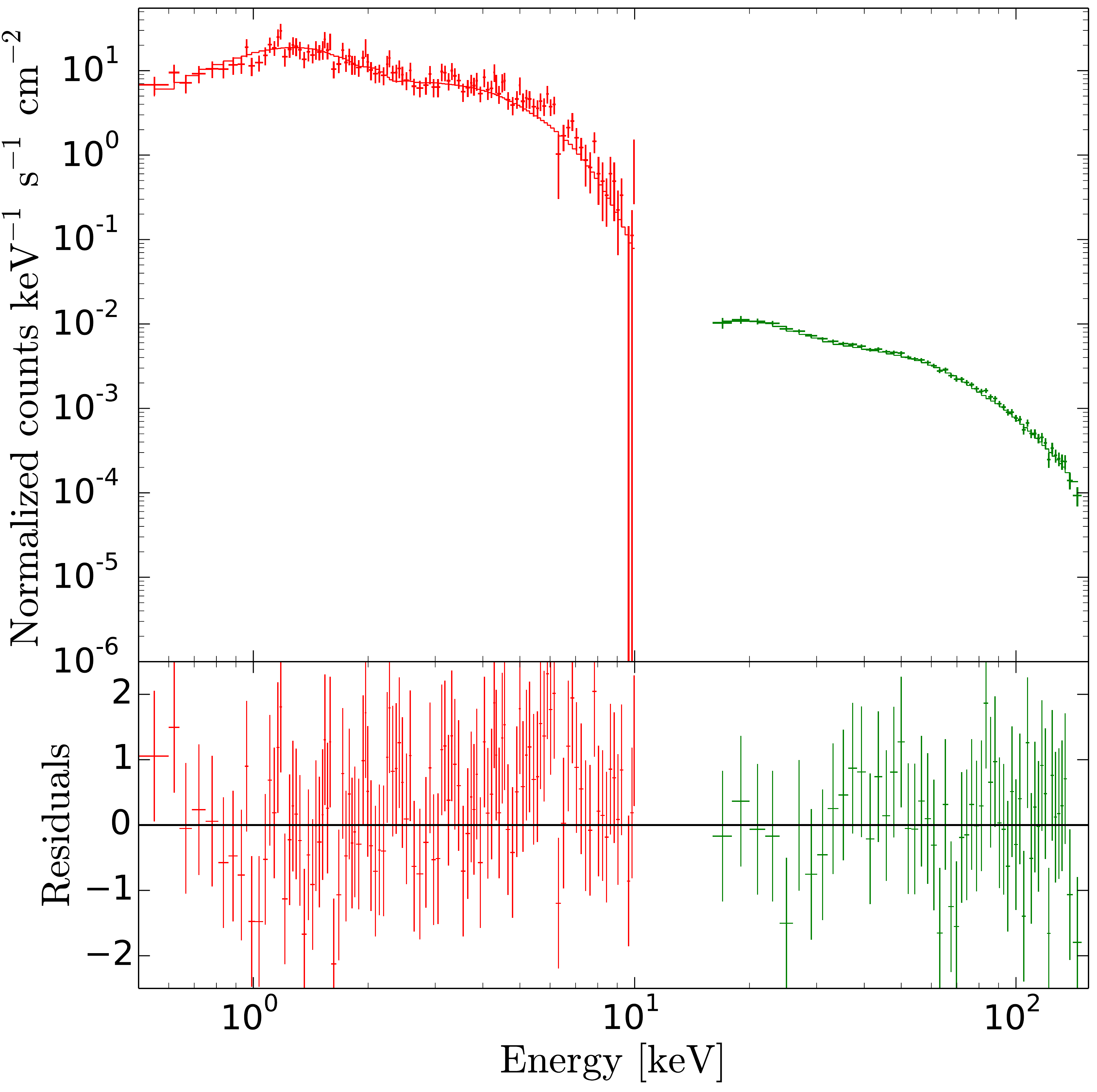} \\
\includegraphics[width = 0.40\textwidth]{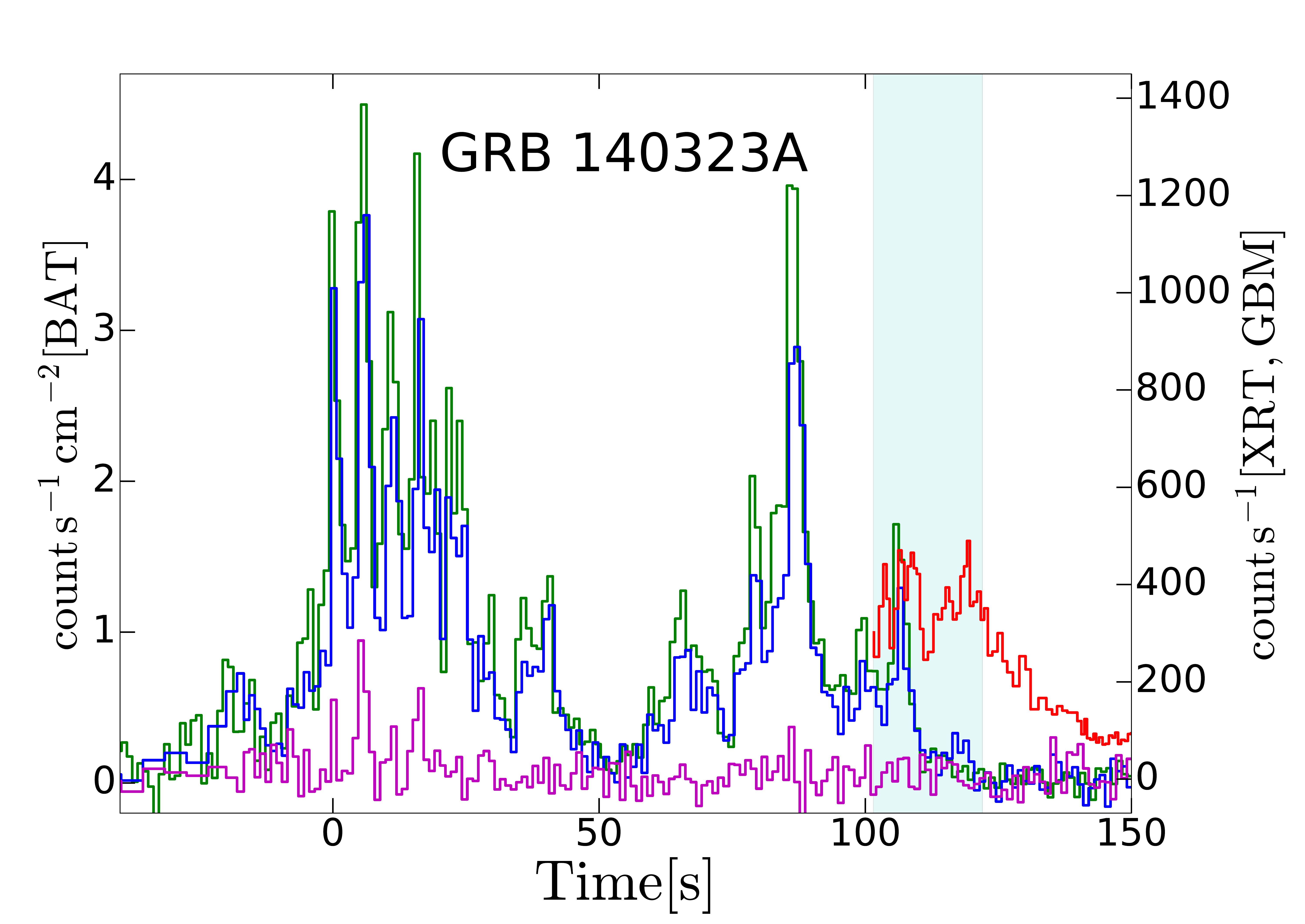}
\includegraphics[width = 0.30\textwidth]{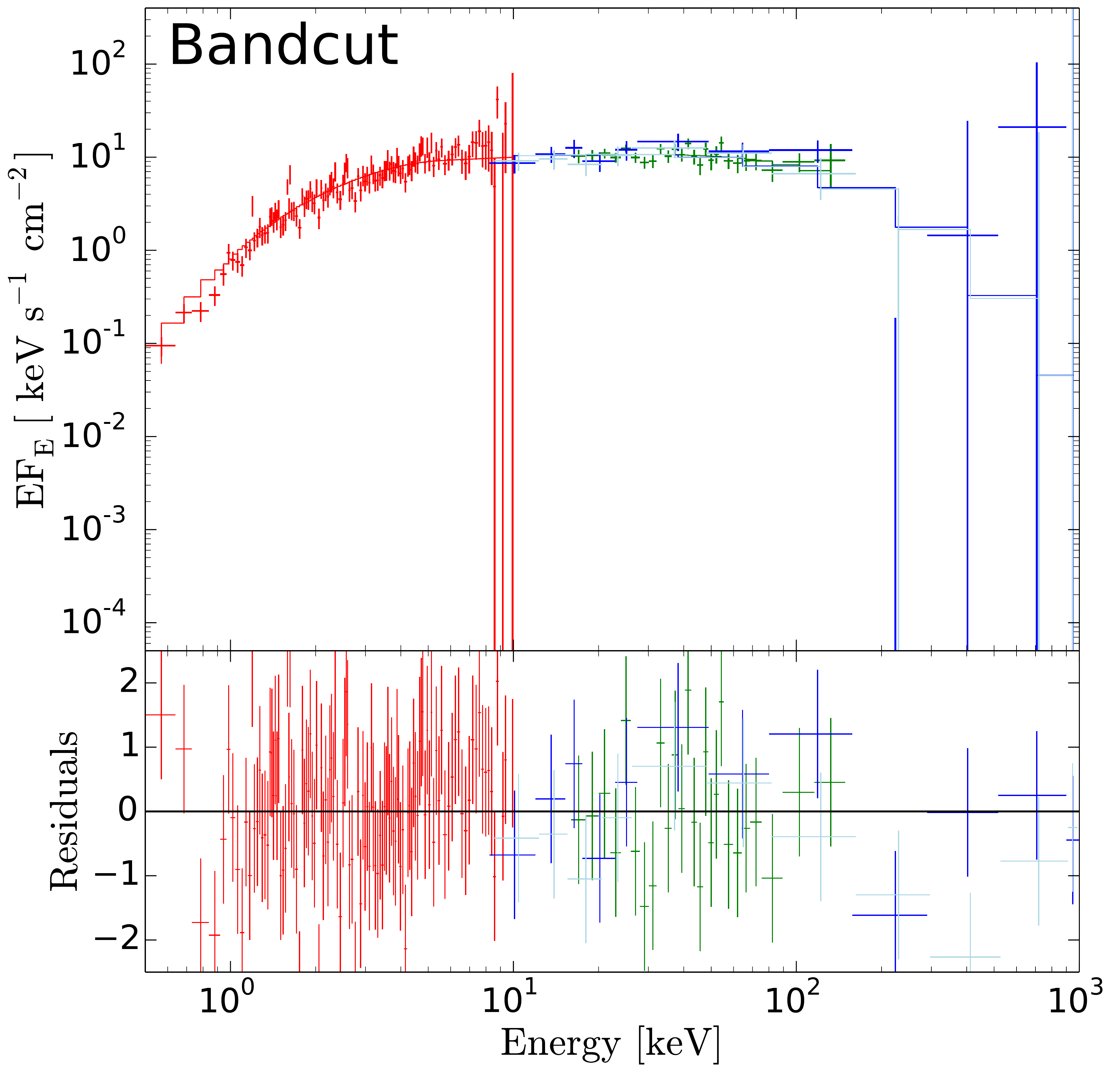} 
\includegraphics[width = 0.30\textwidth]{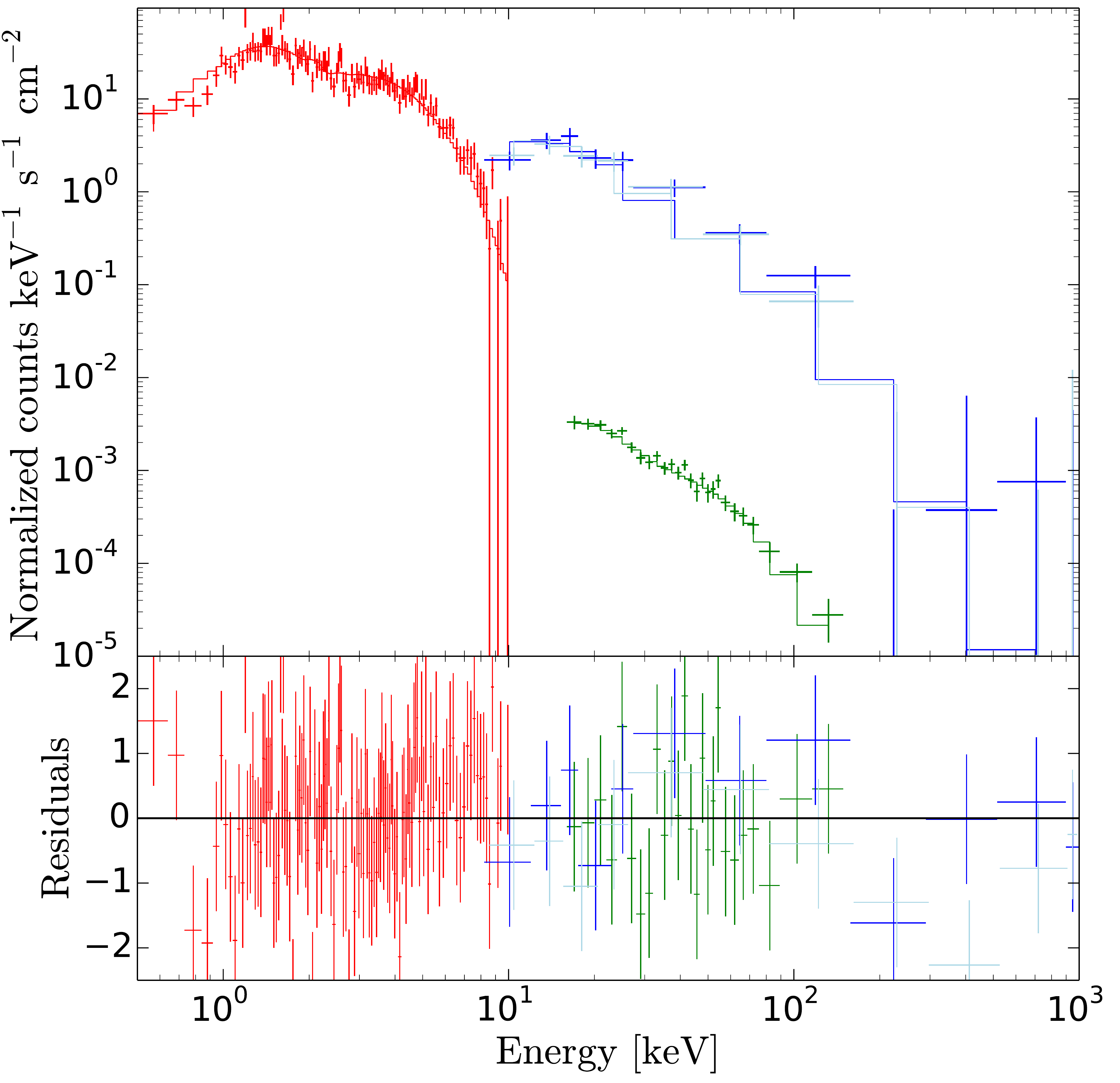} 
\end{figure}

\begin{figure}\ContinuedFloat
\includegraphics[width = 0.40\textwidth]{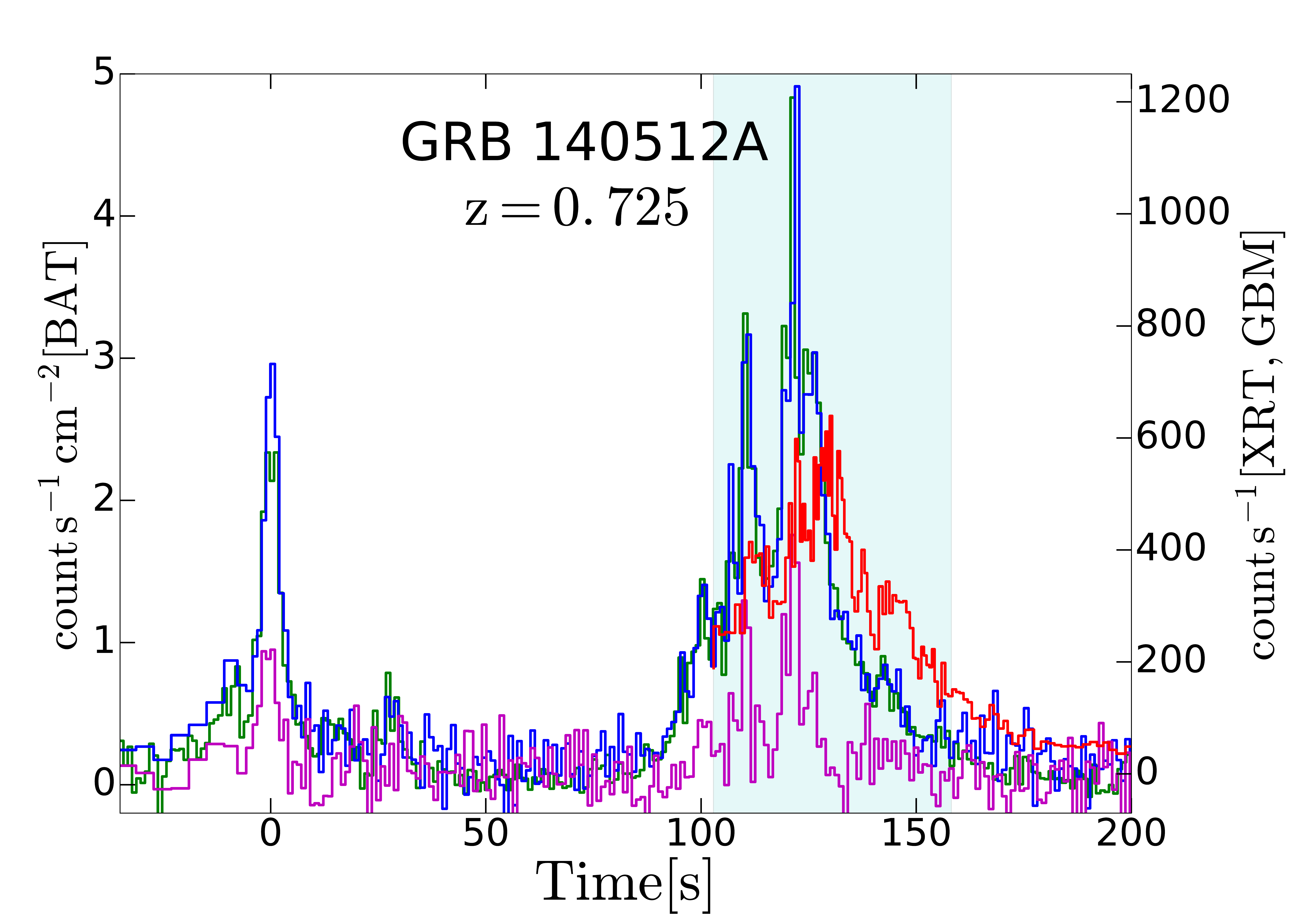} 
\includegraphics[width = 0.30\textwidth]{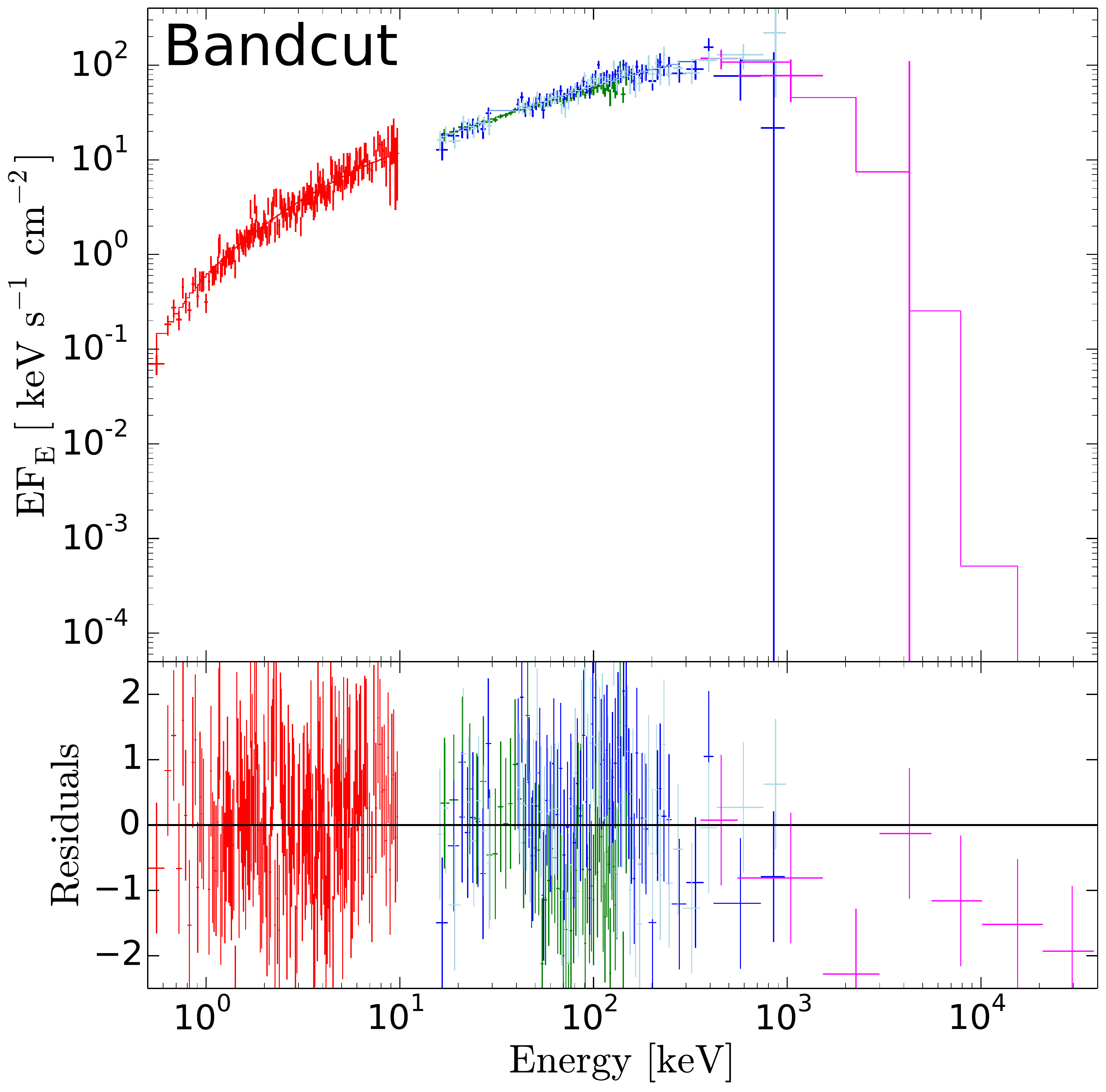} 
\includegraphics[width = 0.30\textwidth]{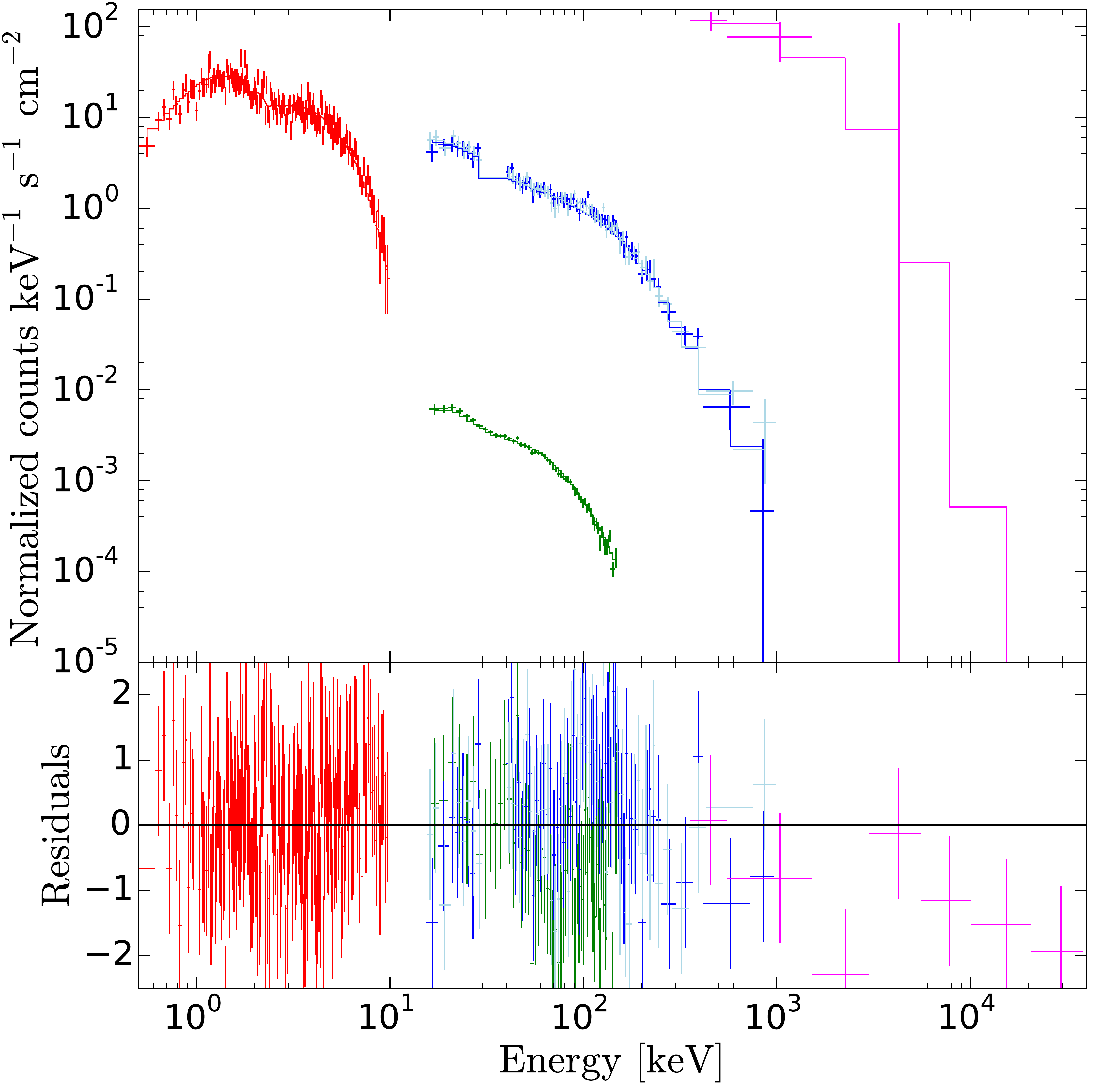} \\
\includegraphics[width = 0.40\textwidth]{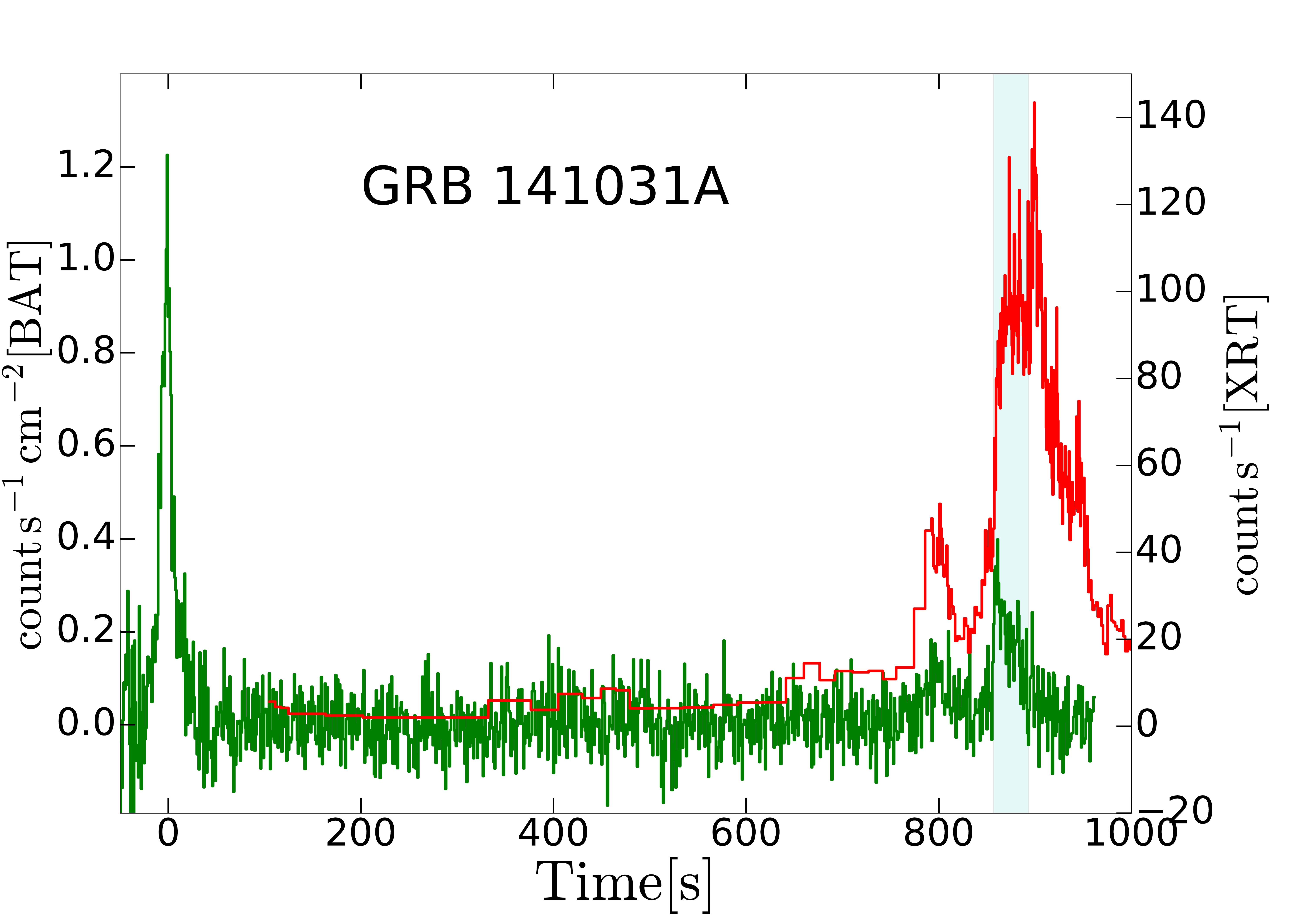} 
\includegraphics[width = 0.30\textwidth]{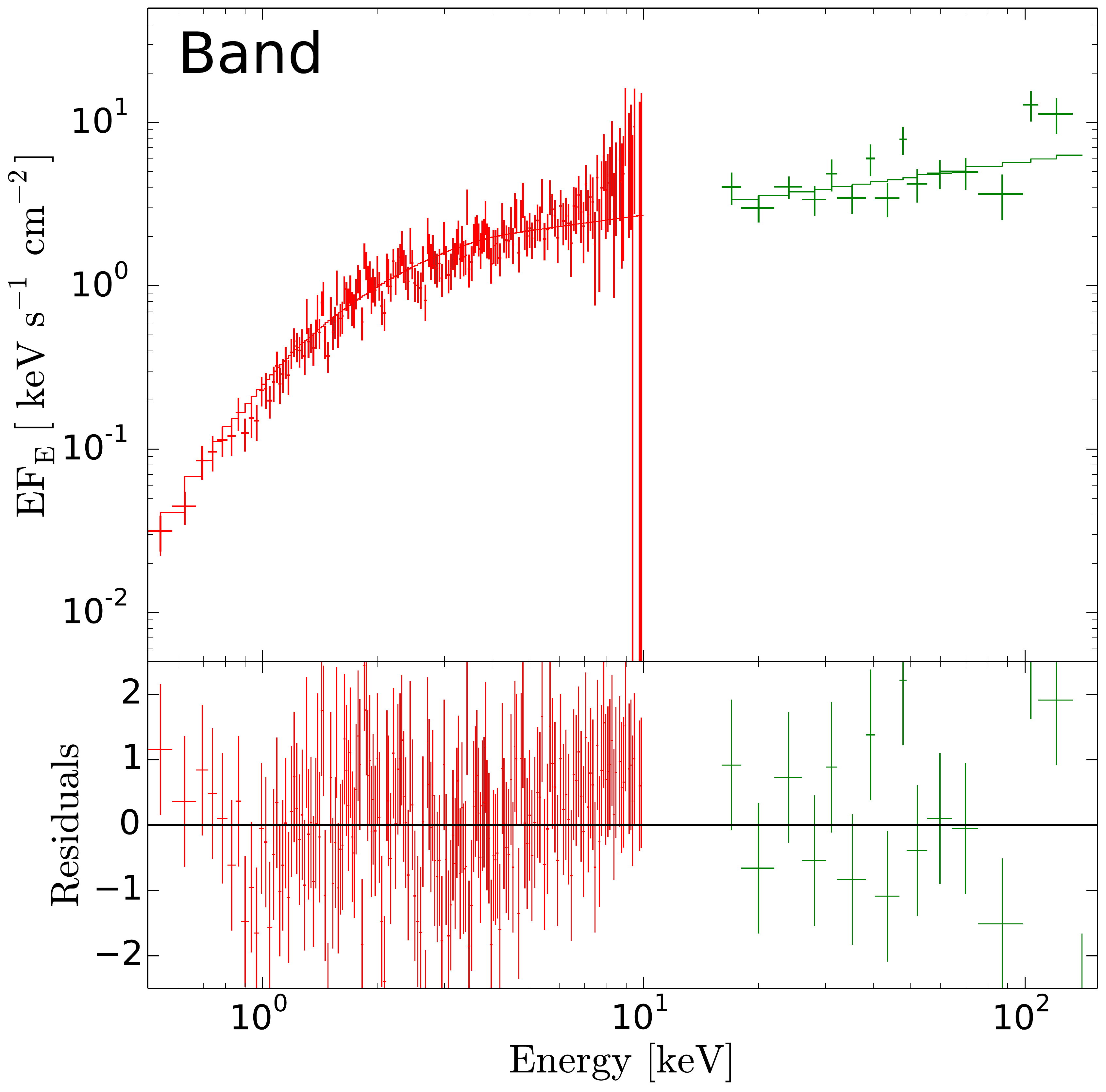} 
\includegraphics[width = 0.30\textwidth]{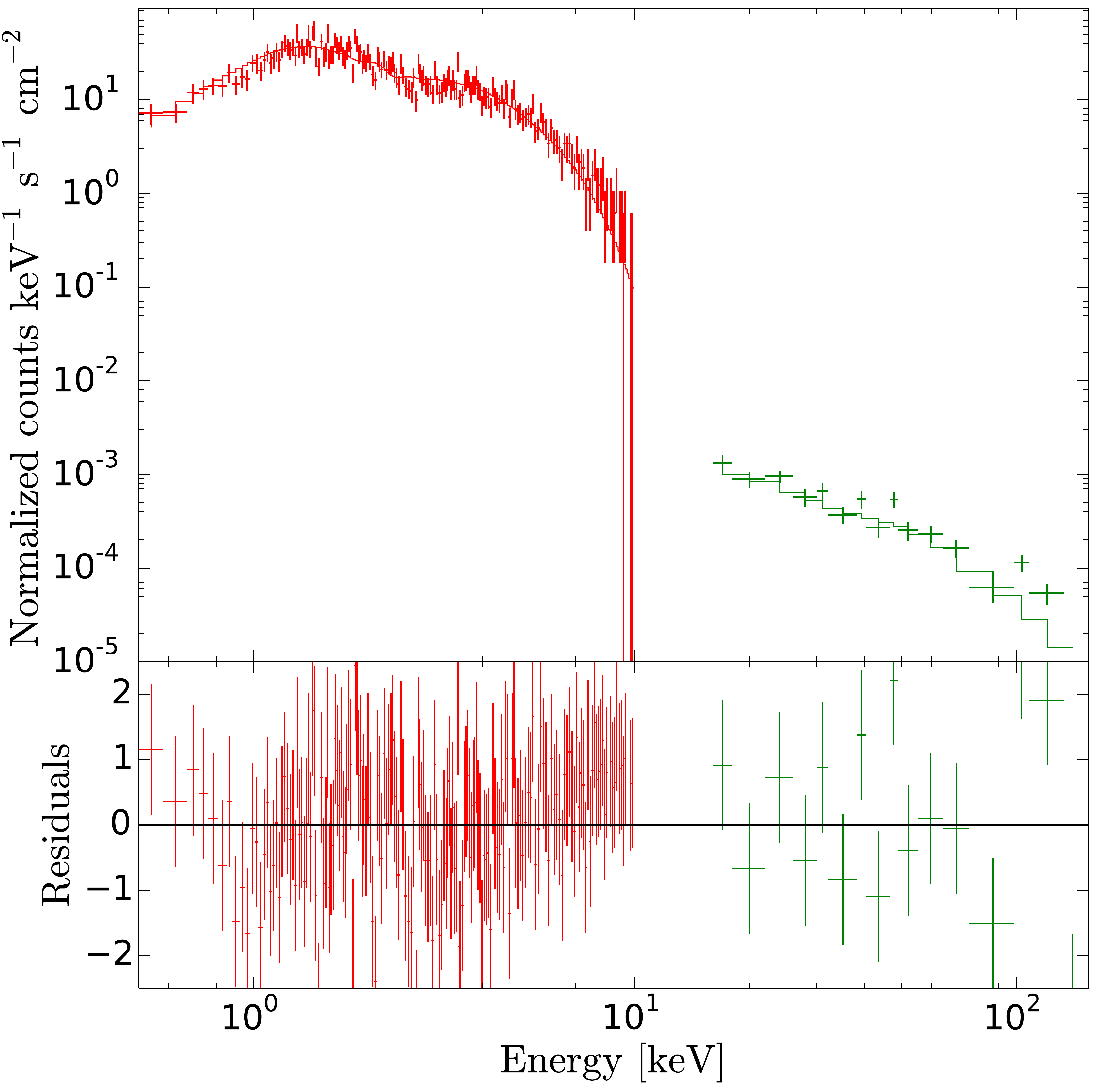} \\
\includegraphics[width = 0.40\textwidth]{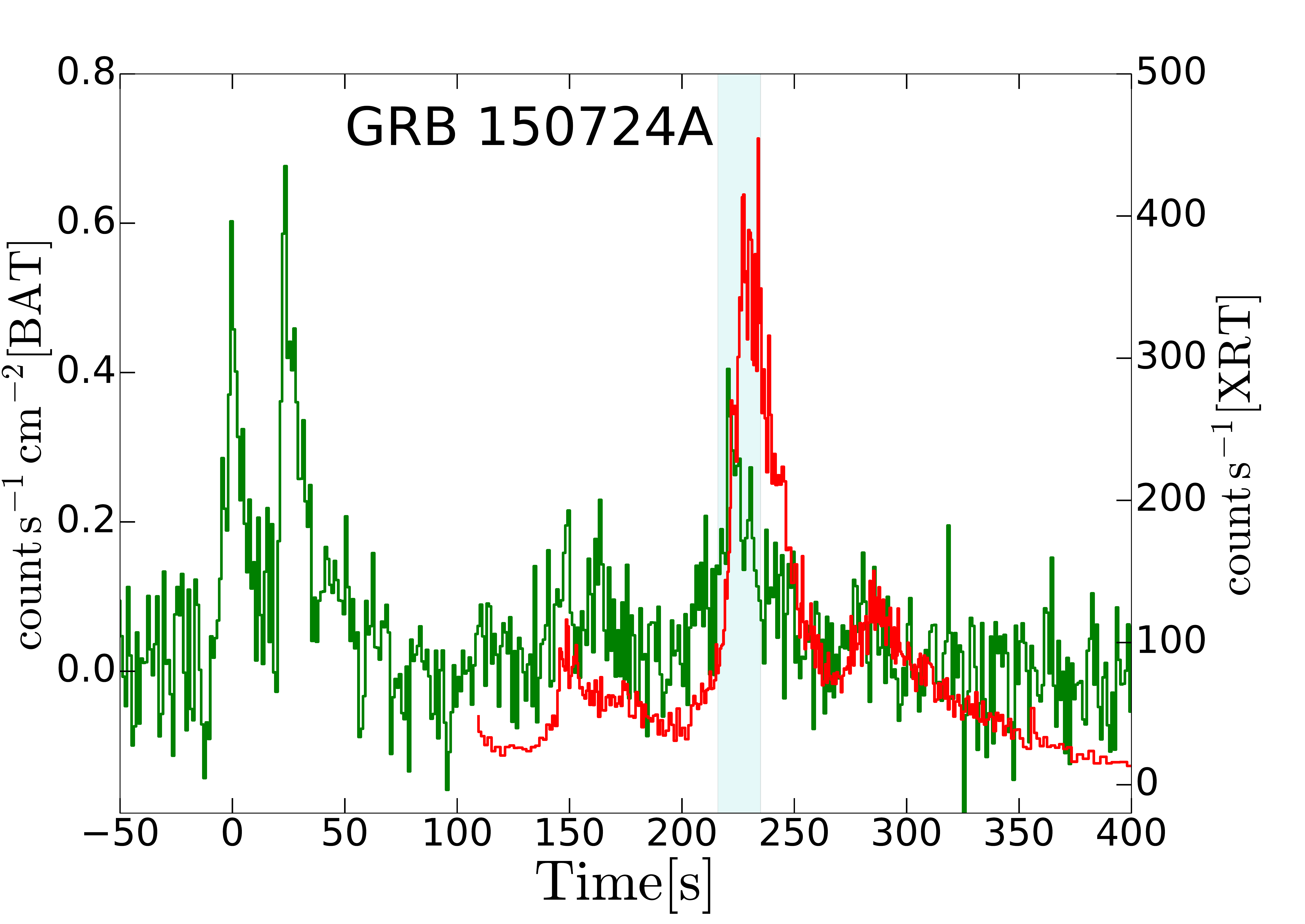} 
\includegraphics[width = 0.30\textwidth]{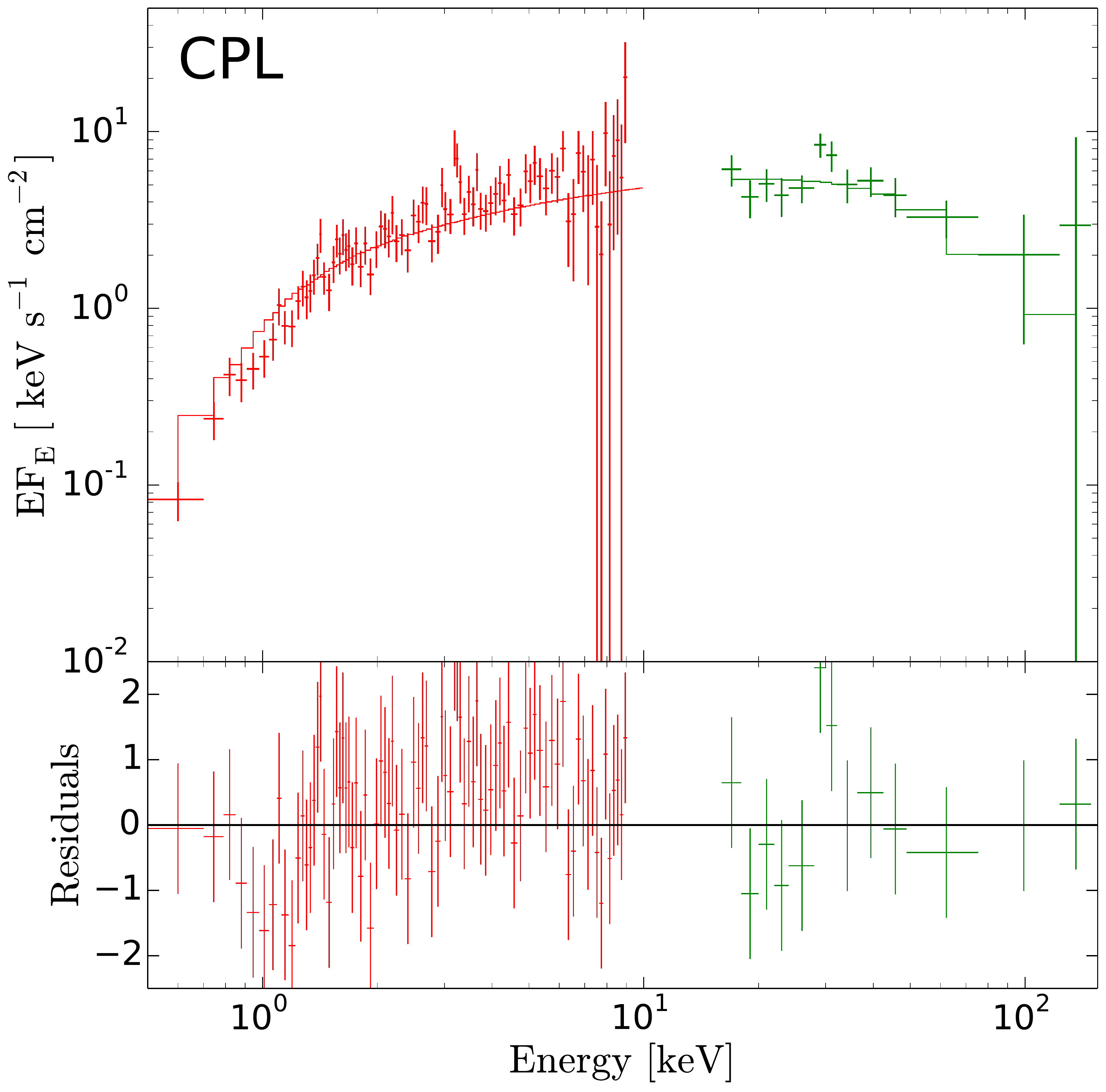} 
\includegraphics[width = 0.30\textwidth]{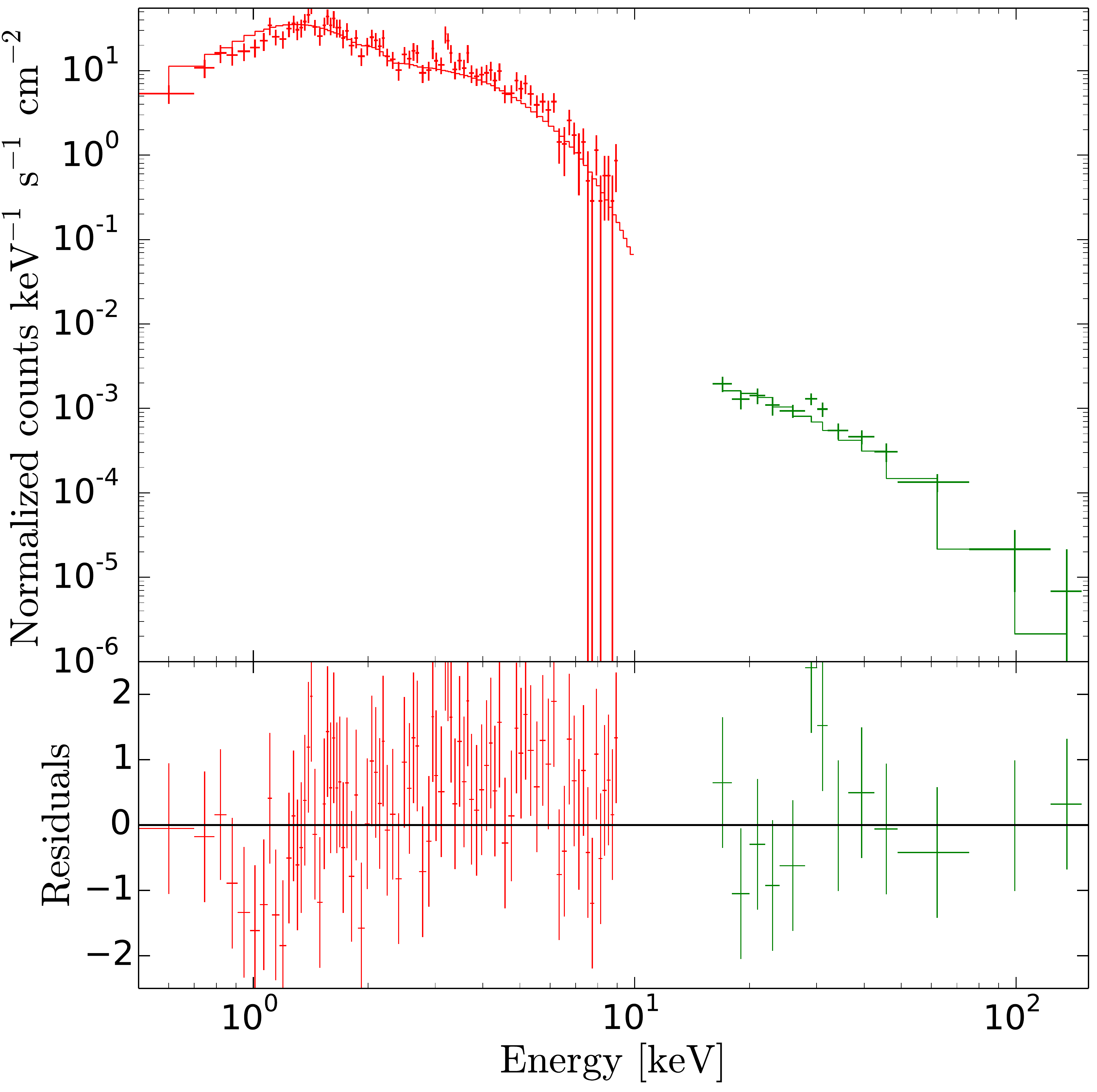} 
\end{figure}

\begin{figure}\ContinuedFloat
\includegraphics[width = 0.40\textwidth]{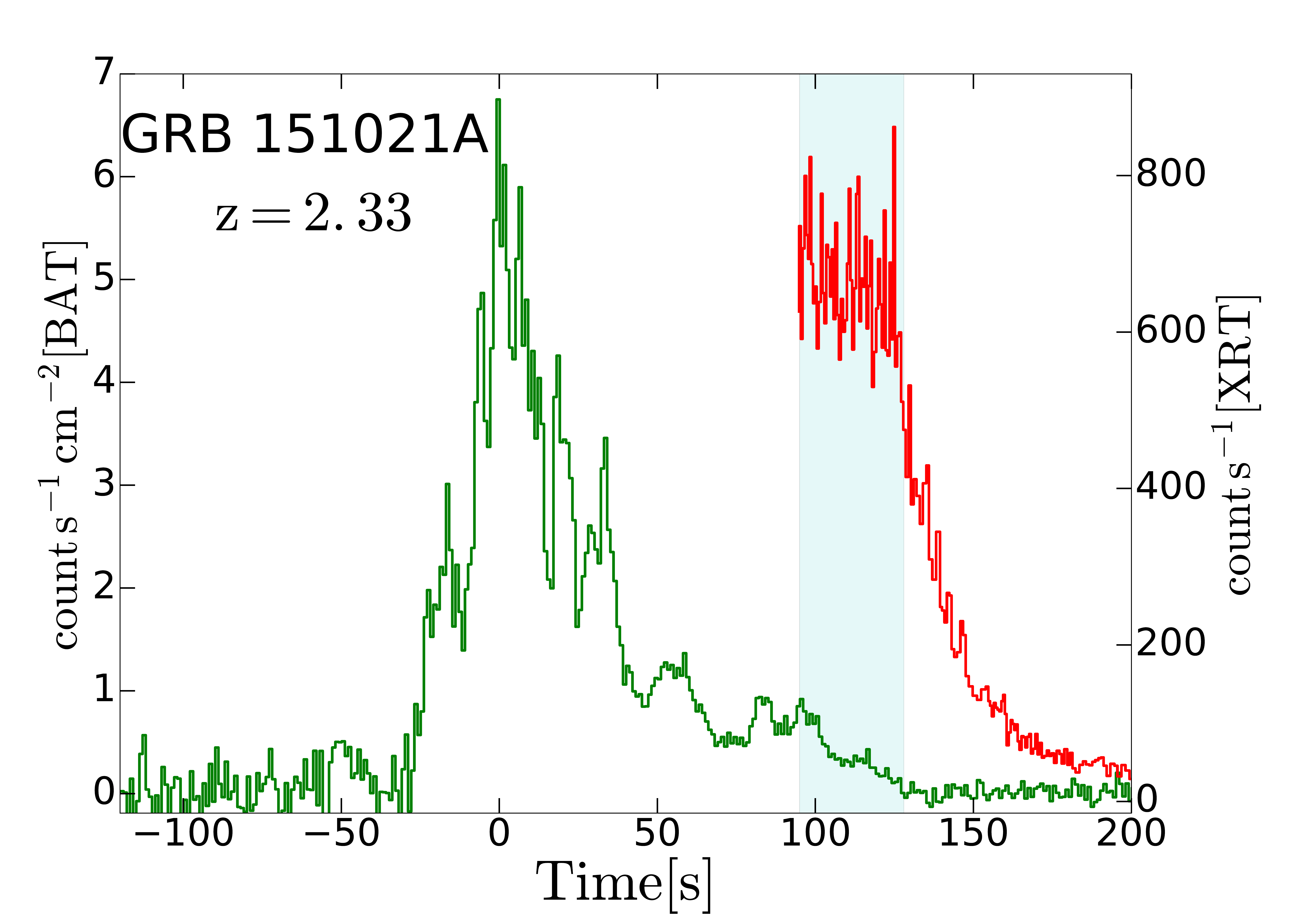}
\includegraphics[width = 0.30\textwidth]{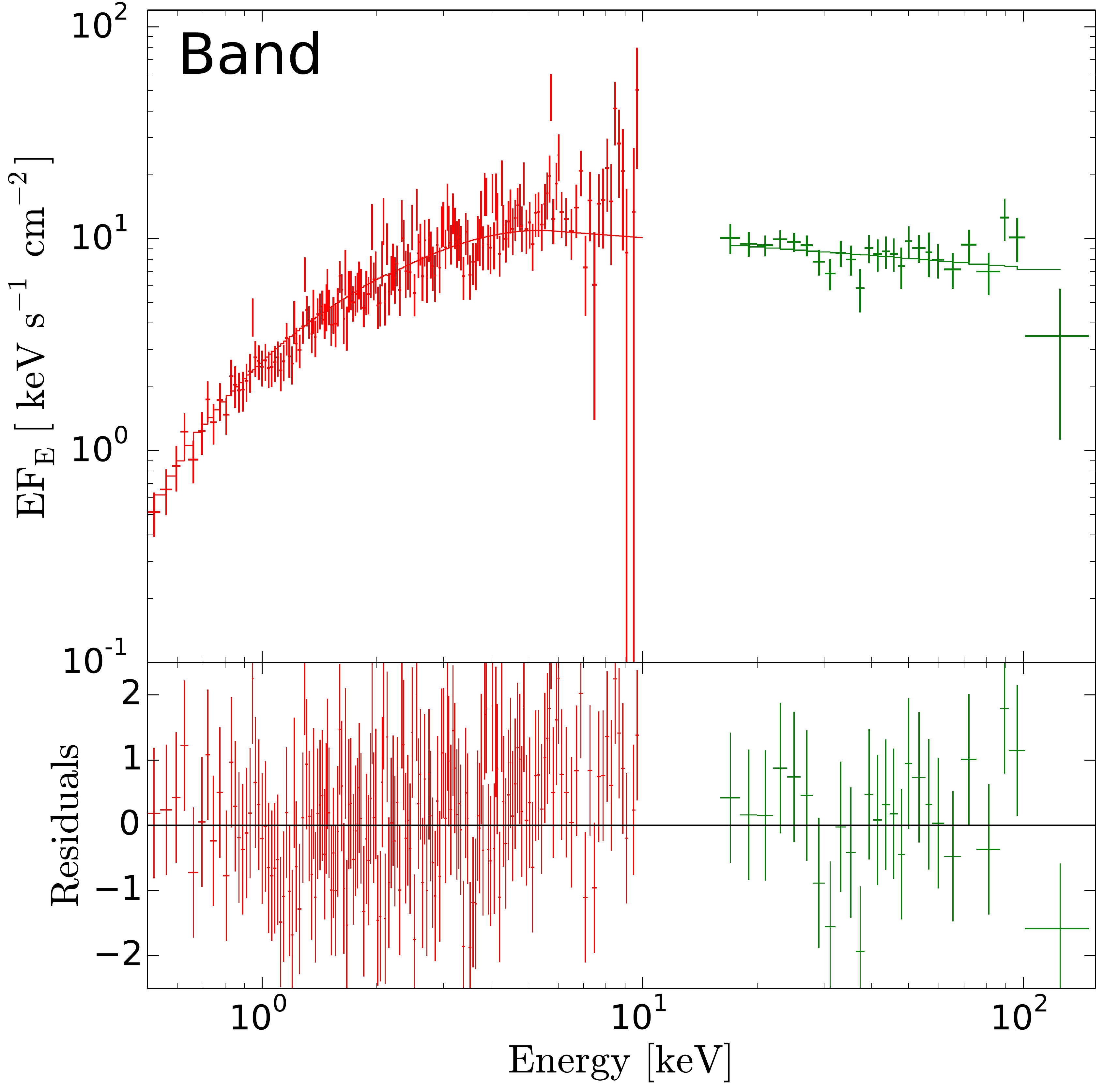} 
\includegraphics[width = 0.30\textwidth]{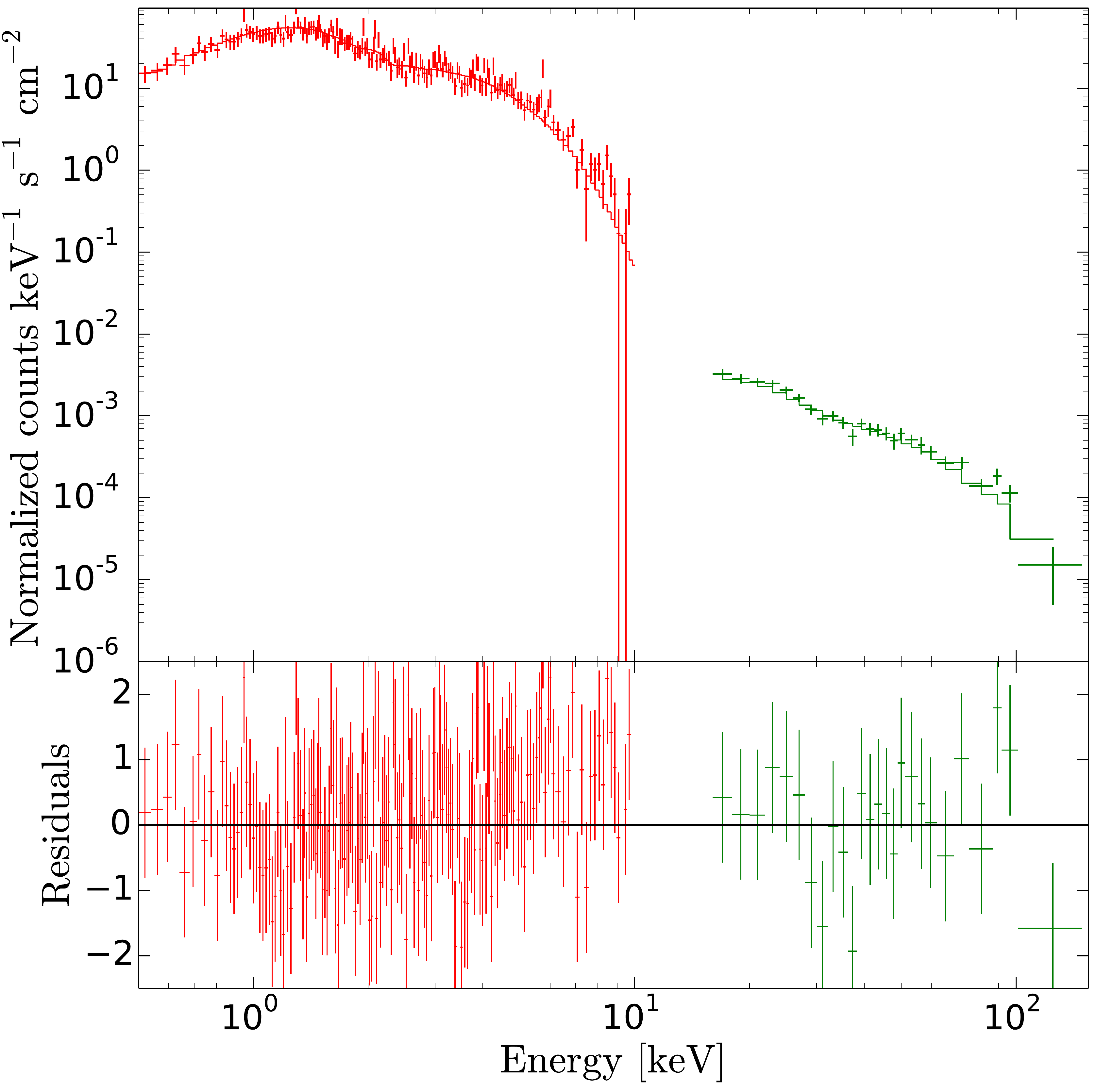} 
\end{figure}

\end{appendix}

\end{document}